\documentclass[12pt,a4paper]{book}
\usepackage{amsmath,amssymb,amsfonts,bm,textcomp,titlesec}
\usepackage{graphicx,subfigure,color,cite,etoolbox,epsfig}
\usepackage{setspace,changepage}
\usepackage[toc,page,titletoc]{appendix}
\usepackage{enumitem,environ,tikz}
\usepackage{slashed}
\usetikzlibrary{shapes,arrows,matrix}
\tikzstyle{process} = [rectangle, minimum width=3cm, minimum height=1cm, text centered, draw=black, fill=orange!30]
\tikzstyle{arrow} = [thick,->,>=stealth]
\usepackage{tabularx,colortbl,fancyhdr}
\usepackage[margin=10pt,font=small,labelfont=bf]{caption}
\graphicspath{{Figures/}}
\apptocmd{\thebibliography}{\csname phantomsection\endcsname\addcontentsline{toc}{chapter}{\bibname}}{}{}

\titleformat{\chapter}[display]
{\normalfont\bfseries\filcenter}
{\LARGE\MakeUppercase{\chaptertitlename} \thechapter}
{1ex}
{\titlerule[2pt]
\vspace{2ex}%
\LARGE}
[\vspace{1ex}%
{\titlerule[2pt]}]

\usepackage[colorlinks=true,linktocpage=true,linkcolor=blue,citecolor=blue]{hyperref}
\usepackage[margin=3.0cm]{geometry}
\def\be{\begin{eqnarray}}
\def\ee{\end{eqnarray}}

\def\nn{\nonumber\\}

\def \Slash{\slash \!\!\!\!}
\def \del{\partial}

\def\nn{\nonumber\\}

\def\[{\left[}
\def\]{\right]}  

\author{By\\Ch. Aminul Islam}

\begin{document}

\begin{center}
 
{\Large {\bf {\LARGE 
{ \doublespacing
Study of hot and dense nuclear matter in effective QCD model}}}

}

\vskip 0.70cm
{\bf {\em By}} 
\vskip -0.2cm
{\bf {\large CH. AMINUL ISLAM}}

{\bf {\large PHYS05201004001}}
\vskip 0.5cm
{\bf {\large Saha Institute of Nuclear Physics}}
\vfill
\vfill
\vskip 2.8cm
{\bf {\em {\large A thesis submitted to the
\vskip 0.05cm
Board of Studies in Physical Sciences
\vskip 0.05cm
In partial fulfillment of requirements
\vskip 0.05cm
For the Degree of 
}}}
\vskip 0.05cm
{\large{\bf{DOCTOR OF PHILOSOPHY}}}
\vskip 0.1cm
{\bf {\em of}}
\vskip 0.1cm
{\bf {\large HOMI BHABHA NATIONAL INSTITUTE}}
\vfill
\includegraphics[height=3.5cm, width=3.5cm]{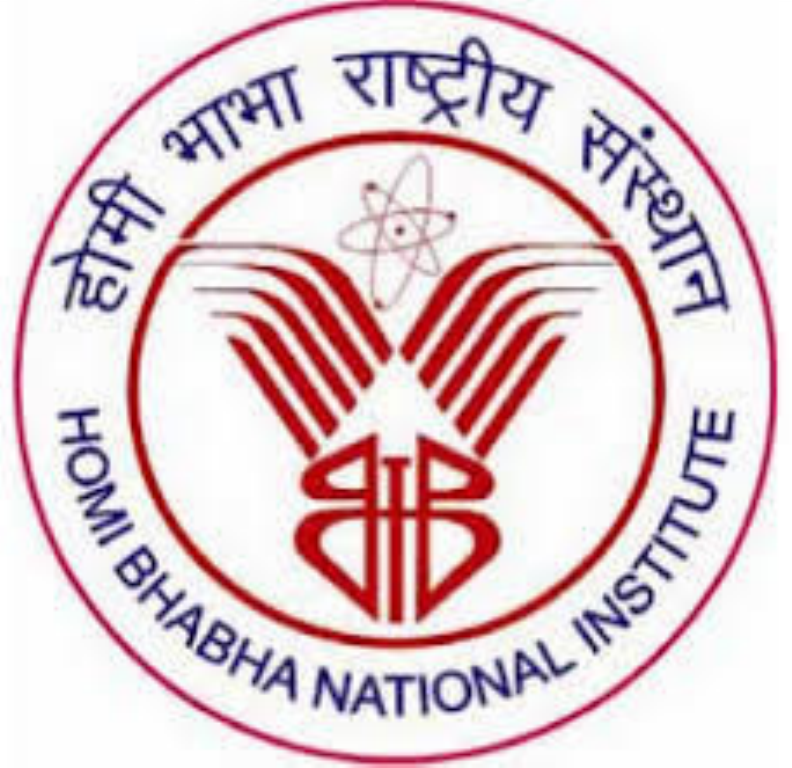}
\vfill
{\bf {\large August, 2016}}
\vfill
\end{center}

\frontmatter
\pagestyle{plain} 
\centerline{{\bf{\LARGE Homi Bhabha National Institute\footnote{This page is to be included only for final submission after successful completion of viva voce.}}}}
\vskip 0.3cm
\centerline{{\bf {\large Recommendations of the Viva Voce Committee}}}
\vskip 0.3cm

{\onehalfspacing 
As members of the Viva Voce Committee, we certify that we have read the
dissertation prepared by {\bf CH. AMINUL ISLAM} entitled {\bf Study 
of hot and dense nuclear matter in effective QCD model} and
recommend that it maybe accepted as fulfilling the dissertation
requirement for the Degree of Doctor of Philosophy.

}
\vskip .8cm
\underline{\hspace{12.0cm}} Date:
\vskip -0.2cm 
Chair - Prof. Asit Kumar De
\vskip .8cm
\underline{\hspace{12.0cm}} Date:
\vskip -0.2cm 
Guide/Convener - Prof. Munshi Golam Mustafa
\vskip .8cm
\underline{\hspace{12.0cm}} Date:
\vskip -0.2cm 
Member 1 - Prof. Palash Baran Pal
\vskip .8cm
\underline{\hspace{12.0cm}} Date:
\vskip -0.2cm 
Member 2 - Prof. Debades Bandyopadhyay
\vskip .8cm
\underline{\hspace{12.0cm}} Date:
\vskip -0.2cm 
External Examiner - Prof. Hiranmaya Mishra
\vskip .80cm
% \vfill
%
{\onehalfspacing 
\hspace{0.7cm} Final approval and acceptance of this dissertation is
contingent upon the candidate's submission of the final copies of the
dissertation to HBNI.
\vskip -0.2cm
\hspace{0.7cm} I hereby certify that I have read this dissertation
prepared under my direction and recommend that it may be accepted as
fulfilling the dissertation requirement.

}

{\doublespacing 
\vskip .2cm 

{\bf Date:} 
\vskip -0.0cm 
{\bf Place:} \hspace{7cm} Guide:{\underline{ \hspace{5.0cm}}}
\vskip -0.4cm 
\hspace{9.65cm}Prof. Munshi Golam Mustafa

\newpage

\centerline{}

\newpage
%\cleardoublepage
%
\centerline{{\bf {\large STATEMENT BY AUTHOR}}}
\vskip 1.00cm
%
%  \doublespacing
%
This dissertation has been submitted in partial fulfillment of
requirements for an advanced degree at Homi Bhabha National Institute
(HBNI) and is deposited in the Library to be made available to borrowers
under rules of the HBNI.
\vskip 0.6cm
Brief quotations from this dissertation are allowable without special
permission, provided that accurate acknowledgement of source is made.
Requests for permission for extended quotation from or reproduction of
this manuscript in whole or in part may be granted by the Competent
Authority of HBNI when in his or her judgment the proposed use of the
material is in the interests of scholarship. In all other instances,
however, permission must be obtained from the author.
\vskip 1.5cm

$~$\hspace{11cm}Ch. Aminul Islam
\newpage
%\cleardoublepage
%
~
\vskip 1.2cm
\centerline{{\bf{\large{DECLARATION}}}}
\vskip 1.2cm
I, hereby declare that the investigation presented in the thesis has been
carried out by me. The work is original and has not been submitted
earlier as a whole or in part for a degree / diploma at this or any
other Institution / University.
\vskip 2.5cm
\rightline{Ch. Aminul Islam \hspace{0.9cm}}
\newpage
%\cleardoublepage
\centerline{{\bf{\large{List of Publications arising from the thesis}}}}

{\bf{\Large Journal}}
\begin{enumerate}

\item $``$Vector meson spectral function and dilepton rate in the presence of strong entanglement
effect between the chiral and the Polyakov loop dynamics"
     \\Chowdhury Aminul Islam, Sarbani Majumder and Munshi G. Mustafa
     \\{\bf Phys. Rev. D 92, 096002(2015)},
     \href{http://arxiv.org/abs/1508.04061}{[arXiv:1508.04061 [hep-ph]].}

\item $``$Vector meson spectral function and dilepton production rate in a hot and dense medium within an effective QCD approach"
     \\Chowdhury Aminul Islam, Sarbani Majumder, Najmul Haque and Munshi G. Mustafa
     \\{\bf JHEP 02 (2015) 011},
     \href{http://arxiv.org/abs/arXiv:1411.6407}{[arXiv:1411.6407 [hep-ph]].}

\item $``$The consequences of $SU(3)$ colorsingletness, Polyakov Loop and $Z(3)$ symmetry on a quark-gluon gas"
  \\Chowdhury Aminul Islam,  Raktim Abir,  Munshi G. Mustafa,  Rajarshi Ray and  Sanjay K. Ghosh
  \\{\bf J.\ Phys.\ G 41 (2014) 025001},
  \href{http://arxiv.org/abs/arXiv:1208.3146}{[arXiv:1208.3146 [hep-ph]].}  

\end{enumerate}

 {\bf{\Large Chapters in books and lectures notes}}
 \hspace{1cm} N. A. \\
\newpage
 {\bf{\Large Conferences}}
\begin{itemize}
\item $``$Vector meson spectral function and dilepton rate in an effective mean field model"
       \\Chowdhury Aminul Islam, Sarbani Majumder, Najmul Haque and Munshi G. Mustafa
       \\ XXI DAE-BRNS HEP Symposium, IIT Guwahati, December 2014
    \\Springer\ Proc.\ Phys. 174 (2016) 31-35,
     \href{http://arxiv.org/abs/1508.04288}{[arXiv:1508.04288 [hep-ph]].}
\end{itemize}

 {\bf{\Large Others}}
\begin{itemize}
 \item $``$Electromagnetic spectral properties and Debye screening of a strongly magnetized hot medium"
      \\Aritra Bandyopadhyay, Chowdhury Aminul Islam and Munshi G. Mustafa
      \\{\bf Phys. Rev. D 94, 114034(2016)},
    \href{http://arxiv.org/abs/1602.06769}{[arXiv:1602.06769 [hep-ph]].}
\end{itemize}
\vskip 3cm
\rightline{Ch. Aminul Islam \hspace{0.9cm}}
\newpage
\centerline{{\bf {\large DEDICATIONS}}}
\vskip 5cm
\hskip 5cm
\hspace{3cm}{\large To my abba and ma,}

\vspace{-0.3cm}
\hspace{7.83cm} {\large who value honesty above all.}

\newpage
%\cleardoublepage
~
\vskip -.50cm
\centerline{{\bf{\large ACKNOWLEDGEMENTS}}}
\vskip 0.30cm

First of all I would like to express my gratitude to my supervisor Munshi Golam Mustafa, without his constant push this thesis would not have been a reality. He was there all the time bearing all of my stupidity, ignorance and sometimes even my audacity. After these six years of association, the supervisor-student relationship has evolved into smooth friendship which reflects in my cellphone call list where his name has morphed into 'Munshi da' from being 'Munshi Sir'.

If academically I have to acknowledge anyone after my supervisor , it is Sarbani Majumder. Things would have been different without her presence. I also take the opportunity here to express my gratitude to all of my collaborators. I would like to thank all the past and present members in Munshi da's group for their all types of help, which in many occasions go beyond academic activities. It is indeed the brotherly behavior of the seniors and junior, which eased up my PhD life. I am also thankful to Samir Mallik and Aritra Bandyopadhyay for their important comments and suggestions while some parts of this manuscript were being written.

I thank all the scholars \textendash~including seniors, juniors and batchmates, academic and non-academic members of SINP, particularly those of Theory Division for their direct and indirect contributions in my efforts. The vibrant atmosphere of Theory Division is indeed admired. I am grateful to the staff members of both the canteens, particularly the cooks of MSA-I, for serving us the food more or less 24$\times$7.

PhD days are also full of stress and anxiety and you can do away with them if you are surrounded with friends. I had a lot of them and probably it will take some pages more to name them all. The friendly ambiance was always there even if I talked with someone for a single time during my whole stay. The stay here could not have been more joyous and was full of fun \textendash~either out of learnings or out of pure fun. I convey my sincere gratitude to all of them for being there. In particular, I wholeheartedly acknowledge all the members of 363 and presently 3319 for making my PhD life something to cherish on for the rest of my life.  All the boarders of MSA-I who has a stay overlapping with me, however small it be, are gratefully acknowledged for the happiness we shared and for all those invaluable lessons which go beyond the scope of a mere PhD degree. I thank heartily the Post MSc batch-2013, who are staying in MSA-I, for those unimaginable crazy moments. You guys are amazing and I will surely miss you. I 
should take this opportunity to apologize to some of my friends who were very close to me and thus in many occasions had to bear with my bossy attitude. In life, sometimes you meet people with whom your frequency really matches and Rajani Raman is one of them during my PhD life. He worked sometimes as a confession box and sometimes as an agony aunt for me, which helped our friendship to grow with time and promises to persist beyond this academic adventure.

In every part of your life you are bound to meet people whom you get influenced by and my PhD life was no exception. I am fortunate that I got Palash Baran Pal as my teacher. I used to visit him frequently with my doubts, queries and thoughts and his immense knowledge about things made each of the conversation, which in many occasions went beyond physics, refreshing and enlightening. I also acknowledge my friend Dipankar Das for boosting my scientific temperament and modifying the scientific attitudes I previously possessed. Senior-cum-friend Arindam Mazumdar really helped me to grow a sense of completeness about life and enforces me to ponder over many important aspects of it, which had been simply invisible to me before I met him. 

I would like to acknowledge all of my family members for their constant support and help towards me. I am thankful to {\it mejo chachima} (aunt) for her care and love for me. Above all I am grateful to my loving abba and ma who had to compromise with their dreams so that I can chase mine.

At last but not least, I am grateful to people of India \textendash~without their investment I would not be even thinking to write this acknowledgement.

}

\cleardoublepage
\phantomsection
% \addcontentsline{toc}{chapter}{Contents}
\tableofcontents

{\doublespacing

\chapter*{Synopsis}
\addcontentsline{toc}{chapter}{Synopsis}

It is an established fact that quantum chromodynamics (QCD) is the fundamental theory of strong interaction and its phase structure is very rich particularly at high temperature and/or high density. As for example there exists a phase transition where normal hadronic matter transforms into a state in which the degrees of freedom are the liberated quarks and gluons. The thermalized state of such deconfined quarks and gluons is known as quark gluon plasma (QGP)~\cite{Muller:1983ed,Heinz:2000bk}. Apart from this there also exist many other exotic phases like quarkyonic phase, color superconducting phase etc~\cite{Fukushima:2010bq}. If we leave aside the immense physical significance of these phase structures, the interesting intricacies alone make the study of QCD a pursue-worthy aspect of natural phenomena. Here in this thesis, only the QGP phase has been explored. Exploration of QGP is expected to shed lights on some less-known important large scale physical phenomena such as evolution of early universe, 
properties of 
neutron star etc. For example, strongly interacting hot nuclear matter is believed to have existed after a few microseconds ($\sim10^{-5}$ s) of the big bang when the temperature was of the order of $10^{12}$ K ($\sim$200 MeV). Also a highly dense matter, much more dense than the ordinary nuclear matter ($\sim10^{18}\rm {kg/m^{3}}$), is believed to exist in the core of neutron star, which in turn can affect its spin and magnetic properties.

Now there are always two aspects of studying any physical phenomena. One is the experimental part and other is the theoretical one. To study experimentally we need to have access to QGP. But the existence of QGP in astrophysical phenomena so far discussed are by far remote in space and time and thus cannot be accessed. This leads to the terrestrial based experiments known as heavy ion collisions (HIC) through which QGP can be formed and thus studied. At present Relativistic Heavy Ion Collider (RHIC) at Brookhaven National Laboratory (BNL)~\cite{Arsene:2004fa,Back:2004je,Adams:2005dq,Adcox:2004mh} and the Large Hadron Collider (LHC) at the European Organization for Nuclear Research (CERN)~\cite{Carminati:2004fp,Alessandro:2006yt} are operational. These experiments have already produced huge amount of data, the analysis of which has led to the indication of creation of the predicted QGP phase. Also some future facilities are coming up, such as Facility for Antiproton and Ion Research (FAIR) at the 
Gesellschaft f\"{u}r Schwerionenforschung (GSI)~\cite{Senger:2004jw,Friman:2011zz} and Nuclotron-based Ion Collider fAcility (NICA) at  Joint Institute for Nuclear Research (JINR), Dubna~\cite{Toneev:2007yu,Sissakian:2009ru}. The goals of these upcoming experiments are basically to complement the operational ones by exploring different regions of QCD phase diagram (particularly the high density regions) which, so far, remain unexplored and also by corroborating many of the findings of those.

On the other side of the coin there are many theoretical tools to study the properties of QGP. The coupling constant of QCD being large at small energies (large length) its study becomes a theoretical challenge specially in those regime because of the nonperturbative nature. There exists a first principle QCD method known as Lattice QCD (LQCD) which is completely a numerical technique and has some difficulties at finite baryon density~\cite{Wilson:1973jj,Wilson:1974sk,Kogut:1982ds,Creutz:1983ev}. Analytical method such as perturbative QCD (PQCD) works when the coupling constant is sufficiently small~\cite{Gribov:1984tu}. To deal with the high temperature QCD, a resummation technique known as hard thermal loop (HTL) perturbation theory~\cite{Braaten:1990wp,Andersen:1999fw,Haque:2013sja}, has been invented. To circumvent all these difficulties, shortcomings of LQCD and PQCD, effective QCD models have been designed. Since the hot and dense matter created in HIC is supposed to be non-perturbative in nature, the 
use of such effective models becomes particularly useful for the purpose.

Eventually there exist many effective QCD models which are being extensively used. There are Nambu\textendash Jona-Lasinio (NJL) model~\cite{Nambu:1961tp,Nambu:1961fr} and its Polyakov loop extended version, PNJL model~\cite{Fukushima:2003fw,Ratti:2005jh}; linear sigma model (LSM) (sometimes also called quark-meson model)~\cite{GellMann:1960np} and also its Polyakov loop extended version, PLSM~\cite{Schaefer:2007pw}; functional methods like Dyson-Schwinger Equation (DSE)~\cite{Bjorken_Drell(book):RQF}, matrix model~\cite{Meisinger:2001cq}, different quasiparticle models~\cite{Peshier:1995ty}, color singlet (CS) model~\cite{Turko:1981nr,Elze:1985wv,Auberson:1986ft,Mustafa:1993np} just to name a few. The PNJL model has been further improved by considering the entanglement between chiral and deconfinement dynamics, which is termed as EPNJL model~\cite{Sakai:2010rp}. The study of the properties of QGP in this 
thesis is based on the NJL, PNJL and 
EPNJL models and also on the color singlet model.

Correlation function (CF) is one of the fruitful and heavily used mathematical tool in high energy physics. There can be different types of CFs. A frequently used one is the propagator for different types of fields. Here we will mainly focus on current-current CFs, viz. vector meson current CF in a thermal background. These CFs and their spectral representation reveal dynamical properties of many particle system and many of the hadron properties.

Such properties in vacuum are very well studied in QCD~\cite{Davidson:1995fq}. The presence of a stable mesonic state is understood by the delta function like peak in the spectral function (SF). For a quasiparticle in the medium, the $\delta$-like peak is expected to be smeared due to the thermal width, which increases with the increase in temperature. At sufficiently high temperature and density, the contribution from the mesonic state in the SF will be broad enough so that it is not very meaningful to speak of it as a well defined state any more. The temporal CF is related to the response of the conserved density fluctuations due to the symmetry of the system. On the other hand, the spatial CF exhibits information on the masses and width. So, in a hot and dense medium the properties of hadrons, viz. response to the fluctuations, masses, width, compressibility etc., will be affected. Hence, hadron properties at finite temperature and density are also encoded in the structure of its CF and the 
corresponding spectral representation, which may reflect the degrees of freedom  around the phase transition point and thus the properties of the deconfined, strongly interacting matter. As for example, the spectral representation of the vector current-current correlation can be indirectly accessible by high energy heavy-ion experiments as it is related to the differential thermal cross section for the production of lepton pairs~\cite{Kapusta_Gale(book):1996FTFTPA,Lebellac(book):1996TFT}. Moreover, in the limit of low energies (small frequencies), various transport coefficients of the hot and dense medium can be determined from the spatial spectral representation of the vector channel correlation.

Here in this dissertation the vector meson current-current CF have been explored with and without the influence of isoscalar-vector interaction~\cite{Islam:2014sea}. The inclusion of isoscalar-vector (also called simply vector) interaction in heavy-ion physics is important for study of the spectral property like dilepton rate at non-zero chemical potential. On the other hand, in nuclear astrophysics the formation of stars with quark matter core depends strongly on the existence of a quark vector repulsion.

It has been more or less accepted in the community that the QGP formed in the HIC is a strongly interacting one~\cite{Adare:2006ti,Aamodt:2010pa}, also known as sQGP. Experimentally measured dilepton spectrum~\cite{Adare:2009qk,Reichelt:2015oze}, which is considered as one of the direct signature, at low invariant mass has been found to be higher than what was predicted by most of the theoretical methods. This excess can be attributed to the strongly coupled nature of the created hot and dense matter~\cite{Gale:2014dfa}. In one of the works within effective model~\cite{Islam:2014sea} in this thesis we have found that the dilepton rate is indeed enhanced for the sQGP as compared to the Born rate (leading order perturbative rate) in a weakly coupled QGP. Here the sQGP has been obtained and tuned by the inclusion of a background gauge field, namely the Polyakov loop field. We further compared our findings with the available Lattice data. Response of the conserved density fluctuations, namely the quark number 
susceptibility (QNS) has 
also been studied in the presence of the vector interaction.

While talking about the transition from normal hadronic matter to QGP, the main interest revolves around two phase transitions - one is the chiral transition  and the other one is the deconfinement transition. If they do not coincide, exotic phases such as the constituent quark phase~\cite{Cleymans:1986cq} or the quarkyonic phase~\cite{McLerran:2007qj} may occur. So, an important question on the QCD thermodynamics is whether the chiral symmetry restoration and the confinement-to-deconfinement transition happen simultaneously or not. We note that chiral and deconfinement transitions are conceptually two distinct phenomena. Though lattice QCD simulation has confirmed that these two transitions occur at the same temperature~\cite{Fukugita:1986rr} or almost at the same temperature~\cite{Aoki:2006br}. Whether this is a mere coincidence or some dynamics between the two phenomena are influencing each other is not well understood and has become an area of active research and exploration. To understand the reason 
behind 
this  coincidence a conjecture has been proposed in~\cite{Sakai:2010rp} through a strong correlation or entanglement between the chiral condensate ($\sigma$) and the Polyakov loop expectation value ($\Phi$) within the PNJL model. Such generalization is known as EPNJL model.

We consider the idea of the EPNJL model and re-explore the vector spectral function and the spectral property such as the dilepton production rate~\cite{Islam:2015koa} previously studied in~\cite{Islam:2014sea}. Because of this strong entanglement between $\Phi$ and $\sigma$, the coupling strengths run with the temperature and chemical potential. This running has interesting implications on the dilepton rate which has been explored in this dissertation. The importance of inclusion of vector interaction has already been discussed, however its inclusion also poses some problems in the fluctuation of conserved density associated with the symmetry, namely the QNS. This issue has also been discussed in details.

In another effort, we assume the hot and dense matter (QGP) to be made of a non-interacting quarks, antiquarks and gluons with the underlying symmetry of $SU(3)$ color gauge theory~\cite{Islam:2012kv}. This assumption holds particularly for a QGP which is weakly coupled. We showed that this type of simple quantum statistical description exhibits very interesting features: the $SU(3)$ color singlet has $Z(3)$ symmetry through the normalized character in the fundamental representation of $SU(3)$. This character becomes equivalent to an ensemble of Polyakov loop ($\Phi$). Furthermore, it was concluded that this $\Phi$ can be taken as an order parameter for color-confinement to color-deconfinement phase transition.

}

{\onehalfspacing

% \addcontentsline{toc}{chapter}{List of Publications}
% \input{./text/publications}

}

{\doublespacing

\cleardoublepage
\phantomsection
\addcontentsline{toc}{chapter}{List of Figures}
\listoffigures

\phantomsection
\addcontentsline{toc}{chapter}{List of Tables}
\listoftables

\mainmatter
\pagestyle{fancy}
\fancyhead{}
\fancyfoot{}
\lhead{\leftmark}
\setlength{\headheight}{14.5pt}
\cfoot{\thepage}
\chapter{Introduction}
\label{chapter:introduction}
% Title Page

\section{Prelude}
\label{sec:prelude}

It has been the foremost priority of science, particularly physical science, to explain the universe and everything within in terms of some basic building blocks. This way of looking at things started long back with the ancient Greek and Indian philosophers trying to understand matter in terms of five fundamental building blocks. But back then the idea was much more philosophical. This doctrine got the boost with the proposition of atoms by John  Dalton and really flourished into its modern form after the discovery of electron by J. J. Thomson in 1897. In modern day's terminology, elementary particle physics is the branch of science which deals with such fundamental questions.

So far, we have understood that there exist two groups of fundamental particles: fermions ((anti)leptons, (anti)quarks) and bosons (gauge bosons and Higgs boson). But it's not only the types and numbers of fundamental particles which matter the most, how these particles interact with each other is also a matter of great importance. Such investigations also belong to the realm of elementary particle physics. Until now four fundamental interactions have been discovered, which govern different types of particle dynamics and related phenomena. Out of these four the quantized version of gravity is yet to be achieved and it is not included in the standard model (SM) of particle physics which describes the other three interactions in their quantized forms. Some of the basic characteristics of these known four interactions are summarized in table~\ref{table:fund_inter}.

\begin{table}
\begin{center}
\begin{tabular}{lccl}
\hline
Interaction     & Theory                        & Mediators         & Strength   \\ \hline \vspace{0.15cm}
Strong          & Quantum chromodynamics        & Gluons            & 1          \\ \vspace{0.15cm}
Electromagnetic & Quantum electrodynamics       & Photons           & $10^{-2}$  \\ \vspace{0.15cm}
Weak            & Quantum flavordynamics        & $\rm W^{\pm}, Z$  & $10^{-13}$ \\ \vspace{0.15cm}
Gravity         & \begin{tabular}[c]{@{}l@{}}Quantum gravity \\ (hypothesized)\end{tabular}  & \begin{tabular}[c]{@{}l@{}} \hspace{0.5cm} Graviton\\ (hypothesized)\end{tabular} & $10^{-38}$   \\ \hline
\end{tabular}
\end{center}
\caption{Fundamental interactions, their mediators and strength along with the corresponding theory.}
\label{table:fund_inter}
\end{table}

Because of their richness and intricacies each of these interactions and related phenomena have become branches of particle physics their own. We, in this dissertation, will be mainly dealing with phenomena related to strong interaction. So, in the next section we discuss some of the basic properties of quantum chromodynamics (QCD), the theory of strong interaction.

\section{Quantum chromodynamics}
\label{sec:QCD}

Because of its success in explaining many of the strong interaction phenomena, QCD is believed to be the theory of strong interaction~\cite{Halzen_Martin(book),Muta(book)}. Particularly, much of the support for QCD derives from its ability to produce results against the experimental verification at high energies (short distances)~\cite{Gross:1973id,Politzer:1973fx}. It is a Yang-Mills theory with the $SU(3)$ color gauge group involving the strong interactions of (anti)quarks and gluons. The non-abelianness of the theory makes it fundamentally very different from abelian theory like quantum electrodynamics (QED). Main difference being the coupling of gluons among themselves, which leads to the property known as asymptotic freedom~\cite{Gross:1973id,Politzer:1973fx}. Before we discuss a bit more about the asymptotic freedom let us introduce the Lagrangian of QCD.

As already mentioned, QCD is a $SU(3)_c$ gauge theory with the (anti)quarks and gluons belonging to its fundamental and adjoint representation, respectively. A quark of a particular flavor comes with three colors, whereas there are eight types of gluons. The Lagrangian has the following form,

\begin{eqnarray}
 {\mathcal L} &=& \bar{\psi}(i\Slash D-m)\psi -\frac{1}{4}F^{\mu\nu}_aF_{\mu\nu}^a,
 \label{eq:qcd_lag}
\end{eqnarray}

with the field strength tensor being

\begin{eqnarray}
 F_{\mu\nu}^a &=& \del_{\mu}A_\nu^a-\del_{\nu}A_\mu^a+g f_{bca}A_\mu^bA_\nu^c;
 \label{eq:qcd_fld_tens}
\end{eqnarray}

where the covariant derivative is defined as $ D_\mu = \del_\mu-igT_aA_{\mu a}$ and $A_\mu^a$ is the non-abelian gauge field with the color index $a$, $g$ is the QCD coupling constant, $T_a$ are the generator of $SU(3)_c$ group with $[T_a,T_b]=if_{abc}T_c$, $f_{abc}$ being the structure constant of the group. The generators are further expressed in terms of Gell-Mann matrices $\lambda_a$ as $T_a=\frac{\lambda_a}{2}$.

The third term in equation (\ref{eq:qcd_fld_tens}) makes the QCD behaving a whole lot different from QED. It is this term through which the gluons can interact among themselves. This leads to the phenomenon known as asymptotic freedom which is discussed in the next paragraph. Before we go into that we should be aware of the fact that apart from this local gauge symmetry QCD also bears some global symmetries which are the backbone of different effective QCD models at low energies. These symmetries are discussed in details in the subsection~\ref{ssec:sym_qcd}.

\begin{figure}[hbt]
\subfigure[]
{\includegraphics[scale=0.5]{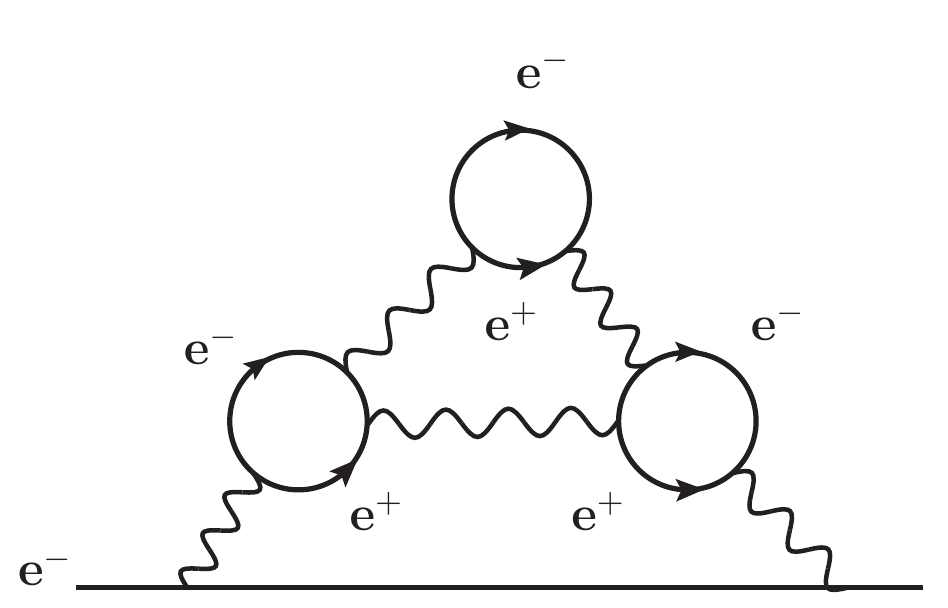}
\label{fig:QED_vac}}
\subfigure[]
{\includegraphics[scale=0.5]{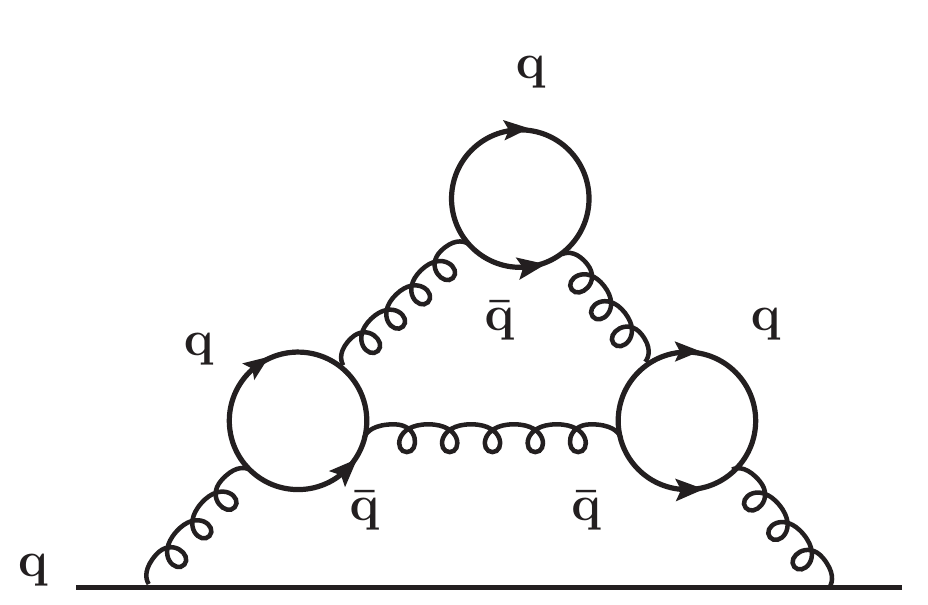}
\label{fig:QCD_vac_q}}
\subfigure[]
{\includegraphics[scale=0.5]{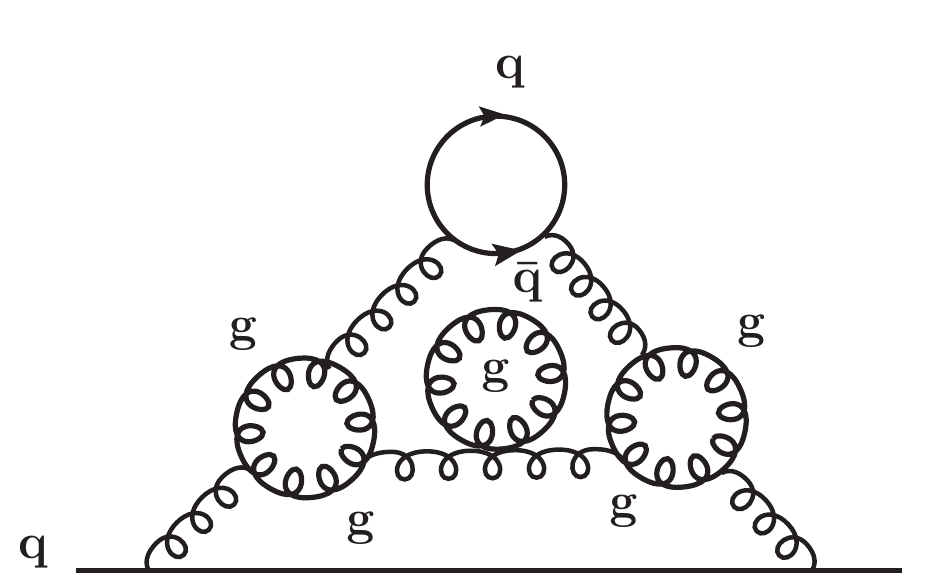}
\label{fig:QCD_vac_g}}
\caption{Schematic representations of vacuum polarization diagrams for QED in (a) and QCD in (b) \& (c). (b) is analogous to (a); the extra contribution for QCD, shown in (c), arises due to the gluon self interaction.}
\label{fig:vac_pola}
\end{figure}

The vacuum of QCD behaves differently from that of QED because of the gluons' self interaction. This has been elucidated appropriately in figure~\ref{fig:vac_pola}. From figure~\ref{fig:QED_vac} it is understood that due to vacuum fluctuation an electron is considered to be surrounded by positrons and thus its electric charge gets screened. So, as we probe closer to the electron by increasing the energy, we experience more of the total charge of the electron (it reminds us of the dielectric medium). Thus in QED the coupling strength increases with the increase of energy or decreasing distance.

\begin{figure} [!htb]
\center
\includegraphics[scale=0.5]{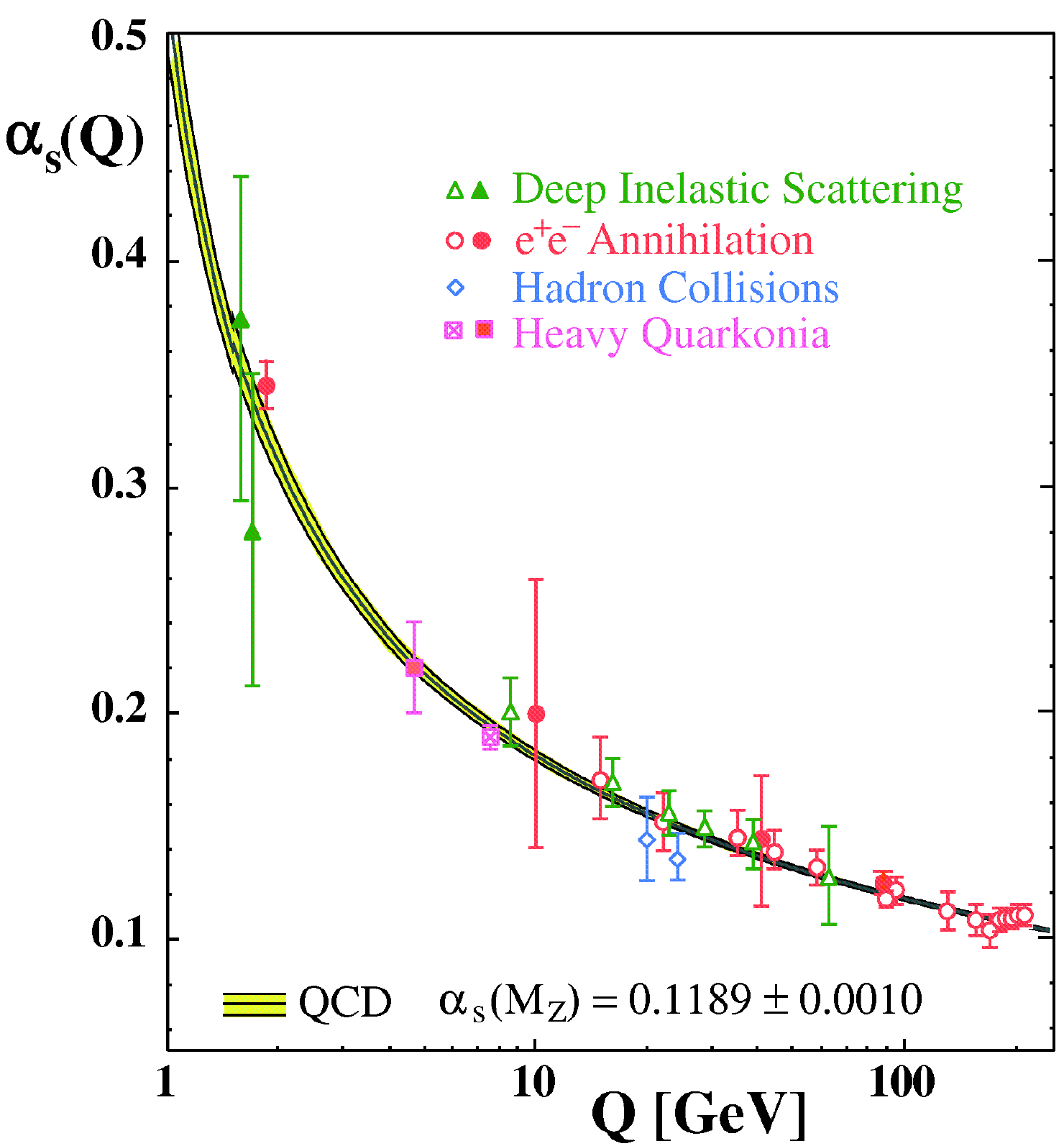}
\caption{Running of QCD coupling constant.}
\label{fig:QCD_run_coup}
\end{figure}

On the other hand the same effect is there for quarks in QCD vacuum as shown in figure~\ref{fig:QCD_vac_q}, though here we are talking about color charge. But that is not all - there is another scenario shown in figure~\ref{fig:QCD_vac_g} which arises because of gluon-gluon interaction. This makes the whole picture different and in QCD vacuum the color charge gets antiscreened. So, in QCD the coupling constant decreases as the probing energy is increased. This is known as asymptotic freedom. The running of the coupling constant in QCD is further demonstrated in figure~\ref{fig:QCD_run_coup}. The predictions from the theory matches well with the experimental findings. This experimental validation gives confidence in QCD as the theory of strong interaction~\cite{Bethke:2006ac}.

Apart from asymptotic freedom there is another property of strong interaction known as color confinement~\cite{Halzen_Martin(book),Muta(book)}, which signifies that no color charged particles such as (anti)quarks, gluons can be isolated and thus can never be directly observed{\footnote{In this sense the hypothesis of color charge, which was introduced to comply with Pauli's exclusion principle, is a strange assumption; because for the first time in subatomic physics it introduces a difference between particles that are identical in all observable ways, viz. between two $d$ quarks of different color.}}. The reasons for confinement is yet to be understood and there is no analytic proof that tells us that QCD should be confining. In other words, the QCD Lagrangian in (\ref{eq:qcd_lag}) gives appropriate descriptions when the degrees of freedom are quarks and gluons at high energies (short distances), but fails to tell us how those quarks and gluons get confined into hadrons. Now there are some way out 
to this through numerical methods like lattice QCD (LQCD), which is a first principle calculation or through effective QCD models based on symmetries of QCD, that try to mimic QCD as closely as possible. These methods will be discussed briefly in the subsection~\ref{ssec:theo_methods}.   

In the next subsection we briefly discuss about the symmetries of QCD along with their physical implications.

\subsection{Symmetries of QCD}
\label{ssec:sym_qcd}

We know that the $SU(3)_c$ color gauge symmetry is the most important local symmetry of QCD. The global symmetries of QCD are also important particularly while building up the effective models. We begin the discussion by considering vanishing current quark masses. In that limit the left and right handed quarks decouple and the QCD Lagrangian possesses $U(N)_L\times U(N)_R$ symmetry (we no more talk about the $SU(3)_c$ symmetry which is always there). The left and right handed fields are defined using the chiral projection operators,
\begin{align}
 \psi_L=\frac{1-\gamma_5}{2}\psi;~~~ \psi_R=\frac{1+\gamma_5}{2}\psi. 
\end{align}
From group properties we can decompose $U(N)$ into $SU(N)\times U(1)$. We can further add and subtract the generators of left and right handed quarks to obtain a new set of generators which correspond to the vector and axial vector symmetries~\cite{Vogl:1991qt}. The vector and axial vector symmetries are defined under global $U(1)$ transformations as,
\begin{align}
 \psi\rightarrow e^{i\theta}\psi,~~~ \psi\rightarrow e^{i\gamma_5\theta}\psi
\end{align}
respectively; $\theta$ being the global parameter. 

\begin{figure}

\hspace{2.0cm}
\begin{tikzpicture}[node distance=2.6cm,line width=0.8pt]
 
 \node(pro1) [process] {$SU(3)_c\times U(3)_L\times U(3)_R$};
 \node(pro2) [process, below of=pro1] {$ SU(3)_c\times SU(3)_L\times U(1)_L\times SU(3)_R\times U(1)_R$};
 \node(pro3) [process, below of=pro2] {$SU(3)_c\times SU(3)_V\times U(1)_V\times SU(3)_A\times U(1)_A$};
 \node(pro4) [process, below of=pro3] {$SU(3)_c\times SU(3)_V\times U(1)_V\times SU(3)_A$};
 \node(pro5) [process, below of=pro4] {$SU(3)_c\times SU(3)_V\times U(1)_V$};
 \node(pro6) [process, below of=pro5] {$SU(3)_c\times SU(3)_f\times U(1)_V$};
 \node(pro7) [process, below of=pro6] {$SU(3)_c\times U(1)_V$};

 \draw  (pro1) -- (0,-1.15);
 \draw [arrow] (0,-1.4) -- node[yshift=14]{Decomposition into subgroups} (pro2);
 
 \draw  (pro2) -- (0,-3.78);
 \draw [arrow] (0,-4.17) -- node[yshift=14]{Change of representations} (pro3);
 
 \draw  (pro3) -- (0,-6.33);
 \draw [arrow] (0,-6.6) -- node[yshift=14]{Anomalous breaking of $U(1)_A$} (pro4);

 \draw  (pro4) -- (0,-8.9);
 \draw [arrow] (0,-9.2) -- node[yshift=14.2]{Spontaneous CSB} (pro5);
  
 \draw  (pro5) -- (0,-11.48);
 \draw [arrow] (0,-11.8) -- node[yshift=14]{Nonzero but degenerate current quark masses} (pro6);
  
 \draw  (pro6) -- (0,-14.07);
 \draw [arrow] (0,-14.42) -- node[yshift=14.5]{Nonzero and nondegenerate current quark masses} (pro7);
\end{tikzpicture}
\caption{Symmetries of QCD and their physical implications with the three lightest quark flavors up ($u$), down ($d$) and strange ($s$).}
\label{fig:sym_qcd}
\end{figure}
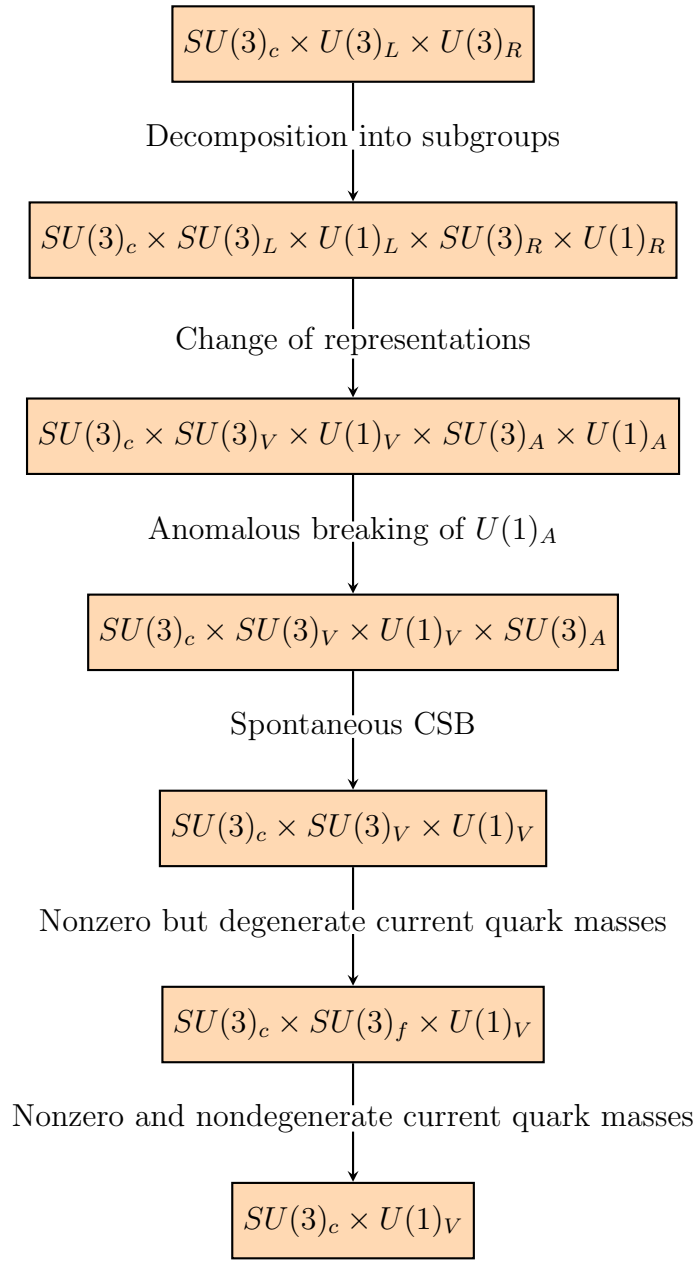 

The $U(1)_A$ symmetry is anomalously broken which is known as axial anomaly. The chiral symmetry $SU(N)_V\times SU(N)_A$ is spontaneously broken into $SU(N)_V$. This is known as spontaneous chiral symmetry breaking (CSB). For two flavor the pion triplet are the Goldstone bosons, whereas for three flavor, the pion octet. If chiral symmetry was not spontaneously broken then there would exist in the meson and baryon spectrum a mirror multiplet with opposite parity to each isospin multiplet. But no such chiral multiplet does exist. This spontaneously broken chiral symmetry is supposed to be restored at high temperature and/or density.
 
Once we introduce the nonzero but degenerate current quark masses the vector symmetry $SU(N)_V$ can be recognized as the symmetry of $N$ flavors, namely $SU(N)_f$. When we consider the quark masses to be nondegenerate, $SU(N)_V$ is explicitly broken and what remains is the symmetry $SU(N)_c\times U(1)_N$. The symmetry $U(1)_N$ is responsible for the conservation of baryon number, which is a consequence of Noether's theorem. In the figure~\ref{fig:sym_qcd} these symmetries are well explained through a flow chart~\cite{Roessner:2006dt}. There we consider the three lightest flavors, $u$, $d$ and $s$ which are the only accessible quark flavors in the energy range of interest in this dissertation.

QCD is also known for its rich phase structure, particularly at high temperature and/or high density associated with the symmetries and their breaking discussed so far. In the next subsection we briefly touch upon the phase structure of QCD.

\subsection{Phase structure of QCD}
\label{ssec:phase_struc_QCD}

The phase diagram of hot and/or dense system of quarks and gluons predicted by the QCD has invited a lot of serious theoretical as well as experimental investigations for last few decades. The first prototype of the QCD phase diagram was conjectured in \cite{Cabibbo:1975ig} where it looked very simple as shown in figure~\ref{fig:phase_dia_old}; with the passage of time more and more investigations culminated in a very complicated looking phase diagram with many exotic phases~\cite{Fukushima:2010bq} as displayed in figure~\ref{fig:phase_dia_new}. We discuss some of its characteristics in the following paragraphs.

So far we have discussed about confining states. There is also the possibility of creating a deconfined state in which the quarks and gluons are liberated. It can be understood intuitively by considering nucleons in a nucleus which can be pictured as system of quark bags. Now if we keep on increasing the density by putting more and more nucleons, the distance among the quarks and gluons keeps on decreasing. Thus there will be a certain critical value of density after which a quark or gluon will no longer be able to recognize which nucleons it belongs to and thus will be moving freely in a volume much larger than the hadronic/nucleonic volume. We can also create the free state of quarks and gluons by increasing the temperature, since increasing temperature leads to pion production which is equivalent to increasing the density. Thermalized state of such deconfined quarks and gluons is known as quark gluon plasma (QGP)~\cite{Muller:1983ed,Heinz:2000bk}, 
which is a relativistic plasma since the thermal velocities of quarks (light flavor) and gluons in it are relativistic. Its constituents are color ionized and it also screens the color charge - the reason why it is called a plasma. How this state of QGP can be formed is discussed in the subsection~\ref{ssec:exp_methods}. Chiral phase transition (vide figure~\ref{fig:sym_qcd}) is another interesting aspect of the QCD phase diagram which has been also at the center of attention.  

\begin{figure}[hbt]
\subfigure[]
{\includegraphics[scale=0.5]{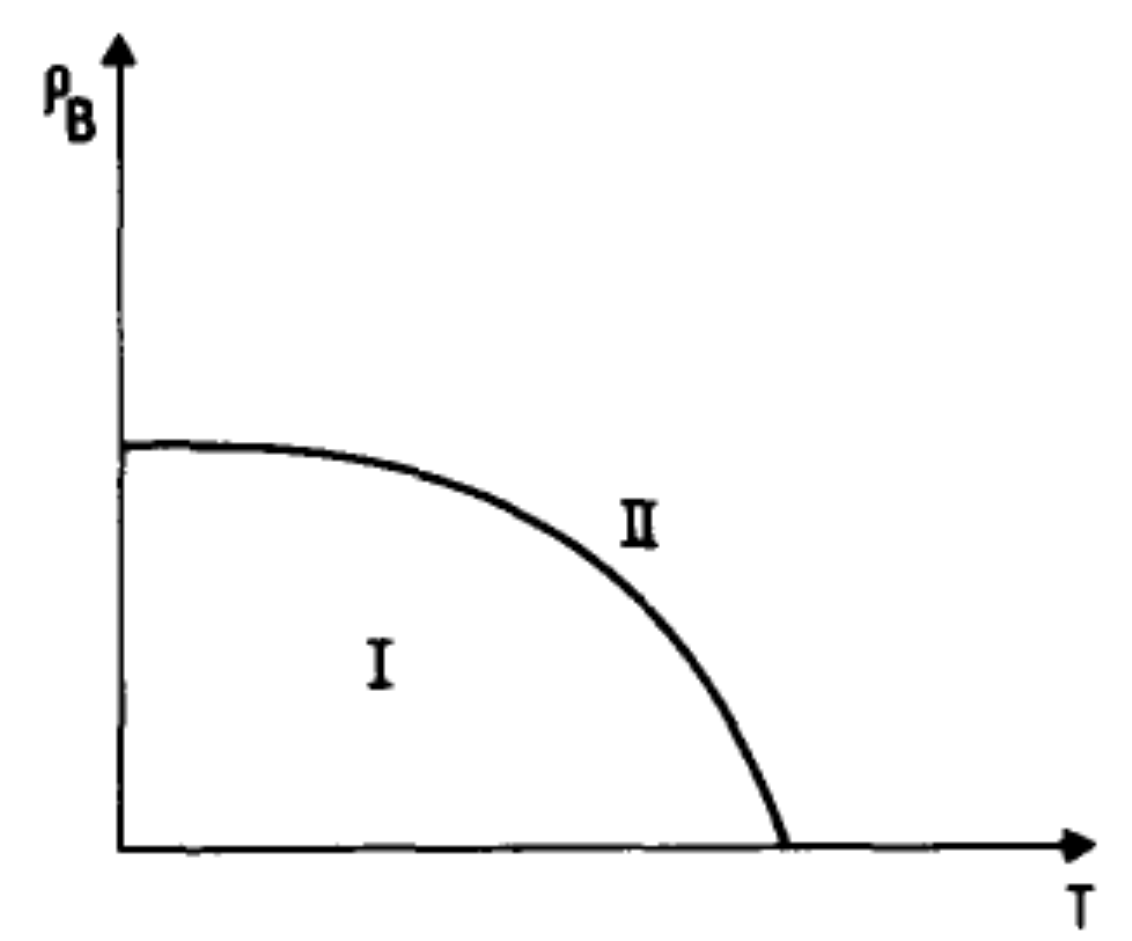}
\label{fig:phase_dia_old}}
\subfigure[]
{\includegraphics[scale=0.33]{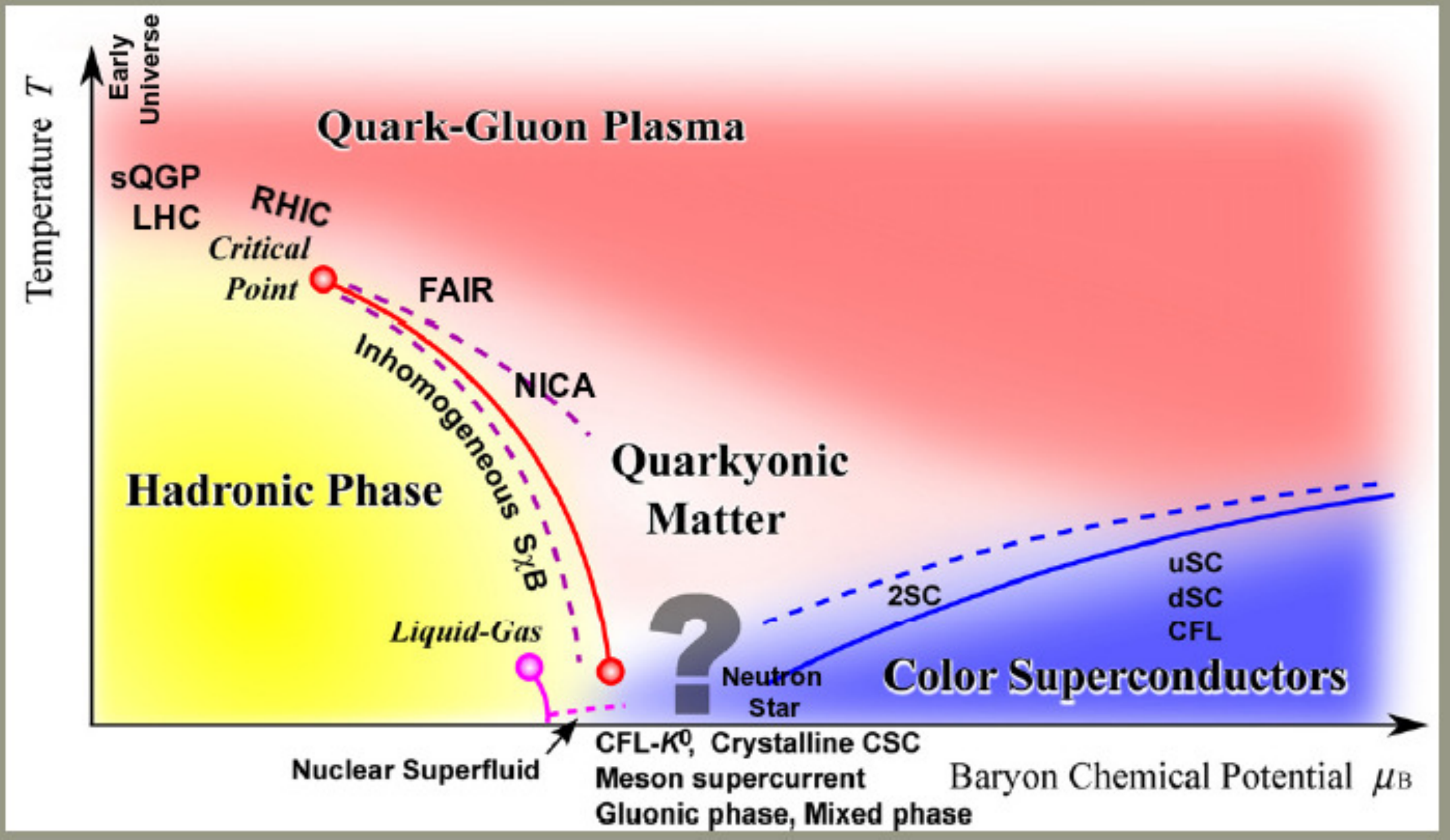}
\label{fig:phase_dia_new}}
\caption{Phase diagram of QCD: (a) The first prototype and (b) it's state of the art avatar.}
\label{fig:QCD_phase_dia}
\end{figure}

It is important to know the order of a transition, since it could reveal the characteristics of the underlying mechanism. For QCD phase diagram in the $T-\mu$ plane, the transition is supposed to be a crossover at zero density. On the other hand, at low temperature and high density the transition is speculated to be first order~\cite{Scavenius:2000qd}. So the first order transition line starting on the chemical potential axis ends up in a critical point at finite $T$ and $\mu$~\cite{Fodor:2001pe}. This is a widely accepted scenario of the QCD phase diagram and the experimental search for this critical point is on~\cite{Mohanty:2009vb}.

If the deconfinement and chiral phase transitions do not coincide, exotic phases such as the constituent quark phase~\cite{Cleymans:1986cq,Kouno:1988bi} or the quarkyonic phase, a confined but chiral symmetric phase~\cite{McLerran:2007qj,Hidaka:2008yy}, may occur. A color superconducting as well as color-flavor-locked phase at zero or small temperature and high density have also been conjectured. There are many more to this whole story, the details of which can be found in the reference~\cite{Fukushima:2010bq}.

In this dissertation we will be dealing with the deconfinement and chiral phase transitions particularly in connection with QGP.

\section{Purpose of studying QGP}
\label{sec:pur_stu_QGP}

If we leave aside the interesting intricacies of studying QCD phase structure, its immense physical significance makes the study of QGP along with the other phases, a pursue-worthy aspect of natural phenomena. Particularly, exploration of QGP is expected to shed lights on some less-known important large scale physical phenomena such as evolution of early universe, properties of neutron star etc.

As far as our present understanding is concerned Big Bang model is the most reliable one regarding the creation and evolution of the universe~\cite{Karen_Fox(book)}. According to this model the universe expanded from a very high density and high temperature state. It went through many phases and ultimately got into its present-day form. When the inflation stopped around $\sim10^{-11}$ s after the birth the universe is supposed to be in a phase filled with QGP. This phase lasts up to $\sim10 \mu \rm{s}$ when the temperature was of the order of $10^{12}$ K ($\sim$ 200 MeV), quarks and gluons get confined into hadrons~\cite{Riordan:2006df}. So understanding of QGP will really help us to peep into the past and comprehend the evolution of the universe in a better way. 

Neutron star is the another astrophysical entity which provides us the motivation for studying QGP. The ongoing research predicts about the structure of the neutron star~\cite{Akmal:1998cf} and it has been speculated that a highly dense matter, much more dense than the ordinary nuclear matter ($\sim 0.125\, {\rm GeV}/{\rm fm}^3 \,{\rm or}\, 10^{18}\, \rm{kg/m^{3}}$), to exist in the core of neutron star, which in turn can affect its spin and magnetic properties. 

The existence of QGP in astrophysical phenomena so far discussed are by far remote in space and time and thus cannot be accessed. But to study experimentally we need to have access to QGP. This leads to the terrestrial based experiments known as heavy ion collisions (HIC) through which QGP can be formed and thus studied. In the next section, this is what we discuss along with the theoretical tools to study QCD and its phase structure, particularly the QGP.

\section{Methods of studying QCD}
\label{sec:methods_stu_QCD}

There are always two aspects of studying any physical phenomena. One is the experimental part and other is the theoretical one. Whereas the history of science suggests that theoretical ideas can really predict new physical phenomena and dictate to the experiment what it should look for, the ultimate role is played by the experiments by deciding the fate of any theoretical advancement and choosing the path of scientific progress. So it is very important that these two go hand in hand. In the following subsections we will discuss in short the tools, experimental (in subsection~\ref{ssec:exp_methods}) as well as theoretical (in subsection~\ref{ssec:theo_methods}), for investigating QCD and many facets of its phase structure.

\subsection{Experimental methods}
\label{ssec:exp_methods}

In recent years, a tremendous effort has been devoted to study how QCD behaves in unusual conditions, particularly the creation of QGP in the laboratory. In all these experiments heavy ions are accelerated to relativistic speeds in order to achieve extreme conditions required for the creation of such a short-lived phase. These extreme conditions are achieved by increasing the energy of the colliding beam. At present Relativistic Heavy Ion Collider (RHIC) at Brookhaven National Laboratory (BNL)~\cite{Arsene:2004fa,Back:2004je,Adams:2005dq,Adcox:2004mh}and the Large Hadron Collider (LHC) at the European Organization for Nuclear Research (CERN)~\cite{Carminati:2004fp,Alessandro:2006yt} are operational. Also some future facilities are coming up, such as Facility for Antiproton and Ion Research (FAIR) at the Gesellschaft f\"{u}r Schwerionenforschung (GSI)~\cite{Senger:2004jw,Friman:2011zz} and Nuclotron-based Ion Collider fAcility (NICA) at  Joint Institute for Nuclear Research (JINR), Dubna~\cite{Toneev:
2007yu,Sissakian:2009ru}. The goals of these upcoming experiments are basically to complement the operational ones by exploring different regions of QCD phase diagram (particularly the high density regions) which, so far, remain unexplored and also by corroborating many of the findings of those. Figure~\ref{fig:phase_dia_new} shows different experiments exploring different parts of the phase diagram in $T-\mu$ plane.

Many observables, which would work as signal of QGP, are proposed and measured in these experiments~\cite{Yagi:2005yb,CY_Wong(book)}. Among them multiplicity vs average $\rm{p_T}$, jet quenching, elliptic flow, electromagnetic emissions (photon and dilepton production), $J/\psi$ suppression, strangeness enhancement, fluctuations etc have been extensively studied. In this thesis we have investigated the electromagnetic spectral function and its spectral properties \textendash~the dilepton production rate and the conserved density fluctuations, namely the quark number susceptibility (QNS).

In QGP a quark can interact with an antiquark to produce a virtual photon and that virtual photon can decay into a lepton-antilepton pair (so called dilepton). These dileptons having a mean free path larger than the system size, leave the fireball almost without any interaction. Thus they carry almost undistorted information of the early times when they are produced in the deconfined hot and dense matter~\cite{Alam:1996fd,Alam:1999sc}. But dileptons are produced from every stages of HIC, thus it becomes difficult to use them as a probe of QGP unless the total dilepton production rate is known. On the other hand study of fluctuations and correlations in HIC reveals important characteristics of the system particularly about the nature of the phase transitions. Critical opalescence arising from fluctuations in second order phase transitions is a well known example. We also get to know about the effective degrees of freedom of the system and how it behaves to any external perturbation~\cite{Koch:2008ia}.

So far various diagnostic measurements have been performed, which indicate a strong hint for the creation of a strongly coupled QGP (not a weakly interacting gas of quarks and gluons)~\cite{Shuryak:2008eq} within a first few fm/$c$ of the collisions through the manifestation of hadronic final states. As for example the measurements at RHIC BNL~\cite{Reichelt:2015oze,Adare:2009qk,Adler:2006yt,Adare:2006ti,Adcox:2001jp,Adcox:2002au,Adler:2003kg, Chujo:2002bi,Adare:2006nq,Abelev:2006db} and the new data from LHC CERN~\cite{Aamodt:2010pb,Aamodt:2010pa,Aamodt:2010jd,Aad:2010bu,Chatrchyan:2011sx}, all support for the existence of a strongly interacting QGP (sQGP).

\subsection{Theoretical methods}
\label{ssec:theo_methods}

On the other side of the coin there are theoretical tools to study the properties of QGP. The coupling constant of QCD being large at small energies (large length) its study becomes a theoretical challenge specially in those regime because of the nonperturbative nature of the theory.

So far we do not have any analytical way to tackle this nonperturbative nature of the theory, but there exists a first principle QCD method known as Lattice QCD (LQCD) which is completely a numerical technique~\cite{Wilson:1973jj,Wilson:1974sk,Kogut:1982ds,Creutz:1983ev}. If we believe that QCD is the theory of strong interaction, then LQCD, being a first principle method, is a numerical experiment. It can provide accurate results for zero as well as nonzero temperature (popularly known as finite temperature) QCD within some systematic errors depending on the lattice size and lattice spacing. But this method suffers from a serious difficulty at finite chemical potential, known as the infamous sign problem~\cite{Karsch:2001cy}. However, there are some ways out to it, which can only be applied in a limited region of the phase diagram~\cite{Aarts:2008wh}.  

Analytical method such as perturbative QCD (PQCD) works well when the coupling constant is sufficiently small at high energies~\cite{Gribov:1984tu,Ioffe:2005ym}. To deal with the high temperature QCD, hard thermal loop (HTL) perturbation theory, a resummation technique, has been invented~\cite{Kalashnikov:1982sc,Andersen:1999fw,Andersen:1999sf,Andersen:1999va,Haque:2010rb}\footnote{Naively, one would expect from asymptotic freedom that perturbative technique will be valid at high temperature and/or high density. But problem of infrared divergence, which arises due to the presence of massless particles, plagues the perturbative calculations at high temperature. The problem is solved, at least in the electric sector, by using the resummation scheme known as HTL. In this technique, a class of loop diagrams for which the loop momenta are of the order of the temperature are resummed, which contribute to a given order. Based on this approximation an improved perturbation theory, known as HTL perturbation theory, 
is developed by performing an expansion around a system of massive quasiparticles generated through thermal fluctuations.}. In a recent study using this HTL perturbation theory various thermodynamic quantities have been calculated and shown to agree well with the available lattice data~\cite{Haque:2013sja}.

To circumvent all these difficulties, shortcomings of LQCD and PQCD, effective QCD models have been designed. Since the hot and dense matter created in HIC is supposed to be non-perturbative in nature, applications of such effective models become particularly useful for the purpose. Eventually there exist many effective QCD models which are being extensively used. There are Nambu\textendash Jona-Lasinio (NJL) model~\cite{Nambu:1961tp,Nambu:1961fr} and its Polyakov loop extended version, PNJL model~\cite{Fukushima:2003fw,Ratti:2005jh}; Linear Sigma Model (LSM)~\cite{GellMann:1960np} and also its Polyakov loop extended version, PLSM~\cite{Schaefer:2007pw}; functional methods like Dyson-Schwinger Equation (DSE)~\cite{Bjorken_Drell(book):RQF}, matrix model~\cite{Meisinger:2001cq}, different quasiparticle models~\cite{Peshier:1995ty}, color singlet (CS) model~\cite{Turko:1981nr,Elze:1985wv,Auberson:1986ft,Mustafa:1993np} just to name a few. The PNJL model has been further improved by considering the entanglement 
between chiral and 
deconfinement 
dynamics, which is termed as EPNJL model~\cite{Sakai:2010rp}. The study of the properties of QGP in this thesis is based on the NJL, PNJL and EPNJL models and also on the CS model, which have been discussed a bit elaborately in the next chapter~\ref{chapter:prelim}. 

A large part of the works covered in this dissertation is based on the correlation function (CF), current-current CF to be precise. This CF, sometimes, is also called correlator. So to maintain the consistency, in the next section we develop some basic ideas about CF and its spectral representation.

\section{Correlation function and its physical importance}
\label{sec:cf_and_sr}

Dynamical behaviors of many particle systems are studied by employing an external perturbation, which disturbs the system slightly from its equilibrium state, and thus assessing the spontaneous responses/fluctuations of the system to this external perturbation. In general, these responses/fluctuations are related to the CFs through fluctuation-dissipation theorem~\cite{Forster(book):1975HFBSCF,Callen:1951vq}.

CF and its spectral representation are  extensively used mathematical tools applied in almost every branch of physics. Generically it describes how microscopic variables co-vary with response to one another. Thus there can be different types of CFs. As for example, a frequently used CF in high energy physics is the propagator for different types of fields. Here, in these series of works, we will mainly focus on current-current CFs, viz. vector meson current CF in a thermal background. These CFs and their spectral representation reveal dynamical properties of many particle system and many of the hadronic properties~\cite{Forster(book):1975HFBSCF,Callen:1951vq,Kubo:1957mj}. Such properties in vacuum are very well studied in QCD~\cite{Davidson:1995fq}.

While propagating through the hot and dense medium, the vacuum properties of any particle get modified due to the change of its dispersion properties in the medium. These changes are reflected in its CFs~\cite{Kapusta_Gale(book):1996FTFTPA,Lebellac(book):1996TFT}. Also the existence of resonances or bound states in the plasma state can be investigated through the study of CFs. The presence of a stable mesonic state is understood by the delta function like peak in the spectral function (SF). For a quasiparticle in the medium, the $\delta$-like peak is expected to be smeared due to the thermal width, which increases with the increase in temperature. At sufficiently high temperature and density, the contribution from the mesonic state in the SF will be broad enough so that it is not very meaningful to speak of it as a well defined state any more. On the other hand, the inverse of equal-time correlation lengths (screening masses) characterize the long-range properties of a thermal system~\cite{Nadkarni:1986as,
Chakraborty:2011uw,Xu:2011ud}. For example, these screening masses specify the infrared sensitivity of various thermodynamic quantities and also the spectral properties of a given system.

The temporal CF is related to the response of the conserved density fluctuations due to the symmetry of the system. On the other hand, the spatial CF exhibits information on the masses and width. Now, in a hot and dense medium the properties of hadrons, viz. response to the fluctuations, masses, width, compressibility etc., will be affected. Hence, hadron properties at finite temperature and density are also encoded in the structure of its CF and the corresponding spectral representation, which may reflect the degrees of freedom  around the phase transition point and thus the properties of the deconfined, strongly interacting matter.

As for example, the spectral representation of the vector current-current correlation can be indirectly accessible by high energy heavy-ion experiments as it is related to the differential thermal cross section for the production of lepton pairs~\cite{Kapusta_Gale(book):1996FTFTPA,Lebellac(book):1996TFT}. These lepton pairs are considered as good signal of QGP, since they leave the fireball with minimum interaction once they are produced. But dilepton pairs are produced in many stages of the HICs and it becomes difficult to identify the stage of the collisions from which the pairs are coming out. Thus it is appropriate to talk about the total rate of lepton pairs produced all over the range of the collision time. In this dissertation, though, we will mainly focus on the dilepton pairs arising from the QGP phase. Moreover, in the limit of low energies (small frequencies), various transport coefficients of the hot and dense medium can be determined from the spatial spectral representation of the vector channel 
correlation.

On the other hand correlations and fluctuations of conserved charges, such as electric and baryonic numbers, are supposed to be good signals of the deconfinement phase transition~\cite{Asakawa:2000wh,Jeon:2000wg,Hatta:2003wn}. Here in this thesis we are specifically interested in the quark number density fluctuation. It can be related with the temporal part of the current-current CF through the fluctuation-dissipation theorem and also can be calculated from the thermodynamic pressure. From the thermodynamic point of view susceptibilities measure the fluctuations. For quark number density fluctuation it is the QNS which is used as a probe for the quark-hadron phase transition. 

 Further details of CF and its spectral representation along with the mathematical details are presented in the upcoming chapter~\ref{chapter:prelim}. In the next section we sketch the scope of the thesis.

\section{Scope of the thesis}
\label{sec:sc_thesis}

In chapter~\ref{chapter:prelim} we briefly review the basic ingredients of various effective QCD models (viz. NJL, PNJL, EPNJL and CS) along with the features of CF and some of their spectral properties, which have extensively been used in the thesis.

In chapter~\ref{chapter:JHEP} we explore the vector meson current-current CF with and without the influence of isoscalar-vector (I-V) interaction in NJL and PNJL models~\cite{Islam:2014sea}. As a spectral property we have computed the dilepton rate which is found to be enhanced in sQGP as compared to the Born rate (leading order perturbative rate) in a weakly coupled QGP. Here the sQGP has been obtained and tuned by the inclusion of a background gauge field, namely the Polyakov loop (PL) field. We further compared our findings with the available Lattice data.

In chapter~\ref{chapter:PRD} we consider the idea of the EPNJL model and re-explore the vector spectral function and the spectral property such as the dilepton production rate~\cite{Islam:2015koa} previously studied in~\cite{Islam:2014sea} as discussed in chapter~\ref{chapter:JHEP}. Because of the strong entanglement between PL field and chiral condensate, the coupling strengths run with the temperature and chemical potential. This running has interesting implications on the dilepton rate which has been explored. 

The Euclidean vector correlator and also the response of the conserved density fluctuations related with the temporal vector correlator, have been studied in the chapter~\ref{chapter:Susc}. We have considered both the scenarios, i.e. presence and absence of the vector (I-V) interaction. The inclusion of the vector interaction also brings forth some intriguing issues in the fluctuation of conserved density, namely the QNS. This has also been discussed in details.

In another effort (chapter~\ref{chapter:JPG}), we assume the hot and dense matter (QGP) to be made of a non-interacting quarks, antiquarks and gluons with the underlying symmetry of $SU(3)$ color gauge theory~\cite{Islam:2012kv}. This assumption holds particularly for a QGP which is weakly coupled. We showed that this type of simple quantum statistical description exhibits very interesting features: the $SU(3)$ color singlet has $Z(3)$ symmetry through the normalized character in the fundamental representation of $SU(3)$. This character becomes equivalent to an ensemble of PL fields which can be exploited as an order parameter for color-confinement to color-deconfinement phase transition.

\chapter{Some preliminaries}
\label{chapter:prelim}
In this chapter we develop the basic ideas of effective QCD models which have been later used in the present thesis. There exists a plethora of such models. But following the requirement of this thesis we will only shed light on NJL, PNJL and EPNJL models. Then we will also give a brief review on color-singlet (CS) model. Some of the preliminaries of correlation functions (CF) and their spectral representations have also been dealt along with their physical significance, particularly how they are associated with dilepton rate and susceptibility.

\section{Effective QCD models}
\label{sec:eff_models}

As has been mentioned in the introduction, QCD is the theory of strong interaction and it shows some interesting properties different from QED. Asymptotic freedom is one of them which allows the perturbative method to be applied in the range of large momentum transfer. But as the momentum transfer becomes smaller and smaller the applicability of such method loses its justification due to the increase of coupling constant and eventually fails completely when the coupling constant becomes sufficiently large. Thus we cannot describe the hadrons and their properties just by starting from QCD Lagrangian. It is a strange situation, since we have the Lagrangian which effectively carries all the dynamical informations of the system one needs to know but it is very difficult to extract useful informations from that. Therefore several methods have been developed to study QCD. 

Lattice gauge theory is the only one of them which actually tries to solve the QCD Lagrangian starting from first principle. It is a numerical method and can provide accurate results for zero as well as nonzero temperature QCD within some systematic errors depending on the lattice size and lattice spacing. But this method suffers from a serious difficulty at finite chemical potential, known as the infamous sign problem, which restricts its applicability in that regime~\cite{Karsch:2001cy}. Thus it becomes necessity to look for some other way to study QCD \textendash \, at least some of its low energy properties if not all. 

Now from our knowledge of atomic physics, nuclear physics or solid state physics we know that sometimes it is really helpful to replace a many-body system by an effective one-body description to extract the essential features. The logic behind such assumption is that many-body system can only be exactly solved if it is simplified. So the lesson from this simplification is that it is sensible to replace a complicated (mathematically intractable) theory like QCD by some effective ones (mathematically tractable) which carries its essential features. There are many such effective models which are being extensively used (vide subsection~\ref{ssec:theo_methods} in the introduction). Here we will discuss only some of them. These models can be categorized in two types. In one type there are NJL, PNJL and EPNJL models whether in the other one there is CS model. Let us move on to the discussion by beginning with the NJL model.

\subsection{Nambu\textendash Jona-Lasinio (NJL) model}
\label{ssec:njl_model}

Historically NJL model was introduced as a pre-QCD theory to describe the strong interaction among the nucleons with the degrees of freedom being the nucleons and mesons~\cite{Nambu:1961tp,Nambu:1961fr}. After the discovery of the quarks~\cite{GellMann:1964nj,Zweig:1981pd,Zweig:1964jf} the theory has been reinterpreted with quark degrees of freedom~\cite{'tHooft:1976fv,Callan:1977gz}. The symmetries of QCD discussed in subsection~\ref{ssec:sym_qcd} in the introduction are the backbone of NJL model. Among all chiral symmetry plays the most crucial role. It is not only all these symmetries but also how they are broken characterize QCD and thus are needed to be considered. This is well included in NJL model in which dynamical mass generation is realized through spontaneous breaking of chiral symmetry. 

Since NJL model was formulated before the discovery of quarks, the idea of confinement is not included in it. It is a drawback of the model but many questions particularly related to hadronic physics, where chiral symmetry is the relevant feature, can be answered using it. In principle starting from QCD one can derive the NJL Lagrangian by integrating out the gluonic degrees of freedom, which leaves the quarks interaction to be a local four point one. Then there arises another problem due to this local interaction - the NJL model is not a renormalizable theory. There are many schemes to regularize the theory, which have been elaborately discussed in the reference~\cite{Klevansky:1992qe}. Here in our work we use the three momentum cut-off scheme, which is the most popular one. In the following few paragraphs we briefly discuss intricacies of the model. The details can be 
found in any of these references~\cite{Vogl:1991qt,
Klevansky:1992qe,Hatsuda:1994pi,Buballa:2003qv}.    

\subsubsection{Two flavor NJL model in vacuum}
\label{sssec:njl_vac}

The NJL model is designed to work below the momentum scale of QCD, $\Lambda_{QCD}\approx0.3~\rm{GeV}$. Within this scale, inclusion of only three lightest flavors $u$, $d$ and $s$ remains relevant. We make further simplification and consider a two flavor NJL model with up and down quark masses being the same. The nonzero quark masses break the $SU(2)_A$ symmetry explicitly but equality of them preserves the isospin one, $SU(2)_V$. First we consider only the scalar and pseudoscalar-isovector interaction terms. The corresponding Lagrangian is:
\begin{eqnarray}
\mathcal{L}_{\rm{NJL}} = \bar{\psi}(i\gamma_{\mu}\partial^{\mu}-m_0)\psi
+ \frac{G_{S}}{2}[(\bar{\psi}\psi)^{2}+(\bar{\psi}i\gamma_{5}\vec{\tau}\psi)^{2}]
\label{eq:njl_lag},
\end{eqnarray}
where, $m_0 = $diag$(m_{u},m_{d})$ with $m_{u}=m_{d}$ and $\vec{\tau}$'s are Pauli matrices; $ G_{S}$ is the coupling constants of local scalar type four-quark interaction. Even for a massless system ($m_0=0$), the Lagrangian is not invariant under $U(1)_A$ transformation due to the interaction term. This is expected from the axial anomaly.

\subsubsection{Bare quarks to constituent quarks and mesons}
\label{sssec:qua_mes}
{\bf Hartree-Fock approximation}\\
The idea of spontaneous chiral symmetry breaking, one of the most important features of NJL model, was borrowed from the BCS (Bardeen, Cooper and Schrieffer) theory~\cite{Bardeen:1957kj}. It has been implemented through a so-called mass gap equation. The mass gap arises from the quark self energy which is calculated through the Dyson equation within Hartree-Fock or Hartree approximation. The local four point interaction renders the Hartree-Fock and Hartree approximations as equivalent, since Hartree (direct) term and the Fock (exchange) term become indistinguishable because of this contraction of the interaction to a point~\cite{Buballa:2003qv}. The equation is diagrammatically represented in figure~\ref{fig:dys_eq_har} and can be mathematically written as 
 \begin{eqnarray}
 M = m_0+iG_S\int \frac{d^4p}{(2\pi)^4}{\rm Tr}\,S(p),
 \label{eq:dys_eq}
 \end{eqnarray}
 where $S(p) = (\,\Slash p-M+i\epsilon)$ is the dressed quark propagator and $M$ is the dressed (constituent) quark mass. Once the traces over the Dirac, flavor and color spaces are performed, we are left with
 \begin{eqnarray}
 M = m_0+4iG_SN_cN_f\int \frac{d^4p}{(2\pi)^4}\frac{M}{p^2-M^2+i\epsilon},
 \label{eq:dys_eq_tr}
 \end{eqnarray}
 where $N_f~(=2)$ and $N_c~(=3)$ are the number of flavors and number of colors respectively. It is evident that $M$ is different from $m_0$ even for chiral limit ($m_0=0$). This is more so for a strong coupling constant ($G_S$). 
\begin{figure} [!htb]
\center
\includegraphics[scale=0.7]{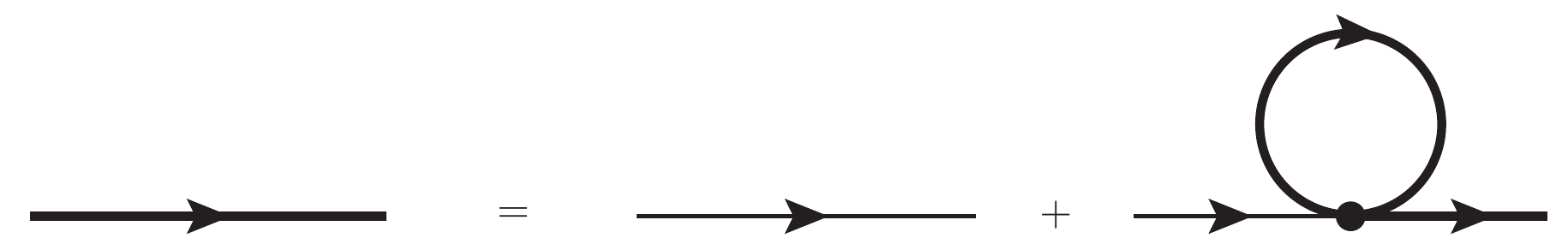}
\caption{Dyson equation for the quark self energy in Hartree approximation. The thick and thin lines signify the dressed and bare quark propagators respectively.}
\label{fig:dys_eq_har}
\end{figure} 
Under the mean field approximation the quark condensate ($\sigma$), also known as chiral condensate, is given as 
 \begin{eqnarray}
 \sigma = \langle \bar{\psi}\psi\rangle = -i\int \frac{d^4p}{(2\pi)^4}{\rm Tr}\,S(p).
 \label{eq:qua_con}
 \end{eqnarray} 
 The pseudoscalar condensate (pion condensate) $\langle \bar{\psi}i\gamma_5\vec\tau\psi\rangle$ becomes zero under the approximation. Using (\ref{eq:qua_con}) in (\ref{eq:dys_eq}) we obtain more familiar expression of the mass gap equation, 
\begin{equation}
 M=m_0-G_S\sigma.
\label{eq:mass_gap}
\end{equation} 

{\bf Random phase approximation}\\
The meson mass spectrum can be reproduced with the dressed quarks and antiquarks in NJL model~\cite{Oertel:2000jp,Davidson:1995fq}. For two flavor case these are the pions which are the Goldstone bosons and thus massless in the chiral limit ($m_0=0$). For nonzero current quarks masses they become massive. Here the mesons are imagined to be collective excitations of a pair of quark-antiquark, which are moving together. To obtain the mass of such compound objects Bethe-Salpeter equation is solved in ring approximation or random phase approximation (RPA) (figure~\ref{fig:bet_sal_eq}). The quark-antiquark $T$-matrix is calculated from that as,
\begin{eqnarray}
 T_M(q^2) = \frac{iG_S}{1-G_S\Pi_M(q^2)},
 \label{eq:tmat_rpa}
\end{eqnarray}
with 
\begin{eqnarray}
 \Pi_M(q^2) = i\int \frac{d^4p}{(2\pi)^4}{\rm Tr}[\mathcal{O}_M S(p+q)\mathcal{O}_MS(p)],
 \label{eq:qua_ant_pol}
\end{eqnarray} 
where ${M}$ depends on the type of mesonic channel considered. For the pion channel $\mathcal{O}_{\pi_a}=i\gamma_5\tau_a$ with $a=1,2,3$. After performing the traces and then using (\ref{eq:mass_gap}) we get,
\begin{figure} [!htb]
\center
\includegraphics[scale=0.55]{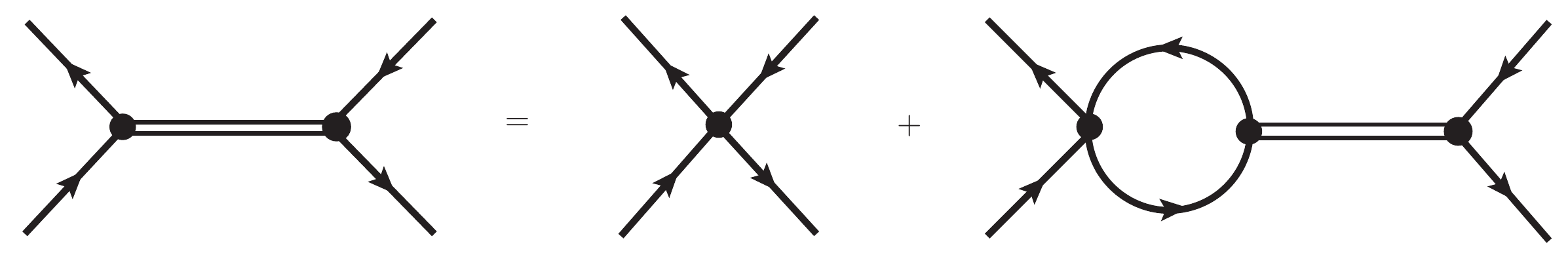}
\caption{Bethe-Salpeter equation in random phase approximation.}
\label{fig:bet_sal_eq}
\end{figure} 
\begin{eqnarray}
 \Pi_{\pi_a}(q^2) = \frac{1}{G_S}\Big(1-\frac{m_0}{M}\Big)-2iN_cN_fq^2I(q^2),
 \label{eq:pion_pol}
\end{eqnarray} 
where
\begin{eqnarray}
 I(q^2) = i\int \frac{d^4p}{(2\pi)^4}\frac{1}{[(p+q)^2-M^2+i\epsilon][p^2-M^2+i\epsilon]}.
 \label{eq:pol_int}
\end{eqnarray} 
Further use of $1-G_S\Pi_{\pi_a}(q^2=m_\pi^2)=0$ leads to an explicit constraint on $m_\pi^2$ 
\begin{eqnarray}
 m_\pi^2 = -\frac{m_0}{M}\frac{1}{2iG_SN_cN_fI(m_\pi^2)}.
 \label{eq:pion_mass}
\end{eqnarray}  
The pion decay constant ($f_\pi$) is calculated by using the pion to axial current matrix element as shown in figure~\ref{fig:pion_decay}. The pion is connected to the vacuum through the axial current.
\begin{eqnarray}
 if_\pi q^\mu\delta_{ab} = g_{\pi\bar qq}\int \frac{d^4p}{(2\pi)^4}{\rm Tr}[i\gamma^\mu\gamma_5\frac{\tau_a}{2}S(p+q)i\gamma_5\tau_bS(p)].
 \label{eq:mat_el_fpi}
\end{eqnarray}
A few more mathematical manipulations leads to the expression~\cite{Klevansky:1992qe}
\begin{eqnarray}
 f_\pi^2 = -8i\frac{N_c}{N_f}M^2I(0).
 \label{eq:pion_dec_con}
\end{eqnarray}

\begin{figure} [!htb]
\center
\includegraphics[scale=0.8]{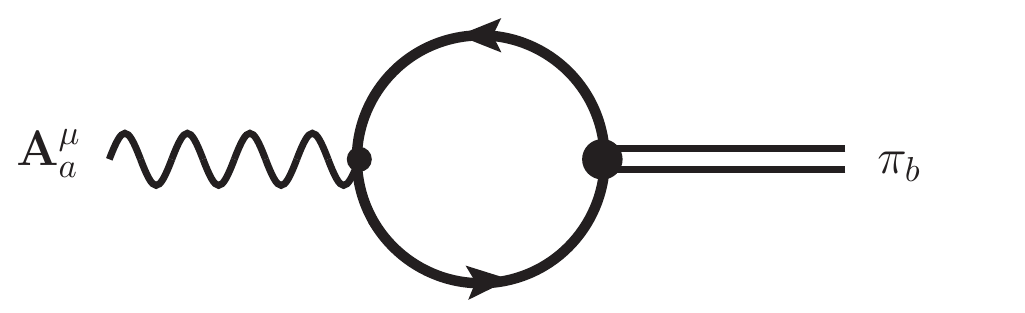}
\caption{Diagram for calculating pion decay constant. The double line represents the pion, whereas the curly line represents axial current. They are connected via a quark-antiquark loop.}
\label{fig:pion_decay}
\end{figure}

{\bf Regularization}\\
The local four point interaction makes the NJL model a non-renormalizable theory. There are integrals such as equations (\ref{eq:dys_eq_tr}), (\ref{eq:qua_con}), (\ref{eq:pol_int}) and (\ref{eq:mat_el_fpi}) which are diverging. So to get finite results these integrals need to be regularized. There are many schemes to do that all of which have some merits and demerits~\cite{Klevansky:1992qe}. We prefer to work with the three momentum cut-off ($\Lambda$). Though introduction of such cut-off breaks the Lorentz invariance of the theory, but as it retains the analytical structure (this becomes particularly useful while dealing with the Matsubara formalism in presence of medium) and also being very easy to imply makes its use advantageous. Thus $\Lambda$ becomes an important parameter of NJL model. In the next subsection we discuss how its value is determined along with $G_S$ and $m_0$ by fitting some known quantities.

\subsubsection{Parameter fitting}
\label{sssec:par_fit}

We have three unknown parameters $G_S,\,\Lambda~ {\rm and}~ m_0$ in the model. We use equations (\ref{eq:qua_con}), (\ref{eq:pion_mass}) and (\ref{eq:pion_dec_con}) to determine the values of the unknowns. The pion mass is an input for the equation (\ref{eq:pion_mass}) whereas pion decay constant is used in (\ref{eq:pion_dec_con}). These two quantities are known quite accurately. Regarding the value of condensate, which is used in (\ref{eq:qua_con}), there are some uncertainties. Its value (range) is determined using QCD sum rules as well as LQCD. The details about the parameter fitting and the associated intricacies can be found in the references~\cite{Buballa:2003qv,Roessner:2006dt}. We use the parameter set from the reference~\cite{Ratti:2005jh}, which is given in the table~\ref{table:para_fit}.

\begin{table}[h]
\begin{center}
\begin{tabular}{ccc|ccc}
\hline
 & Input &  &  & Output &  \\ \hline
$m_\pi$ [MeV] & $f_\pi$ [MeV] & $\langle \bar{\psi}\psi\rangle^\frac{1}{3}$ [MeV] & $G_S$ [GeV] & $\Lambda$ [GeV] & $m_0$ [GeV] \\ \hline
139.3 & 92.3 & 251 & 10.08 & 0.651 & 0.005 \\ \hline
\end{tabular}
\end{center}
\caption{Values of the three unknown parameters ($G_S,\,\Lambda~ {\rm and}~ m_0$) fitted from three given values of physical quantities ($m_\pi,~f_\pi~ {\rm and}~ \langle \bar{\psi}\psi\rangle^\frac{1}{3}$).}
\label{table:para_fit}
\end{table}

\subsubsection{NJL model in presence of medium}
\label{sssec:njl_med}

{\bf Mean field Lagrangian}\\
So far we have discussed how to describe different vacuum properties, such as constructing mesonic states, using NJL model. Now we want to use it to study strong interaction phenomena in presence of medium. Keeping that in mind we would like to construct an effective mean field Lagrangian. We linearize the interaction terms in the Lagrangian (\ref{eq:njl_lag}): $(\bar{\psi}\psi)^{2}\approx2\langle\bar\psi\psi\rangle(\bar\psi\psi)-\langle\bar\psi\psi\rangle^2$. There is no counterpart for the pseudoscalar term as the corresponding condensate term is zero. A more formal way of getting the effective mean field Lagrangian is to bosonize~\cite{Klevansky:1992qe} the model to write it in terms of auxiliary field like $\sigma$. The mean field NJL Lagrangian reads
\begin{eqnarray}
\mathcal{L}_{\rm{NJL_{MF}}} = \bar{\psi}(i\gamma_{\mu}\partial^{\mu}-m_0+G_S\sigma)\psi
- \frac{G_{S}}{2}\sigma^2.
\label{eq:njl_lag_mean}
\end{eqnarray}

{\bf Thermodynamic potential}\\
Thermodynamic potential carries all the necessary informations of a system. Once it is known, all the other thermodynamical quantities can be estimated from it. That is why we are motivated to construct it for studying the system in presence of medium. First we get the action from the Lagrangian. From it we obtain the partition function (PF). Once we know the PF we can get the potential using the relation: $\Omega(T,\mu)=-\frac{T}{V}ln{\cal Z}$, where ${\cal Z}$ is the PF at temperature ($T$) and chemical potential ($\mu$) at a given volume ($V$). Then different thermodynamic quantities can be calculated from $\Omega$.

We apply the techniques of thermal field theory, here the Matsubara formalism~\cite{Kapusta_Gale(book):1996FTFTPA,Lebellac(book):1996TFT}, to introduce the medium. In presence of hot and dense medium the gap equation gets modified to
 \begin{eqnarray}
 M = m_0+2G_SN_cN_f\int \frac{d^3p}{(2\pi)^3}\frac{M}{E_p}\big(1-n_p(T,\mu)-\bar{n}_p(T,\mu)\big).
 \label{eq:dys_eq_tr_med}
 \end{eqnarray}
Here $E_p=\sqrt{{\vec p}^2+M^2}$ is the energy of a quark having constituent mass or the dynamical mass $M$. $n_p$ and $\bar{n}_p$ are the thermal distribution functions for quarks and antiquarks respectively,  
\begin{eqnarray}
 n_p(T,\mu) =\frac{1}{e^{(E_p-{\mu})/T)}+1}; ~~~ \bar{n}_p(T,\mu) = \frac{1}{e^{(E_p+{\mu})/T)}+1}.
\end{eqnarray}
At any nonzero values of $T$ and/or $\mu$ the effective mass gets reduced as compared to that in vacuum. This happens because of the thermal distribution functions. For sufficiently high $T$ and/or $\mu$, $M$ approaches zero. We can get the vacuum result in (\ref{eq:dys_eq_tr}) back if we put $n_p=\bar{n}_p=0$ in (\ref{eq:dys_eq_tr_med}), provided we perform the integration over the temporal component in (\ref{eq:dys_eq_tr}). 

With all these in hand, the thermodynamic potential in mean field approximation is written as:
 \begin{eqnarray}
\Omega_{\rm{NJL}}&=&\frac{G_{S}}{2}\sigma^2-
2N_fN_c\int_{\Lambda}\frac{d^3p}{(2\pi)^3}E_p \nonumber\\
 &-&2N_fN_cT\int\frac{d^3p}{(2\pi)^3} \left[{\rm{ln}}(1+e^{-(E_p-{\mu})/T)})+{\rm{ln}}(1+e^{-(E_p+{\mu})/T)}) \right].
 \label{eq:njl_pot}
 \end{eqnarray}
We should mention here that the chiral condensate ($\sigma$) depends on both $T$ and $\mu$, though it has not been, neither will be explicitly mentioned anywhere in this dissertation. There are both explicit and implicit dependence, the implicit dependence arises through the constituent quark mass~(\ref{eq:dys_eq_tr_med}). The potential in the equation~(\ref{eq:njl_pot}) can be specifically divided in three parts. The first term is known as condensation energy, the second term as the zero point energy and the third term as thermal quark energy. The chiral condensation energy contributes negatively to the pressure and as one increases the temperature or density, contribution becomes less significant and eventually vanishes at sufficient higher values of $T$ and $\mu$. The other two terms contribute positively to the pressure. Particularly the contribution of zero point energy, which understandably does not depend on $T$ and $\mu$, largely affects the model output.

{\bf Region of validity of the model}\\
Effective models are called effective because they are valid within a given energy range and cannot be applied for arbitrary energy scale. In this sense all the established theoretical models are effective \textendash~the standard model being an epitome. So is for effective models like NJL one. But being a model which needs to be regularized the validity of such models have some arbitrariness depending on the choice of the regularization scheme. Specifically, results depend heavily on the value of the cut-off parameter and are not reliable if any of the external parameter attains value beyond that cut-off value. That is why, while dealing a system at zero temperature the value of the external momentum should not cross the cut-off value and while discussing system in presence of medium values of $T$ and $\mu$ should always be kept under that value to obtain legitimate results.

\subsubsection{NJL model with isoscalar-vector interaction}
\label{sssec:njl_iv_int} 
 
Apart from the scalar and pseudoscalar channel, many more terms can be added to the Lagrangian (\ref{eq:njl_lag}), provided they are allowed by the QCD symmetries~\ref{fig:sym_qcd}. One such term is the isoscalar-vector(I-V) channel. This term particularly becomes important for system at nonzero densities. A major part of this thesis deals with such scenario and further details of the reasons for inclusion of I-V term is discussed in chapters~\ref{chapter:JHEP},~\ref{chapter:PRD} and~\ref{chapter:Susc}. The NJL Lagrangian with I-V term is,

\begin{eqnarray}
\mathcal{L}_{\rm{NJL}} = \bar{\psi}(i\gamma_{\mu}\partial^{\mu}-m_0)\psi
+ \frac{G_{S}}{2}[(\bar{\psi}\psi)^{2}+(\bar{\psi}i\gamma_{5}\vec{\tau}\psi)^{2}]
- \frac{G_{V}}{2}(\bar{\psi}\gamma_{\mu}\psi)^2
\label{eq:njl_lag_vec},
\end{eqnarray}
where $G_V$ is the strength of the I-V interaction. It is another parameter in the theory. However, its value is difficult to fix within the model formalism, since this quantity should be fixed using the $\rho$ meson mass which, in general, happens to be higher than the maximum energy scale $\Lambda$ of the model~\footnote{Some efforts have also been made to estimate the value of $G_V$ mainly by fitting lattice data through two-phase model, which is not very conclusive. Interested readers are referred to Refs.~\cite{Steinheimer:2014kka,Sugano:2014pxa} and references therein.}. So, we consider the vector coupling constant $G_V$ as a free parameter and different choices are considered as ${G_V}=x \times {G_S}$, where $x$ is chosen from $0$ to $1$ appropriately.

There is an additional condensate due to the added interaction term, known as quark number density. It is given as $n=\langle \bar{\psi}\gamma^0 \psi\rangle$~\cite{Buballa:2003qv,Kashiwa:2006rc}. Before writing the corresponding thermodynamic potential we have to linearize the I-V term: $(\bar{\psi}\gamma_\mu\psi)^{2}\approx2\langle\bar\psi\gamma_0\psi\rangle(\bar\psi\gamma_0\psi)-\langle\bar\psi\gamma_0\psi\rangle^2$. With this the potential in presence of the I-V term is written as 
\begin{eqnarray}
\Omega_{\rm{NJL}}&=&\frac{G_{S}}{2}\sigma^2-\frac{G_{V}}{2}n^2-
2N_fN_c\int_{\Lambda}\frac{d^3p}{(2\pi)^3}E_p \nonumber\\
 &-&2N_fN_cT\int\frac{d^3p}{(2\pi)^3} \left[{\rm{ln}}(1+e^{-(E_p-\tilde{\mu})/T)})+{\rm{ln}}(1+e^{-(E_p+\tilde{\mu})/T)}) \right],
 \label{eq:njl_pot_vec}
\end{eqnarray}
with the modified quark chemical potential ($\tilde{\mu}$) related to vector condensate $n$ through the relation
\begin{equation}
 \tilde{\mu}=\mu-G_V n.
 \label{eq:mu_tilde}
\end{equation}
In equation~(\ref{eq:njl_pot_vec}) there is an extra term (the second term) than what we have in equation~(\ref{eq:njl_pot}). This term with the vector condensate ($n$), however contributes positively to the pressure and increases as one increases the external parameters like $T$ and $\mu$. Now this thermodynamic potential is minimized with respect to the mean fields $\sigma$ and $n$ to extract informations about them. This is done by solving the following gap equation numerically
\begin{align}
 \frac{\partial \Omega_{\rm NJL}}{\partial X}=0,
\end{align}
where $X$ represents $\sigma$ and $n$.

\subsection{PNJL Model}
\label{ssec:pnjl_model}

Apart from the chiral symmetry breaking, which is incorporated in the NJL model, there is another important property of QCD - known as confinement. The confinement effect is completely missing in the NJL model, since the gluons are integrated out there. This is why below the transition temperature the quark densities are overestimated in NJL model, since the gluons plays the utmost role in the confinement dynamics. It has been long strived to take into account the confinement effect along with the chiral dynamics to mimic QCD as closely as possible, particularly in the nonperturbative regime. PNJL model has been designed to encompass such characteristics, in which the confinement transition is combined with the chiral transition. Here in the following few paragraphs we try to depict the essence of the model. 

\subsubsection{Polyakov loop}
\label{sssec:pl}
The calculations with the pure gluons in lattice are much more accurate and cheaper as compared to those with the inclusion of quark dynamics. As gluons can interact among themselves, a pure gluonic confined system is possible - known as glueballs. While discussing the thermodynamic behaviour of such system, specially the phase transition, the order parameter becomes the most relevant quantity. The confinement-deconfinement transition of a pure gluonic system is well explored using LQCD. This transition can be shown to be related with the spontaneous breaking of $Z(N_c)$ symmetry, the center symmetry of $SU(N_c)$~\cite{Weiss:1980rj}.   

To describe the deconfinement transition an order parameter can be constructed from the Wilson loop \begin{eqnarray}
W(x)={\mathcal P}\, \exp[i\oint_c dx^\mu{\mathcal A}_\mu(x)],
\label{eq:pre:wil_loop}
\end{eqnarray}
through the Polyakov loop (PL)~\cite{Pisarski:2000eq,Dumitru:2001xa,Dumitru:2002cf,Dumitru:2003hp} which is the timelike Wilson line given as
\begin{eqnarray}
L(\vec x)={\mathcal P}\, \exp\left[i\int_0^\beta dx^4 A_4(\tau,\vec x)\right],
\label{eq:pol_loop}
\end{eqnarray}
with $A_4=iA^0$ is the temporal component of the gauge field and $\mathcal{P}$ indicates the path-ordering in imaginary time $\tau$ which runs from $0$ to $\beta = 1/T$. The normalized trace of PL with respect to fundamental representation is known as PL field and  is given as
\begin{eqnarray}
 \Phi =\frac{1}{N_c} {\mbox{tr}_c}L(x),
 \label{eq:pol_loop_tr}
\end{eqnarray}
where the $\Phi$ is complex and transforms under the global $Z(N_c)$ symmetry as a field with charge one as  $\Phi \rightarrow e^{{i2\pi j}/{N_c}}\Phi$ with $j=0,\cdots, (N_c-1)$. In the Polyakov gauge, i.e. for static and diagonal $A_4$, the PL matrix can be expressed through a diagonal representation~\cite{Fukushima:2003fw}. Also the normalized trace over the PL in (\ref{eq:pol_loop_tr}) is related with the free energy of a heavy static quark $q$ as $\Phi =\exp[-\beta F_q(T)]$ and is equal to that of antiquark, $F_{\bar q}$ (i.e.\,$\Phi = \bar\Phi$). This acts as an order parameter for confinement-deconfinement transition for a pure gauge theory.
For pure gluonic system the transition is of first order type. For confined phase the free energy diverges and $\Phi$ becomes $0$, whereas for deconfined phase at high temperature $\Phi$ attains nonzero value.

Given the role of an order parameter for 
pure gauge~\cite{Pisarski:2000eq,Dumitru:2001xa,Dumitru:2002cf,Dumitru:2003hp}, if $\Phi =0$ the $Z(N_c$) is unbroken and there is no ionization of $Z(N_c)$ charge, which is the confined phase below a certain temperature. At high temperature the symmetry is spontaneously broken,  $\Phi\ne 0$ corresponds to a deconfined phase of gluonic plasma and there are $N_c$ different equilibrium states distinguished by the phase $2\pi j/N_c$.

But once the quarks are included the free energy no more diverges in the confined phase and $\Phi$ never becomes strictly zero in that regime. We also note that the $Z(N_c)$ symmetry is explicitly broken in presence of dynamical quark, yet it can be considered as an approximate symmetry and $\Phi(\bar\Phi)$ can still provide useful information as an order parameter for deconfinement transition which essentially becomes a crossover~\cite{Weiss:1981ev} with the inclusion of quarks. 
At finite density $\Phi$ and $\bar\Phi$ become unequal~\cite{Fukushima:2008wg} and use of traced PL as an order parameter becomes more restricted.

\subsubsection{The Polyakov loop potential}
\label{sssec:pl_pot}

Now our goal is to write down an ansatz for an effective potential which will incorporate the gluon dynamics. We have in our hands the PL fields ($\Phi$, $\bar\Phi$) which act as the order parameter for deconfinement transition. For pure gluonic system $\Phi$ = $\bar\Phi$ and this can be considered as the starting point. The $Z(N_c)$ symmetry provides further guidelines while guessing such effective potential. Different forms of the potential are available in the literature. We use the form given in~\cite{Ratti:2005jh}. But that is not all - that form is further augmented with a Vandermonde term to keep $\Phi(\bar\Phi)$ within the domain [$0,1$]~\cite{Ghosh:2007wy}. The Vandermonde term arises from invariant Haar measure of $SU(3)_c$. It is further elaborated in~\cite{Islam:2012kv}. With all these considerations the effective PL potential~\cite{Ghosh:2007wy} reads as
\begin{eqnarray}
 \frac{{\mathcal U^\prime}(\Phi,\bar{\Phi},T)}{T^4} = \frac{{\mathcal U}(\Phi,\bar{\Phi},T)}{T^4}-\kappa\ln[J(\Phi,{\bar \Phi})]
 \label{eq:pre:pot_vdm}
\end{eqnarray}with
\begin{eqnarray}
 \frac{{\mathcal U}(\Phi,\bar{\Phi},T)}{T^4} = 
    -\frac{b_2(T)}{2}\Phi\bar{\Phi} -
    \frac{b_3}{6}(\Phi^3+{\bar{\Phi}}^3) +
    \frac{b_4}{4}(\bar{\Phi}\Phi)^2,
\label{eq:pre:pot_gauge}
\end{eqnarray}
where $b_2(T) = a_0 + a_1\left(\frac{T_0}{T}\right) + a_2\left(\frac{T_0}{T}\right)^2 + a_3\left(\frac{T_0}{T}\right)^3$, $T_0$ ($=270\, \rm{MeV}$) is the transition temperature for pure gauge theory taken from lattice calculation~\cite{Boyd:1996bx} and the values of the coefficients $a_i$ and $b_i$ are taken from~\cite{Ratti:2005jh}. The value of $\kappa$ is tuned so as to reproduce the lattice result~\cite{Petreczky:2007ny}. Their values have been tabulated in~\ref{table:par_pnjl}. The Vandermonde determinant $J(\Phi,{\bar \Phi})$ is given as \cite{Islam:2012kv,Ghosh:2007wy}
\begin{eqnarray}
J[\Phi, {\bar \Phi}]=\frac{27}{24\pi^2}\left[1-6\Phi {\bar \Phi}+
4(\Phi^3+{\bar \Phi}^3)-3{(\Phi {\bar \Phi})}^2\right].
\end{eqnarray}
In the high temperature limit ($T\rightarrow\infty$), $(\Phi,\bar\Phi)$ becomes unity and the equation (\ref{eq:pre:pot_gauge}) reproduces the free gluonic pressure $(16. \frac{\pi^2}{90}) T^4$ (Stefan-Boltzmann limit).

\begin{table}
\begin{center}
\begin{tabular}{cccccccc}
 \hline
 & $a_0$ & $a_1$ & $a_2$ & $a_3$ & $b_3$ & $b_4$ & $\kappa$\\
 \hline
  & 6.75 & -1.95 & 2.625  & -7.44  & 0.75 & 7.5 & 0.1\\
 \hline
\end{tabular}
\end{center}
\caption{Parameter set used in this work for the PL potential and the Vandermonde term.}
\label{table:par_pnjl}
\end{table}

\subsubsection{Thermodynamic potential}
\label{sssec:pnjl_pot}

We have learned so far how to build up the effective PL potential to describe the deconfinement transition. Now we want to understand how it can be realized starting from a Lagrangian. First of all, while considering the PL potential, only the spatially constant temporal gauge field is included which acts as a background field. So there is no spatial fluctuation of the gauge field. This is incorporated at the level of Lagrangian by introducing a covariant derivative $D^\mu=\partial^\mu-ig{\mathcal A}^\mu_a\lambda_a/2$ with ${\mathcal A}^\mu = \delta_{\mu 0} A^{0}$ and $A^0 = -iA_4$. Inclusion of such background field reduces the $SU(3)_c$ local gauge symmetry into a global one. 

We have eight degrees of freedom represented by eight Gell-Mann matrices $\lambda_a$'s. But in the PL potential we have only two degrees of freedom, $\Phi$ and $\bar\Phi$. Thus we need to get rid of the other six. This can be achieved by taking into account only the diagonal matrices $\lambda_3$ and $\lambda_8$ by rotating the elements of $SU(3)_c$ in the Polyakov gauge~\cite{Fukushima:2003fw}. Thus with the inclusion of covariant derivative and the PL potential the two flavor PNJL Lagrangian becomes
\begin{eqnarray}
{\mathcal L}_{\rm PNJL} &=& \bar{\psi}(i\Slash D-m_0+\gamma_0\mu)\psi +
\frac{G_S}{2}[(\bar{\psi}\psi)^2+(\bar{\psi}i\gamma_5\vec{\tau}\psi)^2]\nonumber\\
&-& {\mathcal U}(\Phi[A],\bar{\Phi}[A],T).
\label{eq:pnjl_lag}
\end{eqnarray}
Using the mean field approach described in subsection~\ref{sssec:qua_mes} the corresponding potential can be obtained as
\begin{eqnarray}
 {\Omega}_{\textrm{PNJL}} &=& {\mathcal U}(\Phi,{\bar \Phi},T) + \frac{ G_S}{2} \sigma^2 \nonumber\\
 &-&2N_fT\int \frac{d^3p}{(2\pi)^3} \ln \left[1+ 3\left(\Phi +{\bar \Phi}
 e^{-(E_p-\mu)/T} \right)e^{-(E_p-\mu)/T} + e^{-3(E_p-\mu)/T} \right ]  \nonumber \\
 &-&  2N_fT\int \frac{d^3p}{(2\pi)^3} \ln \left[1+ 3\left({\bar \Phi} + \Phi
 e^{-(E_p+\mu)/T} \right)e^{-(E_p+\mu)/T} + e^{-3(E_p+\mu)/T} \right ] \nonumber \\ 
 &-&\kappa T^4 \ln[J(\Phi,{\bar \Phi})] 
  -2N_fN_c\int_{\Lambda}\frac{d^3p}{(2\pi)^3}E_p\ .
 \label{eq:pnjl_pot}
\end{eqnarray}
Comparing between the equations~(\ref{eq:pnjl_pot}) and~(\ref{eq:njl_pot}) we observe that there are two new terms in~(\ref{eq:pnjl_pot}), whereas the expressions of thermal quark part gets modified and those of the condensation and zero point energy remain the same. Here one should be careful and understands that with the same expressions these two terms will now contribute differently, since ultimately coupled gap equations will be solved to extract information about the mean fields. The thermal quark energy part is now regulated with the presence of PL fields. The two new terms \textendash~the pure gluonic part and the Vandermonde term have already been explained.

\subsubsection{PNJL model with isoscalar-vector interaction}
\label{sssec:pnjl_iv_int}
Following the section~\ref{sssec:njl_iv_int} it is straightforward to include the I-V interaction in PNJL model. The Lagrangian becomes 
\begin{eqnarray}
 {\mathcal L}_{\rm PNJL} &=& \bar{\psi}(i\Slash D-m_0+\gamma_0\mu)\psi +
\frac{G_S}{2}[(\bar{\psi}\psi)^2+(\bar{\psi}i\gamma_5\vec{\tau}\psi)^2]
- \frac{G_V}{2}(\bar{\psi}\gamma_{\mu}\psi)^2\nonumber\\
&-& {\mathcal U}(\Phi[A],\bar{\Phi}[A],T)
\label{eq:pnjl_lag_vec}
\end{eqnarray}
and the corresponding thermodynamic potential is
\begin{eqnarray}
 {\Omega}_{\textrm{PNJL}} &=& {\mathcal U}(\Phi,{\bar \Phi},T) + \frac{ G_S}{2} \sigma^2 -\frac{ G_V}{2} n^2 
 \nonumber\\
 &-&2N_fT\int \frac{d^3p}{(2\pi)^3} \ln \left[1+ 3\left(\Phi +{\bar \Phi}
 e^{-(E_p-\tilde \mu)/T} \right)e^{-(E_p-\tilde \mu)/T} + e^{-3(E_p-\tilde \mu)/T} \right ]  \nonumber \\
 &-&  2N_fT\int \frac{d^3p}{(2\pi)^3} \ln \left[1+ 3\left({\bar \Phi} + \Phi
 e^{-(E_p+\tilde \mu)/T} \right)e^{-(E_p+\tilde \mu)/T} + e^{-3(E_p+\tilde \mu)/T} \right ] \nonumber \\ 
 &-&\kappa T^4 \ln[J(\Phi,{\bar \Phi})] 
  -2N_fN_c\int_{\Lambda}\frac{d^3p}{(2\pi)^3}E_p.
 \label{eq:pnjl_pot_vec}
\end{eqnarray}
Here the potential is minimized with respect to the fields $\sigma$, $n$, $\Phi$ and $\bar\Phi$ to get informations about the mean fields themselves. The gap equation for the potential is given by
\begin{align}
 \frac{\partial \Omega_{\rm PNJL}}{\partial X}=0,
\end{align}
where $X$ represents $\sigma$, $n$, $\Phi$ and $\bar\Phi$.

\subsection{EPNJL Model}
\label{ssec:epnjl_model}

An important question on the QCD thermodynamics is whether the chiral symmetry restoration and the confinement-to-deconfinement transition happen simultaneously or not. We note that chiral and deconfinement transitions are conceptually two distinct phenomena. Though lattice QCD simulation has confirmed that these two transitions occur at the same temperature \cite{Fukugita:1986rr} or almost at the same temperature \cite{Aoki:2006br}. Whether this is a mere coincidence or some dynamics between the two phenomena are influencing each other is not understood yet and is matter of intense current research exploration. 

To understand the reason behind this  coincidence a conjecture has been proposed in~\cite{Sakai:2010rp}  through a strong correlation or entanglement between the chiral condensate ($\sigma$) and the PL expectation value ($\Phi$) within the PNJL model. Usually, in PNJL model, there is a weak correlation between the chiral dynamics $\sigma$ and the confinement-deconfinement dynamics  $\Phi$ that is in-built through the covariant derivative between quark and gauge fields as discussed in the subsection~\ref{sssec:pnjl_pot}. With this kind of weak correlation the coincidence between the chiral and deconfinement crossover transitions~\cite{Sakai:2009dv,Ghosh:2006qh,Mukherjee:2006hq,Ghosh:2007wy,Ratti:2005jh,Deb:2009ng} can be described but it requires  some fine-tuning of parameters, inclusion of the scalar type eight-quark interaction for zero chemical potential $\mu$ and the vector-type four-quark interaction for imaginary $\mu$. This reveals 
that there may be a stronger correlation 
between $\Phi$ and $\sigma$ than that in the usual PNJL model associated through the covariant derivative between quark and gauge fields. Also, some recent analyses~\cite{Braun:2009gm,Kondo:2010ts} of the exact renormalization-group (ERG) equation~\cite{Wetterich:1992yh} suggest a strong entanglement interaction between $\Phi$ and $\sigma$ in  addition to the original entanglement through the covariant derivative. Based on this the two-flavor PNJL model is further generalized~\cite{Sakai:2010rp} by considering the effective four-quark scalar type interaction with the coupling strength that depends on the PL field $\Phi$. The effective vertex in turn generates entanglement interaction between $\Phi$ and $\sigma$. Such generalization  of the PNJL model is known as Entangled-PNJL (EPNJL) model~\cite{Sakai:2010rp}. 

\subsubsection{Modifying the scalar vertex ($G_S$)}
\label{sssec:mod_sca_ver}

The weak correlation between the chiral ($\sigma$) and the deconfinement  ($\Phi$ and $\bar \Phi$) dynamics in PNJL model has been modified into a strong one so that chiral and deconfinement transitions coincide. This is achieved by introducing a strong entanglement interaction between $\Phi$ and $\sigma$ through an effective scalar type four-quark interaction with the coupling strengths that depend on the Polyakov field. The Lagrangian in EPNJL will be the same as that in (\ref{eq:pnjl_lag}) except that now the coupling constants $G_S$ will be replaced by the effective ones $\tilde G_S(\Phi)$. The effective vertex $\tilde{G}_S(\Phi)$ generates entanglement interaction between $\Phi$ and $\sigma$ and its form is chosen~\cite{Sakai:2010rp} to preserve chiral and $Z(3)$ symmetry,
\begin{equation}
 \tilde{G}_S(\Phi)= G_S[1-\alpha_1\Phi\bar\Phi-\alpha_2(\Phi^3+\bar\Phi^3)],
 \label{eq:entangle_Gs}
\end{equation}
We  note that for $\alpha_1 =\alpha_2=0$, $\tilde{G}_S(\Phi)= G_S$: the EPNJL model reduces to PNJL model. Also at $T=0$, $\Phi={\bar \Phi}=0$ (confined phase), then $\tilde{G}_S= G_S$.

In EPNJL model, $\alpha_1$ and $\alpha_2$ are two new parameters, which are to be fixed from the lattice QCD data. The values of other parameters are taken to be as those in PNJL model~\cite{Islam:2014sea} except the value of $T_0$, which is taken as 190 MeV. The thermodynamic potential $\Omega_{\rm{EPNJL}}$ in EPNJL model can be obtained from (\ref{eq:pnjl_pot}) by replacing $G_S$ with $\tilde{G}_S(\Phi)$.

\subsubsection{In presence of I-V interaction ($G_V$)}
\label{sssec:epnjl_iv_int}

The inclusion of vector interaction is also done in the similar fashion. The coupling strength $G_V$ is replaced by the effective vertex $\tilde{G}_V$, which depends on the PL field. Here also the ansatz is chosen to maintain the $Z(3)$ and chiral symmetry~\cite{Sugano:2014pxa}. It is given as,
\begin{equation}
 \tilde{G}_V(\Phi)= G_V[1-\alpha_1\Phi\bar\Phi-\alpha_2(\Phi^3+\bar\Phi^3)].
 \label{eq:entangle_Gv}
\end{equation}
 Due to the reason already mentioned in 
the previous subsection, here again the strength of the vector interaction is taken in terms of the value of $G_S$ as $G_V=x\times G_S$, 
which on using (\ref{eq:entangle_Gv}) reduces to
\begin{equation}
\tilde{G}_V(\Phi)=x \times G_S[1-\alpha_1\Phi\bar\Phi-\alpha_2(\Phi^3+\bar\Phi^3)]=x \times \tilde{G}_S(\Phi).
\label{eq:entangle_Gv_tilde}
\end{equation}
Now it is easy to obtain the corresponding Lagrangian and the thermodynamic potential. One just needs to replace $G_S$ and $G_V$ by ${\tilde G}_S$ and ${\tilde G}_V$ respectively, in (\ref{eq:pnjl_lag_vec}) and (\ref{eq:pnjl_pot_vec}).

\subsection{Color singlet model}
\label{ssec:col_sin_model}

We often employ statistical thermodynamical description to study many-particle system. Particularly it is known to be very useful to describe system of quantum gas, such as, electrons in metal, blackbody photons 
in a heated cavity, phonons at low temperature, neutron matter in neutron stars, etc. There are conservation laws which act as guiding principles while developing such statistical models. These conservation laws can be related to the associated symmetries. Symmetries can be, in terms of physical interpretations, of two types - external and internal. Conservation of energy, linear momentum, angular momentum etc are reflections of the external symmetries that the system obeys. Also there are conservation of electric charges, baryon numbers, color charges etc which are to be associated with the internal symmetries of the system.

Here we are interested in describing a system of quarks and gluons using statistical methods. For that $SU(3)_c$ is the major internal symmetry that needs to be taken care of. Many QCD related phenomena such as condensation problems, critical phenomena etc have been studied using such methods~\cite{Turko:1981nr}. The thermodynamics of QGP and the related phase transition are also extensively studied~\cite{Gorenstein:1982ua,Skagerstam:1983gv,Ansari:1992qd,Zakout:2007nb,Mustafa:1993np}. Of course these phenomena should be investigated within the framework of QCD. But due to the nonperturbative nature of QCD, which leads it to be difficult to deal with (vide subsection~\ref{ssec:theo_methods}), use of such methods becomes indeed insightful~\cite{Auberson:1986ft}.

We construct a partition function (PF) for a quantum gas of quarks, antiquarks and gluons, which is restricted by the assumption of a color singlet projection to conform with the $SU(3)_c$ symmetry. This projection allows only the color singlet physical states to exist~\cite{Abir:2009sh}. Such projection is performed employing group theoretical projection method~\cite{Redlich:1979bf,Turko:1981nr}. Because of the projection many physical properties gets modified as compared to the unprojected one, particularly the deconfinement phase transition~\cite{Dey:1989fx}. This method is particularly useful for describing a system of weakly interacting quarks and gluons at high temperature and/or high density. We call such model as color singlet (CS) model~\cite{Turko:1981nr,Elze:1985wv,Auberson:1986ft,Mustafa:1993np}. 

QGP in finite volume has also been treated in a canonical color single PF~\cite{Elze:1983du}. But while describing the relativistic plasma the use of canonical PF is no more justified, since there are particle creations and annihilations. Thus the introduction of grand canonical PF becomes mandatory~\cite{Turko:1981nr,Auberson:1986ft}. 

In the following subsection we describe how to get a color singlet PF for a quantum gas of massless quarks, antiquarks and gluons.

\subsubsection{Color singlet partition function}
\label{sssec:col_sin_par_fn}

In thermal equilibrium the statistical behaviour of a quantum gas is studied through a density matrix in an appropriate ensemble as
\begin{equation}
{\rho(\beta)} =  {\rm {exp}} (-\beta
{\hat H}) \ \ , 
\end{equation}
where $\beta=1/T$ is the inverse of temperature and ${\hat H}$ is the Hamiltonian of a physical system. The corresponding PF for a quantum gas having a finite volume can be written as
\begin{equation}
{\cal Z} = {\rm{Tr}} \left (  e^{-\beta {\hat H}} \right ) 
= \sum_n \left \langle n \left |  e^{-\beta {\hat H}} 
\right | n \right \rangle \ \ , 
\end{equation}
where $|n\rangle $ is a many-particle state in the full Hilbert space ${\cal H}$. Now, the full Hilbert space contains states which should not contribute to a desired configuration of the system. One can restrict those states from contributing to the PF by defining a reduced ensemble for a desired configuration through the use of projection operator as 
\begin{equation}
{\cal Z} = {\rm{Tr}} \left ( {\hat {\cal P}} e^{-\beta 
{\hat H}} \right ) = \sum_n \left \langle n \left | 
{\hat {\cal P}} e^{-\beta {\hat H}} 
\right | n \right \rangle \ \ . 
\label{eq:par_fn_sum}
\end{equation}
Now, let  ${\cal G}$  be a symmetry group with unitary representation ${\hat U}(g)$ in a Hilbert space ${\cal H}$. The group theoretical projection operator~\cite{Weyl(book):TCG} ${\hat {\cal P}}$ for a desired configuration is defined as ${\hat{\cal P}}_j = d_j \int_{\cal G} {\rm d}\mu (g) \chi^{\star}_j(g) {\hat U}(g)$ , where $d_j$ and $\chi_j$ are, respectively, the dimension and the character of the irreducible representation $j$ of ${\cal G}$ and ${\rm d}\mu (g)$ is the invariant Haar measure. The symmetry group associated with the color singlet configuration is $SU(N_c)$ and $d_j=1$ and $\chi_j=1$. Now the color singlet PF for the system becomes,
\begin{equation}
{\cal Z}_S =\int_{SU(N_c)} {\rm d}\mu (g) {\rm{Tr}} 
\left ( {\hat U}(g) 
{\rm {exp}} (-\beta {\hat H}) \right ) \ \ .
\label{eq:par_fn_int} 
\end{equation}
The invariant Haar measure~\cite{Weyl(book):TCG,Auberson:1986ft} is expressed in terms of the distribution of eigenvalues of $SU(N_c)$ as
\begin{eqnarray}
\int {\rm d}\mu (g)\!\! = \!\!
 \frac{1}{N_c!}\! \left ( \prod^{N_c}_{l=1} 
\! \int\limits_{-\pi}^\pi \frac{{\rm d}\theta_l}{2\pi}\right )
\delta\Big(\sum_i\theta_i\Big)\!\! \prod_{i>j}\left |e^{i\theta_i}- e^{i\theta_j}\right |^2 , 
\label{eq:pre:haar_measure}
\end{eqnarray}
where the square of the product of the differences of the eigenvalues is known as the Vandermonde determinant. The class parameter $\theta_l$ obeys $\sum^{N_c}_{l=1}\theta_l=0 \ ({\rm{mod}} 2\pi)$ ensuring  the requirement of unit determinant in $SU(N_c)$. This also restricts that the $SU(3)$ has only two parameter abelian subgroups associated with two diagonal generators, which would completely characterize the ${\hat U}(g)$. Apart from the color singlet restriction the conservation of baryon number is also taken care through the introduction of baryon chemical potential. It is included in the definition of Hamiltonian that we are going to use.

Now, the Hilbert space ${\cal H}$ of a composite system has a structure of a tensor product of the individual Fock spaces as ${\cal H} = {\cal H}_q \otimes {\cal H}_{\bar q} \otimes {\cal H}_g$. The PF  in (\ref{eq:par_fn_int}) decomposes in respective Fock spaces as
\begin{eqnarray}
{\cal Z}_S & =&
\int_{SU(N_c)}  
 {\rm d}\mu (g) \, {\rm{Tr}}  \left ( {\hat U}_q 
 e^{-\beta {\hat H}_q} \right )   
 {\rm{Tr}}  \left ( {\hat U}_{\bar q}
 e^{-\beta {\hat H}_{\bar q}} \right ) 
\, {\rm{Tr}}  \left ( {\hat U}_g
 e^{-\beta {\hat H}_g} \right ),
\label{eq:par_fn_deco}
\end{eqnarray}
where the various  $U_i(g)$ act as link variables that link, respectively, the quarks, antiquarks and spatial gluons in a given state of the physical system. In each Fock space there exists a basis that diagonalizes both operators as long as ${\hat H}_i$ and ${\hat U}_i$ commute. Performing the traces in (\ref{eq:par_fn_deco}) using the standard procedure~\cite{Kapusta_Gale(book):1996FTFTPA}, the projected PF in Hilbert space becomes
\begin{eqnarray}
{\cal Z}_S\!\! &=&\!\!\! \int_{SU(N_c)}
\!\!\!\! \!\! \!\!\! {\rm d}\mu (g) \ \ e^{\Theta_q + \Theta_{\bar q} + \Theta_g}
=\int_{SU(N_c)}{\hspace*{-0.15in}}\!\!\! {\rm d}\mu (g) \ \ e^{\Theta_p }
\ \ ;
\label{eq:pre:par_fn_final}
\end{eqnarray}
with
\begin{eqnarray}
\Theta_p 
&=& \Theta_q + \Theta_{\bar q} + \Theta_g \nonumber \\ 
&=&
2 N_f\sum_\alpha {\rm{tr}_c}  \ln \left (1+R_q
e^{-\beta (\epsilon_{q}^\alpha -\mu_q)}
\right ) 
+ 2N_f \sum_\alpha {\rm{tr}_c}  
\ln \left (1+R_{\bar q}
e^{-\beta (\epsilon_q^\alpha +\mu_{ q})}
\right ) 
\nonumber \\
&& 
- 2\sum_\alpha {\rm{tr}_c}  
\ln \left (1-R_g
e^{-\beta \epsilon_g^\alpha}\right ),
\label{eq:theta}
\end{eqnarray}
where $\epsilon_i^\alpha= \sqrt{(p^\alpha_i)^2+m^2_i}$, $R_{q ({\bar q})}$ are the finite dimensional diagonal matrices in the basis of the color space in fundamental representation and $R_{g}$ is that in adjoint representation. Also the quark flavor ($N_f$), their spin and the chemical potential $\mu$, and the polarization of gluons are introduced.

Once we know the PF, many properties of QCD can be studied using this CS model. As for example the phase structure resulting from HICs - specifically the deconfinement phase transition, the hadron formation temperature, baryon and meson density of states etc are particularly well investigated.

\subsubsection{Unprojected partition function}
\label{sssec:unp_par_fn}

It is straightforward to obtain the unprojected PF from (\ref{eq:par_fn_sum}) by putting ${\hat {\cal P}}=1$. Then the colorsingletness of the physical states is lost and the PF is given as
\begin{equation}
{\cal Z} = {\rm{Tr}} \left ( e^{-\beta 
{\hat H}} \right ) = \sum_n \left \langle n \left | 
 e^{-\beta {\hat H}} 
\right | n \right \rangle \ \ . 
\end{equation}
From here the unprojected PF for quarks, antiquarks and gluons can be written as
\begin{eqnarray}
{\cal Z}_U & =&
{\rm{Tr}}  \left (e^{-\beta {\hat H}_q} \right )   
 {\rm{Tr}}  \left (e^{-\beta {\hat H}_{\bar q}} \right ) 
\, {\rm{Tr}}  \left (e^{-\beta {\hat H}_g} \right ).
 \label{eq:par_fn_deco_un}
\end{eqnarray}
Performing the traces in (\ref{eq:par_fn_deco_un}) the unprojected PF in Hilbert space becomes
\begin{eqnarray}
{\cal Z}_U\!\! &=&\!\!\!  e^{\Theta_q + \Theta_{\bar q} + \Theta_g} = e^{\Theta_p }
\ \ ;
\end{eqnarray}
with
\begin{eqnarray}
\Theta_p 
&=& \Theta_q + \Theta_{\bar q} + \Theta_g \nonumber \\ 
&=&
2 N_f\sum_\alpha  \ln \left (1+e^{-\beta (\epsilon_{q}^\alpha -\mu_q)}
\right ) 
+ 2N_f \sum_\alpha \ln \left (1+e^{-\beta (\epsilon_q^\alpha +\mu_{ q})}
\right ) 
\nonumber \\
&& 
- 2\sum_\alpha \ln \left (1-e^{-\beta \epsilon_g^\alpha}\right ),
%\label{eq:theta}
\end{eqnarray}
where the symbols have the same meaning as in the previous subsection.

\section{Correlation function and its spectral representation}
\label{sec:cor_sp_rep}

Correlation functions (CF) is a mathematical tool which is extensively used in almost every branch of physics. Particularly it is ubiquitous in high energy physics, condensed matter physics, astronomy etc. It is basically a measure of how microscopic variables change with response to one another. In many occasions this mutual response is ensemble averaged. When the CF is between the same variables at two different space-time points, it is know as autocorrelation function and that between different variables is known as cross-correlation function. In quantum field theory CFs are also known as correlators. Propagator is one such example. The term Green's function is also, sometimes, used in place of correlator. 

The physical significance of CFs in context to this dissertation have already been discussed in the section~\ref{sec:cf_and_sr} in the introduction. In the following subsection we develop the mathematics of CF and its spectral representation. Then we briefly describe how to obtain the dilepton multiplicity and quark number susceptibility (QNS) from the spectral function (SF). We focus on the two point autocorrelation functions and our language is being that of high energy physics for obvious reason.

\subsection{Mathematical formulation}
\label{ssec:math_form}

In general the CF in coordinate space is given by
\begin{eqnarray}
{\cal G}_{AB}(t,{\vec x})\equiv 
{\cal T} \langle{\hat A}(t,{\vec x}){\hat B}(0,{\vec 0})\rangle
= \int \frac{d\omega}{2\pi} \int \frac{d^3{ q}}{(2\pi)^3}
 \ \ e^{i\omega t-i{\vec q}\cdot{\vec x}} \ \ 
{\cal G}_{AB}(\omega,{ \vec q}),
\end{eqnarray}
where ${\cal T}$ is the time-ordered product of the two operators $\hat A$ and $\hat B$, and the four momentum $Q\equiv (\omega, {\vec q})$ with $q=|{\vec q}|$.

By taking the Fourier transformation one can obtain the momentum space CF as
\begin{eqnarray}
{\cal G}_{AB}(\omega,{\vec q})
&=& \int dt \int d^3{\vec x} \ \
{\cal G}_{AB}(t,{\vec x}) 
\ \ e^{-i\omega t+i{\vec q}\cdot{\vec x}} \ \ .
\end{eqnarray}
We are specifically interested in current-current CFs. The thermal meson current-current correlator in Euclidean time $\tau \in $ [$0,\ \beta=1/T$] is given as \cite{Karsch:2000gi}
\begin{eqnarray}
{\cal G}^E_{M}(\tau, {\vec x})&=&\langle {\cal T}(J_M(\tau, 
{\vec x})
J_M^\dagger(0, {\vec 0}))\rangle_\beta \nonumber\\ 
&=&T\sum_{n=-\infty}^{\infty} \int
\frac{d^3q}{(2\pi)^3} \ e^{-i(\omega_n\tau+{\vec q}
\cdot {\vec x})}\ {\cal G}^E_{M}(i\omega_n, {\vec q}),
\label{eq:cur_cor_fn}
\end{eqnarray}
where the mesonic current is defined as $J_M={\bar {\psi}} (\tau,{\vec x})\Gamma_M \psi(\tau,{\vec x})$, with $\Gamma_M$ =$ 1, \gamma_5, \gamma_\mu,$ $\gamma_\mu\gamma_5$ for scalar, pseudoscalar, vector and pseudovector channel respectively. The momentum space correlator  ${\cal G}^E_{M}(i\omega_n, \vec q)$ at the discrete Matsubara modes $\omega_n=2\pi nT$ can be obtained as
\begin{eqnarray}
{\cal G}^E_{M}(i\omega_n, \vec q) =
- \int_{-\infty}^{\infty} d\omega \ \ 
\frac{\sigma_{M}(\omega, \vec q)}{i\omega_n-\omega} .
\label{eq:cor_fn_dis} 
\end{eqnarray}
For a given mesonic channel $H$, we replace the SF $\sigma_M(\omega, \vec q)$ by $\sigma_H(\omega, \vec q)$
which can be obtained from the imaginary part of the momentum space Euclidean correlator in (\ref{eq:cor_fn_dis}) by analytic continuation as
\begin{eqnarray}
\sigma_H(\omega, \vec q)&=&\frac{1}{\pi} {\mbox{Im}}
\ {\cal G}^E_H(i\omega_n=\omega+i\epsilon,\vec q).
\label{eq:pre:spectral}
\end{eqnarray}
For a given mesonic channel, $H=00,ii$ and $V$ stand for temporal, spatial and vector SF, respectively. The vector SF is expressed in terms of temporal and spatial components as ${\rm\sigma_V}=\sigma_{00}-\sigma_{ii}$.

Using (\ref{eq:cur_cor_fn}) and (\ref{eq:cor_fn_dis}) one obtains (vide appendix~\ref{app:cor_sp_rep}) the spectral representation of the thermal CF in Euclidean time but at a fixed momentum $\vec q$ as
 \begin{equation}
{\cal G}^E_H(\tau, \vec q)=\int_0^\infty\ d\omega \
\sigma_H (\omega,  \vec q) \
\frac{\cosh[\omega(\tau-\beta/2)]}{\sinh[\omega\beta/2]}. 
\label{eq:cor_tot}
\end{equation}
Spectral properties are heavily studied in LQCD. But because of the discretized space-time there is difficulty in analytic continuation in LQCD the SF can not be obtained directly using (\ref{eq:pre:spectral}). Instead a calculation in LQCD proceeds by evaluating the Euclidean CF. Using a probabilistic application based on maximum entropy method (MEM)~\cite{Nakahara:1999vy,Asakawa:2000tr,Wetzorke:2000ez}, (\ref{eq:cor_tot}) is then inverted  to extract the SF and  thus various spectral properties are computed in LQCD. 

\subsection{Vector spectral function and dilepton rate}
\label{ssec:v_sf_dilepton_rate}

We know that self-energy is nothing but the current-current CF. Now photon self-energy in thermal medium is related with the dilepton production rate at finite temperature~\cite{Weldon:1990iw}. These lepton pairs are created from virtual photons. So the dilepton multiplicity in medium can be calculated using the vector current-current CF. We briefly outline the procedure here; the details can be found in~\cite{Weldon:1990iw}.

Dynamics-wise there are two parts in the expression of dilepton multiplicity - one is the leptonic part and the other one is the photon SF. Then there are terms arising from the kinematics and the availability of phase space. Since the medium created in HICs is visualized to be thermalized, we have to take the average over the initial states along with the final states, where the initial states have to be sampled by the Boltzmann factor. Thus the dilepton multiplicity in a four volume is given by~\cite{Weldon:1990iw},
\begin{eqnarray}
\frac{dR}{d^4x}&=& 2\pi e^2 e^{-\beta \omega}L_{\mu\nu}\rho^{\mu\nu}\frac{d^3p_1}{(2\pi)^3E_1}\frac{d^3p_2}{(2\pi)^3E_2},
\label{eq:pre:dil_rate}
\end{eqnarray}
where the photonic tensor is given by,
\begin{eqnarray}
\rho^{\mu\nu}(\omega,{\vec q}) =-\frac{1}{\pi}\frac{e^{\beta \omega}}{e^{\beta \omega}-1}~\frac{e^2}{q^4}\mathrm{Im}\left[\Pi^{\mu\nu}(\omega,{\vec q})\right]
\end{eqnarray}
and the leptonic part is given by,
\begin{eqnarray}
L_{\mu\nu} &=& \frac{1}{4} \sum\limits_{\mathrm{spins}}\mathrm{tr}\left[\bar{u}(p_2)\gamma_\mu v(p_1)\bar{v}(p_1)\gamma_\nu u(p_2)\right]\nonumber\\
&=& p_{1\mu}p_{2\nu}+p_{1\nu}p_{2\mu}-(p_1\cdot p_2+m_l^2)g_{\mu\nu};
\label{eq:pre:lmunu_unmagnetized}
\end{eqnarray}
$m_l$ being the mass of the lepton. With all these expressions in hand the equation (\ref{eq:pre:dil_rate}) can be further simplified to
\begin{eqnarray}
\frac{dR}{d^4xd^4Q} &=& 2\pi e^2 e^{-\beta \omega}\rho^{\mu\nu}\int\frac{d^3p_1}{(2\pi)^3E_1}\frac{d^3p_2}{(2\pi)^3E_2}\delta^4(p_1+p_2-Q)L_{\mu\nu}.
\label{eq:pre:dpr_d4q_1}
\end{eqnarray}
We further proceed in the calculation by taking the massless assumption which simplifies the expression, that we use in this thesis. So we put $m_l=0$ in (\ref{eq:pre:lmunu_unmagnetized}) and then after few mathematical manipulations we obtain~\cite{Karsch:2000gi}
\begin{equation}
  \frac{dR}{d^4xd^4Q}  =\frac{5\alpha^2} {54\pi^2} \frac{1}{M^2} 
\ \frac{1}{e^{\omega/T}-1} \ \sigma_V(\omega, {{\vec q}}) \ ,
\label{eq:pre:rel_dilep_spec}
\end{equation}
where the invariant mass of the lepton pair is $M^2=\omega^2-q^2$ and 
$\alpha$ is the fine structure constant. We call $\sigma_V$ as the SF function and is given by
\begin{eqnarray}
 \frac{1}{\pi}\mathrm{Im}\Pi^{\mu}_{\mu}(\omega,{\vec q}).
\end{eqnarray}

\subsection{Temporal vector spectral function and response to conserved density fluctuation}
\label{ssec:tem_v_sf_qns}

The quark number susceptibility (QNS), $\chi_q$ measures of the response of the quark number 
density $\rho$ with 
infinitesimal change in the quark chemical  potential, $\mu+\delta\mu$. It is related to
temporal CF through fluctuation-dissipation theorem~\cite{Hatsuda:1994pi,Kunihiro:1991qu} as 
\begin{eqnarray}
\chi_q(T) &=&\!\! \left.\frac{\partial \rho}{\partial \mu}\right |_{\mu=0} \! \!
= \!\!\! \int \ d^4x \ \left \langle J_0(0,{\vec { x}})J_0(0,{\vec { 0}})
\right \rangle_\beta \ \nonumber \\
&=& {\lim_{\vec{ q}\rightarrow 0}}\ \beta \int \frac{d\omega}{2\pi}
\frac{-2}{1-e^{-\omega/T}} {\mbox{Im}}
{\cal G}_{00}(\omega,{{\vec q}})
 =- {\lim_{\vec{ q}\rightarrow 0}} 
{\mbox {Re}}\ {\cal G}_{00}(\omega=0, \vec q),
\label{eq:pre:qns}
\end{eqnarray}
where Kramers-Kronig dispersion relation has also been used. 

The quark number conservation  implies that ${\lim_{\vec{ q}\rightarrow 0}}{\mbox{Im}}
{\cal G}_{00}(\omega,{\vec{ q}}) \propto \delta(\omega)$ and the temporal SF in (\ref{eq:pre:spectral}) becomes
\begin{equation}
\sigma_{00}(\omega, {\vec 0})=\frac{1}{\pi} {\mbox{Im}} {\cal G}_{00}(\omega,{\vec 0})
=-\omega \delta(\omega) \chi_q(T) . \label{eq.corr_chiq}
\end{equation}
The relation of the Euclidean temporal CF and the response to the fluctuation
of conserved number density, $\chi_q$, can be obtained  from (\ref{eq:cor_tot}) as
\begin{equation}
 {\cal G}_{00}^E(\tau T)= -T \chi_q(T), \label{eq.eucl_chiq}
\end{equation}
which is independent of the Euclidean time, $\tau$, but depends on $T$. This result is shown in appendix~\ref{app:flu_qns}

\chapter[Dilepton production]{Dilepton production rate with and without the isoscalar-vector interaction}
\label{chapter:JHEP}

In this chapter the properties of the vector meson current-current correlation function (CF) and its spectral representation are investigated in details with and without isoscalar-vector (I-V) interaction within the framework of NJL and PNJL models at finite temperature and finite density. Then using the vector meson spectral function (SF) we obtained the dilepton production rate. This chapter is based on: {\em Vector meson spectral function and dilepton production rate in a hot and dense medium within an effective QCD approach}, Chowdhury Aminul Islam, Sarbani Majumder, Najmul Haque and Munshi G. Mustafa, {\bf JHEP 1502 (2015) 011}.

%%%%%%%%%%%%%%%%%%%%%%%%%%%%%%%%%%%%%%%%%%%%%%%%%%%%%%%%%%%%%%%%%%%%%%%%%%%                         
\section{Introduction}  

As we have already mentioned many of the hadron properties are embedded in the CF and its spectral representation. As for example informations on the masses and width are encoded in the spatial CF whereas the temporal CF is related to the response of the conserved density fluctuations; the vector current-current CF is related to the differential thermal cross section for the production of lepton pairs~\cite{Weldon:1990iw}. Thus the study of CF and its spectral representation for the hot and dense matter created in HICs becomes pursue-worthy investigations.

At high temperature but zero chemical potential, the structure of the vector CF, determination of thermal dilepton rate and various transport coefficients have been studied using LQCD framework at zero momentum~\cite{Ding:2010ga,Kaczmarek:2011ht,Francis:2011bt,Datta:2003ww,Aarts:2007wj,Aarts:2002cc,Aarts:2002tn,Amato:2013naa,Karsch:2001uw} and also at nonzero momentum~\cite{Aarts:2005hg}, which is a first principle calculation that takes into account the nonperturbative effects of QCD. Such studies for hot and dense medium  are also performed within perturbative techniques like HTL 
approximation~\cite{Karsch:2000gi,Braaten:1990wp,Greiner:2010zg,Chakraborty:2003uw,Chakraborty:2001kx,
Chakraborty:2002yt,Laine:2011xm,Czerski:2008zz,Czerski:2006ct,Alberico:2006wc,Alberico:2004we,Arnold:2000dr,Arnold:2003zc,Burnier:2012ze}, 
and in dimensional reduction~\cite{Laine:2003bd,Hansson:1991kb} appropriate for weakly interacting QGP. Nevertheless,  at RHIC and LHC energies the 
maximum temperature reached is not very far from the phase transition temperature $T_c$ and a hot
and dense matter created in these collisions is nonperturbative in nature (a strongly interacting one, also called sQGP). So, most of the perturbative methods may not be applicable in this temperature domain but 
these methods, however, are very reliable and accurate at very  high 
temperature~\cite{Haque:2013sja,Haque:2014rua,Haque:2013qta,Haque:2012my} for usual weakly
interacting QGP. In effective QCD model 
framework~\cite{Gale:1990pn,Mustafa:1999dt,He:2003jza,He:2002pv,He:2002ii,Kunihiro:1991qu,Oertel:2000jp}, 
several studies have also been done in this direction. 
Mesonic SFs and the  Euclidean  correlator in scalar, pseudoscalar 
and vector channel have been discussed in Refs.~\cite{He:2003jza,He:2002pv,He:2002ii,Kunihiro:1991qu,Oertel:2000jp} 
using NJL model. The NJL model has no information related to the confinement and thus does not have any nonperturbative effect associated with the sQGP above $T_c$. However, the PNJL model~\cite{Fukushima:2003fw,Fukushima:2003fm} contains nonperturbative information through Polyakov 
Loop~\cite{Pisarski:2000eq,Dumitru:2001xa,Dumitru:2002cf,Dumitru:2003hp,Gross:1980br,Gocksch:1993iy,Weiss:1980rj,Weiss:1981ev} 
that suppresses the color degrees of freedom  as the Polyakov loop (PL) expectation value 
decreases when $T\rightarrow T_c^+$. The thermodynamic properties  
of strongly interacting matter have  been studied 
extensively within the framework of NJL and PNJL 
models~\cite{Ghosh:2006qh,Mukherjee:2006hq,Ghosh:2007wy,Ratti:2005jh,Ghosh:2014zra}.  
The properties of the mesonic CFs have also been studied in scalar and pseudoscalar channel
using PNJL model in references~\cite{Hansen:2006ee,Deb:2009ng}. We intend to study here, for the first time 
in PNJL model, the properties of the vector CF and its spectral representation to understand the nonperturbative effect on the spectral properties, {\it e.g.,} the dilepton production rate in a hot and dense matter created in HICs.

Further, the low temperature and high density part of the phase diagram is
still less explored compared to the high temperature one. 
At finite densities the effect of chirally symmetric vector channel
interaction becomes important and it is established that within the NJL or PNJL model
this type of interaction weakens the first order transition line 
\cite{Carignano:2010ac,Kashiwa:2006rc,Sakai:2008ga,Fukushima:2008wg,Fukushima:2008is,Denke:2013jmp}
in contrary to the scalar coupling which tends to favor the appearance of 
first order phase transitions. 
It is important to mention here that the determination of
the strength of vector coupling constant is crucial under model formalism.
It cannot be fixed using vector meson mass as it is beyond
the characteristic energy cut-off of the model.
However, at the same time incorporation of vector interaction
is important if one intends to study the various spectral properties
of the system  at non-zero chemical potential~\cite{Kunihiro:1991qu,Jaikumar:2001jq}
appropriate for FAIR scenario~\cite{Friman:2011zz}.
In this chapter, we study the nonperturbative effect of PL, in the presence as well as 
the absence  of the repulsive I-V interaction,
on the SF, CF  and spectral property ({\it e.g.,} the rate of dilepton production) in a hot and dense matter. The influence of this repulsive vector channel interaction on the correlator and its spectral representation has been obtained using ring resummation. The results are compared with NJL model and the available LQCD data.

This chapter is organized as follows: In section~\ref{sec:c1:ring} we obtain the  vector CF and its spectral properties through the I-V interaction within the ring resummation both in vacuum as well as in a hot and dense medium. We present our findings relevant for a hot and dense medium in section~\ref{sec:c1:result} and finally, conclude in section~\ref{sec:c1:conclusion}.

%%%%%%%%%%%%%%%%%%%%%%%%%%%%%%%%%%%%%%%%%%%%%%%%%%%%%%%%%%%%%%%%%%%%
\section{Vector Meson correlator in Ring Resummation}
\label{sec:c1:ring}

\begin{figure}
\center
 \includegraphics[scale=0.7]{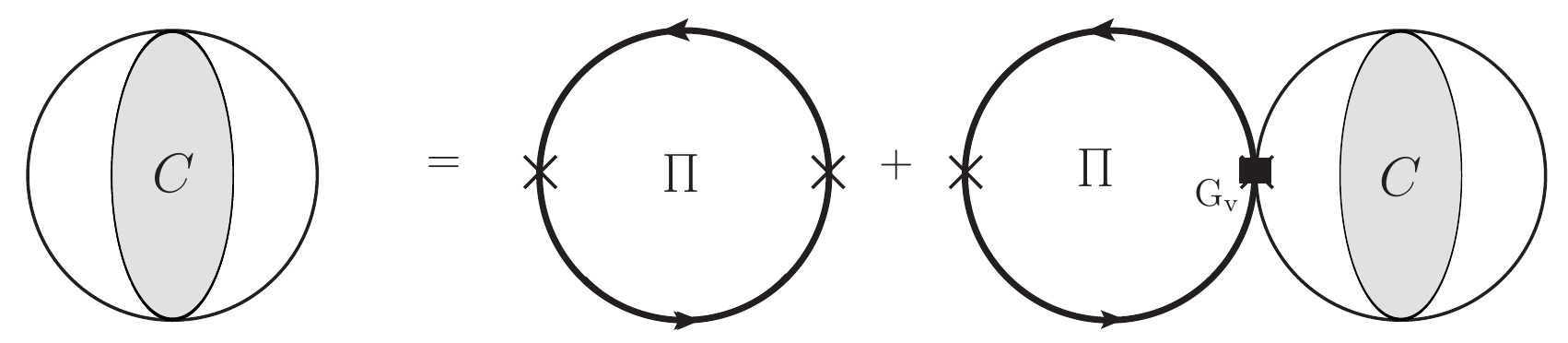}
 \caption{Vector correlator in ring resummation.}
 \label{fig:c1:corr_resum}
\end{figure}
We intend here to consider the I-V interaction. Generally,
from the structure of the interaction one can write the full vector 
channel CF  by a geometric progression of one-loop 
irreducible amplitudes \cite{Davidson:1995fq}.
In the present form of our model Lagrangian
with effective coupling $G_V$ (vide equations (\ref{eq:njl_lag_vec}) and (\ref{eq:pnjl_lag_vec})), the Dyson-Schwinger equation (DSE)
for the vector correlator $C_{\mu\nu}$ within
the ring approximation, as shown in figure~\ref{fig:c1:corr_resum}, reads as
\begin{equation}
C_{\mu\nu} = \Pi_{\mu\nu} + G_V \Pi_{\mu\sigma} C^{\sigma}_{\nu},
\label{eq:c1:C_munu}
\end{equation}
where $\Pi_{\mu\nu}$ is one loop vector correlator (figure~\ref{fig:c1:corr_oneloop}).

\subsection{Ring resummation at zero temperature and chemical potential}
The general properties of a vector CF at vacuum :
\begin{eqnarray}
\Pi_{\mu\nu} (Q^2) &=& \left (g_{\mu\nu}-
\frac{Q_\mu Q_\nu}{Q^2} \right) \Pi(Q^2)
\label{eq:c1:pi_munuk}, \\
C_{\mu\nu}(Q^2)&=& \left (g_{\mu\nu}-
\frac{Q_\mu Q_\nu}{Q^2} \right) C(Q^2),
\label{eq:c1:c_munuk}
\end{eqnarray}
where $\Pi(Q^2)$ and $C(Q^2)$ are 
scalar quantities with $Q\equiv(q_0,{\vec q})$, is the four momentum.

Using (\ref{eq:c1:pi_munuk}) and (\ref{eq:c1:c_munuk}),  (\ref{eq:c1:C_munu})  can
be reduced to a scalar DSE as
\begin{eqnarray}
  C&=& \Pi + G_V\Pi C \nonumber \\
C&=& \frac{\Pi}{1-G_V\Pi}  \label{eq8}
\end{eqnarray}

The general structure of a vector correlator in vacuum  becomes
\begin{equation}
C_{\mu\nu}= \frac{\Pi}{1-G_V\Pi} \left (g_{\mu\nu}-
\frac{Q_\mu Q_\nu}{Q^2} \right). \label{eq:c1:eq9}
\end{equation}
Using (\ref{eq:c1:eq9}) the spectral representation in vacuum can be obtained from (\ref{eq:pre:spectral}).
The vacuum properties of vector meson can be studied using this SF~\cite{Davidson:1995fq}
but we are interested in those at finite temperature and density appropriate for hot and dense
medium formed in heavy-ion collisions and the vacuum, as we will see later,  is in-built therein. 
Below we briefly outline how the vector CF 
and its spectral properties will be modified in a hot and dense medium.

\subsection{Ring resummation at finite temperature and chemical potential}

The general structure of one-loop and resummed  vector correlation
function in the medium ($T \  {\mbox{and}} \  \mu \ne 0$) can be
decomposed~\cite{Kapusta_Gale(book):1996FTFTPA,Lebellac(book):1996TFT,Das(book):FTFT} as :
\begin{eqnarray}
\Pi_{\mu\nu} (Q^2) &=& \Pi_T(Q^2) P^T_{\mu\nu}+\Pi_L(Q^2) P^L_{\mu\nu},
\label{eq:c1:pi_munu_T} \\
C_{\mu\nu} (Q^2) &=& C_T(Q^2) P^T_{\mu\nu}+C_L(Q^2) P^L_{\mu\nu},
\label{eq:c1:c_munu_T}
\end{eqnarray}
where  $\Pi_{L(T)}$ and $C_{L(T)}$ are the respective scalar parts
of $\Pi_{\mu\nu}$ and $C_{\mu\nu}$.
$P_{\mu\nu}^{L(T)}$ are longitudinal (transverse) projection
operators with their well defined properties in the medium and can be chosen as \cite{Das(book):FTFT},
\begin{equation}
P_{\mu \nu}^L=\frac{Q^2}{\tilde Q^2} {\bar U_{\mu}}{\bar U_{\nu}}, ~~~~~~
P_{\mu \nu}^T=\eta_{\mu \nu}-U_\mu U_\nu -\frac{\tilde Q_\mu \tilde Q_\nu}{\tilde Q^2}.
\label{eq:c1:p_munu}
\end{equation}
Here $U_\mu$ is the proper four velocity which in the rest frame of the heat bath
has the form $U_\mu =(1,0,0,0)$. $\tilde{Q}_\mu=(Q_\mu-\omega U_\mu)$ is the four 
momentum orthogonal to $U_\mu$ whereas
$\bar U_\mu=(U_\mu-\omega Q_\mu/Q^2)$ is orthogonal component of $U_\mu$'s with 
respect to the four momentum $Q_\mu$.
Also,  the respective scalar parts, $\Pi_{L(T)}$,
of $\Pi_{\mu\nu}$ are obtained as
\begin{equation}
\Pi_L=-\frac{Q^2}{{q}^2}\Pi_{00} ;~~~~~~~
 \Pi_T=\frac{1}{(D-2)}\left [\frac{\omega^2}{{q}^2}\Pi_{00}-\Pi_{ii}\right ],
 \label{eq.pi_munu}
\end{equation}
where $D$, is the space-time dimension of the theory.
%At finite temperature the DSE for $C_{\mu\nu}$ within ring
%summation reads as
%\begin{equation}
%C_{\mu\nu} = \Pi_{\mu\nu} + G_V \Pi_{\mu\sigma} C^{\sigma}_{\nu}.
%\end{equation}

Using (\ref{eq:c1:pi_munu_T}) and (\ref{eq:c1:c_munu_T}) in (\ref{eq:c1:C_munu}) one obtains
\begin{align}
C_T(Q^2) P^T_{\mu\nu}+C_L(Q^2) P^L_{\mu\nu} = \left [\Pi_T
+G_V \Pi_T C_T \right] P^T_{\mu\nu}
+\left [\Pi_L+G_V \Pi_L C_L \right] P^L_{\mu\nu}.
\end{align}
Now, the comparison of coefficients on both sides leads to two scalar DSEs:  
One, for transverse mode, reads as
\begin{eqnarray}
%C_T&=& \Pi_T + G_V\Pi_T C_T \nonumber \\
C_T&=& \frac{\Pi_T}{1-G_V\Pi_T}, 
\label{eq:c1:cT}
\end{eqnarray}
and the other one, for the longitudinal mode, reads as:
\begin{eqnarray}
 C_L&=& \frac{\Pi_L}{1-G_V\Pi_L} 
 \label{eq:c1:cL}
\end{eqnarray}
Let us first write the temporal component of the resummed correlator:
 it is clear from (\ref{eq:c1:p_munu}) that $P_{00}^T=0$. So we have
 from (\ref{eq:c1:c_munu_T})
\begin{eqnarray}
 C_{00}&=&\frac{\Pi_L}{1-G_V \Pi_L}P_{00}^L
 = \frac{\Pi_{00}}{1+G_V\frac{Q^2}{{q}^2}\Pi_{00}}, \label{eq:c1:c00}
\end{eqnarray}
and  the imaginary part of the temporal component of $C_{00}$ is obtained as
\begin{equation}
 {\rm Im C_{00}}=\frac{{\rm Im \Pi_{00}}}
 {\Big[ 1- G_V \Big(1-\frac{\omega^2}{{q}^2}\Big){\rm Re \Pi_{00}}\Big]^2+
 \Big[G_V (1-\frac{\omega^2}{{q}^2})
 {\rm Im \Pi_{00}}\Big ]^2} . \label{eq:c1:C00}
\end{equation}
%\par
The spatial component of the resummed correlator ($C_{ii}$) can be written as:
\begin{equation}
 C_{ii}= C_T P_{ii}^T+C_L P_{ii}^L  =\frac{\Pi_T}{1-G_V \Pi_T}P_{ii}^T+ \frac{\Pi_L}
 {1-G_V \Pi_L}P_{ii}^L . 
\end{equation}
Using (\ref{eq:c1:cT}) and (\ref{eq:c1:cL}), it becomes for $D=4$
\begin{eqnarray}
% C_T'&=& C_T P_{ii}^T \nonumber\\
 C_{ii}&=& \frac{\Pi_{ii}-\frac{\omega^2}{q^2}\Pi_{00}}{1-
 \frac{G_V}{2}(\frac{\omega^2}{ q^2}\Pi_{00}-\Pi_{ii})}
 + \frac{\frac{\omega^2}{ q^2}\Pi_{00}}
 {1+G_V(\frac{Q^2}{ q^2})\Pi_{00}} \nonumber \\
 &=& C_T' + C_L'.
\end{eqnarray}
The imaginary part of the spatial vector correlator can be obtained as
\begin{eqnarray}
 {\mbox{Im}}C_{ii}= {\mbox{Im}}C_T' + {\mbox{Im}}C_L'.
 \label{eq:c1:Cii}
\end{eqnarray}
where
\begin{equation}
 {\rm Im}C_T'=\frac{{\rm{Im}}\Pi_{ii}-\frac{\omega^2}
 {{q}^2}{\rm{Im}}\Pi_{00}}
 {\left[1+\frac{G_V}{2}{\rm{Re}}\Pi_{ii}-\frac{G_V}{2}
 \frac{\omega^2}
 {{q}^2}{\rm{Re}}\Pi_{00}\right]^2
 +\frac{G_V^2}{4}\Big[{{\rm{Im}}\Pi_{ii}-\frac{\omega^2}
 {{q}^2}\rm{{Im}}\Pi_{00}}\Big]^2}, \label{eq:c1:CTprime}
\end{equation}
and,
\begin{equation}
 {\rm Im}C_L'=\frac{\frac{\omega^2}{{q}^2}\rm{Im}\Pi_{00}}
 {\Big[ 1- G_V \Big(1-\frac{\omega^2}{{q}^2}\Big){\rm Re \Pi_{00}}\Big]^2+
 \Big[G_V (1-\frac{\omega^2}{{q}^2})
 {\rm Im \Pi_{00}}\Big ]^2} = \frac{\omega^2}{{q}^2}{\rm Im}C_{00}.
%{\left(1+G_V\frac{\omega^2}{\vec{q}^2}{\rm{Re}}\Pi_{00}-G_V{\rm{Re}}
%\Pi_{00}\right)^2
%+G_V^2\left(\frac{\omega^2}{\vec{q}^2}-1\right)^2\Big({\rm{Im}}\Pi_{00}\Big)^2}.
\label{eq:c1:CLprime}
\end{equation}
Following (\ref{eq:pre:spectral}), the resummed vector SF can be written as
\begin{equation}
\sigma_V=\frac{1}{\pi}\Big[{\rm Im}C_{00}-{\rm Im}C_{ii}\Big ].
\label{eq:c1:spectral_resum}
\end{equation}
%where $R$ stands for resummation in the ring approximation.

%%%%%%%%%%%%%%%%%%%%%%%%%%%%%%%%%%%%%%%%%%%%%%%%%%%%%%%%%%%%%%%%%%%%%%%%%%%%%%%%%%%%%%%%%%%%%%
%\section{Formalism}
%\label{sc.formalism}
\subsection{Vector correlation function in one-loop}
\label{ssec:resum_corr}

\begin{figure} [!htb]
\center
\includegraphics[scale=0.4]{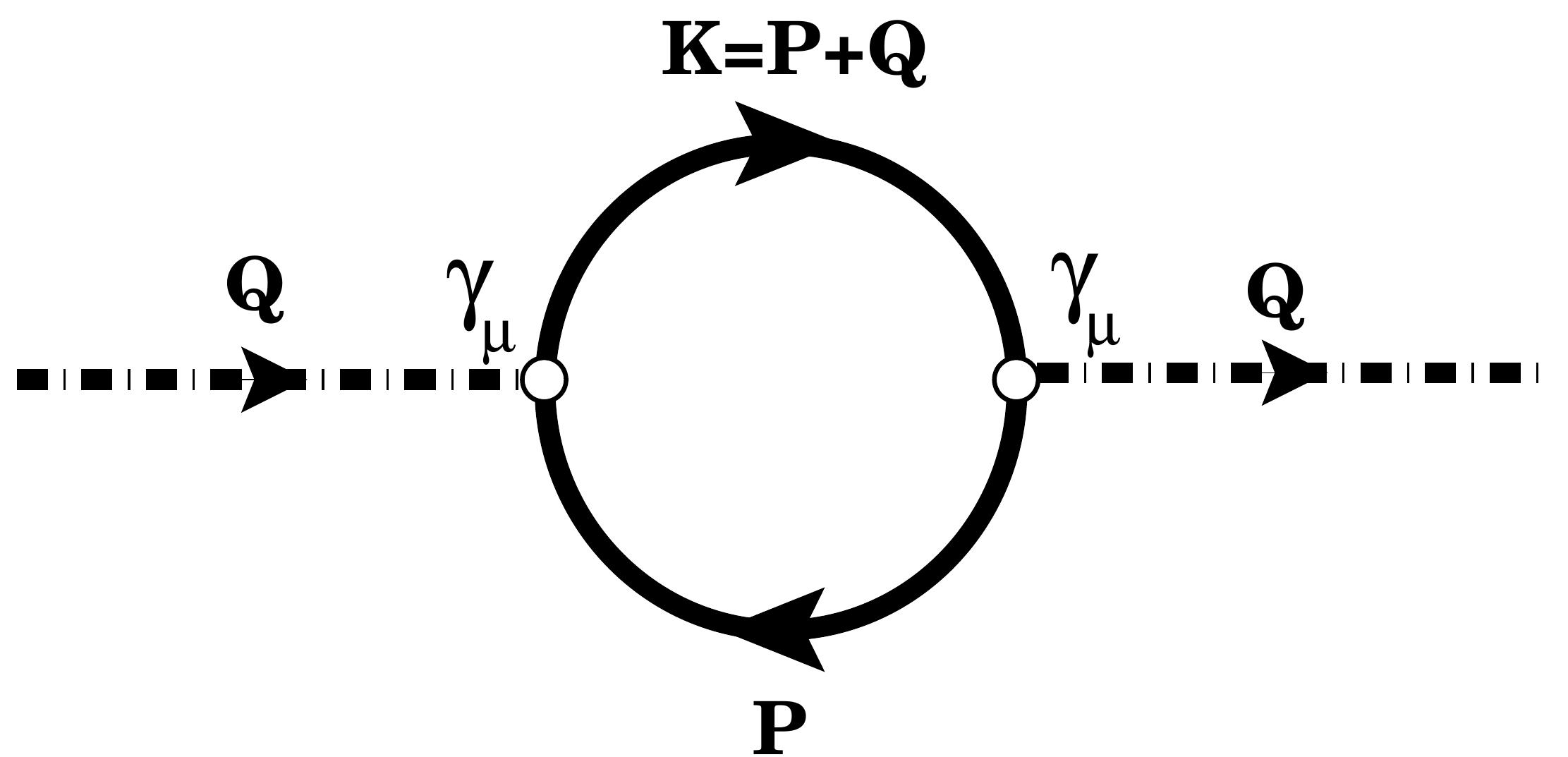}
\caption{Vector channel correlator at one-loop.}
\label{fig:c1:corr_oneloop}
\end{figure}

The current-current correlator in vector channel
at one-loop level (figure~\ref{fig:c1:corr_oneloop}) can be written as
\begin{equation}
 \Pi_{\mu\nu}(Q)= \int\frac{d^4 P}{(2\pi)^4} {\rm{Tr}}_{D,c}
 \left[\gamma_\mu S(P+Q) \gamma_\nu S(P)\right],
 \label{eq:c1:pi_munu}
\end{equation}
where ${\textrm {Tr}}_{D,c}$ is trace over Dirac and color indices, respectively.
We would like to compute this in effective models, viz. NJL and PNJL models. 

The NJL quark propagator in Hartree approximation is given as
\begin{equation}
{S_{\textrm{NJL}}}(L)= \left[\slashed{l}-m_0 +\gamma_0\tilde\mu+G_S\sigma\right ]^{-1}
= \left[\slashed{l}-M_f +\gamma_0{\tilde\mu}\right ]^{-1},
\label{eq.mod_prop_inv}
\end{equation}
whereas for PNJL it reads as
\begin{equation}
S_{\textrm{PNJL}}(L)=\left[\slashed{l} -M_f +\gamma_0{\tilde \mu} -i\gamma_0{\cal A}_4\right]^{-1} 
\label{eq.mod_prop_pnjl_inv}
\end{equation}
where the four momentum, $L\equiv(l_0,{\vec l}\hspace{0.05cm})$. The gap equation for constituent quark mass $M_f$ and the modified quark chemical 
potential $\tilde \mu$ due to vector coupling $G_V$ are given, respectively, in (\ref{eq:mass_gap}) 
and (\ref{eq:mu_tilde}). Now in contrast to NJL one, 
the presence of background temporal gauge field ${\cal A}_4$ will make a connection 
to the PL field $\Phi$ 
\cite{Pisarski:2000eq,Dumitru:2001xa,Dumitru:2002cf,Dumitru:2003hp,Gross:1980br,Gocksch:1993iy,Weiss:1980rj,Weiss:1981ev}. 
While performing the frequency sum and color trace in (\ref{eq:c1:pi_munu}), the thermal distribution
function in PNJL case will be different from that of  NJL one.

For convenience, we will calculate the one-loop vector correlation in NJL model.  
This NJL CF, as discussed, can easily be generalized to PNJL one by replacing 
the thermal distribution functions~\cite{Hansen:2006ee,Ghosh:2014zra} as
\begin{align}
 f(E_p-\tilde\mu)=\frac{\bar\Phi e^{-\beta(E_p-\tilde \mu)}
 +2{\Phi}e^{-2\beta(E_p-\tilde \mu)}+e^{-3\beta(E_p-\tilde \mu)}}
 {1+3\bar\Phi e^{-\beta(E_p-\tilde\mu)}+3{\Phi}e^{-2\beta(E_p-\tilde\mu)}
 +e^{-3\beta(E_p-\tilde\mu)}}, \nonumber\\
 f(E_p+\tilde\mu)=\frac{\Phi e^{-\beta(E_p+\tilde \mu)}
 +2\bar{\Phi}e^{-2\beta(E_p+\tilde \mu)}+e^{-3\beta(E_p+\tilde \mu)}}
 {1+3\Phi e^{-\beta(E_p+\tilde\mu)}+3\bar{\Phi}e^{-2\beta(E_p+\tilde\mu)}
 +e^{-3\beta(E_p+\tilde\mu)}}, \label{eq:c1:dist_pnjl}
\end{align}
which at $\Phi ({\bar \Phi})=1$ reduces to the thermal distributions for NJL or free as the cases may be. 
On the other hand for $\Phi ({\bar \Phi})=0$, the effect of confinement
is clearly evident in which three quarks are stacked in a same momentum and color state~\cite{Islam:2012kv}. 

\subsubsection{Temporal Part}
The time-time component of the vector correlator in (\ref{eq:c1:pi_munu}) reads as
\begin{equation}
 \Pi_{00}(Q)=\int{\frac{d^4P}{(2\pi)^4}{\mathrm {Tr}}[\gamma_0 S(K)\gamma_0 S(P)]},
\label{eq.pi00}
\end{equation}
where $K=P+Q$. After some mathematical simplifications (vide appendix~\ref{app:correlator}) we are left with
\begin{align}
 \Pi_{00}(\omega,\vec q) =&\,\, N_c N_f\int\frac{d^3p}{(2\pi)^3}\frac{1}{E_{p}E_{k}}\Bigg\{
\frac{E_{p}E_{k}+M_f^2+\vec{p}\cdot\vec{k}}{\omega+E_{p}-E_{k}} \nonumber\\
& \times \left[f(E_{p}-\tilde \mu)+f(E_{p}+\tilde \mu)
-f(E_{k}-\tilde\mu)-
f(E_{k}+\tilde \mu)\right] \nonumber \\
& + \left(E_{p}E_{k}-M_f^2-\vec{p}\cdot\vec{k}\right)
\left[\frac{1}{\omega-E_{p}-E_{k}}-
\frac{1}{\omega+E_{p}+E_{k}}\right] \nonumber\\
& \times \left[1-f(E_{p}+\tilde\mu)
-f(E_{k}-\tilde\mu)\right]
\Bigg\}. \label{eq:c1total_pi00} 
\end{align}

The real and imaginary parts (vide appendix~\ref{app:correlator}) of the temporal vector correlator are, respectively,  obtained as
\begin{align}
\textrm{Re} \Pi_{00}(\omega,\vec q) =&~ {\large {\textrm  P}}\Bigg [\ N_f N_c \int\frac{d^3p}{(2\pi)^3}\frac{1}{E_pE_k}
\Bigg \{\frac{E_p E_k+M_f^2+\vec{p}\cdot\vec{k}}{\omega +E_p -E_k} \nonumber\\
& \times \left[f(E_{p}-\tilde\mu)+f(E_{p}+\tilde\mu)-f(E_{k}-\tilde\mu)-f(E_{k}+\tilde\mu)\right] \nonumber\\
& +(E_p E_k-M_f^2-\vec{p}\cdot\vec{k})\left(\frac{1}{\omega-E_p-E_k}-\frac{1}{\omega+E_p+E_k}\right) \nonumber\\
& \times \left[1-f(E_p+\tilde\mu)-f(E_k-\tilde\mu)\right]\Bigg\}\Bigg ],
\label{eq:c1:repi00}
\end{align}
as {\large P} stands for principal value, and
\begin{align}
 {\textrm {Im}} \Pi_{00}(\omega,\vec q) =& \lim_{\eta\rightarrow0}\ \frac{1}{2i}
\Big [\Pi_{00}(\omega\rightarrow \omega +i\eta,q)
-\Pi_{00}(\omega\rightarrow \omega -i\eta,q)\Big ] \nonumber \\
=& -\pi N_f N_c \int\frac{d^3p}{(2\pi)^3} 
\frac{1}{E_pE_k}
\Bigg\{ 
(E_p E_k+M_f^2+\vec{p}\cdot\vec{k}) \nonumber \\
& \times \left[f(E_{p}-\tilde\mu)+
f(E_{p}+\tilde\mu)-f(E_{k}-\tilde \mu)-f(E_{k}+\tilde\mu)\right]
\times \delta(\omega+E_p-E_k) \nonumber \\
& + (E_p E_k-M_f^2-\vec{p}\cdot\vec{k})
\left[1-f(E_p+\tilde\mu)-
f(E_k-\tilde \mu)\right] \nonumber \\
& \times [\delta(\omega-E_p-E_k)]\Bigg\}.
\label{eq:c1:im_pi}
\end{align}
It is now worthwhile 
to check some known results in the limit  
$\vec q \rightarrow 0$ and $\tilde\mu=0$,  (\ref{eq:c1:im_pi}) can be written as: 
\begin{align}
\textrm{Im} \Pi_{00}(\omega)=-\pi N_f N_c \int\frac{d^3p}
{(2\pi)^3}\frac{1}{E_p^2}
\left (3E_p^2-3M_f^2-p^2 \right) \left(2f'(E_p)\right)
\left(-\omega \delta\left(\omega \right)\right),
 \label{eq.impi_00_q0_2}
\end{align}
and which, in the limit $M_f=m_0-G_S\sigma=0$, further becomes
\begin{eqnarray}
\textrm{Im} \Pi_{00}(\omega)
&=& -2\pi T^2\omega \delta(\omega).
\label{eq.impi_00_q0_m0}
\end{eqnarray}
The vacuum part in (\ref{eq:c1:repi00}) is now separated as 
\begin{align}
{\textrm {Re}}\Pi_{00}^{\textrm {vac}}(\omega,\vec q) =&~ \frac{N_f N_c}{4\pi ^2} \!\! \int _0^{\Lambda } \!\! p\ dp
\frac{1}{2 E_p q}\left [ 4 p q+6 E_p X_- -6 E_p X_+ 
- Y_- \ln\left | \frac{E_p+ X_- -\omega } {E_p+ X_+ -\omega }\right|\right. \nonumber \\
&\left. + Y_+  \ln\frac{E_p+ X_+ +\omega } {E_p + X_- +\omega }\right] , \label{eq.repivac00} \\
{\textrm{with}}  \, \,  Y_\pm &= (4 E_p^2 \pm 4 E_p \omega +M^2),  \, \,
X_\pm =\sqrt{E_p^2\pm 2 p q+q^2}  \ {\textrm{and}}  \  M^2=\omega^2-q^2.\nonumber
\end{align}
We note that the ultraviolet divergence in the vacuum part is 
regulated by using  a finite three momentum cut-off $\Lambda$. The corresponding matter 
part of (\ref{eq:c1:repi00}) is obtained as
\begin{align}
{\textrm {Re}} \Pi_{00}^{\textrm {mat}}(\omega,\vec q)=&~\frac{N_f N_c}{2\pi ^2}\int _0^{\infty }p \ dp
\big[f(E_p-\tilde\mu)+f(E_p+\tilde\mu) \big]
\left[\frac{\omega }{q}\ln\left |\frac{M^2-4\left(pq+\omega E_p\right)^2}
{M^2-4\left(pq-\omega E_p \right)^2}\right | \right. \nonumber\\
&\left. 
-\left(\frac{4E_p^2 + M^2}{4qE_p}\right)
\ln\left | \frac{\left(M^2-2pq\right)^2-4\omega ^2 E_p^2}{\left(M^2+2pq\right)^2-4\omega^2 E_p^2}\right |
-\frac{2p}{E_p}\right] . \label{eq.repimat00}
\end{align}
The imaginary part in (\ref{eq:c1:im_pi}) can be simplified as
\begin{align}
{\textrm {Im}} \Pi_{00}(\omega,\vec q) = \frac{ N_f N_c}{4\pi } \int _{p_-}^{p_+}
p \ dp \ \frac{4\omega E_p -4E_p^2-M^2}{2E_p q} \ \big [f(E_p-\tilde\mu)+ f(E_p+ \tilde \mu) - 1 \big ]
\label{eq.impimat00}
\end{align}
where the vacuum part does not require any momentum cut-off as 
the energy conserving $\delta$-function 
ensures the finiteness of the limits:
\begin{equation}
p_\pm = {\frac{\omega }{2}\sqrt{1-\frac{4M_f^2}{M^2}}\pm\frac{q}{2}}, \label{eq:c1:limits}
\end{equation}
with a threshold restricted by a step function, $\Theta(M^2-4M_f^2)$.

\subsubsection{Spatial Part}
The space-space component of the vector correlator in (\ref{eq:c1:pi_munu}) reads as
\begin{equation}
 \Pi_{ii}(Q)=\int{\frac{d^4P}{(2\pi)^4}{\mathrm {Tr}}
 [\gamma_i S(K)\gamma_i S(P)]},
\label{eq_pi_ii}
\end{equation}
which can be simplified in the similar way as it is done for temporal component to
\begin{align}
 \Pi_{ii}(\omega,\vec q) =&~ N_c N_f\int\frac{d^3p}{(2\pi)^3}\frac{1}
 {E_{p}E_{k}}\Bigg\{
 \frac{3E_{p}E_{k}-3M_f^2-\vec{p}\cdot\vec{k}}{\omega-E_{p}+E_{k}}
 \nonumber \\
 & \times \left[f(E_k+\tilde\mu)+f(E_k-\tilde\mu) - f(E_p+\tilde\mu)-f(E_p-\tilde\mu)\right]
 \nonumber \\
&+ \left (3E_{p}E_{k}+3M_f^2+\vec{p}\cdot\vec{k}\right ) 
 \left [\frac{1}{\omega-E_p-E_k} - \frac{1}{\omega+E_p+E_k}\right ]
 \nonumber \\
&\times  \left [1-f(E_k+\tilde\mu)-f(E_p-\tilde\mu) \right ]
\Bigg\}. \label{eq.total_pi_ii} 
\end{align}
In the similar way as before the imaginary part can be obtained 
\begin{align}
 {\textrm {Im}} \Pi_{ii}(\omega,\vec q) =& -\pi N_f N_c
 \int\frac{d^3p}{(2\pi)^3} 
 \frac{1}{E_pE_k}
\Bigg\{(3E_{p}E_{k}-3M_f^2-\vec{p}\cdot\vec{k}) \nonumber\\
& \times \left[f(E_k+\tilde\mu)+f(E_k-\tilde\mu) - f(E_p+\tilde\mu)-f(E_p-\tilde\mu)
\right]\delta(\omega+E_p-E_k) \nonumber \\
& + \left (3E_{p}E_{k}+3M_f^2+\vec{p}\cdot\vec{k}\right )
\left [1-f(E_k+\tilde\mu)-f(E_p-\tilde\mu) \right ] \nonumber\\
& \times \delta(\omega-E_p-E_k)\Bigg\},
\label{eq:c1:impi_ii}
\end{align}
whereas the real part can be obtained as
\begin{eqnarray}
\textrm{Re} \Pi_{ii}(\omega,\vec q)\!\!\! &=&\!\!\! {\large {\textrm  P}} \Bigg[ N_f N_c \int
\frac{d^3p}{(2\pi)^3}\frac{1}{E_pE_k}
\Bigg \{\frac{3E_p E_k-3M_f^2-\vec{p}\cdot\vec{k}}
{\omega -E_p +E_k} \nonumber\\
&&\!\!\! \times \left[f(E_{k}-\tilde\mu)+f(E_{k}+\tilde\mu)
-f(E_{p}-\tilde\mu)-f(E_{p}+\tilde \mu)\right] \nonumber\\
&&\!\!\! + (3E_p E_k+3M_f^2+\vec{p}\cdot\vec{k})
\left(\frac{1}{\omega-E_p-E_k}
-\frac{1}{\omega+E_p+E_k}\right) \nonumber\\
&&\!\!\! \times \left[1-f(E_k+\tilde \mu)-f(E_p-\tilde \mu)\right]\Bigg\}\Bigg].
\label{eq:c1:repi_ii}
\end{eqnarray}
At this point we also check the known results in the limit 
$\vec q \rightarrow 0$ and $\tilde\mu=0$,  (\ref{eq:c1:impi_ii}) can be written as: 
\begin{align}
\textrm{Im} \Pi_{ii}(\omega)=& -\pi N_f N_c \int\frac{d^3p}
{(2\pi)^3}\frac{1}{E_p^2}
\left (3E_p^2-3M_f^2-p^2 \right) \left(2f'(E_p)\right)
\left(-\omega \delta\left(\omega \right)\right) \nonumber \\
& - \frac{3}{2\pi\omega}\sqrt{\omega^2-4M_f^2}
\left(\omega^2+2M_f^2\right)\tanh
\left(\frac{\omega}{4T}\right)\Theta(\omega-2M_f),
 \label{eq:c1:impi_ii_q0_2}
\end{align}
and when $M_f=m_0-G_S\sigma=0$, it becomes
\begin{eqnarray}
\textrm{Im} \Pi_{ii}(\omega)
= -2\pi T^2\omega \delta(\omega)-
{\frac{3}{2\pi}}\omega^2 \tanh
\left(\frac{\omega}{4T}\right).
\label{eq.impi_ii_q0_m0}
\end{eqnarray}
As seen both massive and massless cases show a sharp peak due to the delta function at $\omega \rightarrow 0$, which leads to pinch singularity for calculation of transport coefficients.

Now, the vacuum part in (\ref{eq:c1:repi_ii}) is simplified as
\begin{align}
{\textrm {Re}} \Pi^{\textrm{vac}}_{ii}(\omega,\vec q)=&~ \frac{N_f N_c}{4\pi ^2}
\!\! \int _0^{\Lambda }\!\!\!\! p\ dp \
\frac{1}{2 E_p q}\left[-4 p q+10 E_p (X_- - X_+) - Z_-
\ln\left |\frac{E_p+X_--\omega }{E_p+X_+ -\omega }\right | \nonumber \right. \nonumber \\ 
& \left. + Z_+ 
\ln \frac {E_p+X_+ +\omega } {E_p+X_-+\omega }\right] , \label{eq.repivacii}
\end{align}
where $Z_\pm=4p^2 \pm 4 E_p \omega -M^2$. The corresponding matter part is obtained as
\begin{align}
{\textrm {Re}} \Pi_{ii}^{\textrm{mat}}(\omega,\vec q)=&~\frac{N_f N_c}{4\pi ^2}\int _0^{\infty }p\ dp \ 
\big[f(E_p-\tilde\mu)+f(E_p+\tilde\mu)\big]
\left[2\frac{\omega}{q}\ln\left |\frac{M^2-4(pq+E_p \omega)^2}{M^2-4(pq-E_p\omega)^2}\right| \right. \nonumber \\
& \left. +\left(\frac{M^2-4p^2}{2qE_p}\right)
\ln\left |\frac{\left(M^2-2pq\right)^2-4\omega^2E_p^2}{\left(M^2+2pq\right)^2-4\omega^2 E_p^2}\right |+4\frac{p}{E_p}\right] .
\label{eq.repimatii}
\end{align}
Finally, the imaginary part in (\ref{eq:c1:impi_ii}) is simplified as
\begin{align}
{\textrm {Im}} \Pi_{ii}(\omega,\vec q)=\frac{ N_f N_c}{4\pi } \int _{p_-}^{p_+}p \ dp \
\frac{4\omega E_p-4p^2+M^2}{2E_p q}\ \left (f(E_p-\tilde\mu)+ f(E_p+ \tilde\mu) - 1 \right )
\label{eq.impimatii}
\end{align}
where the vacuum part does not need any finite momentum cut-off as stated above.

\section{Results}
\label{sec:c1:result}
\subsection{Gap Equations and Mean Fields}
%%%%%%%%%%%%%%%%%%%%%%%%%%%%%%%%%%%%%%%%%%%%%%%%%%%%%%%%%%%%%%%%%%%%%%%%
\begin{figure} [!htb]
\subfigure[]
{\includegraphics[scale=0.8]{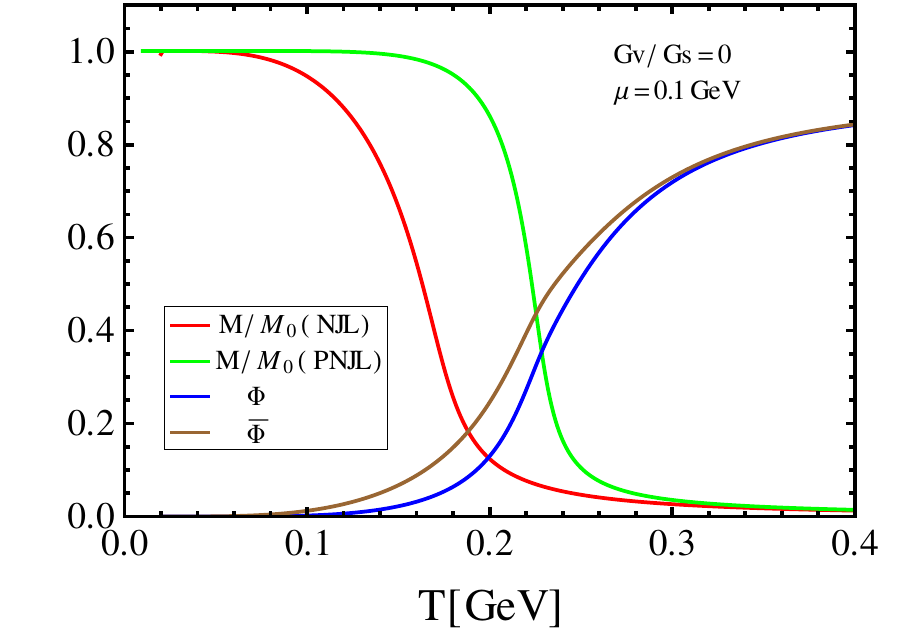}
\label{fig:c1:massfields_gv0}}
\subfigure[]
 {\includegraphics[scale=0.8]{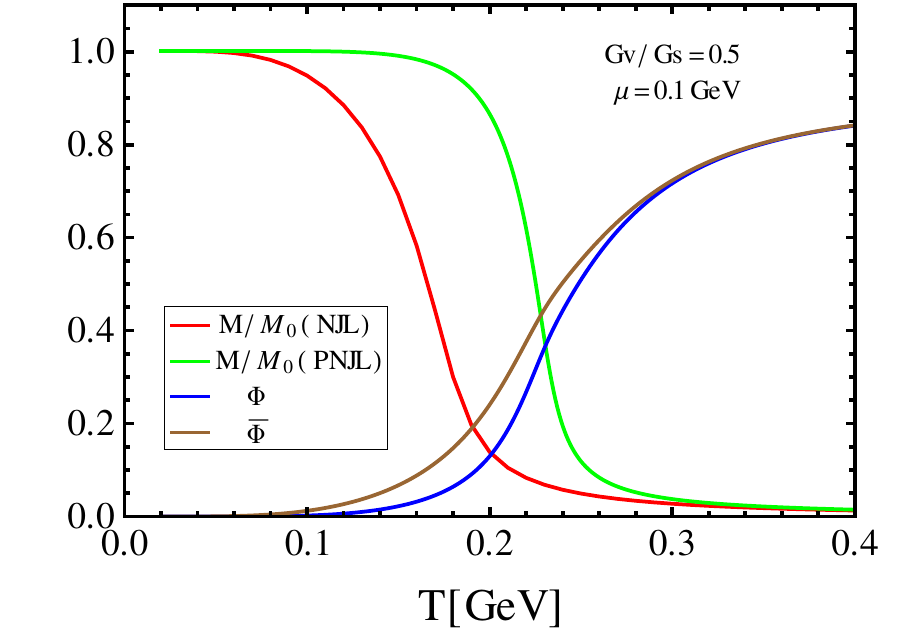}
 \label{fg.massfields_gv05}}
 \caption{Variation of scaled constituent quark mass with zero temperature quark mass 
 and the PL fields ($\Phi$ and $\bar \Phi$) with temperature 
 at chemical potential $\mu=100$ MeV for (a) $G_V/G_S=0$ 
 and (b) $G_V/G_S=0.5$.}
\label{fig:c1:fields&mass_vs_T}
\end{figure}
%%%%%%%%%%%%%%%%%%%%%%%%%%%%%%%%%%%%%%%%%%%%%%%%%%%%%%%%%%%%%%%%%%%%%%%%%%%%%%%%%%%%%

%%%%%%%%%%%%%%%%%%%%%%%%%%%%%%%%%%%%%%%%%%%%%%%%%%%%%%%%
\begin{figure} [!htb]
\subfigure[]
{\includegraphics[scale=0.8]{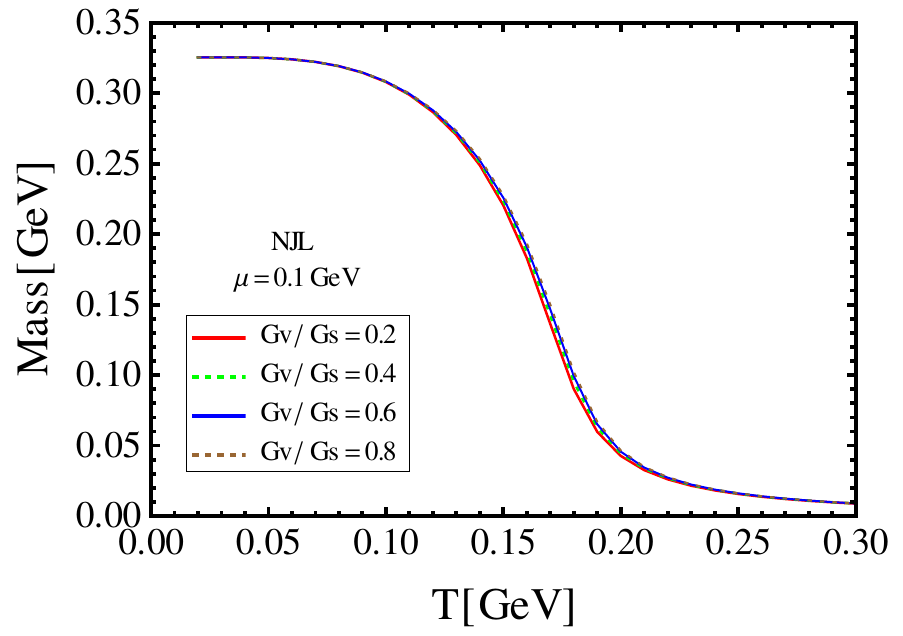}
\label{fg.mass_njl}}
\subfigure[]
 {\includegraphics[scale=0.8]{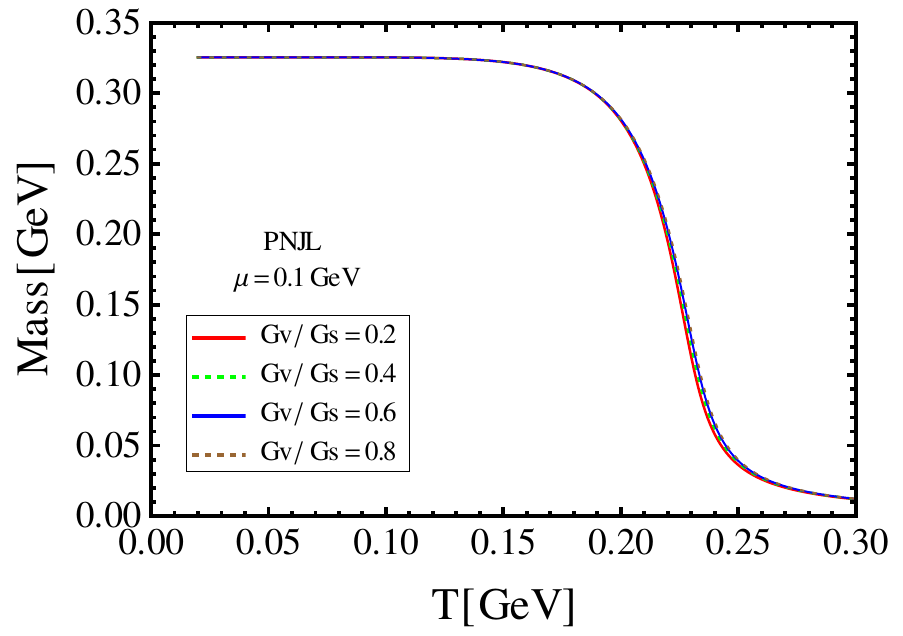}
 \label{fg.mass_pnjl}}
 \caption{Variation of constituent quark mass with temperature in  (a) NJL  and 
(b) PNJL  model for chemical potential $\mu=100$ MeV and a set of
 values for vector coupling $G_V$.}
\label{fig:c1:mass_vs_T}
\end{figure}
%%%%%%%%%%%%%%%%%%%%%%%%%%%%%%%%%%%%%%%%%%%%%%%%%%%%%%%%%%%%%%%%%%%%%%%%%%%%%%%%%%%%%%%%%%

The thermodynamic potentials $\Omega$ for both NJL and PNJL model are extremized 
with respect to the mean fields X, i.e.,
\begin{equation}
 \frac{\partial \Omega}{\partial X}=0
\end{equation}
where, X stands for $\sigma$ and $n$ for NJL model and 
$\Phi, \bar{\Phi},\sigma$ and $n$ for PNJL model.
The value of the parameters, $G_S=10.08 \ {\textrm{GeV}}^{-2}$ and
$\Lambda=0.651$ GeV were taken from literature \cite{Ratti:2005jh}
and $m_0=0.005$ GeV.
However, the value of $G_V$ is difficult to fix within
the model formalism,
since this quantity should be fixed using the $\rho$ meson
mass which, in general, happens to be higher than the
maximum energy scale $\Lambda$ of the model. 
So, we consider the vector coupling constant $G_V$ as a free parameter and
different choices are considered as ${G_V}=x \times {G_S}$, where $x$ is chosen
from $0$ to $1$ appropriately.

%%%%%%%%%%%%%%%%%%%%%%%%%%%%%%%%%%%%%%%%%%%%%%%%%%%%%%%%
\begin{figure} [!htb]
\subfigure[]
{\includegraphics[scale=0.8]{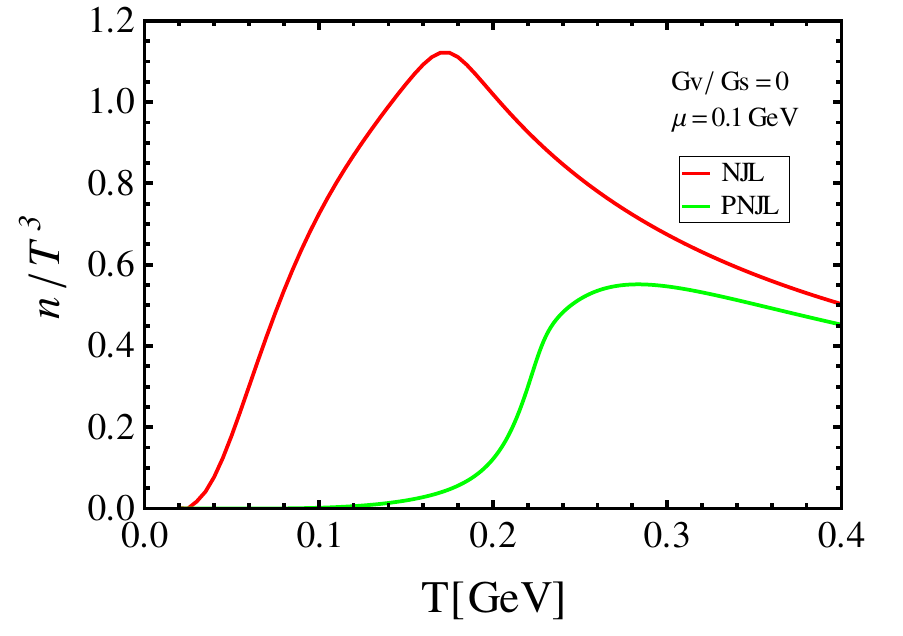}
\label{fig:c1:density_gv0}}
\subfigure[]
 {\includegraphics[scale=0.8]{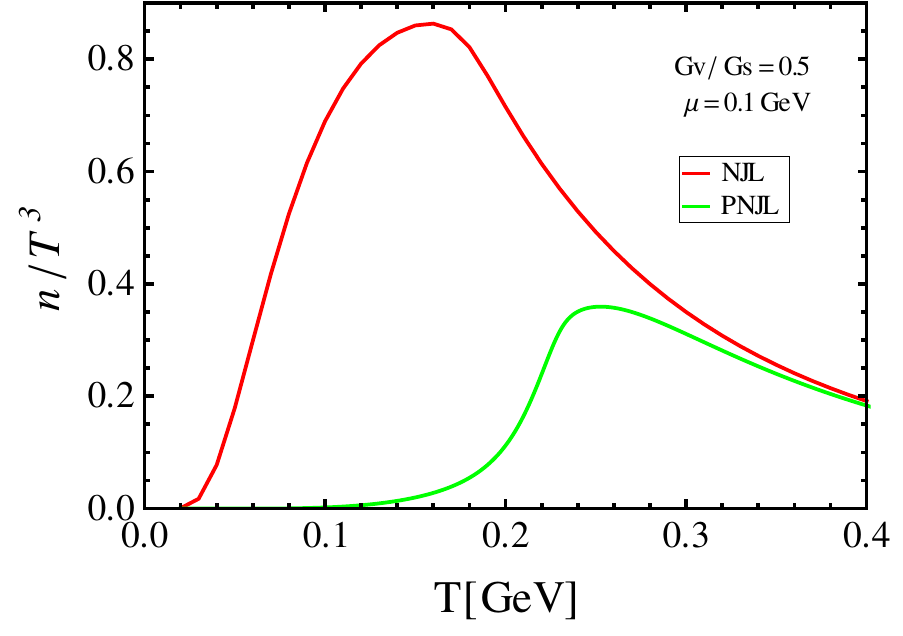}
 \label{fig:c1:density_gv05}}
 \caption{Comparison of the scaled quark number density with $T^3$ 
 as a function of  $T$ in  NJL and PNJL model with chemical potential $\mu=100$ MeV
 for (a) $G_V/G_S=0$ and (b) $G_V/G_S=0.5$ .}
\label{fig:c1:number_density}
\end{figure}
%%%%%%%%%%%%%%%%%%%%%%%%%%%%%%%%%%%%%%%%%%%%%%%%%%%%%%%%

In figure~\ref{fig:c1:fields&mass_vs_T}, a comparison between the scaled quark mass with its zero temperature
value 
($M_f(T)/M_f(0)$) in NJL and PNJL model is displayed as a function of temperature $T$ for two values 
of $G_V (=0 \ {\textrm{and}} \ 0.5G_S)$ with $\mu=100$MeV. It also contains a variation of the PL fields
($\Phi(T)$ and $\bar \Phi(T)$) with $T$. 
In both models the scaled quark mass decreases with increase in $T$ and approaches 
the chiral limit at very high $T$. However, in the temperature range $60\le T({\textrm{MeV}})\le 300$,
the variation of the scaled quark mass is slower in PNJL than that of NJL model. This slow variation is
due to the presence of the confinement effect in it through the PL fields ($\Phi(T)$ and $\bar \Phi(T)$),
as can be seen that the  PL field ($\Phi(T)$) and its conjugate ($\bar \Phi(T)$) increase from zero 
in confined phase and approaches unity (free state) at high temperature. Now, we note that $\Phi={\bar \Phi}$ 
at $\mu=0$ as there is equal number of quarks and antiquarks.  However, because of non-zero chemical potential 
there is an asymmetry in 
quark and antiquark numbers, which leads to an asymmetry in the PL fields $\Phi$ and $\bar \Phi$. 
This asymmetry disappears for $T> 300$ MeV, which is much greater than $\mu$. 
We also note that the fields depend weakly with the variation of the vector coupling $G_V$ akin to that 
of mass as displayed in figure~\ref{fig:c1:mass_vs_T} for $\mu=100$ MeV and $G_V/G_S=0 \ {\textrm {to}} \ 0.8$.

In figure~\ref{fig:c1:number_density} the  number density scaled with $T^3$ for both NJL and PNJL model is 
displayed as a function of $T$ with  $\mu= 100$ MeV for two values of $G_V$. 
 At very high temperature, as seen in figure~\ref{fig:c1:fields&mass_vs_T}, 
the PL fields $\Phi({\bar \Phi})\rightarrow 1$ and masses in both models
become same. So, PNJL model becomes equivalent to NJL model because the thermal distribution 
function becomes equal as can be seen from (\ref{eq:c1:dist_pnjl}) and thus the number density. 
On the other hand, for temperature $T< 400$ MeV and a given $\mu$, the PNJL number density 
is found to be suppressed than that of NJL case as the thermal distribution function in 
(\ref{eq:c1:dist_pnjl}) is suppressed. This is due to the combination of two complementary effects:
(i) the nonperturbative effect through $\Phi$ and ${\bar \Phi}$ 
and (ii) the slower variation of mass in PNJL model, which are clearly evident from 
figure~\ref{fig:c1:fields&mass_vs_T}. It is also obvious from
figure~\ref{fig:c1:density_gv0} and figure~\ref{fig:c1:density_gv05} that 
the presence of vector interaction $G_V$ reduces the number density for both NJL and PNJL model, 
which could be understood due to the reduction of ${\tilde \mu}$ as given in (\ref{eq:mu_tilde}).

\subsection{Vector spectral function and dilepton rate}

The vector SF is proportional to the imaginary part of the vector correlation
function as defined in (\ref{eq:pre:spectral}) or (\ref{eq:c1:spectral_resum}). This imaginary part is restricted by, the energy conservation, $\omega = E_p+E_k$, as can be seen from (\ref{eq:c1:im_pi}) and 
(\ref{eq:c1:impi_ii}). This equivalently leads to a threshold, $M^2\ge 4M_f^2$, which can also be found
from (\ref{eq:c1:limits}). Now for a given $G_V$ and  $T$, the resummed spectral 
function in (\ref{eq:c1:spectral_resum}) picks up continuous contribution  
above the threshold, $M^2> 4M^2_f$, which provides a finite width to a vector meson that decays 
into a pair of leptons.  However, below the threshold $M^2 < 4M^2_f$, the continuous contribution of
the SF in (\ref{eq:c1:spectral_resum}) becomes zero and the decay to dileptons are forbidden. 
But if one analyses it below the threshold, one can find bound state contributions in
SFs.
%which would follow from (\ref{eq.C00}), (\ref{eq.CTprime}) and (\ref{eq.CLprime}). 
When imaginary part  approaches zero, the SF in (\ref{eq:c1:spectral_resum}) becomes 
discrete and can be written as:
\begin{eqnarray}
% \sigma_V^R(\omega,q) &=\atop {M<2M_f} & \frac{1}{\pi} \Big [ \delta \big(F_0(\omega,q)\big) + \delta\big (F_1(\omega,q)\big ) 
% +\delta \big(F_2(\omega,q)\big)  \Big], \nonumber \\
 \sigma_V(\omega,\vec q) &=\atop {M<2M_f} & \frac{1}{\pi} \Big [\delta\big (F_1(\omega,\vec q)\big )\Big], \nonumber \\
\textrm{where} \, \,  \, \, \, \, \, \, \,   
%F_0(\omega, q) &=& {1- G_V \Big[1-\frac{\omega^2}{\vec{q}^2}\Big]{\rm Re \Pi_{00}}}=0 ,  \nonumber \\
 F_1(\omega, \vec q) &=& {1+\frac{G_V}{2}{\rm{Re}}\Pi_{ii}-\frac{G_V}{2} 
 \frac{\omega^2}
 {{q}^2}{\rm{Re}}\Pi_{00}}=0,
% \nonumber \\
% F_2(\omega, q) &=&{1+G_V\frac{\omega^2}{\vec{q}^2}{\rm{Re}}\Pi_{00}-G_V{\rm{Re}} \Pi_{00}}=0.    
\label{eq.delta_spect}
\end{eqnarray}
where only the dominant contribution of (\ref{eq:c1:CTprime}) is considered.
Using the properties of $\delta$-function,
%and the dominant contribution in (\ref{eq.delta_spect}), 
one can write
\begin{eqnarray}
 \sigma_V(\omega, \vec q) & = \atop {M<2M_f} & \frac{1}{\pi} \frac{\delta(\omega-\omega_0)}
 {\left | dF_1(\omega,\vec q)/d\omega\right |_{\omega=\omega_0}},   \label{eq:c1:discrete_spect}
\end{eqnarray}
which corresponds to a sharp $\delta$-function peak at $\omega=\omega_0$. However, we are interested here 
in continuous contribution $M> 2M_f$, which are discussed below.

%%%%%%%%%%%%%%%%%%%%%%%%%%%%%%%%%%%%%%%%%%%%%%%%%%%%%%%%%%%%%%%
\begin{figure} [!htb]
\subfigure[]
{\includegraphics[scale=0.8]{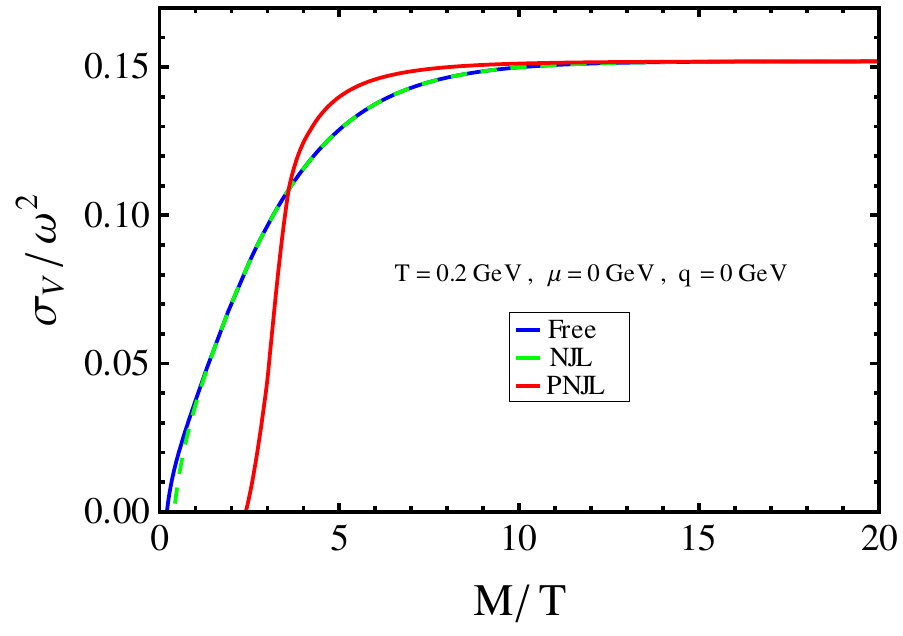}
\label{fg.free_mu0_t2}}
\subfigure[]
{\includegraphics[scale=0.8]{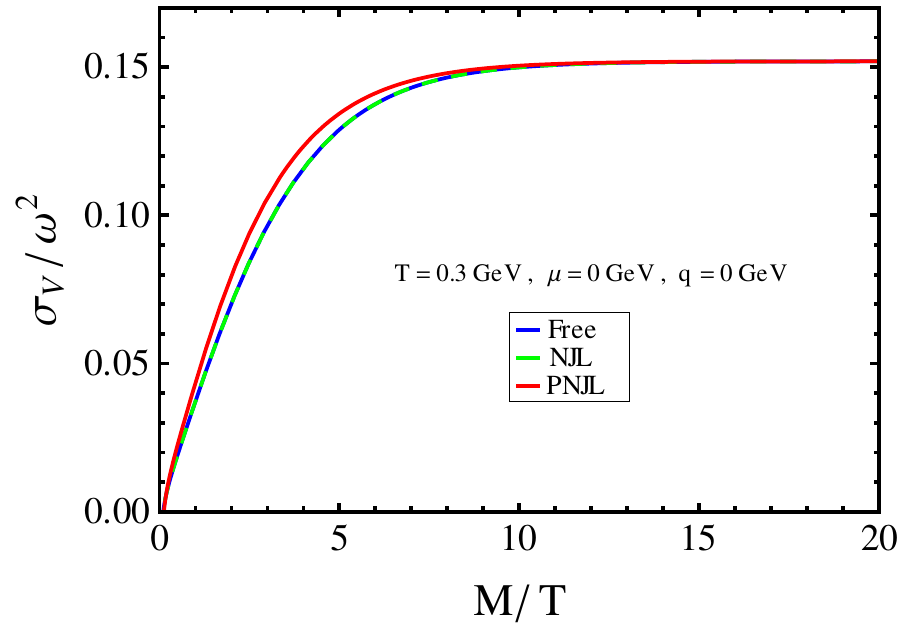}
 \label{fg.free_mu0_t3}}
\caption{Scaled vector SF $\sigma_V/\omega^2$ as a function of scaled 
invariant mass, $M/T$, in NJL and PNJL model with external momentum 
$q=0$, quark chemical potential $\mu=0$ and $G_V/G_S=0$ for (a) $T=200$ MeV and (b) $T=300$ MeV.}
\label{fig:c1:spect_gv0_q0_mu0}
\end{figure}
%%%%%%%%%%%%%%%%%%%%%%%%%%%%%%%%%%%%%%%%%%%%%%%%%%%%%%%%%%%%%%%%%%%%%%%%%%

%%%%%%%%%%%%%%%%%%%%%%%%%%%%%%%%%%%%%%%%%%%%%%%%%%%%%%%%%%%%%%%%%%%%%%%%%
\begin{figure} [!htb]
\subfigure[]
{\includegraphics[scale=0.8]{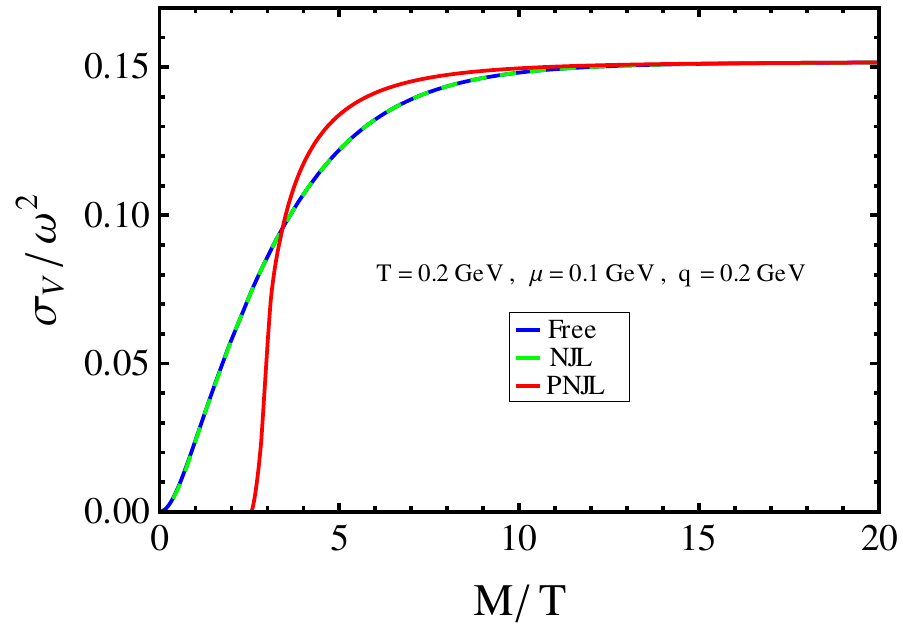}
\label{fg.free_mu1_t2}}
\subfigure[]
{\includegraphics[scale=0.8]{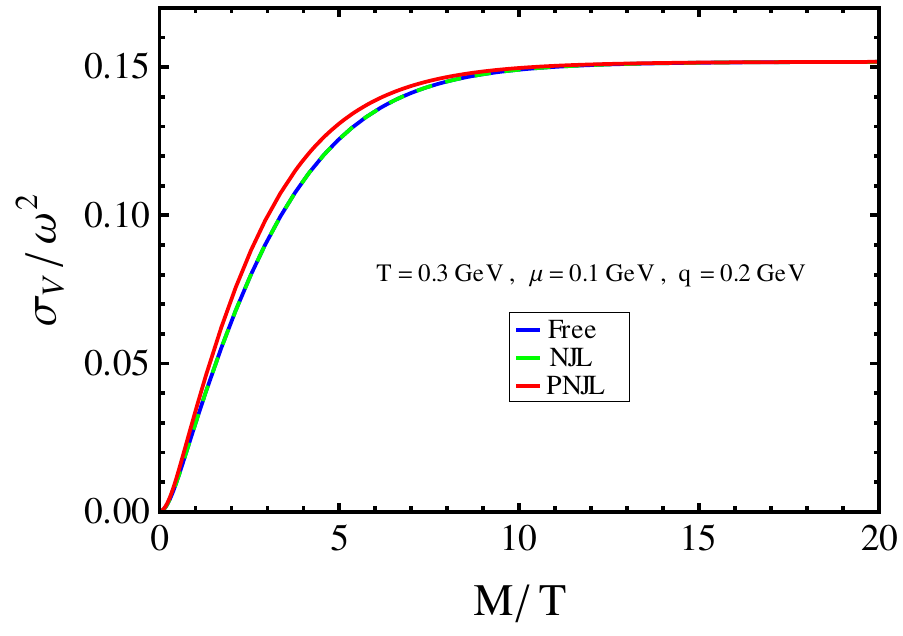}
\label{fg.free_mu1_t3}}
\caption{Scaled vector SF $\sigma_V/\omega^2$ as a function of scaled 
invariant mass, $M/T$, in NJL and PNJL model with external momentum 
$q=200$ MeV, quark chemical potential $\mu=100$ MeV and $G_V/G_S=0$ for (a) $T=200$ MeV and (b) $T=300$ MeV}
\label{fig:c1:spect_gv0_q02_mu0}
\end{figure}
%%%%%%%%%%%%%%%%%%%%%%%%%%%%%%%%%%%%%%%%%%%%%%%%%%%%%%%%%%%%%%%%%%%%%%%%%%
 
\subsubsection{Without vector interaction ($G_V=0$)}

With no vector interaction ($G_V=0$), the SF in (\ref{eq:c1:spectral_resum}) 
is solely determined by  the imaginary part of the one loop vector self energies 
$\Pi_{00}(\omega,\vec q)$  and $\Pi_{ii}(\omega,\vec q)$.
figure~\ref{fig:c1:spect_gv0_q0_mu0} displays a comparison of vector SF
with zero external momentum ($\vec q=0, \ \ M=\omega$) in NJL and PNJL model for $T=200$ MeV and $300$ MeV, 
when there is no vector interaction ($G_V=0$). 
Now, for $T=200$ MeV (left panel) the SF in PNJL model has larger
threshold than NJL  model because the quark mass in PNJL model is much
larger than that of NJL one (see figure~\ref{fig:c1:fields&mass_vs_T}). Also the 
PNJL SF dominates over that of NJL one, because of the presence of
nonperturbative effects due to PL fields $\Phi$ and ${\bar \Phi}$. At
higher values of $T~(=300~\rm {MeV})$ (right panel), the threshold becomes almost same due to the 
reduction of mass effect in PNJL case whereas the nonperturbative effects 
at low $M/T$ still dominate. The reason is the following: at zero external momentum and zero
chemical potential the SF is proportional to $[1-2f(E_p)]$ (apart from the mass
dependent prefactor) as can be seen from the second
term of (\ref{eq:c1:impi_ii_q0_2}). In PNJL case the thermal distribution function, $f(E_p)$, 
is more suppressed due to the suppression of color degrees of freedom than NJL at moderate
values of $T$, so the weight factor $[1-2 f(E_p)]$ is larger than NJL case and 
causing an enhancement in the SF . All these features also 
persist at non-zero chemical
potential and external momentum as can also be seen from figure~\ref{fig:c1:spect_gv0_q02_mu0}.

%%%%%%%%%%%%%%%%%%%%%%%%%%%%%%%%%%%%%%%%%%%%%%%%%%%%%%%%%%%%%%%%%%%%%%%%%
\begin{figure} [!htb]
\vspace*{-0.0in}
\begin{center}
\includegraphics[scale=1.0]{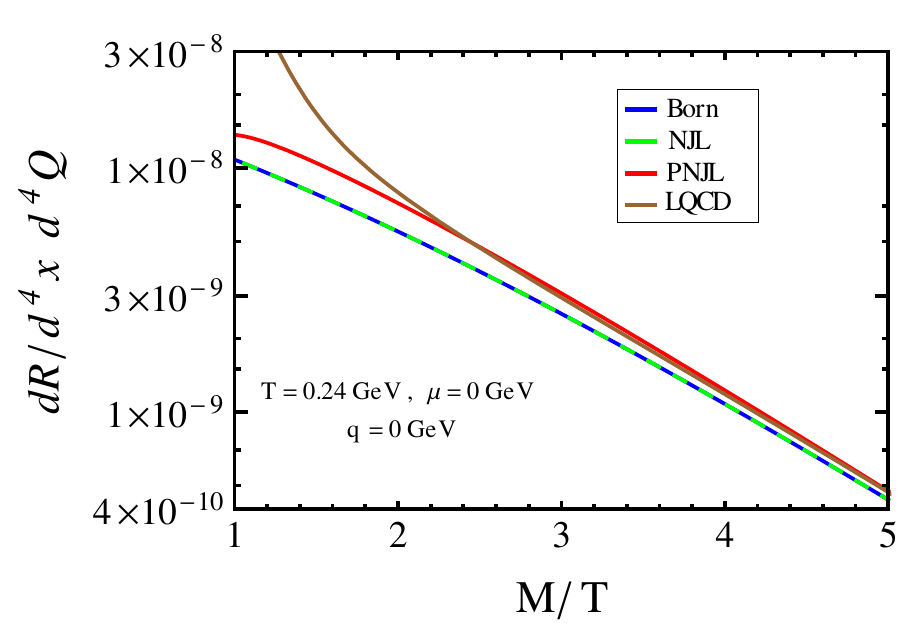}
\end{center}
\caption{Comparison of  dilepton rates  as a function of  $M/T$  
for $T=240$ MeV with external momentum 
$q=0$, quark chemical potential $\mu=0$ and $G_V/G_S=0$. The LQCD rate is from Ref.~\cite{Ding:2010ga}.}
\label{fig:c1:dilep_gv0_q0_mu0}
\end{figure}
%%%%%%%%%%%%%%%%%%%%%%%%%%%%%%%%%%%%%%%%%%%%%%%%%%%%%%%%%%%%%%%%%%%%%%%%%%

At this point it is important to note that for 
$T>250$ MeV the mass in NJL model almost approaches current
quark mass (see figure~\ref{fig:c1:massfields_gv0}) and can be considered as a free case 
since there is no vector interaction present ($G_V=0$). Nevertheless, the PNJL case is 
different  because of the presence of the nonperturbative confinement effect
through the PL fields.  The PNJL model can suitably describe a sQGP~\cite{Fukushima:2003fw,Fukushima:2003fm,Gale:2014dfa} 
scenario having nonperturbative effect due to the suppression of color degrees of freedom 
compared to NJL vis-a-vis free case above the deconfinement temperature.

The above features of the SF  in a sQGP with no vector interaction 
will be reflected  in the dilepton rate which is related to the SF,  as given 
in (\ref{eq:pre:rel_dilep_spec}). In figure~\ref{fig:c1:dilep_gv0_q0_mu0}, the dilepton rate is displayed
as a function of scaled invariant mass $M$ with $T$. As already discussed, at this temperatures 
the quark mass in NJL approaches current quark mass faster than PNJL, 
thus the dilepton rates for Born and NJL cases become almost the same. However, the dilepton rate 
in PNJL model is enhanced than those of Born or NJL case. This in turn suggests that the 
nonperturbative dilepton production rate is higher in a sQGP than the Born rate in 
a weakly coupled QGP. The dilepton
rate is also compared with that from LQCD result~\cite{Ding:2010ga} within a quenched approximation.
It is found to agree well for $M/T\ge 2$, below which it differs from LQCD rate. 
We try to understand this as follows:  the SF in LQCD 
is extracted using maximum entropy method from Euclidean vector CF by inverting (\ref{eq:cor_tot}),
which requires an ansatz for the SF. Using a free field SF as an ansatz, 
the SF in a  quenched approximation of QCD was obtained earlier~\cite{Karsch:2001uw}
by inverting (\ref{eq:cor_tot}), which was then approaching zero in the limit $M/T \rightarrow 0$.
So was the first lattice dilepton rate~\cite{Karsch:2001uw} at low $M/T$ whereas it was oscillating around 
the Born rate for $M/T>3$. Now, in a very recent LQCD calculation~\cite{Ding:2010ga} with larger size, 
while extracting the SF using maximum entropy method from Euclidean vector CF, an ansatz for the SF, a Briet-Wigner for low  $M/T $ plus a free field one for $M/T\ge 2$,  has been used. The ansatz of Briet-Wigner at  low $M/T$ pushes up the SF and so is the recent dilepton 
rate in LQCD below $M/T \le 2$. However, no such ansatz is required in thermal QCD and we can 
directly calculate the SF without any uncertainty by virtue of the analytic continuation.

%%%%%%%%%%%%%%%%%%%%%%%%%%%%%%%%%%%%%%%%%%%%%%%%%%%%%%%%%%%%%%%%%%%%%%%%%
\begin{figure} [!htb]
\vspace*{-0.0in}
\subfigure[]
{\includegraphics[scale=0.8]{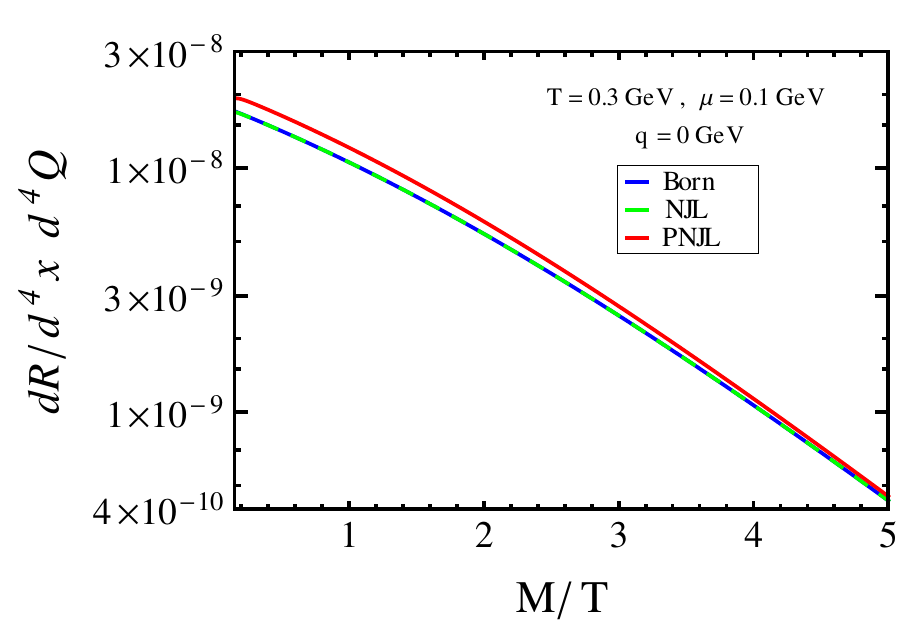}
\label{fg.free_dilepq0mu}}
\subfigure[]
{\includegraphics[scale=0.8]{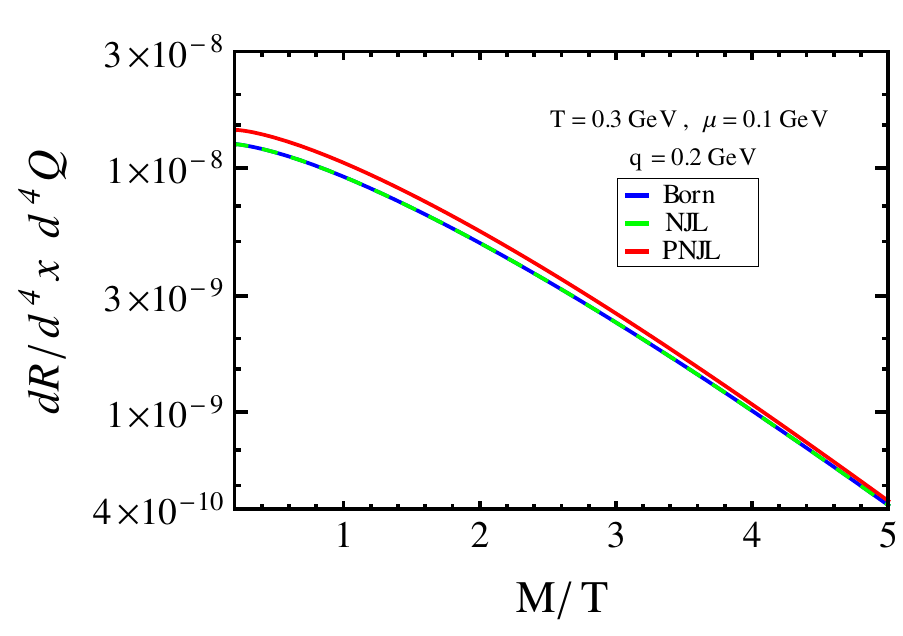}
\label{fg.free_dilepq2mu}}
\caption{Different dilepton rates  as a function of  $M/T$ 
at  $T=300$ MeV, $\mu=100$ MeV and $G_V/G_S=0$ for (a) $q=0$ (b) $q=200$ MeV}
\label{fig:c1:dilep_gv0_mu1}
\end{figure}
%%%%%%%%%%%%%%%%%%%%%%%%%%%%%%%%%%%%%%%%%%%%%%%%%%%%%%%%%%%%%%%%%%%%%%%%%%
In figure~\ref{fig:c1:dilep_gv0_mu1} the dilepton rate is also displayed at $T=300$ MeV, 
non-zero chemical potential ($\mu=100$ MeV) and external momentum ($q=0$ and $200$ MeV).
We note that this information could also be indicative for future LQCD 
computation of dilepton rate at  non-zero $\mu$ and $q$.
The similar feature of sQGP as found in figure~\ref{fig:c1:dilep_gv0_q0_mu0} is 
also seen here but with a quantitative difference especially due to higher $T$,
which could be understood from figure~\ref{fig:c1:massfields_gv0}.

\subsubsection{With vector interaction ($G_V\ne 0$)}

%%%%%%%%%%%%%%%%%%%%%%%%%%%%%%%%%%%%%%%%%%%%%%%%%%%%%%%%%%%%%%%%%%%%%%%%%
\begin{figure} [!htb]
\vspace*{-0.0in}
\subfigure[]
{\includegraphics[scale=0.8]{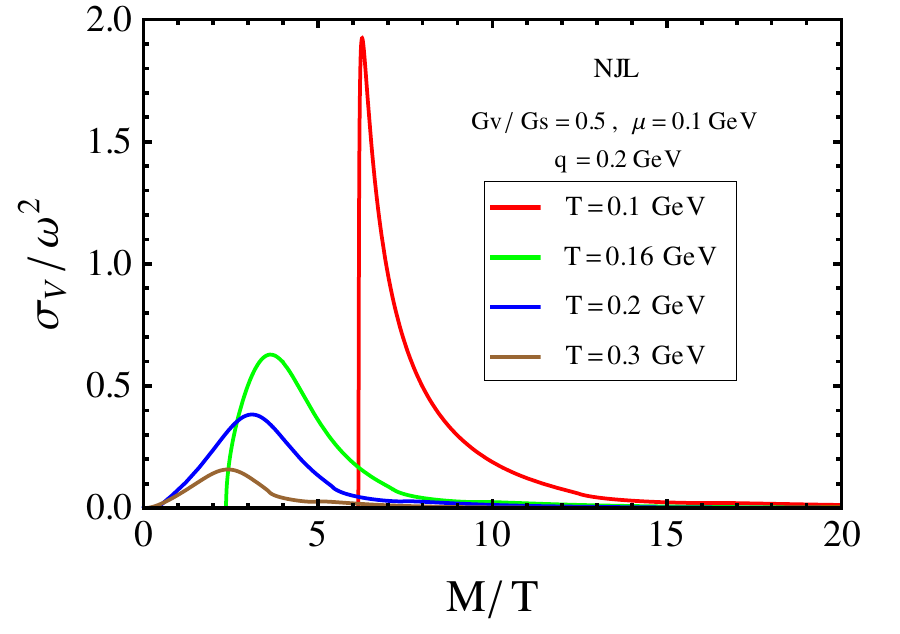}
\label{fg.njl_spect1q2mu1}}
\subfigure[]
{\includegraphics[scale=0.8]{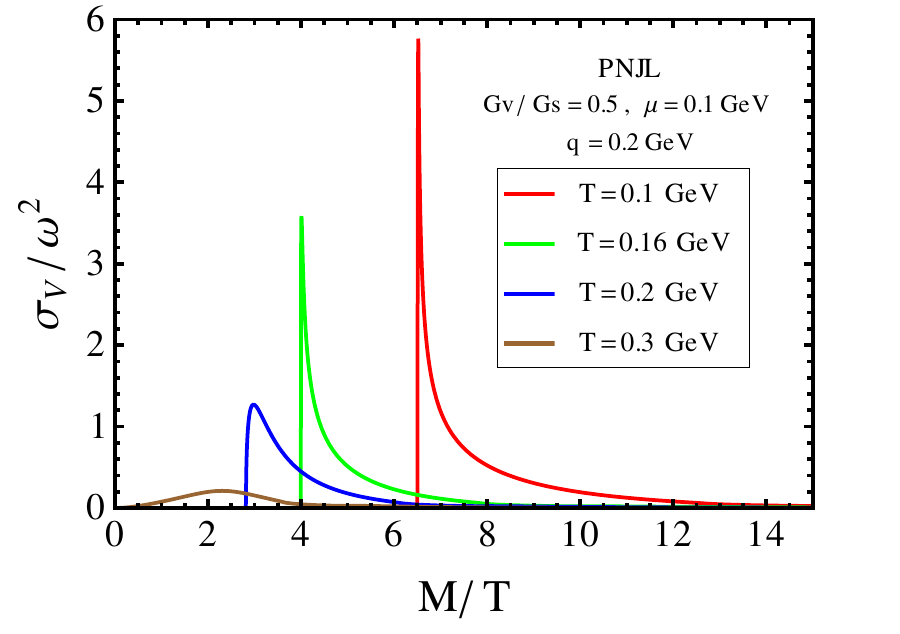}
\label{fg.pnjl_spect1q2mu1}}
\caption{Scaled spectral function  as a function of  $M/T$ in (a) NJL and (b) PNJL model 
for a set of $T$ with 
$\mu=100$ MeV,  $q=200$ MeV and $G_V/G_S=0.5$. Note the difference in $y$-scale.}
\label{fig:c1:spect_njl_gv05}
\end{figure}
%%%%%%%%%%%%%%%%%%%%%%%%%%%%%%%%%%%%%%%%%%%%%%%%%%%%%%%%%%%%%%%%%%%%%%%%%%

%%%%%%%%%%%%%%%%%%%%%%%%%%%%%%%%%%%%%%%%%%%%%%%%%%%%%%%%%%%%%%%%%%%%%%%%%
\begin{figure} [!htb]
\vspace*{0.1in}
\subfigure[]
{\includegraphics[scale=0.8]{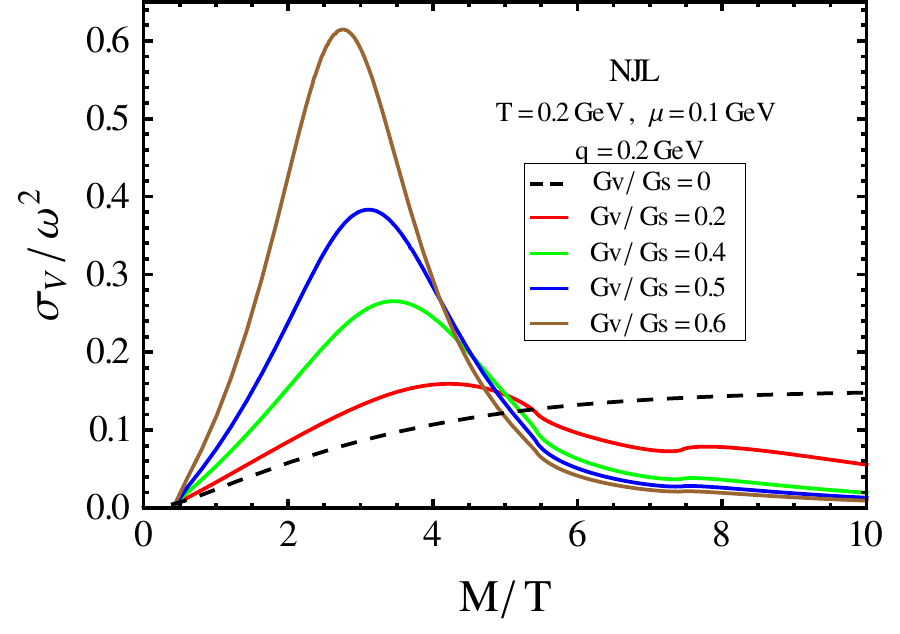}
\label{fg.njl_spect2q2mu1}}
\subfigure[]
{\includegraphics[scale=0.8]{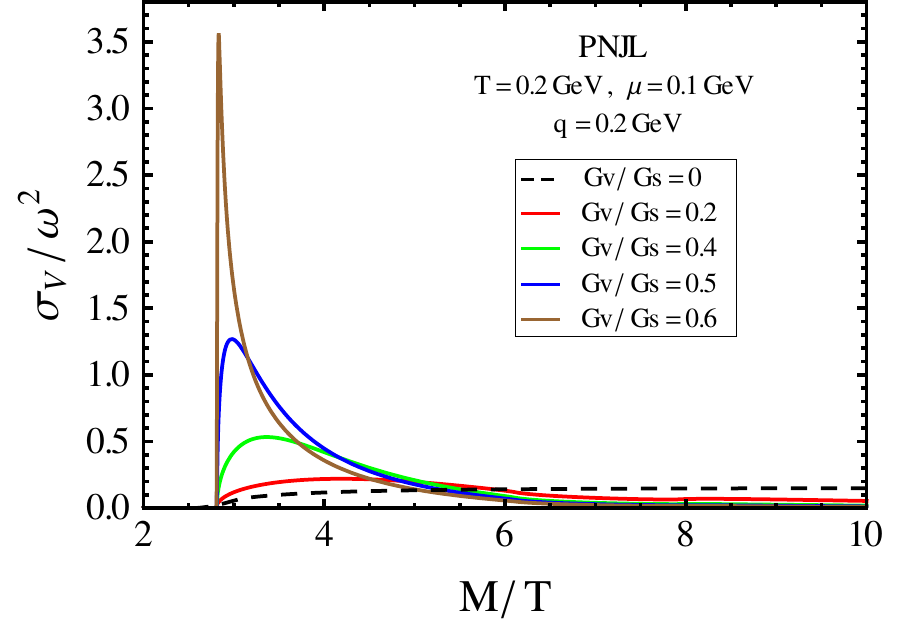}
\label{fg.pnjl_spect2q2mu1}}
\caption{Scaled spectral function  as a function of  $M/T$ 
for (a) NJL  and (b) PNJL model with $T=200$ MeV, 
$\mu=100$ MeV,  $q=200$ MeV and a set of values of $G_V/G_S=0,\, 0.2,\, 0.4, \, 0.5
\, \, {\textrm{and}} \, \, 0.6$.}
\label{fig:c1:spect_t2_q2_mu1}
\end{figure}
%%%%%%%%%%%%%%%%%%%%%%%%%%%%%%%%%%%%%%%%%%%%%%%%%%%%%%%%%%%%%%%%%%%%%%%%%%
%%%%%%%%%%%%%%%%%%%%%%%%%%%%%%%%%%%%%%%%%%%%%%%%%%%%%%%%%%%%%%%%%%%%%%%%%
\begin{figure} [!htb]
\vspace*{-0.0in}
\subfigure[]
{\includegraphics[scale=0.8]{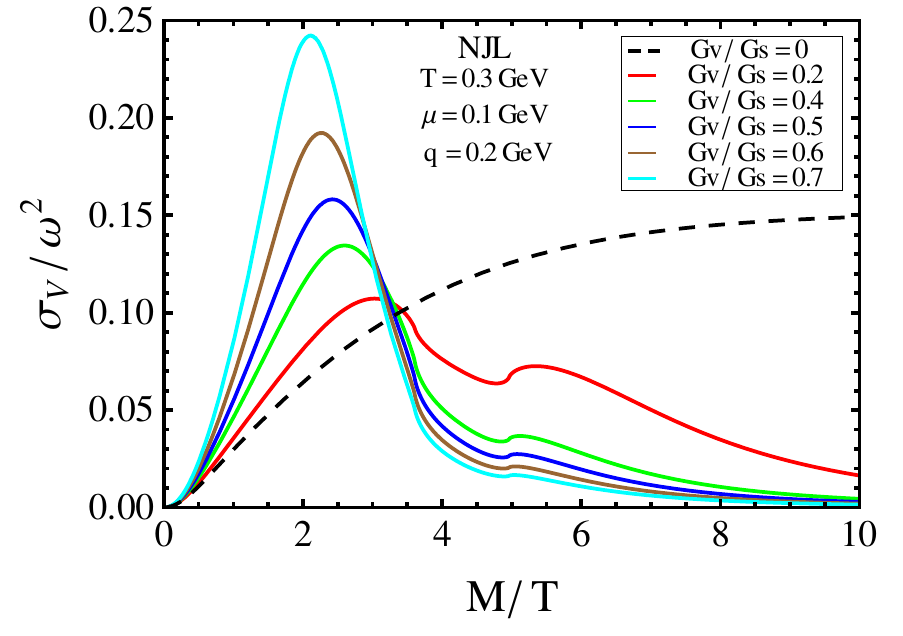}
\label{fg.njl_spect3q2mu1}}
\subfigure[]
{\includegraphics[scale=0.8]{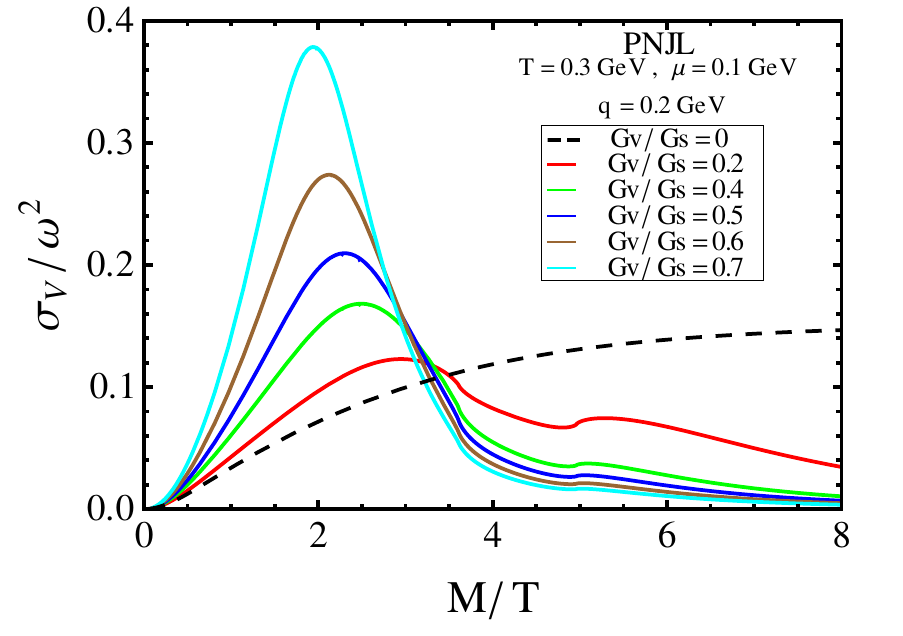}
\label{fg.pnjl_spect3q2mu1}}
\caption{Scaled spectral function  as a function of  $M/T$ 
for (a) NJL  and (b) PNJL model with $T=300$ MeV, 
$\mu=100$ MeV,  $q=200$ MeV and a set of values of $G_V/G_S=0, \ 0.2,\, 0.4, \, 0.5, 
\, 0.6 \, \, {\textrm{and}} \, \, 0.7$.}
\label{fig:c1:spect_t3_q2_mu1}
\end{figure}
%%%%%%%%%%%%%%%%%%%%%%%%%%%%%%%%%%%%%%%%%%%%%%%%%%%%%%%%%%%%%%%%%%%%%%%%%%
\begin{figure} [!htb]
\vspace*{-0.0in}
\subfigure[]
{\includegraphics[scale=0.8]{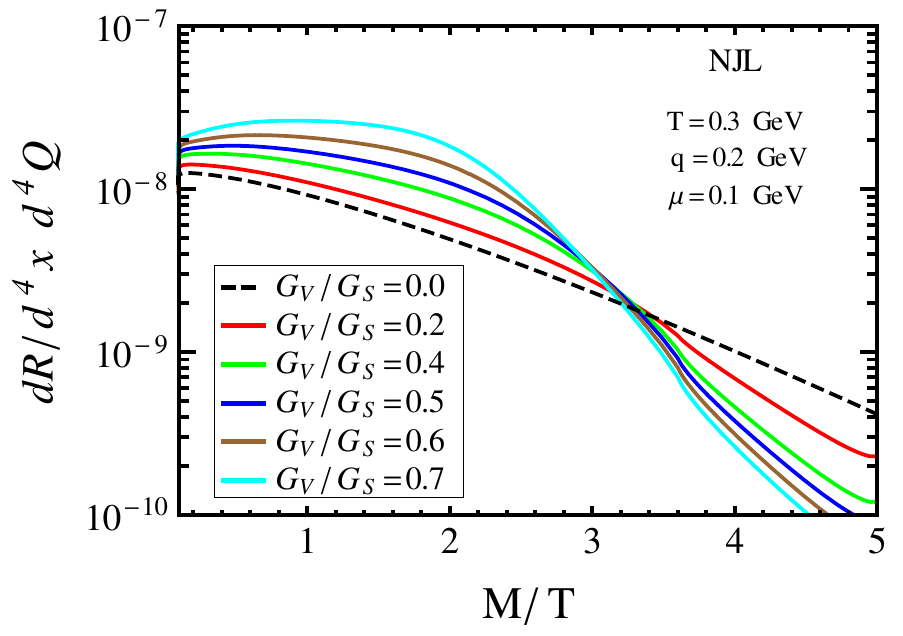}
\label{fg.njl_dilepq0t3}}
\subfigure[]
{\includegraphics[scale=0.8]{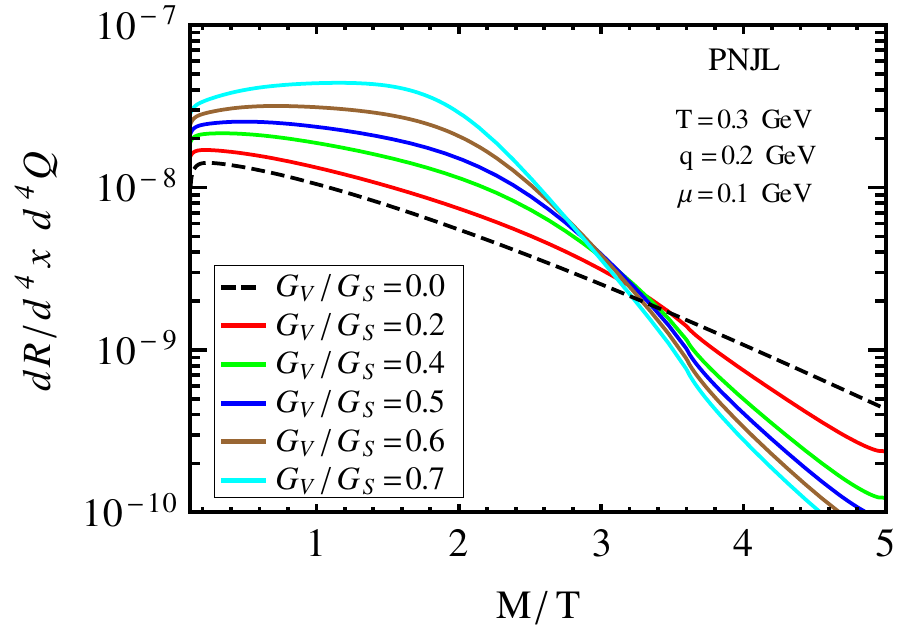}
\label{fg.pnjl_dilepq0t3}}
\caption{Dilepton rates  as a function of  $M/T$  (a) NJL and (b) PNJL model
for a set of values of $G_V/G_S$  with $T=300$ MeV with external momentum 
$q=200$ MeV, quark chemical potential $\mu=100$ MeV.}
\label{fig:c1:dilep_gv_t3}
\end{figure}
%%%%%%%%%%%%%%%%%%%%%%%%%%%%%%%%%%%%%%%%%%%%%%%%%%%%%%%%%%%%%%%%%%%%%%%%%%%%

In figure~\ref{fig:c1:spect_njl_gv05} the SF for $G_V/G_S=0.5$ with $q=200$ MeV and 
$\mu=100$ MeV in NJL (left panel) and PNJL (right panel) model is displayed. 
At $T=100 < T_c \sim 160$ MeV~\cite{Islam:2012kv,Bazavov:2011nk,Petreczky:2012ct,Borsanyi:2010bp}
the SF  above the respective threshold, $M> 2M_f$, starts with a large value because the
denominator in (\ref{eq:c1:CTprime}) is very small compared to those in (\ref{eq:c1:C00}) and (\ref{eq:c1:CLprime}).
This is due to the two reasons: (i) the first term in the denominator involving real parts of $\Pi$ has zero 
below the threshold that corresponds to a sharp $\delta$-like peak as discussed in (\ref{eq:c1:discrete_spect}), 
thus it also becomes a very small number just above the threshold and 
(ii) the second term involving imaginary parts start building up, which is also very small. However, the increase in 
$T$ causes the SF to decrease due to mutual effects of denominator 
(involving both real and imaginary parts of $\Pi$)  and numerator (involving only imaginary parts of $\Pi$). 
On the other hand, with the increase in $T$, the threshold in NJL case reduces quickly as the quark mass 
decreases faster whereas it reduces slowly for PNJL case because the PL fields experience a 
slow variation of the quark mass. So, the  vector meson in NJL model acquires a width earlier
than the PNJL model due to suppression of color degrees of freedom in presence of PL
fields. As seen, for NJL model at $T= T_c \sim 160$ MeV the sharp peak like structure gets 
a substantial width than  PNJL model. This suggests that the vector meson retains its bound properties 
at and above $T_c$ in PNJL  model in presence of $G_V$ along with the nonperturbative effects 
through PL fields.

In figures~\ref{fig:c1:spect_t2_q2_mu1} and \ref{fig:c1:spect_t3_q2_mu1} we present the dependence of 
the SF on the vector interaction in QGP for a set of values of the coupling 
$G_V$ in NJL (left panel) and PNJL (right panel) model, respectively,
for $T=200$ and  $T=300$ MeV. In both cases the spectral strength increases 
with that of $G_V$. Nevertheless, the strength of the  SF in PNJL case at a 
given $T$ and $G_V$ is always stronger than that of NJL model. This suggests that
the presence of the vector interaction further suppresses the color degrees of freedom 
in addition to the PL fields.
The dilepton rates corresponding to $T=300$ MeV are  also displayed in figure~\ref{fig:c1:dilep_gv_t3}, which
show an enhancement at low $M/T$ compared to $G_V=0$ case. The enhancement in PNJL case
indicates that more lepton pairs will be produced at low mass ($M/T < 4$) in sQGP 
with vector interaction, which would be appropriate for the hot and dense matter likely to be
produced at FAIR energies.

\section{Conclusion}
\label{sec:c1:conclusion}

In the present work, the behavior of the vector meson 
CF and its spectral representation have been studied 
within the effective model framework, viz. NJL and PNJL models.
PNJL model contains additional nonperturbative information through
PL fields than NJL model. In addition to this nonperturbative 
effect of PL, the repulsive I-V interaction
is also considered. The influence of such interaction on the correlator 
and its spectral representation in a hot and dense medium has been obtained 
using ring resummation known as 
Random Phase Approximation. The incorporation of vector interaction 
is important, in particular, for various spectral properties of the system 
at non-zero chemical potential. However, the value of the vector coupling
strength is difficult to fix from the mass scale which is higher than the maximum 
energy scale $\Lambda$ of the effective theory. So, we have made different choices of this 
vector coupling strength to understand qualitatively its effect on the various quantities
we have computed.

In absence of the I-V interaction, the static SF and the CF in NJL model become quantitatively equivalent to those of free field theory. In case of PNJL these quantities are different from both free and NJL case because of the presence of the nonperturbative PL fields that suppress the color degrees of freedom in the deconfined phase just above $T_c$. This suggests that some nontrivial correlation 
exist among the color charges in the deconfined phase. As an important consequence, the nonperturbative 
dilepton production rate is enhanced in the deconfined phase compared to the leading 
order perturbative rate. We note that the nonperturbative rate with zero chemical potential agree  well with the available LQCD data in quenched approximation. We also discussed the rate in presence of finite chemical potential and external momentum which could provide useful information if, in future, LQCD computes them at finite chemical potential and external momentum.

In presence of the I-V interaction, appropriate for hot but very dense medium 
likely to be created at FAIR GSI, it is found that the color degrees of freedoms are, further, suppressed
up to a moderate value of the temperature above the critical temperature implying a stronger correlation 
among the color charges in the deconfined phase. The CF, SF and its spectral
property, {\it e.g.}, the low mass dilepton rate are strongly affected in PNJL case than NJL case. Finally, some of our results presented in this work can be tested  when LQCD computes them, in future, with the inclusion of the dynamical fermions.

\chapter[Entanglement and dilepton rate]{Dilepton production rate with the entangled vertex}
\label{chapter:PRD}

In this chapter we re-explore our earlier study, discussed in previous chapter, on the vector meson spectral function (SF) and  
its spectral property in the form of dilepton rate in a two-flavour PNJL model in presence  of a strong entanglement
between the chiral and Polyakov loop (PL) dynamics. It is based on: {\em Vector meson spectral function and dilepton rate in
the presence of strong entanglement effect between the chiral and the Polyakov loop dynamics}, Chowdhury Aminul Islam,
Sarbani Majumder and Munshi G. Mustafa, {\bf Phys. Rev. D 92, 096002 (2015) }.

\section{Introduction}

The phase diagram of hot and/or dense system of quarks and gluons predicted by the QCD has invited a 
lot of serious theoretical  investigations for last few decades. The first prototype of the QCD phase 
diagram was conjectured in \cite{Cabibbo:1975ig} where it looked very simple; with the passage of time
more and more investigations culminated in a very complicated looking phase diagram with many exotic 
phases \cite{Fukushima:2010bq}. Nevertheless, the interest mainly revolved around  two phase 
transitions - one is the chiral phase transition  and the other one is the deconfinement transition.
If they do not coincide, exotic phases such as the constituent quark phase~\cite{Cleymans:1986cq,Kouno:1988bi} or the
quarkyonic phase~\cite{McLerran:2007qj,Hidaka:2008yy} may occur. So, an important question on the QCD thermodynamics 
is whether the chiral symmetry restoration and the confinement-to-deconfinement transition happen simultaneously or not. 
We note that chiral and deconfinement transitions are conceptually two distinct phenomena. Though LQCD simulation 
has confirmed that these two transitions occur at the same temperature \cite{Fukugita:1986rr} or almost at the 
same temperature \cite{Aoki:2006br}. Whether this is a mere coincidence or some dynamics between the two phenomena 
are influencing each other is not understood yet and is matter of intense current research exploration. 

To understand the reason behind this  coincidence a conjecture has been proposed in 
the article \cite{Sakai:2010rp}  through a strong correlation or entanglement between the 
chiral condensate ($\sigma$) and the PL expectation value ($\Phi$) 
within the PNJL model. Usually, in PNJL model, there is 
a weak correlation between the chiral dynamics $\sigma$ and the confinement-deconfinement 
dynamics  $\Phi$, which is in-built through the covariant derivative between quark and gauge fields.
With this kind of weak correlation the coincidence between the chiral and deconfinement 
crossover transitions~\cite{Sakai:2009dv,Ghosh:2006qh,Mukherjee:2006hq,Ghosh:2007wy,Ratti:2005jh,Deb:2009ng}
can be described but it requires  some fine-tuning of parameters, inclusion of the scalar type
eight-quark interaction for zero chemical potential $\mu$ and the vector-type four-quark interaction
for imaginary $\mu$. This reveals that there may be a stronger correlation 
between $\Phi$ and $\sigma$ than that in the usual PNJL model associated through the
covariant derivative between quark and gauge fields. Also, some recent analyses~\cite{Braun:2009gm,Kondo:2010ts} 
of the exact renormalization-group (ERG) equation~\cite{Wetterich:1992yh} suggest a strong entanglement 
interaction between $\Phi$ and $\sigma$ in  addition to the original entanglement through the covariant
derivative. Based on this the two-flavor PNJL model is further generalized~\cite{Sakai:2010rp} 
by considering the effective four-quark scalar type interaction with the coupling strength that 
depends on the PL field $\Phi$. The effective vertex in turn generates entanglement interaction 
between $\Phi$ and $\sigma$. Such generalization  of the PNJL model is known as EPNJL model~\cite{Sakai:2010rp}. This EPNJL model has been used to study the location of 
the tricritical point at real isospin chemical potential~\cite{Sakai:2010rp} and
on the location of the critical endpoint at real quark-number chemical potential~\cite{Sakai:2010rp,Friesen:2014mha,Sugano:2014pxa}.
It has also been used to study~\cite{Restrepo:2014fna} the effect of dynamical generation of a repulsive vector contribution to the quark pressure.
The EPNJL model has further been generalized to the three-flavor phase diagram~\cite{Sasaki:2011wu} as a function of light- and 
strange-quark masses for both zero and imaginary quark chemical potential.

The properties of the vector current CF and its spectral representation in the deconfined phase have been studied to understand the nonperturbative effect on the vector current spectral properties, e.g., the dilepton production rate in LQCD framework~\cite{Ding:2010ga}. In the previous chapter~\ref{chapter:JHEP}, within the PNJL model, we have analysed~\cite{Islam:2014sea} the effect of isoscalar-vector (I-V) interaction on the vector meson SF and spectral property (such as, dilepton production rate\footnote{We also note that both the dilepton and real photon rate have been computed in a matrix model of QGP by considering only the confinement effect~\cite{Gale:2014dfa,Hidaka:2015ima} and taking into account both the confinement and chiral symmetry breaking effects~\cite{Satow:2015oha}.} in a hot and dense medium.
In this present chapter, we consider the idea of the EPNJL model in which the effective vertex  generates a strong entanglement interaction 
between the chiral condensate $\sigma$ and the PL field $\Phi$ to re-explore  the vector SF and the spectral property such as the dilepton production rate previously studied in \cite{Islam:2014sea}. Because of this strong entanglement 
between $\Phi$ and $\sigma$, the coupling strengths run with the temperature and chemical potential. First we study the characteristics 
of mean fields with various constraints: with and without the I-V interaction in both PNJL and EPNJL models. Then we further 
demonstrate the effect of the entanglement on vector meson SF and dilepton rate.

This chapter is organized as follows: in section~\ref{sec:c2:eff_qcd_model} we briefly outline the usual PNJL model and extend it with the entanglement effect, namely the EPNJL model. In section~\ref{sec:c2:spect} we write the expression for the vector SF and its various spectral properties following our earlier calculation in~\cite{Islam:2014sea}. In section~\ref{sec:c2:results} we discuss our results and finally we conclude in section~\ref{sec:c2:concl}.
\section{Parameter fitting in effective QCD models}
\label{sec:c2:eff_qcd_model}

\subsection{PNJL Model}
\label{ssec:c2:pnjl}

We start with the two flavour PNJL model Lagrangian with I-V interaction~\cite{Islam:2014sea} given in equation (\ref{eq:pnjl_lag_vec}) and the corresponding thermodynamic potential is obtained as in (\ref{eq:pnjl_pot_vec}). Different parameters of the models are fitted as described in the section~\ref{sec:eff_models}. 

In the pure gauge theory the Polyakov potential is fitted to lattice QCD that yields a first order phase transition at $T_0=270$ MeV. 
With this value of $T_0$ for zero chemical potential we get, for 2-flavour case, almost a coincidence between the chiral and deconfinement 
transitions\footnote{We note that the chiral transition temperature $T_\sigma$ is obtained from the peak position of 
the  $\partial\sigma/\partial T$ whereas the deconfinement transition temperature 
$T_\Phi$ is that from the  $\partial \Phi/\partial T$.} 
($T_\sigma = 233$ MeV and $T_\Phi=228$ MeV). Thus the two transitions almost coincide [e.g., figure~\ref{fig:c2:Tsigma_Tphi_PNJL}]
but at a value higher than the range provided by 
the 2-flavour\footnote{It is worth mentioning here that the chiral transition temperature is found 
to be $ T_c= (154\pm 9)$ MeV in the recent (2+1) flavour LQCD computations by HotQCD collaboration~\cite{Bazavov:2011nk}.
In (2+1) flavor QCD the chiral order parameter contains both 
the light quark condensate and the strange quark condensate. 
Only the former is used to define the chiral transition temperature, as the strange condensate varies
very smoothly~\cite{Bazavov:2013yv}. Now, the behaviour of the light quark condensate
in (2+1) flavour and 2-flavour QCD will be similar
if the light quark masses are similar but  will be
different at quantitative level as it leads to two different chiral transition 
temperatures simply because one has {\it two different scales} in the theory.   
The value $ T_c= (154\pm 9)$ MeV was extracted~\cite{Bazavov:2011nk}  entirely in reference to the chiral 
phase transition for (2+1) flavour QCD. Further, we also note that the Wuppertal-Budapest 
collaboration~\cite{Borsanyi:2010bp} has also extracted three 
somewhat different values of $T_c$ ranging from $147$ MeV
to $157$ MeV, depending on the chiral observables considered for the purpose.
Since we  restrict our calculation only to 2-flavour case, we stick to the corresponding  
$T_c=(173\pm8$) MeV as extracted for 2-flavour case in LQCD simulation\cite{Karsch:2001cy,Karsch:2000kv}.} 
lattice QCD \cite{Karsch:2001cy,Karsch:2000kv} which is  
$T_\sigma\approx T_\Phi\approx (173\pm8)$ MeV. In 
Ref~\cite{Ratti:2005jh} the value of $T_0$ was changed to 190 MeV but keeping all the other parameters  same and obtained 
a lower value of $T_\sigma$ ($\approx200$ MeV) and $T_\Phi$ ($\approx170$ MeV) [e.g., figure~\ref{fig:c2:Tsigma_Tphi_PNJL(T0=190)}]. 
Taking the average of the two while defining $T_c$ 
gives a value almost within the range provided by the lattice QCD but then the coincidence is lost.
Here in this chapter we work with the same Polyakov potential but the entanglement between the chiral and deconfinement mechanism is introduced in the next subsection~\ref{ssec:c2:epnjl}.

\subsection{EPNJL}
\label{ssec:c2:epnjl}

We introduce the entanglement effect through effective vertices. The Lagrangian in EPNJL model will be the same as that in (\ref{eq:pnjl_lag_vec}) except that now the coupling constants $G_S$ and $G_V$ will be replaced by the effective ones $\tilde G_S(\Phi)$ and $\tilde G_V(\Phi)$. The forms of the effective vertices are chosen~\cite{Sakai:2010rp,Sugano:2014pxa} to preserve chiral and $Z(3)$ symmetry as given by
\begin{equation}
 \tilde{G}_S(\Phi)= G_S[1-\alpha_1\Phi\bar\Phi-\alpha_2(\Phi^3+\bar\Phi^3)]  , \label{eq:c2:entangle_Gs}
\end{equation}
and
\begin{equation}
 \tilde{G}_V(\Phi)= G_V[1-\alpha_1\Phi\bar\Phi-\alpha_2(\Phi^3+\bar\Phi^3)]  . \label{eq:c2:entangle_Gv}
\end{equation}
We  note that for $\alpha_1 =\alpha_2=0$, 
$\tilde{G}_S(\Phi)= G_S$ and $\tilde{G}_V(\Phi)= G_V$, the EPNJL model reduces to PNJL model. Also
at $T=0$, $\Phi={\bar \Phi}=0$ (confined phase), then $\tilde{G}_S= G_S$ and $\tilde{G}_V= G_V$. Due to the reason already mentioned in the section~\ref{sssec:njl_iv_int}, here again the strength of the vector interaction is taken in terms of the value of $G_S$ as $G_V=x\times G_S$, 
which on using (\ref{eq:c2:entangle_Gv}) reduces to 
\begin{equation}
\tilde{G}_V(\Phi)=x \times G_S[1-\alpha_1\Phi\bar\Phi-\alpha_2(\Phi^3+\bar\Phi^3)]=x \times \tilde{G}_S(\Phi) . \label{entangle_Gv_tilde}
\end{equation}
Now in EPNJL model, $\alpha_1$ and $\alpha_2$ are two new parameters, which are
to be fixed from the lattice QCD data. The thermodynamic potential $\Omega_{\rm{EPNJL}}$ in EPNJL model can be obtained from (\ref{eq:pnjl_pot_vec}) by replacing $G_S$ with $\tilde{G}_S(\Phi)$ and $G_V$ with $\tilde{G}_V(\Phi)$. For the EPNJL model we take same values of the parameters 
as those in PNJL model~\cite{Islam:2014sea} except the value of $T_0$, which is taken as 190 MeV. 
Then we fix the values of parameters $\alpha_1$ and $\alpha_2$  so as  to reproduce the coincidence of chiral and deconfinement 
transitions within the range given by lattice QCD data at zero chemical potential \cite{Karsch:2001cy,Karsch:2000kv} and it 
is found that ($\alpha_1$,~$\alpha_2$)= (0.1, 0.1). We further  mention that the coincidence of $T_\sigma$ and $T_\Phi$ are preserved [e.g., figure~\ref{fig:c2:Tsigma_Tphi_epnjl}] within the parameter region $\alpha_1$, $\alpha_2$ $\approx$ 0.10 $\pm$ 0.05. Note that the values $\alpha_1$ and $\alpha_2$ in our model differ from that of reference~\cite{Sakai:2010rp} because of the choice of different PL potential. We chose the form of the potential as given in reference~\cite{Ratti:2005jh} whereas that used in reference~\cite{Sakai:2010rp} is taken from reference~\cite{Roessner:2006xn}. It is also noteworthy that the two forms of PL potentials are consistent with each other in the validity domain of the model~\cite{Fukushima:2008wg}.

\section{Vector meson spectral function and dilepton rate}
\label{sec:c2:spect}

The resummed vector meson SF in presence of I-V interaction within ring approximation~\cite{Islam:2014sea} is written as
\begin{equation}
\sigma_V(\omega, {\vec q})=\frac{1}{\pi}\Big[{\rm Im}C_{00}(\omega, {\vec q})-{\rm Im}C_{ii}(\omega, {\vec q})\Big ],
\label{eq:c2:spectral_resum}
\end{equation}
where $Q\equiv(\omega, {\vec q})$, the four momentum of the vector meson. The imaginary part of the temporal ($C_{00}$) and spatial ($C_{ii}$) components of the resummed correlator are given as 
\begin{equation}
 {\rm Im C_{00}}=\frac{{\rm Im \Pi_{00}}}
 {\Big[ 1- {\tilde G}_V(\Phi) \Big(1-\frac{\omega^2}{{q}^2}\Big){\rm Re \Pi_{00}}\Big]^2+
 \Big[{\tilde G}_V(\Phi) (1-\frac{\omega^2}{{q}^2})
 {\rm Im \Pi_{00}}\Big ]^2},\label{eq.C00}
\end{equation}
and
\begin{align}
 {\mbox{Im}}C_{ii}=\frac{{\cal F}}
 {\left[1+\frac{{\tilde G}_V(\Phi)}{2}{\rm{Re}}\Pi_{ii}-\frac{{\tilde G}_V(\Phi)}{2}
 \frac{\omega^2}
 {{q}^2}{\rm{Re}}\Pi_{00}\right]^2
 +\frac{{\tilde G}_V^2(\Phi)}{4}\big[{\cal F}\big]^2}
 +\frac{\omega^2}{{q}^2}{\rm Im}C_{00}, \label{eq.CTprime}
\end{align}
respectively, which are obtained by replacing $G_V$ with the effective vertex $\tilde{G}_V(\Phi)$ in equations (\ref{eq:c1:C00}) and (\ref{eq:c1:Cii}) with ${\cal F}={\rm{Im}}\Pi_{ii}-\frac{\omega^2}
 {{q}^2}{\rm{Im}}\Pi_{00}$. The various expressions for one-loop self-energies, $\Pi_{00}$ and $\Pi_{ii}$, are explicitly computed in the previous chapter~\ref{chapter:JHEP}. The dilepton rate is obtained from the vector SF using the expression given in (\ref{eq:pre:rel_dilep_spec}).

\section{Results}
\label{sec:c2:results}

\subsection{Mean Fields}
\subsubsection{Without the isoscalar-vector interaction ($G_V=0$)}

The gap equation for the thermodynamic potential is 
\begin{eqnarray}
\frac{\partial \Omega_{(E)PNJL}}{\partial X}=0, \label{eq:c2:gap_equation}.
\end{eqnarray}
The thermodynamic potential is minimized with respect to mean fields $X$; with $X$ 
representing $\sigma,~ \Phi, ~\bar{\Phi}$ and $n$. In this section we compare the variations of the mean fields in PNJL model with that 
of EPNJL one without the effect of I-V interaction i.e. $G_V = 0$. As discussed in subsection~\ref{ssec:c2:pnjl}  the scalar 
type four-quark coupling strength ($G_S$)
in NJL/PNJL model is fixed along with three momentum cutoff $\Lambda$ and bare quark mass $m_0$ to reproduce known zero temperature 
chiral physics in the hadronic sector. We note that in principle it should depend on the parameters $T$ and $\mu$ but it is not usually considered in NJL model~\cite{Klevansky:1992qe,Hatsuda:1994pi}. 
However, in PNJL model the PL field ($\Phi$) is related to the temporal gluon 
which should make $G_S$ to depend on $\Phi$. But this dependence is also neglected 
in the same spirit~\cite{Fukushima:2003fw}. So, the value of $G_S$ remains 
fixed as represented by solid line in figure~\ref{fig:c2:couplings_Gv0}.

Now we pay attention to the features of EPNJL model in figure~\ref{fig:c2:couplings_Gv0}. As soon as one introduces the $\Phi$ dependence in the scalar coupling strength through (\ref{eq:c2:entangle_Gs}) in EPNJL model, it ($\tilde{G}_S$) becomes dependent on both $T$ and $\mu$. This running is due to  the gap equation in (\ref{eq:c2:gap_equation}), which is solved in a self-consistent manner for different mean fields.  As can be seen the increase in $T$ causes  $\tilde{G}_S$ to decrease for a given $\mu$ and the decrease becomes faster as one increases $\mu$. This can be understood from (\ref{eq:c2:entangle_Gs}) as  for a given $T$ if  one increases $\mu$, the PL fields ($\Phi$ and $\bar \Phi$) increase  and thus  $\tilde{G}_S$ decreases. figure~\ref{fig:c2:mass_fields_Gv0_Mu0} displays the temperature dependence of 
the scaled constituent quark mass and PL fields for both PNJL and EPNJL models at $\mu =0$. Here we  
mention that for $\mu=0$, $\Phi=\bar\Phi=|\Phi|$ \cite{Ratti:2005jh}. It clearly shows 
a considerable change in the chiral condensate $(\sigma=\langle\bar\psi\psi\rangle)$ 
and the PL fields in EPNJL model as compared to those in  PNJL model. For nonzero chemical potential similar behaviour of $\sigma$ and $\Phi, \bar\Phi$ is also observed. This is obviously 
due to the running of the coupling ${\tilde G}_S$ which is arising due to the entanglement effect as shown in Fig~\ref{fig:c2:couplings_Gv0}.

\begin{figure}[hbt]
\subfigure[]
{\includegraphics[scale=0.8]{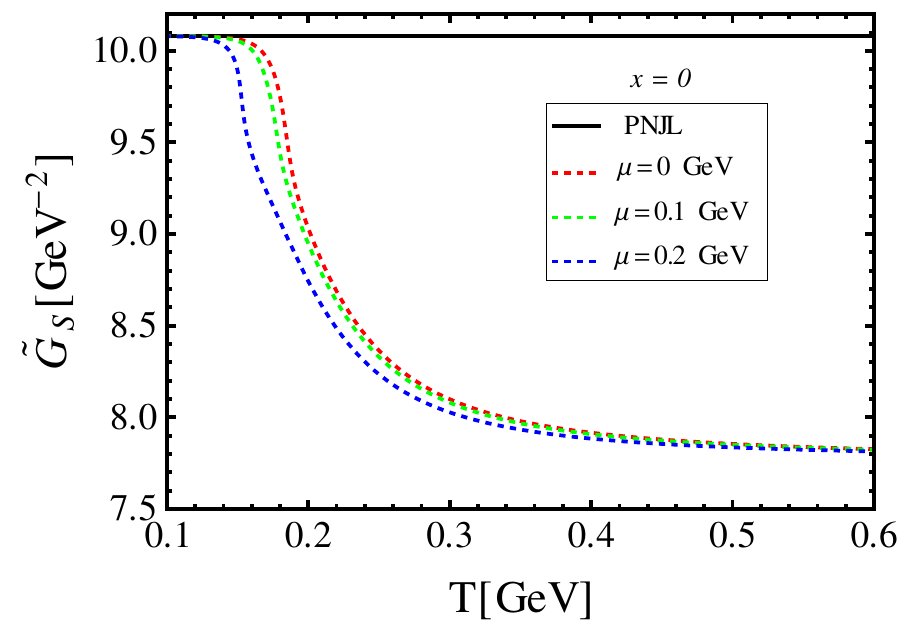}
\label{fig:c2:couplings_Gv0}}
\subfigure[]
{\includegraphics[scale=0.785]{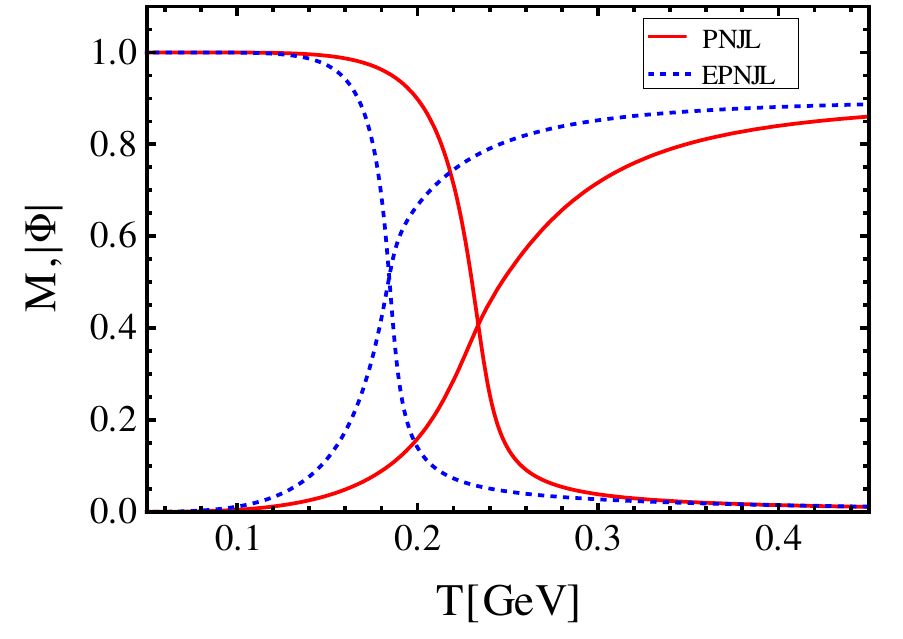}
\label{fig:c2:mass_fields_Gv0_Mu0}}
\caption{ Variation of (a) scalar type four-quark coupling strength ${\tilde G}_S(\Phi)$ with temperature $T$ 
for different values of $\mu$ and (b) the constituent quark mass scaled with its zero temperature value and Polyakov loop fields with 
$T$  for $\mu=0$ for both PNJL (solid lines) and EPNJL (dotted lines) model.}
\label{coupling_mass_fields_Gv0_Mu0}
\end{figure}

Figure~\ref{fig:c2:Tsigma_Tphi_Gv0_Mu0} displays the variations of $\frac{\partial \sigma}{\partial T}$ and 
$\frac{\partial\Phi}{\partial T}$ with the temperature at $\mu=0$ for various model conditions as discussed 
in subsections~\ref{ssec:c2:pnjl} and ~\ref{ssec:c2:epnjl} in details. We note that  $T_\sigma$ and $T_\Phi$ coincide for EPNJL model 
at $\approx 184$ MeV (e.g, figure~\ref{fig:c2:Tsigma_Tphi_epnjl}), which is almost within the range, $T_c=(173\pm8)$ MeV,  
given by the two flavour lattice  QCD~\cite{Karsch:2001cy,Karsch:2000kv}.

\begin{figure}[hbt]
\subfigure[]
{\includegraphics[scale=0.52]{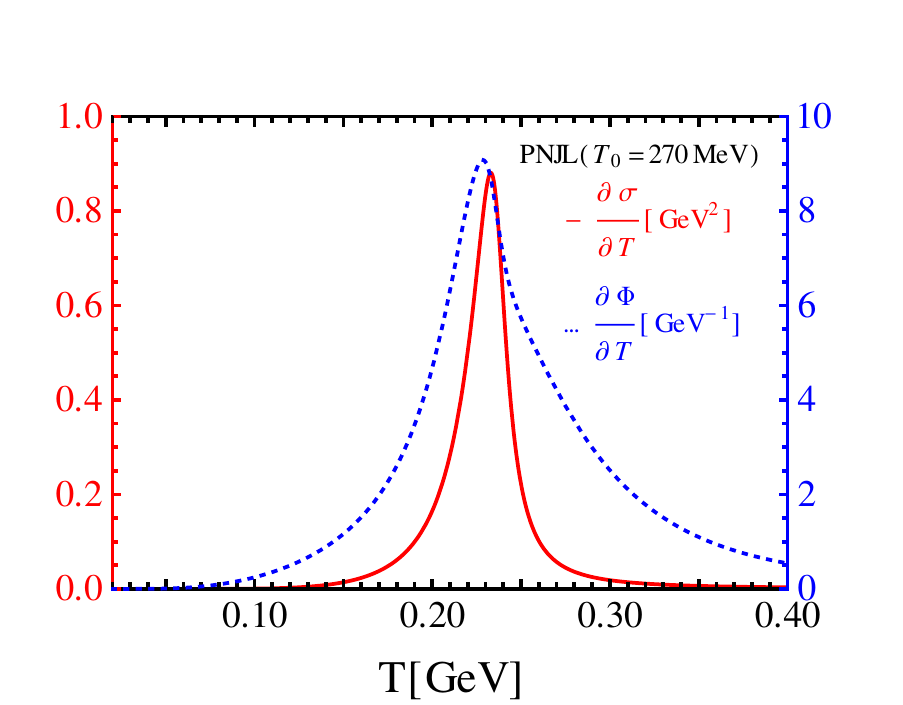}
\label{fig:c2:Tsigma_Tphi_PNJL}}
\subfigure[]
{\includegraphics[scale=0.52]{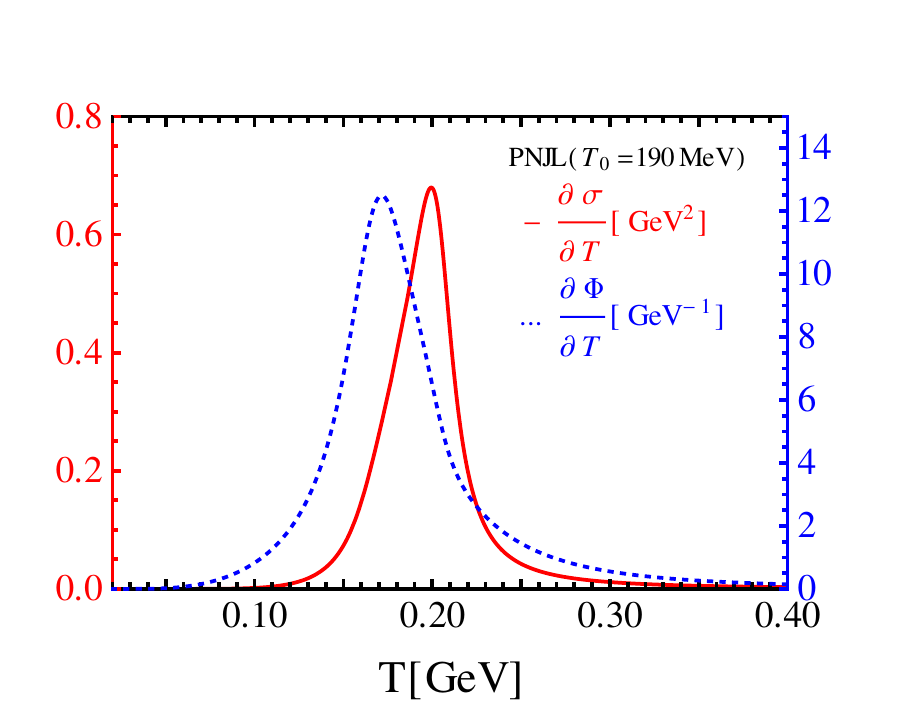}
\label{fig:c2:Tsigma_Tphi_PNJL(T0=190)}}
\subfigure[]
{\includegraphics[scale=0.52]{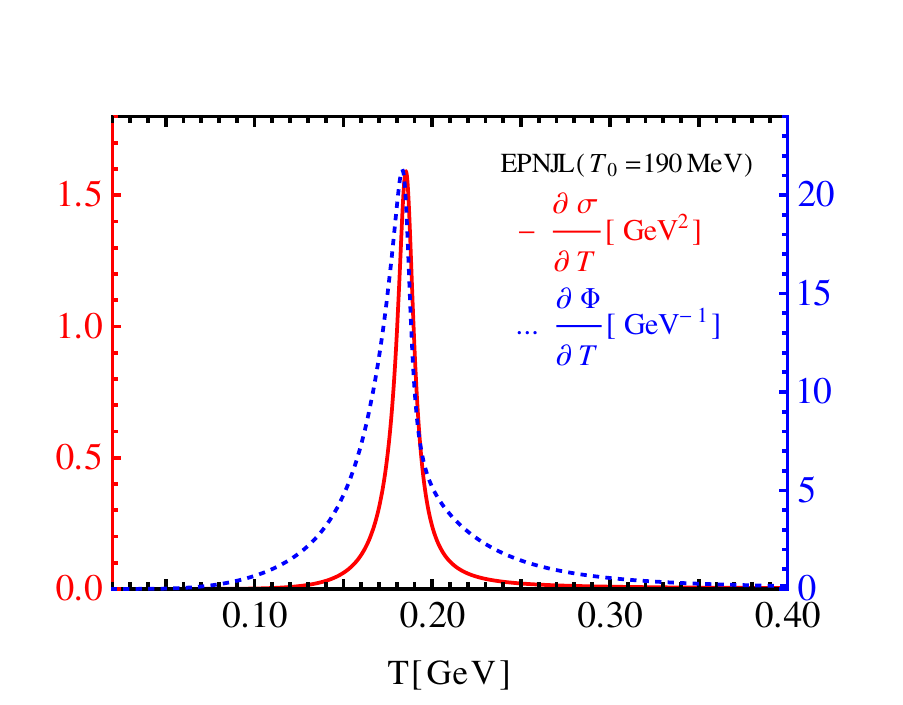}
\label{fig:c2:Tsigma_Tphi_epnjl}}
\caption{Plot of $\partial \sigma/\partial T$ and  
$\partial\Phi/\partial T$  as function of $T$ with $\mu=0$ 
for (a) PNJL model with $T_0=270$ MeV in~\cite{Ghosh:2007wy,Ratti:2005jh},
(b) PNJL model with $T_0=190$ MeV in reference~\cite{Ratti:2005jh} and
(c) present calculation in EPNJL model with $T_0=190$ MeV. For details it is referred to text 
in subsecs.~\ref{ssec:c2:pnjl} and ~\ref{ssec:c2:epnjl}, respectively.}
\label{fig:c2:Tsigma_Tphi_Gv0_Mu0}
\end{figure}

We note that once $\mu$ is introduced in the system the transition temperatures 
(both chiral and deconfinement) get reduced, which is expected. 
Now for a  given $T$ and $\mu\ne 0$,  $\Phi \ne \bar\Phi$ \cite{Dumitru:2005ng} generates
two separate but close values of inflection points leading to different  $T_\Phi$ and $T_{\bar\Phi}$. 
In that case one can take the average of $T_\Phi$ and $T_{\bar\Phi}$ as the deconfinement transition 
temperature. For $\mu=150$ MeV, we found  $T_\Phi=166$ MeV and 
$T_{\bar\Phi}=160$ MeV  and the average of the them ($163$ MeV)  is very close to 
the value of $T_\sigma=167$ MeV. With the  increase of $\mu$  the transition temperatures further get reduced; 
for example at $\mu=200$ MeV, $T_\Phi=153$ MeV, $T_{\bar\Phi}=151$ MeV and $T_\sigma= 153$ MeV.

\begin{figure}[hbt]
\begin{center}
 \includegraphics[scale=0.8]{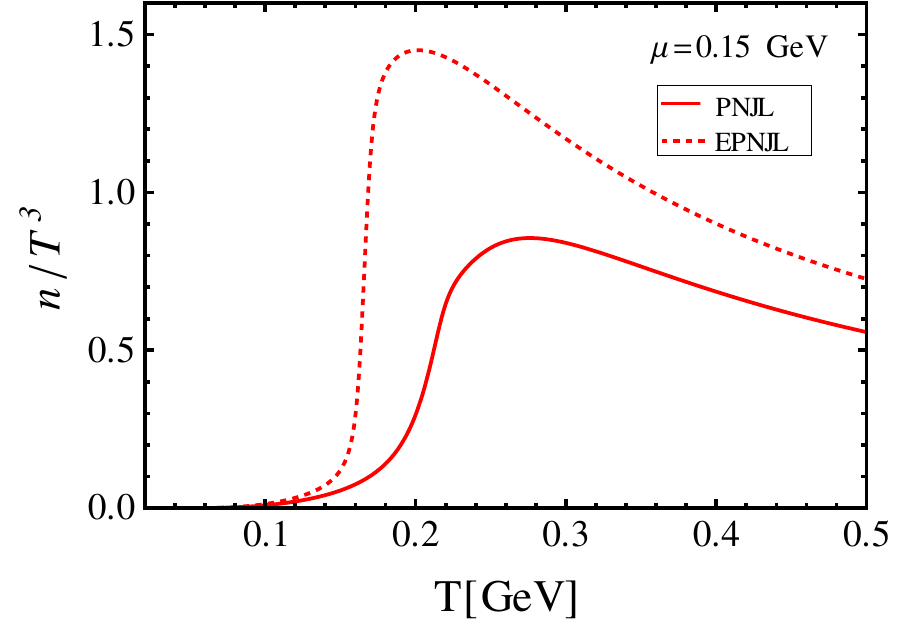}
\caption{Comparison of variations of scaled quark number density  between PNJL and EPNJL model for  $\mu=0.15$ GeV.}
\label{fig:c2:numberdensity_Gv0}
\end{center}
\end{figure}

We now discuss the differences in quark number density in EPNJL model with that of the PNJL one. 
In figure~\ref{fig:c2:numberdensity_Gv0} we observe that for temperature beyond 150 MeV the quark number density rises 
very sharply for EPNJL model as compared to PNJL one. This can be understood from figure~\ref{fig:c2:mass_fields_Gv0_Mu0} 
in which the value of PL field rises very sharply beyond $T=150$ MeV for EPNJL model. 
This indicates that the PL field provides a strong correlation among the quarks at low $T$ whereas 
the strength of the correlation among the quarks decreases when the value of the PL field increases 
at high $T$ and we have more and more free quarks in the system for EPNJL model as compared to the PNJL one.

\subsubsection{With the isoscalar-vector interaction ($G_V\neq0$)}

Now we deal with the same set up but the I-V interaction (${\tilde G}_V$) is turned on 
through (\ref{eq:c2:entangle_Gv}). In EPNJL model both couplings in (\ref{eq:c2:entangle_Gs}) and (\ref{eq:c2:entangle_Gv}) 
are entangled  and run with $T$ and $\mu$ by virtue of the gap equation in (\ref{eq:c2:gap_equation}).
We choose three different values of the strength of the I-V interaction to demonstrate its 
effects within the EPNJL model. These values are taken in terms of $\tilde{G}_S$ and the reason for which is 
already mentioned in the section~\ref{sec:c2:eff_qcd_model}.  

%in the EPNJL model 
%$\tilde{G}_S$ runs with the temperature and so does $\tilde{G}_V$.

\begin{figure}[hbt]
\subfigure[]
{\includegraphics[scale=0.8]{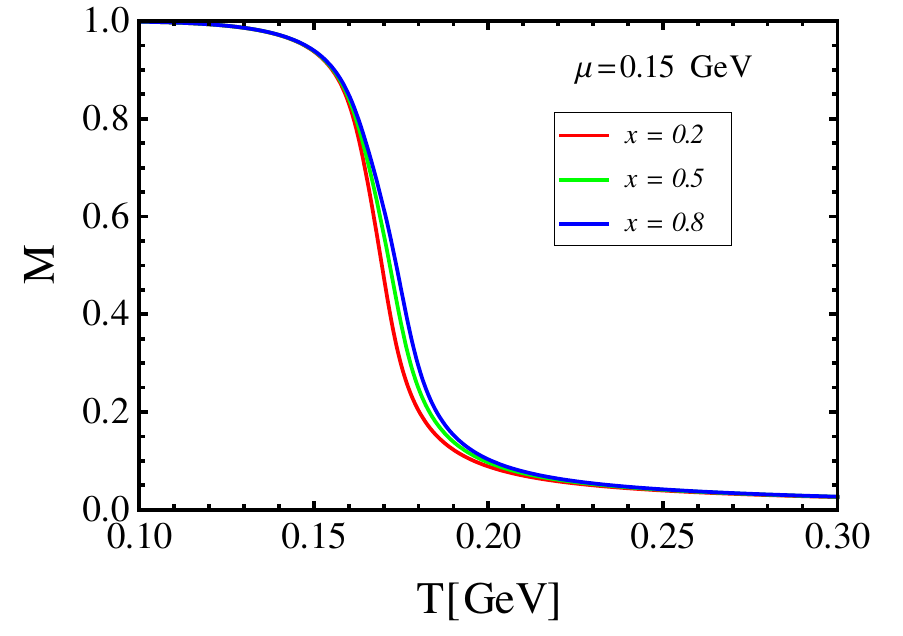}
\label{fig:c2:mass_differentGv}}
\subfigure[]
{\includegraphics[scale=0.785]{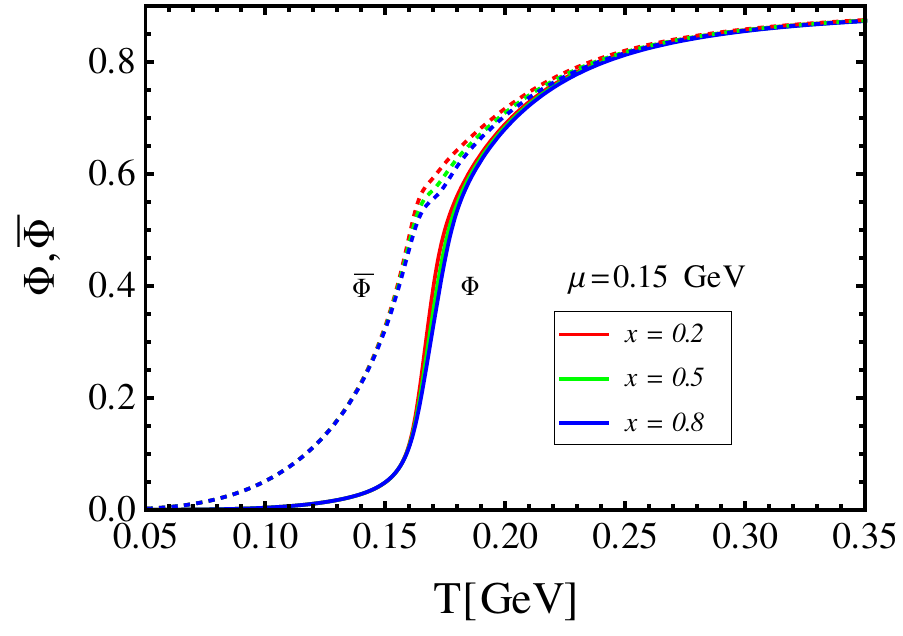}
\label{fig:c2:fields_differentGv}}
\caption{Variations of (a) scaled constituent quark mass and (b) Polyakov loop fields with temperature for 
three different values of $G_V$ at $\mu=0.15$ GeV in EPNJL model.}
\label{mass_fields_differentGv}
\end{figure}

In figure~\ref{fig:c2:mass_differentGv} the variation of the scaled constituent quark mass 
is shown for $\mu = 150$ MeV.  As one increases the strength of the vector interaction the rate of mass 
variation with the temperature  becomes slower.  
Since the couplings run in the EPNJL model the effect of the vector interaction is more prominent 
than that of the PNJL model with fixed values of couplings~\cite{Islam:2014sea}.
In the right panel 
(figure~\ref{fig:c2:fields_differentGv}) the variations of the PL fields with temperature at $\mu = 150$ MeV are shown. 
We observe that with the increase of the value of $G_V$ the rate of increase of PL fields with temperature 
decreases. The differences in the constituent quark masses or the PL fields for different values of $G_V$ are however
more prominent within the temperature range $165\leq T(\mathrm {MeV}) \leq210$.

We have already discussed the effects of chemical potential on the transition temperatures in the previous section. Here in 
Table~\ref{table_Tsigma_Tphi} we present the variations of 
the transition temperatures by the inclusion of the vector interaction. It shows that as we increase 
the strength of the vector interaction for the same chemical potential, the values of $T_\sigma$ and as well as the average of $T_\Phi$ 
and $T_{\bar\Phi}$ increase \cite{Friesen:2014mha}.

\begin{table}
\begin{center}
\begin{tabular}{cc|cccccc}
 \hline
 & Values of $G_V$ and $\mu$ && $T_\Phi$ && $T_{\bar\Phi}$ & $\frac{T_\Phi+T_{\bar\Phi}}{2}$ & $T_\sigma$\\
 \hline
  & $x=0$,~~ $\mu=150$ MeV && 166 && 160  & 163  & 167\\
 \hline
  & $x=0.2$, $\mu=150$ MeV && 167 && 160  & 163.5  & 170\\
 \hline
  & $x=0.5$, $\mu=150$ MeV && 168 && 159  & 163.5  & 173\\
 \hline
  & $x=0.8$, $\mu=150$ MeV && 169 && 159  & 164  & 175\\
 \hline
 
\end{tabular}
\end{center}
\caption{Values of $T_\sigma$, $T_\Phi$ and $T_{\bar\Phi}$ for different values of $G_V$ and $\mu=150$ MeV.}
\label{table_Tsigma_Tphi}
\end{table}

\begin{figure}[hbt]
\subfigure[]
{\includegraphics[scale=0.8]{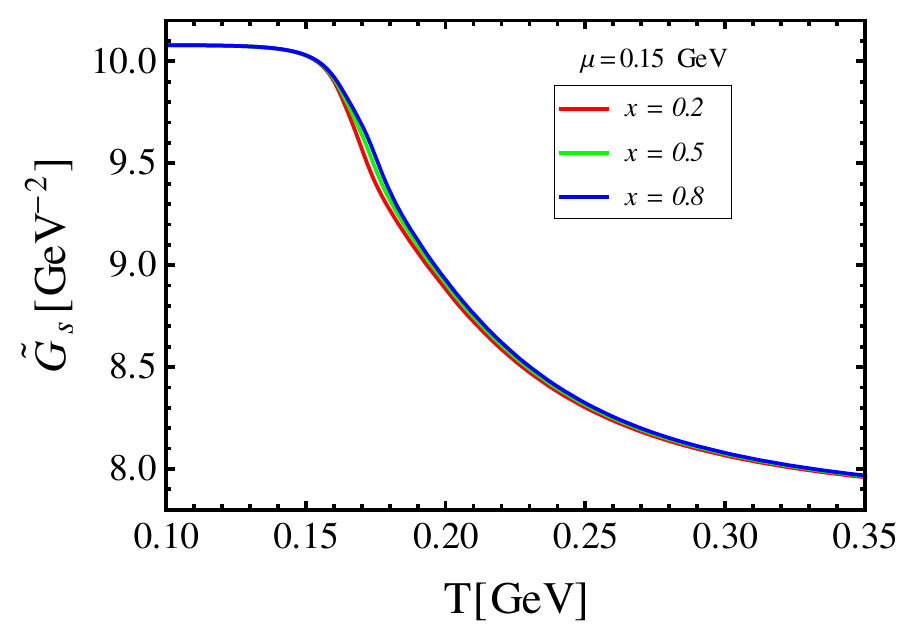}
\label{fig:c2:couplings_differentGv}}
\subfigure[]
{\includegraphics[scale=0.785]{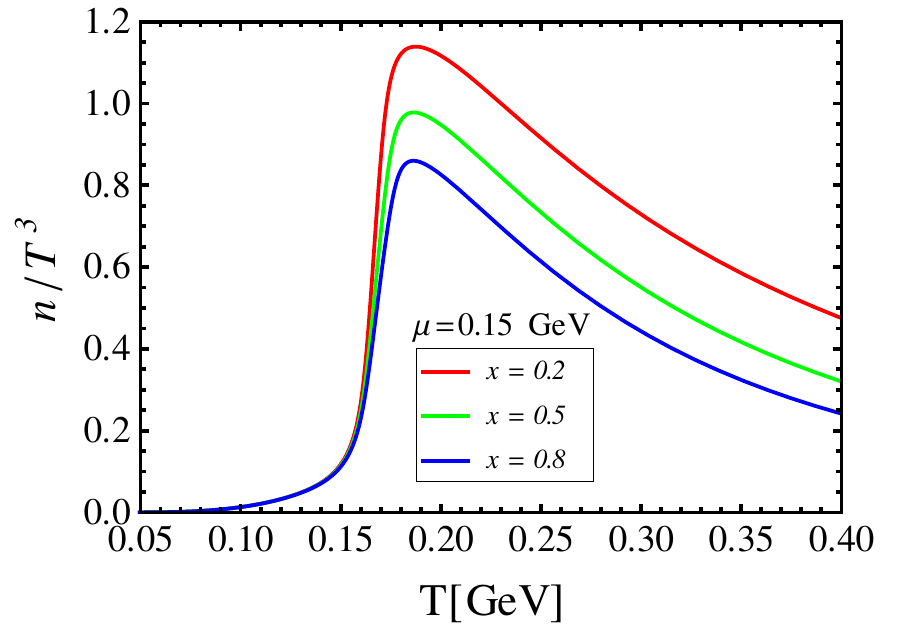}
\label{fig:c2:numberdensity_differentGv}}
\caption{Variations of  (a) scalar type four-quark coupling strength  and (b) scaled number density with 
temperature for three different values of $G_V$ at $\mu=0.15$ GeV in EPNJL model.}
\label{numberdensity_couplings_differentGv}
\end{figure}

In figure~\ref{fig:c2:couplings_differentGv} the variation of $\tilde{G}_S$ with temperature at ${\mu}=150$ MeV is shown for 
different value of $G_V$. It is found that the value of $\tilde{G}_S$ increases 
as the strength of the vector interaction increases for a given value of temperature and chemical potential. 
This can be understood from figure \ref{fig:c2:fields_differentGv} where the values of PL fields decrease
with increase of $G_V$. This in turn  leads to an enhancement of $\tilde{G}_S$ according to (\ref{eq:c2:entangle_Gs}).
In figure~\ref{fig:c2:numberdensity_differentGv} the variation of scaled quark number density with temperature is displayed for
same ${\mu}$ and  $G_V$ as in figure~\ref{fig:c2:couplings_differentGv}. For a given temperature and chemical potential the number density
is found to decrease with the increase of $G_V$. This is because the number of free quarks in the system is reduced
since the correlation among quarks increases due to the decrease  of PL fields  with  the increase of the couplings.

\subsection{Vector spectral function and dilepton rate}
\subsubsection{Without the isoscalar-vector interaction ($G_V=0$)}
Now we will be discussing the entanglement effect on the SF vis-a-vis the dilepton rates without the inclusion 
of the vector interaction but considering only the scalar type interaction. In  figure~\ref{fig:c2:spectral_T0.25_Mu0} the SFs 
with zero external momentum ($q$) and zero chemical potential ($\mu$) i.e. $q=\mu=0$, for PNJL and EPNJL model along with the free 
case are displayed  whereas those in  figure~\ref{fig:c2:spectral_T0.25_Mu0.2} are for $q=\mu=200$ MeV. The corresponding dilepton rates are 
shown in figure~\ref{fig:c2:dilepton_T0.25_Gv0}. Due to the entanglement effect through scalar type interaction the SF vis-a-vis dilepton 
rate for EPNJL model gets suppressed  compared to PNJL model but is still higher than the Born rate. This could be understood in the following way.  
\begin{figure}[hbt]
\subfigure[]
{\includegraphics[scale=0.8]{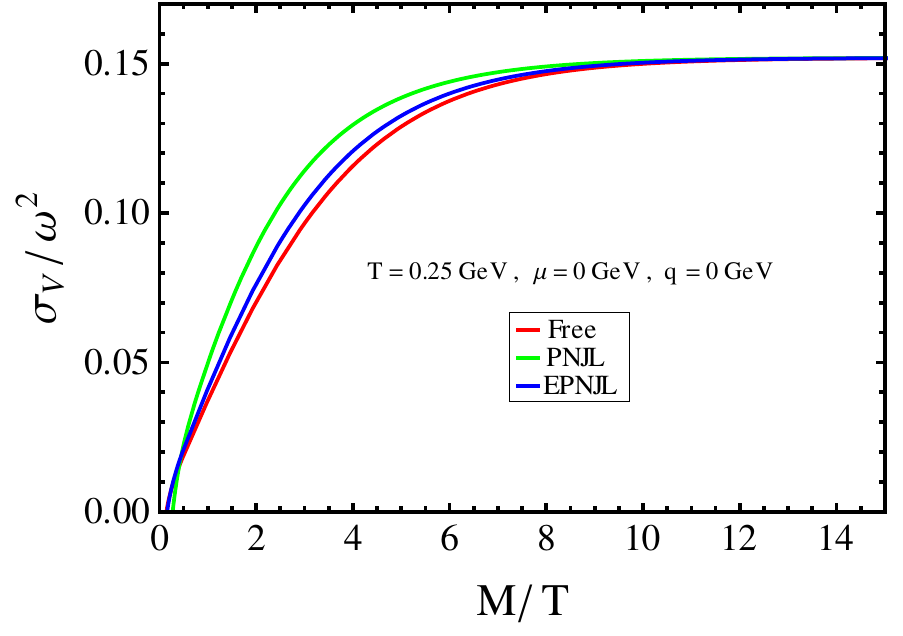}
\label{fig:c2:spectral_T0.25_Mu0}}
\subfigure[]
{\includegraphics[scale=0.8]{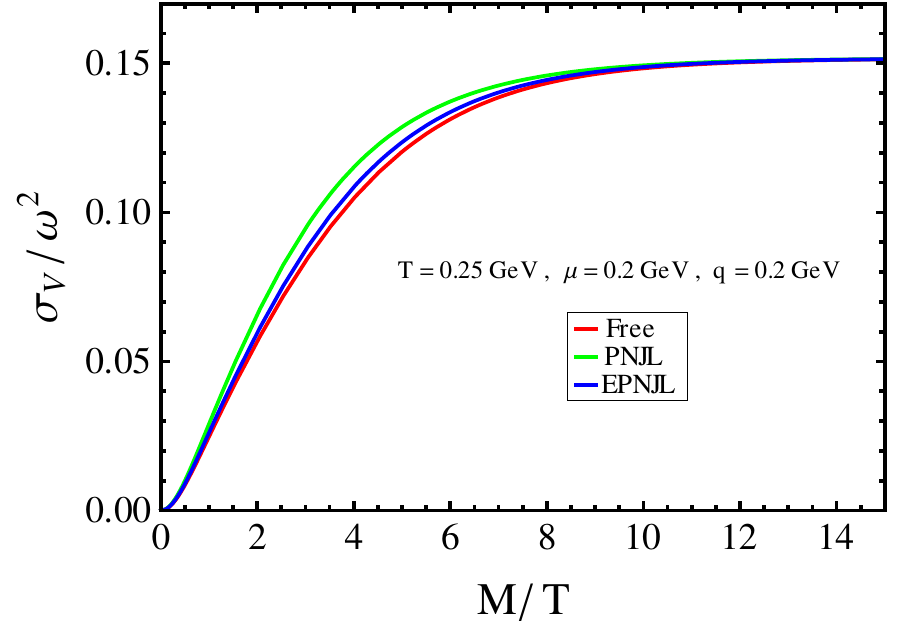}
\label{fig:c2:spectral_T0.25_Mu0.2}}
\caption{Scaled SF in PNJL and EPNJL model are compared with the free case for 
(a) $\mu=0$ and the three momentum $q=0$  and (b) $\mu=q=0.2$ GeV at $T=0.25$ GeV with $x=0$.}
\label{spectral_T0.25_Gv0}
\end{figure}
\begin{figure}[hbt]
\subfigure[]
{\includegraphics[scale=0.8]{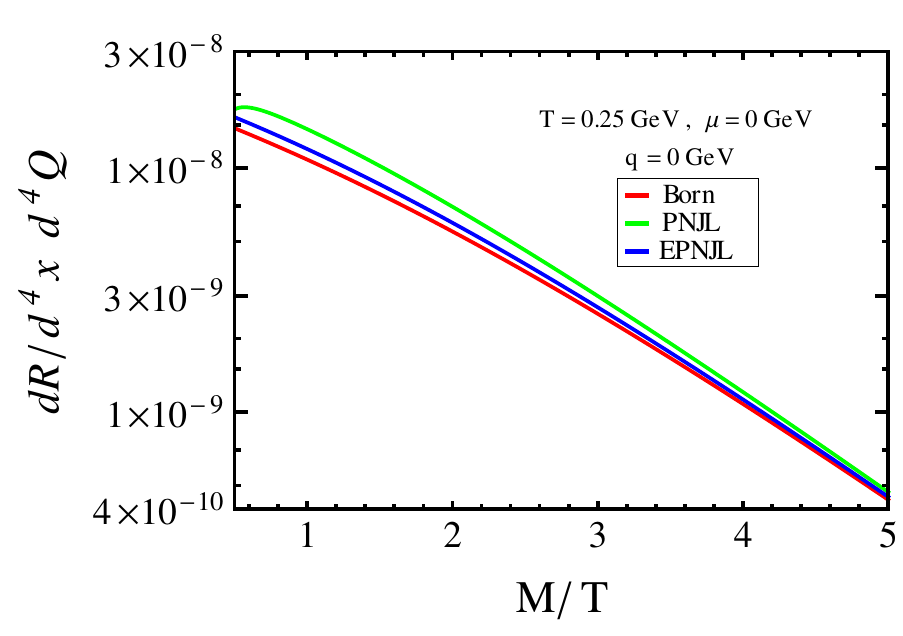}
\label{dilepton_T0.25_Mu0}}
\subfigure[]
{\includegraphics[scale=0.8]{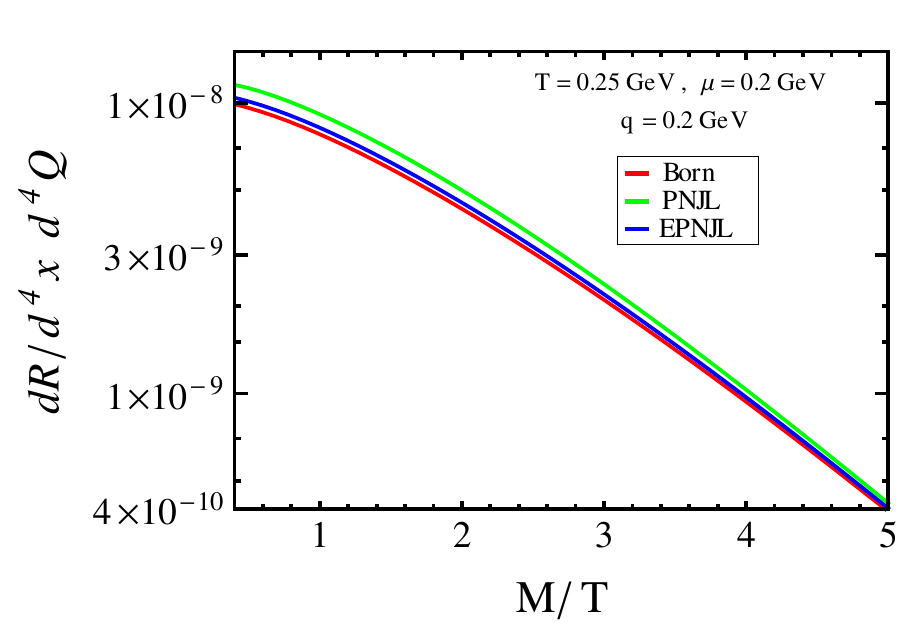}
\label{dilepton_T0.25_Mu0.2}}
\caption{Dilepton rate as a function of $M/T$ at $T=0.25$ GeV for PNJL and EPNJL model at (a) $\mu=q=0$ and (b) $\mu=q=0.2$ GeV with $x=0$.
The leading order perturbative dilepton (Born) rate is also shown.}
\label{fig:c2:dilepton_T0.25_Gv0}
\end{figure}
Usually the color degrees of freedom are suppressed in PNJL model due to the nonperturbative effect of 
the PL field that causes an enhancement~\cite{Islam:2014sea} of the dilepton rate compared to the Born one. As soon as the entanglement effect is introduced through the scalar type interaction that relatively enhances the color degrees of freedom in the system due to the running in ${\tilde G}_S$ as evident from figure~\ref{fig:c2:numberdensity_Gv0}, hence the dilepton rate is reduced 
compared to that in PNJL model.

\subsubsection{With the isoscalar-vector interaction ($G_V\neq0$)}

\begin{figure}[hbt]
\subfigure[]
{\includegraphics[scale=0.8]{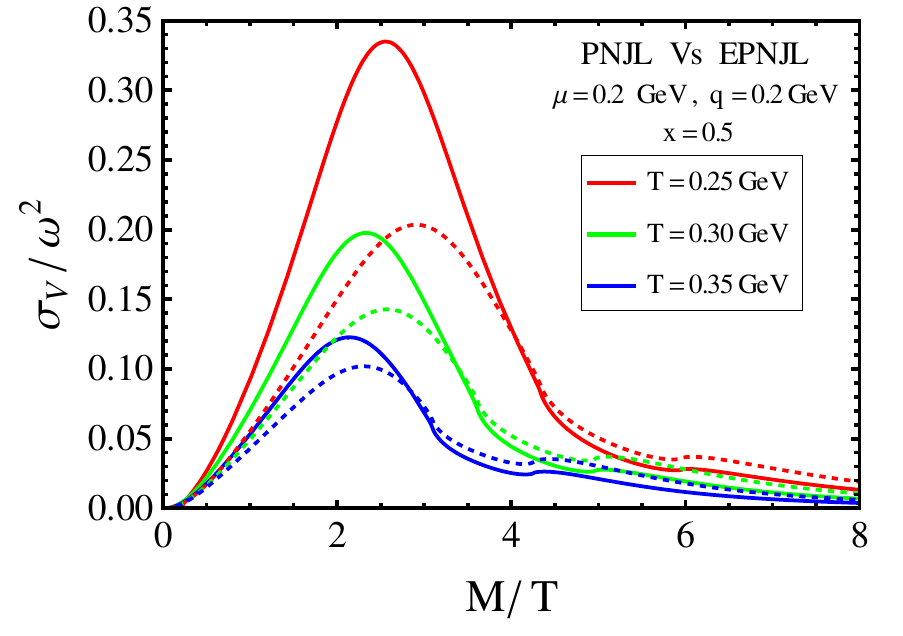}
\label{fig:c2:spectral_differentT}}
\subfigure[]
{\includegraphics[scale=0.78]{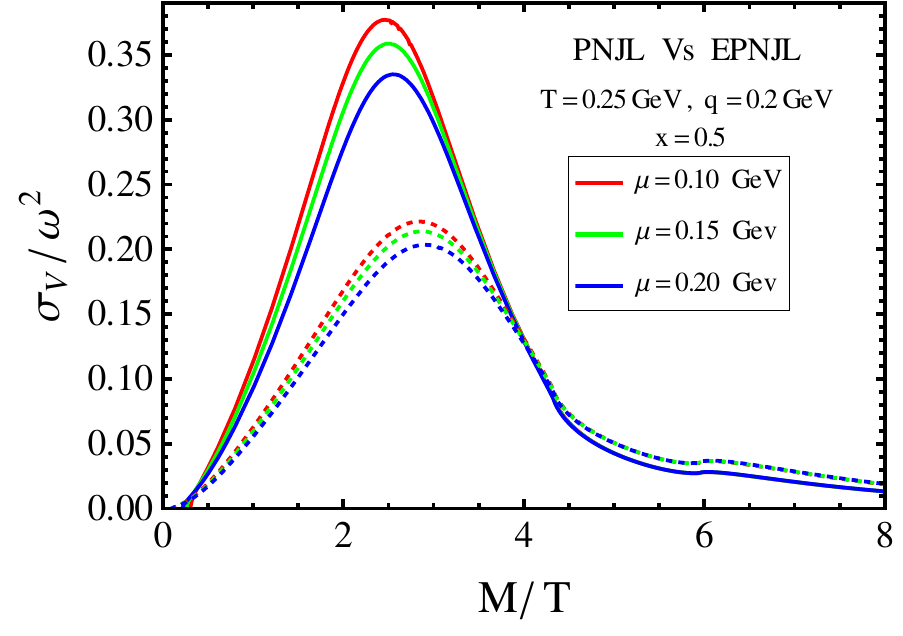}
\label{fig:c2:spectral_differentMu}}
\caption{Comparison between scaled SFs in PNJL (solid lines) and EPNJL (dotted lines) model for 
(a) a given chemical potential but different temperatures and (b) a given temperature but different chemical potentials 
at $x = 0.5$.}
\label{spectral_differentTMu}
\end{figure}

The free SF, in general, has a peak that appears at infinite value of $M$. This is also true for
four-quark scalar type interaction as seen above. However, in presence of I-V 
interaction $G_V$ the peak appears at finite $M$ in the resummed SF in (\ref{eq:c2:spectral_resum}) for given $G_V$ and $T$:
(a) below the kinematic threshold, $M <2M_f$, the resummed SF has a $\delta$-like peak due to the pole 
that can lead to bound state information of the vector meson and
(b) above the threshold  $M > 2M_f$ the resummed SF  picks
up a continuous contribution along with a somewhat broader peak\footnote{The width of the peak will depend on the value of $T$. If $T$ is around $T_c$ the peak will still be sharp around the threshold~\cite{Islam:2014sea}.}. We here concentrate on the continuous contribution ($M > 2M_f$) 
of the SF above $T_c$ that provides a finite width to a vector meson which decays to lepton pairs.
Now we focus on the effects of entanglement on SF and dilepton rate when the vector interaction is included
in addition to the scalar type interaction. In the left panel (figure~\ref{fig:c2:spectral_differentT}) the scaled SFs at $\mu=200$ MeV
in PNJL (solid line) and EPNJL (dotted line) model  are shown for three different values of $T$. 
In figure~\ref{fig:c2:spectral_differentT} the peak of the vector SF, for a given $T$ and $G_V$, 
is found to be suppressed and shifted to a higher $M$ in EPNJL model compared to PNJL one. This is purely due to the entangled vector 
interaction as the correlation among the quarks in the deconfined states becomes weaker in EPNJL model. 
In particular, the suppression is larger at lower value of $T$ and becomes smaller with the increase of
$T$.

The right panel (figure~\ref{fig:c2:spectral_differentMu}) displays the same quantity for three different values of $\mu$ but at a given
$T=250$ MeV. Comparison with the left panel reveals that the variation of the suppression of the SF due to entanglement is strongly temperature dependent than the chemical potential.

\begin{figure}[hbt]
\subfigure[]
{\includegraphics[scale=0.8]{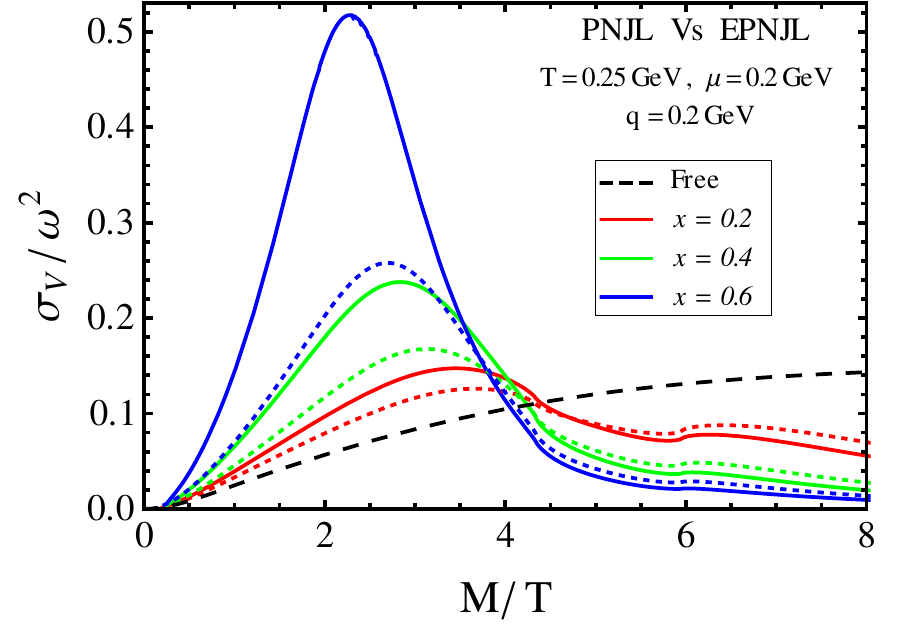}
\label{fig:c2:spectral_differentGv}}
\subfigure[]
{\includegraphics[scale=0.87]{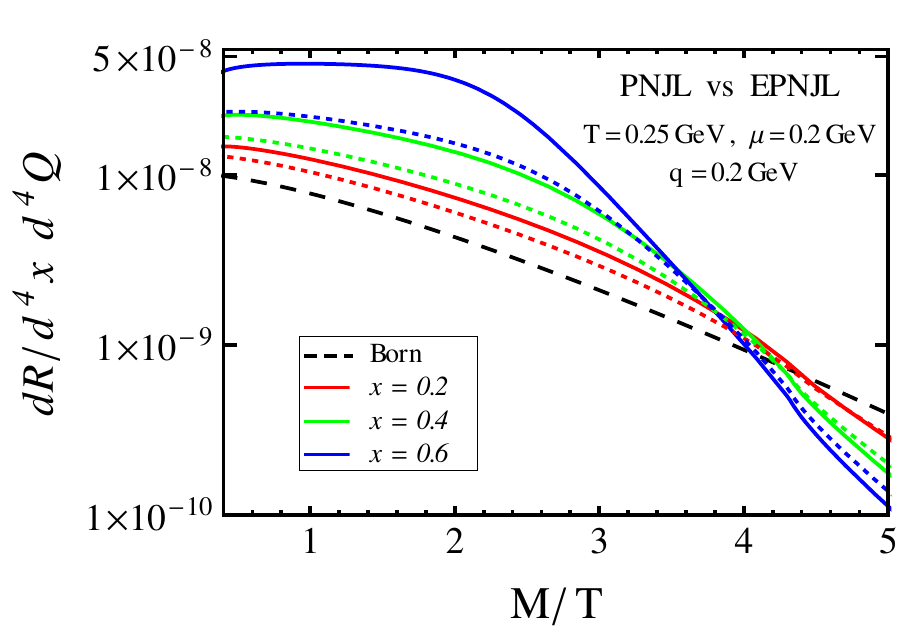}
\label{fig:c2:dilepton_differentGv}}
\caption{Plot of (a) scaled SF and (b) dilepton rate as a function of $M/T$ for PNJL (solid line) and EPNJL (dotted line) 
model for $T = 0.25$ GeV and $\mu=0.2$ GeV at three different choices of $G_V$.}
\label{spectral_dilepton_differentGv}
\end{figure}

In the left panel (figure~\ref{fig:c2:spectral_differentGv}) the SFs in PNJL (solid line) and EPNJL (dotted line) model 
at $T=250$ MeV and $\mu=200$ MeV for three different choices of $G_V$ are compared. As evident for any value of $G_V$ the strength 
of the SF for PNJL model is greater than that in the EPNJL one. In EPNJL model both couplings are strongly entangled 
through the mean fields and as one increases the strength of the vector interaction ($G_V$) that enhances the strength of the both 
running couplings. This in turn provides an enhancement in the strength of the SF that
decays to the dilepton pairs in the medium. The entanglement effect becomes more prominent than that with only the scalar 
type interaction. These features 
are well reflected in the right panel (figure~\ref{fig:c2:dilepton_differentGv}) where the corresponding dilepton rates in 
PNJL and EPNJL models are compared. For a given $G_V$ there is more lepton pairs at low mass in both EPNJL and PNJL model
compared to the leading order (Born) rate. Moreover, EPNJL model produces less lepton pairs than PNJL one. This is due to the entangled 
vector interaction that reduces the correlation among the quarks in the medium. However, as the strength of the vector interaction
increases, there is a relatively more dilepton production in both models.

\section{Conclusions}
\label{sec:c2:concl}

In general PNJL model contains nonperturbative information of confinement/decon-\\finement dynamics 
through the PL fields in addition to the chiral symmetry breaking dynamics. This model 
also employs the coupling of local scalar type four-quark interaction as well I-V 
interaction. The scalar type four-quark coupling strength is fixed along with three momentum cutoff $\Lambda$ and bare quark mass $m_0$ to reproduce known zero temperature chiral physics in the hadronic sector. However, the value of the vector coupling is difficult to fix and there exists ambiguity about 
its value as discussed. Nevertheless, the introduction of vector interaction in heavy-ion physics 
is important for study of the spectral property like dilepton rate at non-zero chemical potential. 
On the other hand, in nuclear astrophysics the formation of stars with quark matter core depends
strongly on the existence of a quark vector repulsion. However, in PNJL model both the couplings are  
considered to be constant in the literature. Since, this model also contains temporal gluons 
these couplings, in principle, should depend on the PL fields. But this dependence is
usually neglected and the correlation of the PL and chiral dynamics  
is a weak one as it arises through the covariant derivative that couples the quark and 
the temporal gauge field in the model. 

In this chapter we have extended the usual PNJL model by introducing a strong entanglement between 
the chiral ($\sigma$) and the PL dynamics ($\Phi$), known as EPNJL model in the literature. 
The strong entanglement has been introduced  via effective four-quark scalar type  
interaction that obeys the centre symmetry, $Z(3)$ of pure $SU(3)$ gauge group. 
Since the PL and chiral fields run with temperature and chemical potential,
the entanglement  makes also those coupling run. This entanglement effect is capable of 
reproducing the coincidence of chiral ($T_\sigma$) and deconfinement ($T_\Phi$) transition 
temperature within the range provided by the 2-flavour lattice data.

% The  spectral behaviour  of two point correlation function in terms of mesonic currents 
% describes the various properties of mesons. The vacuum properties of hadrons are well 
% studied in QCD. However, the vacuum properties of mesons are also affected in the medium due to
% the change of their dispersion properties in a medium. At finite temperature and density
% the hadron properties are also encoded in the structure of its correlation function and 
% corresponding spectral representation. Around the phase transition point the spectral behaviour may reflect 
% the degrees of freedom and thus the properties of the strongly interacting deconfined 
% hadronic matter. 
The SF of the vector current-current correlation is related to 
the production of lepton pairs, which is considered as an important probe of the deconfined hadronic 
matter and has been measured in  high energy heavy-ion experiments~\cite{Adare:2009qk,Adler:2006yt}. 
On the other hand,  at RHIC and LHC energies the maximum temperature reached of a 
hot and dense strongly interacting matter created is not very far from the phase transition 
temperature $T_c$ and is nonperturbative in nature. In LQCD framework the dilepton production 
rate~\cite{Ding:2010ga} at finite temperature but zero chemical potential has also been computed  using a SF obtained from Euclidean CF through a probabilistic method that involves certain uncertainties and intricacies~\cite{Islam:2014sea}. In the previous chapter~\ref{chapter:JHEP}, the influence of the four-quark scalar and I-V interaction without entanglement effect on the SF vis-a-vis the dilepton production was studied within PNJL model. In the present chapter we have updated the SF and the dilepton production rate within the EPNJL model that takes into consideration the entanglement between the PL and the chiral dynamics through scalar and vector interaction. In PNJL model both scalar and vector couplings do not run and the dominance of the PL fields substantially suppresses the color degrees of freedom around the phase transition temperature. On the other hand  EPNJL model introduces a strong entanglement between the chiral and the PL dynamics which relatively enhances 
color degrees of freedom in the deconfined phase compared to the PNJL model. Because of this the strength of the vector SF is suppressed and the peak is shifted to a higher energy compared to that of  PNJL model but the strength is higher than the free one at low energy. Since  the dilepton production  is related to the vector SF, it is also suppressed in EPNJL model compared to the PNJL model but is more compared to the Born rate (leading order perturbative one) in the deconfined phase. This indicates relatively less production of lepton pairs at low energy with entangled vector interaction. However, as the strength of the vector interaction is increased there is a relative increase in the strength of the SF in both EPNJL and PNJL model, which also results in a relatively more production in lepton pairs at low invariant mass.

\chapter[Vector meson correlator and QNS]{Vector meson correlator and the conserved density fluctuation}
\label{chapter:Susc}

In this chapter we compute the Euclidean correlation function (CF) in vector channel and the conserved density fluctuation associated with temporal CF appropriate for a hot and dense medium. The study of such conserved density fluctuation is performed under the influence of isoscalar-vector (I-V) interaction within the ambit of NJL, PNJL and EPNJL models. The whole discussion is based on two articles: partly on {\em Vector meson spectral function (SF) and dilepton production rate in a hot and dense medium within an effective QCD approach}, Chowdhury Aminul Islam, Sarbani Majumder, Najmul Haque and Munshi G. Mustafa, {\bf JHEP 1502 (2015) 011} and {\em Vector meson SF and dilepton rate in the presence of strong entanglement effect between the chiral and the Polyakov loop dynamics}, Chowdhury Aminul Islam, Sarbani Majumder and Munshi G. Mustafa, {\bf Phys. Rev. D 92, 096002 (2015)}.

%%%%%%%%%%%%%%%%%%%%%%%%%%%%%%%%%%%%%%%%%%%%%%%%%%%%%%%%%%%%%%%%%%%%%%%%%%%                         
\section{Introduction}  

Fluctuations of conserved charges are considered as appropriate signals for quark-hadron phase transitions~\cite{Asakawa:2000wh,Jeon:2000wg,Hatta:2003wn}, which basically bring forth the information about the degrees of freedom of the system under consideration. Thus these fluctuations are important to get a physical picture of the different phases of the strongly interacting matter. Considering the thermodynamical point of view, fluctuations of conserved charges are associated with susceptibilities. There can be diagonal and off-diagonal susceptibilities. The diagonal ones provide the correlations among the similar conserved charges whereas the off-diagonal ones present the same but for different conserved charges. These fluctuations and the corresponding susceptibilities are extensively studied in LQCD~\cite{Karsch:2001cy,Gottlieb:1987ac,Gavai:2001ie,Krieg:2013xwa}, HTL perturbation theory~\cite{Blaizot:2001vr,Chakraborty:2001kx,Chakraborty:2003uw,Chakraborty:2002yt,Haque:2014rua,Andersen:2015eoa} and in 
PNJL model~\cite{Ghosh:2006qh,
Roessner:2006xn,Sasaki:
2006ww,Mukherjee:2006hq}. The theoretical investigations are specifically being encouraged by the current bunch of HIC experiments, in particular beam energy scan, which provide more and more detailed accounts on fluctuations. 

We investigate the conserved density fluctuation associated with the temporal CF. Quark number density fluctuation is investigated via the quark number susceptibility (QNS) which is defined as the response of the quark number density with the infinitesimal change in quark chemical potential. As mentioned earlier it can be calculated from the temporal component of the current-current correlator using fluctuation-dissipation theorem~\cite{Hatsuda:1994pi,Kunihiro:1991qu}. We can also obtain it by taking the derivative of the thermodynamic potential with respect to the quark chemical potential. These two methods are equivalent~\cite{Ghosh:2014zra}. The compressibility of a system is directly related with the QNS at finite quark number density. Inclusions of vector mesons become particularly important for a system with finite density.  

At finite density QNS has been already studied in PNJL model~\cite{Ratti:2007jf}. Here we will investigate the QNS at finite density with and without the I-V interaction ($G_V$) for the two lightest flavors. The IV interaction has been considered through the random phase approximation. The effect of the vector interaction on the QNS has been critically reviewed in NJL and PNJL models and its implications for the hot and dense matter created in the HICs are mentioned. Then we further take our calculations forward and estimate the QNS in EPNJL model. We also compute the Euclidean CF in vector channel in NJL and PNJL models. Such correlators have been investigated in details in LQCD~\cite{Ding:2010ga,Hashimoto:1992np,Boyd:1994np} and as well as in HTL perturbative calculations~\cite{Karsch:2000gi,Haque:2011iz,Chakraborty:2001kx,Chakraborty:2003uw,Ding:2016hua}. The correlator can be calculated from the polarization diagram.

% Usually the bare quark propagators are used for the evaluation. In our calculation we replace the bare quark propagator by an effective one which is obtained through the effective model, namely PNJL. We then compare our findings with the available lattice data, which is found to agree well in certain domain. 

The chapter is organized as follows: in section~\ref{sec:c3:math_set_up} we briefly recapitulate the mathematical tools required for calculating the vector correlator and the QNS from the temporal correlator for with and without the I-V interaction. Then in section~\ref{sec:c3:res} we outline the results and finally we conclude in section~\ref{sec:c3:conclusion}.

\section{Mathematical set up}
\label{sec:c3:math_set_up}

The mathematical formula for calculating QNS is given in equation (\ref{eq:pre:qns}) in chapter~\ref{chapter:prelim} through thermodynamic sum rule. There we have both the expressions - one for obtaining it from the thermodynamical pressure and the other from the temporal correlator in the vector channel. We have particularly used the second derivative of pressure with respect to the quark chemical potential. But it can be equivalently estimated from the temporal correlator~\cite{Ghosh:2014zra}. While using the derivative of pressure with respect to the chemical potential to obtain the QNS within the effective mean field models, we have to be careful about the fact that the mean fields implicitly depend on the chemical potential. Thus we can no more take the explicit derivative, rather we have to take the total derivative with respect to the chemical potential, which is given by~\cite{Ghosh:2014zra}
\begin{eqnarray}
\chi_q &=&\frac{d^2 P}{d \mu_q^2}\nonumber\\
 &=&\dfrac{\partial^2  P}
{\partial \mu_q^2}
+\Big[\frac{\partial}{\partial \mu_q}
\Big(\frac{\partial P}{\partial \sigma}\Big)
+\frac{\partial}{\partial \sigma}
\Big(\frac{\partial P}{\partial \mu_q}\Big)\Big]
\cdot\frac{d \sigma}{d \mu_q}
+\frac{\partial P}{\partial \sigma}\cdot
\frac{d^2 \sigma}{d \mu_q^2}
+\frac{\partial^2 P}{\partial \sigma^2}\cdot
\Big(\frac{d \sigma}{d \mu_q}\Big)^2, 
\label{eq:c3:d2pdmu2b}
\end{eqnarray}
with the pressure $P$ obtained from the thermodynamic potential by using $P=-\Omega$. On the other hand, the vector correlator as well as its spatial and temporal components separately can be calculated from equation (\ref{eq:cor_tot}).

\subsection{Resummed correlator in ring approximation}
\label{ssec:c3:res_cor_ring}

The resummed vector correlator is obtained using equation (\ref{eq:cor_tot}), where $\sigma_H(\omega,  \vec q)$ has been replaced by the $\sigma_V$ in (\ref{eq:c1:spectral_resum}). The resummed temporal correlator i.e. the QNS can be obtained in terms of the real part of the temporal CF, $C_{00}$ following the equation (\ref{eq:pre:qns}). The real part of $C_{00}$ is obtained (vide appendix~\ref{app:susc}) from (\ref{eq:c1:c00}) as 
\begin{eqnarray}
{\rm{Re}}C_{00}(\omega,\vec q) = \frac{{\rm{Re}}\Pi_{00}(\omega, \vec q)+G_V\left (\frac{\omega^2}{ q^2}-1\right){\cal I}(\omega,\vec q)}
 {1+2G_V{\left (\frac{\omega^2}{ q^2}-1\right ){\rm{Re}}\Pi_{00}(\omega, \vec q)}
 +\left(G_V(\frac{\omega^2}{ q^2}-1)\right)^2{\cal I}(\omega,\vec q)}, 
 \label{eq_re_c00}
\end{eqnarray}
where ${\cal I}(\omega,\vec q)=\big({\rm{Re}}\Pi_{00}(\omega, \vec q)\big)^2+\big({\rm{Im}}\Pi_{00}(\omega, \vec q)\big)^2$.  Now the resummed QNS in ring approximation becomes
\begin{equation}
\chi^R_{q}(T,\tilde\mu)=-\lim_{\vec{q}\rightarrow 0} {\rm{Re}} C_{00}(0,\vec q) 
=\frac{\chi_q(T,\tilde \mu)}{1+G_V \chi_q(T,\tilde\mu)},
\label{eq:c3:resum_suscp} 
\end{equation}
where we have used that $\lim_{ {\vec q}\rightarrow 0}{\rm{Im}}\Pi_{00}(0, \vec q)=0$ and for one-loop the QNS is given by $\chi_q(T,\tilde\mu)=-\lim_{\vec{q}\rightarrow 0}{\rm{Re}}\Pi_{00}(0, \vec q).$ 

Now, we also note that at $G_V=0$, the resummed susceptibility in ring approximation reduces to that of one-loop. To compute the resummed one (\ref{eq:c3:resum_suscp}), we just need to compute one-loop (vide figure~\ref{fig:c1:corr_oneloop}) vector self-energy within the effective models considered here.

\section{Results}
\label{sec:c3:res}

\subsection{Vector correlator in PNJL model}

The vector CF can be obtained using ({\ref{eq:cor_tot}) in the scaled
Euclidean time $\tau T \in [0, 1]$. We note that
the CF in the $\tau T$ range is symmetrical around $\tau T=1/2$  
due to the periodicity condition in Euclidean time guaranteed by the kernel 
$\cosh[\omega (2\tau T-1)/2T]/\sinh(\omega/2T)$ in 
(\ref{eq:cor_tot}).

\begin{figure} [!htb]
\vspace*{0in}
\begin{center}
\includegraphics[scale=0.9]{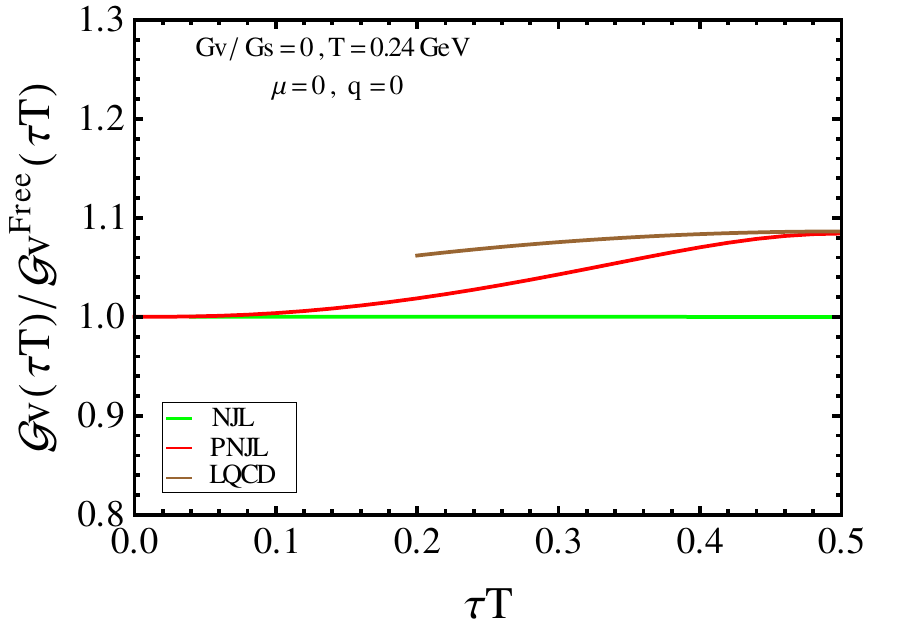}
\end{center}
\caption{Comparison of various scaled Euclidean CF with respect to that of 
free field theory, 
${\cal G}_V(\tau T)/{\cal G}_V^{\textrm{free}}(\tau T)$, as a function of the scaled Euclidean time
$\tau T$ for $T=240$ MeV with external momentum 
$q=0$, quark chemical potential $\mu=0$ and $G_V/G_S=0$. The continuum extrapolated 
LQCD result is from Ref.~\cite{Ding:2010ga}.}
\label{fig:c3:corr_gv0_q0_mu0}
\end{figure}
%%%%%%%%%%%%%%%%%%%%%%%%%%%%%%%%%%%%%%%%%%%%%%%%%%%%%%%%%%%%%%%%%%%%%%%%%%

%%%%%%%%%%%%%%%%%%%%%%%%%%%%%%%%%%%%%%%%%%%%%%%%%%%%%%%%%%%%%%%%%%%%%%%%%%%%%
\begin{figure} [!htb]
\begin{center}
\includegraphics[scale=0.9]{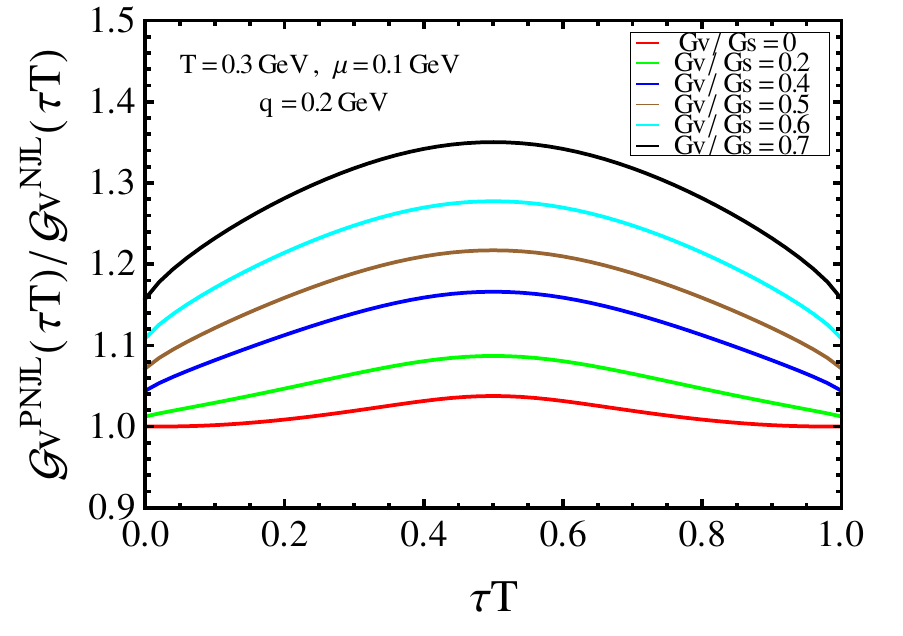}
\end{center}
 \caption{Ratio of Euclidean CF in PNJL model to that of NJL model
 as a function of $\tau T$  at $T=300$ MeV,  $\mu=100$ MeV and $q=200$ MeV with a set of values of $G_V/G_S$.}
\label{fig:c3:corr_pnjl_njl}
\end{figure}
%%%%%%%%%%%%%%%%%%%%%%%%%%%%%%%%%%%%%%%%%%%%%%%%%%%%%%%%%%%%%%%%%%%%%%%%%%%%%%%%%%%%%%%%%%

In figure~\ref{fig:c3:corr_gv0_q0_mu0} a comparison of the ratio of the vector CF to that of 
free one is displayed at $T=240$ MeV, $G_V/G_S=0$, $\mu=0$ and $q=0$ for NJL, PNJL and the continuum extrapolated 
LQCD data~\cite{Ding:2010ga} in quenched approximation. It is plotted in the $\tau T$ range $[0,1/2]$ because 
LQCD data are available for the same range. As seen the NJL case becomes
equal to that of free one as there is no effects from background gauge fields. On the other hand
the PNJL results at $\tau T=0$ and $1$ become similar to those of free case because 
$\sigma_V^{\textrm{PNJL}}(\omega)/\sigma_V^{\textrm{free}}(\omega)=1$ as $\omega \rightarrow \infty$.  
Around $\tau T=1/2 $ it deviates maximum from the free case due to the difference in SF
at small $\omega $ and thus a nontrivial correlation exists among color charges due to the presence 
of the nonperturbative Polyakov loop (PL) fields. This features are consistent with those in
the SF in figure~\ref{fig:c1:spect_gv0_q0_mu0} and the dilepton rate in figure~\ref{fig:c1:dilep_gv0_q0_mu0}. 
In contrary, the CF around $\tau T=1/2$ agrees\footnote{This is true at low $\omega$ with the lattice data as long as we are dealing with the Euclidean correlator. But when we compare the SF at low $\omega$ it differs from that of LQCD. Because of difficulties in performing analytic continuation in lattice calculation the continuous-time SF is obtained from discrete Euclidean correlator, using a finite set of data, through a probabilistic method known as maximum entropy method. Also one needs to rely on some ansatz for the extraction of SF at low energy in such methods. So the lattice computation of SF involves some amount of uncertainties.} better with that of LQCD in 
quenched approximation~\cite{Ding:2010ga}.

In figure~\ref{fig:c3:corr_pnjl_njl}, we display the effect of the vector interaction $G_V$ in
addition to the presence of the PL fields in QGP through the ratio of the correlation 
function in PNJL model to that of NJL model. Here we have displayed the result in the full range
of the scaled Euclidean time $\tau T \in [0,1]$. As discussed the ratio is symmetric around
$\tau T=1/2$ and always stay above unity. The ratio increases with the increase of the strength of the vector interaction.
This is due to the fact that PNJL CF 
is always larger than that of the NJL case since   
$\sigma_V^{\textrm{PNJL}}(\omega,\vec q)/\sigma_V^{\textrm{NJL}}(\omega,\vec q)>1$, and it is even stronger
in particular at small $\omega$ (see figure~\ref{fig:c1:spect_t3_q2_mu1}).  
This indicates that the color 
charges maintain a strong correlation  among them due to the presence of both 
PL fields and the vector interaction.  Thus the vector meson retains its 
bound properties in the deconfined phase.

\subsection{Quark number susceptibility in NJL and PNJL models}

%%%%%%%%%%%%%%%%%%%%%%%%%%%%%%%%%%%%%%%%%%%%%%%%%%%%%%%%%%%%%
\begin{figure} [!htb]
\subfigure[]
{\includegraphics[scale=0.8]{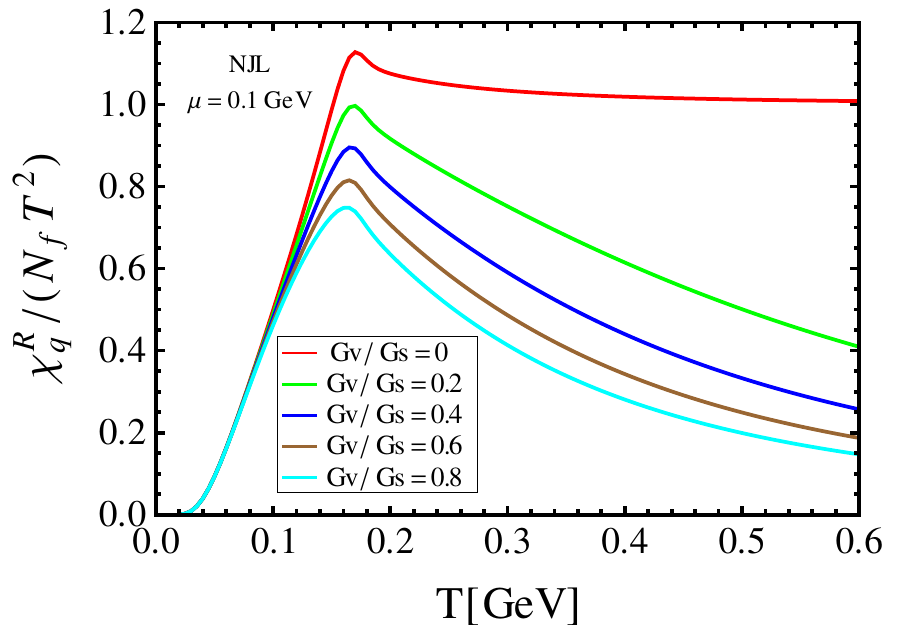}
\label{fig:c3:suscp_njl}}
\subfigure[]
 {\includegraphics[scale=0.8]{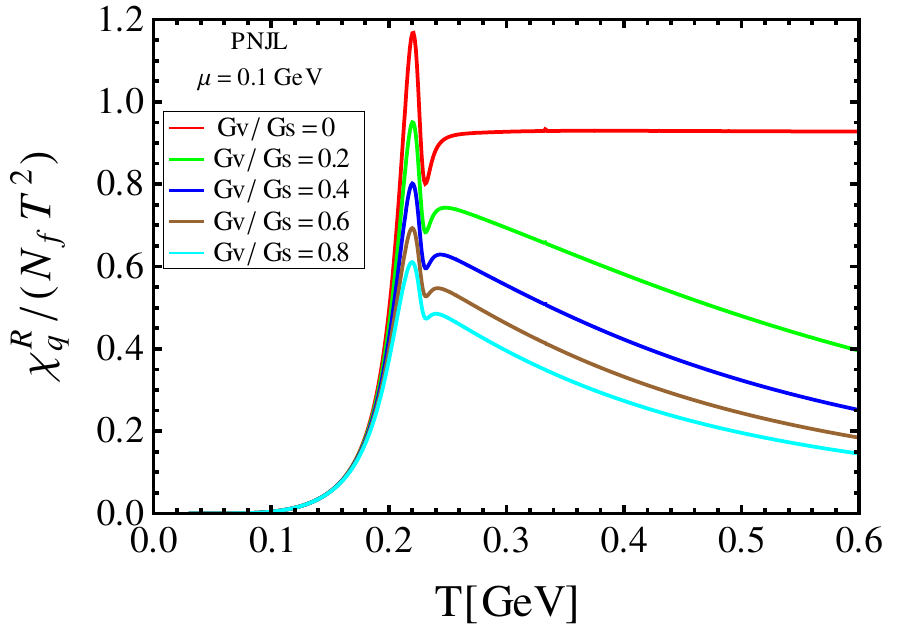}
 \label{fig:c3:suscp_pnjl}}
 \caption{Resummed QNS in (a) NJL model
 and (b) PNJL model at non-zero chemical potential for two flavor $(N_f=2)$.}
\label{fig:c3:sus_njl_pnjl}
\end{figure}
%%%%%%%%%%%%%%%%%%%%%%%%%%%%%%%%%%%%%%%%%%%%%%%%%%%%%%%%%%%%%%%%% 

Now, we can calculate QNS associated with 
the temporal part of the vector SF through the
conserved density fluctuation as given in (\ref{eq:c3:resum_suscp}). 
The resummed susceptibility for a set of values of $G_V$ at finite
quark chemical potential ($\mu=100$ MeV) is shown in  
figure~\ref{fig:c3:sus_njl_pnjl}.
For positive vector coupling $G_V$ the denominator of 
(\ref{eq:c3:resum_suscp}) is always greater than unity and
as a result the resummed susceptibility gets suppressed as 
one increases $G_V$. Since positive $G_V$ implies a repulsive
interaction, the compressibility of the system decreases
with increase of $G_V$, hence the susceptibility as seen from
figure~\ref{fig:c3:sus_njl_pnjl} decreases. We note that 
the QNS at finite $\mu$ shows an important feature around
the phase transition temperature than that at $\mu=0$~\cite{Ghosh:2014zra}. 
This is due to the fact that the mean fields  ($X=\sigma, \ \, \Phi$ and $\bar \Phi$)  
which implicitly depend on $\mu$ contributes strongly as the change 
of these fields are most significant around the transition region. 
This feature could be important in the perspective of FAIR scenario
where a hot but very dense matter is expected to be created.
Using (\ref{eq.eucl_chiq}) one can compute the temporal Euclidean 
CF associated with the QNS as ${\cal G}^E_{00}(\tau T)/T=-\chi^R_q(T)$, which 
does not depend on $\tau$ but on $T$. This study could also provide useful information to
future LQCD calculation at finite $\mu$.

\subsection{Quark number susceptibility in EPNJL model}
\label{ssec:c3:qns_epnjl}

Compressibility of a system is proportional to QNS~\cite{Hatsuda:1994pi}. So at high temperature when the susceptibility increases, the compressibility increases along with it and the system becomes more and more compressible which implies the system being either attractive or weakly repulsive. Once we include the vector interaction, which becomes important specifically for a system with finite quark number density, the susceptibility is governed by the strength of the the vector interaction ($G_V$). Here in our calculation we have assumed a repulsive vector interaction. Thus as we increase its strength the system becomes lesser and lesser compressible and the QNS decreases (figure~\ref{fig:c3:sus_njl_pnjl}).

\begin{figure} [!htb]
\begin{center}
\includegraphics[scale=0.8]{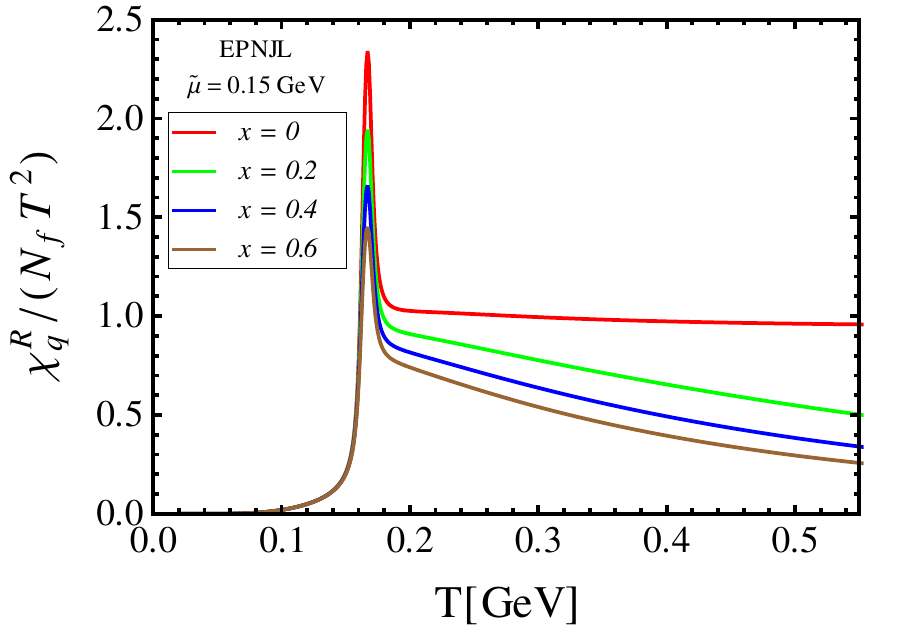}
\caption{Resummed QNS in EPNJL model at non-zero chemical potential for two flavor $(N_f=2)$.}
\label{fig:c3:sus_epnjl}
\end{center}
\end{figure}

Up to this point the line of argument is appropriate, but there is one problem with the results that we have obtained. In general the QNS should decrease with the increase of $G_V$ but then it should approach to Stefan-Boltzmann limit as one increases the temperature. It is obvious from the figure~\ref{fig:c3:sus_njl_pnjl} that this is not the case. The reason is the inclusion of a constant strength of the vector interaction that neither depends on the temperature nor on the chemical potential. But in principle, the coupling strengths ($G_S$ and $G_V$ both) should depend both on $T$ and $\mu$; since in QCD the coupling strength runs. 
 
This problem was supposed to be solved in the EPNJL model, where due to the entangled vertex the coupling strengths run as shown in figures~\ref{fig:c2:couplings_Gv0} and~\ref{fig:c2:couplings_differentGv} in section~\ref{sec:c2:results} of chapter~\ref{chapter:PRD}. But from figure~\ref{fig:c3:sus_epnjl} with running coupling within EPNJL, one could see that the problem is not fully resolved but becomes less severe. 
 
However, this is because of the choice of the ansatzes (vide equations~\ref{eq:entangle_Gs} and~\ref{eq:entangle_Gv}) that we made. At higher temperature $\Phi$ and $\bar\Phi$ go to unity. Then with the values of $\alpha_1$ and $\alpha_2$ we considered the coupling strengths can be reduced to $70\%$, at best, to their respective initial values. Thus with this form of ansatz, the problem in the QNS with the inclusion of the vector interaction remains even in the EPNJL model. One needs to improve on this ansatz.

\section{Conclusion}
\label{sec:c3:conclusion}

In this chapter we have studied the behavior of the vector meson CF and conserved density fluctuation associated with the temporal vector correlator in a hot and dense environment. The investigations are performed with and without the I-V interaction. This interaction is considered using ring resummation known as Random Phase Approximation. 

In absence of the I-V interaction, the CF in NJL model become quantitatively equivalent to those of free field theory. In case of PNJL it is different from both free and NJL ones because of the presence of the PL fields that suppress the color degrees of freedom in the deconfined phase just above $T_c$. This suggests that some nontrivial correlation exists among the color charges in the deconfined phase. We note that the Euclidean CF with zero chemical potential agree in certain domain with the available LQCD data in quenched approximation. We also discussed the same quantity in presence of finite chemical potential and external momentum which could provide useful information if, in future, LQCD computes them at finite chemical potential and external momentum.

In presence of the I-V interaction, we note that the response to the conserved number density fluctuation at finite chemical potential exhibits an interesting characteristic around the phase transition temperature than that at vanishing chemical potential. This is because the mean fields (PL fields and condensates etc.)  depend implicitly on chemical potential and  so their variations are most significant around the transition region, in particular for PNJL model.

We have also discussed the issue with QNS in presence of I-V interaction. Some efforts are also made to resolve the problem by introducing entangled vertex through EPNJL model. As found that the choices for the ansatz (vide equation (\ref{eq:entangle_Gs})) available in the literature for the entangled vertex has not yet fully cured the problem and further work is required in this direction.

\chapter{The consequences of $SU(3)$ colorsingletness}
\label{chapter:JPG}

In this chapter we show, using the quantum statistical mechanics, that the $SU(3)$ color singlet ensemble of 
a quark-gluon gas exhibits a $Z(3)$ symmetry through the normalized character in fundamental 
representation and also becomes equivalent, within a stationary point approximation, to the ensemble 
given by Polyakov loop (PL). We further obtain the PL gauge potential by considering spatial  
gluons along with the invariant Haar measure at each space point. This discussion is based on: {\em The consequences of SU(3) colorsingletness, Polyakov Loop and Z(3) symmetry on a quark-gluon gas}, Chowdhury Aminul Islam, Raktim Abir, Munshi G. Mustafa, Sanjay K. Ghosh and Rajarshi Ray, {\bf J. Phys. G 41, 025001 (2014)}.

\section {Introduction:}

We aim here in this chapter to use color singlet (CS) model, introduced in chapter two~\ref{chapter:prelim} to study the phase structure of strongly interacting matter created in HICs - the deconfinement transition specifically. Within CS model we restrict the partition function (PF) by introducing group theoretical projection operator, which takes care of the colorsingletness of physical states. The conservation of baryon number is also taken care of through the introduction of baryon chemical potential. Once we have the PF it can be used to study different thermodynamical properties of QCD, hadronic density of states etc.

As already mentioned we are not interested in studying the thermodynamics, rather we discuss some of the phenomenological consequences of $SU(3)$ colorsingletness vis-a-vis PL that may lead to the formation of center domains in QGP produced in relativistic HICs. The implications of these center domains in the context of HICs have also been envisaged. This chapter is organized as follows: In section~\ref{sec:c4:ther_pot} we obtain the thermodynamic potential, in section~\ref{sec:c4:pl_nc} we recognize the PL as the normalized character of $SU(3)_c$ and in~\ref{sec:c4:nc_cs} we find out the consequences of center symmetry and finally we conclude in~\ref{sec:c4:con}.

\section{Color singlet ensemble and the thermodynamic potential}
\label{sec:c4:ther_pot}

To start with we have the color singlet PF (\ref{eq:pre:par_fn_final}) which is given as
\begin{eqnarray}
{\cal Z}_S\!\!=\int_{SU(N_c)}{\hspace*{-0.15in}}\!\!\! {\rm d}\mu (g) \ \ e^{\Theta_p }
\ \ ,
\label{eq:c4:par_fn_final}
\end{eqnarray}
with
\begin{eqnarray}
\Theta_p 
&=& \Theta_q + \Theta_{\bar q} + \Theta_g \nonumber \\ 
&=&
2 N_f\sum_\alpha {\rm{tr}_c}  \ln \left (1+R_q
e^{-\beta (\epsilon_{q}^\alpha -\mu_q)}
\right ) 
+ 2N_f \sum_\alpha {\rm{tr}_c}  
\ln \left (1+R_{\bar q}
e^{-\beta (\epsilon_q^\alpha +\mu_{ q})}
\right ) 
\nonumber \\
&& 
- 2\sum_\alpha {\rm{tr}_c}  
\ln \left (1-R_g
e^{-\beta \epsilon_g^\alpha}\right ),
\label{eq:c4:theta}
\end{eqnarray}
where $\epsilon_i^\alpha= \sqrt{(p^\alpha_i)^2+m^2_i}$. Also the quark flavor ($N_f$), their spin and the chemical potential $\mu$, and the polarization of gluons are introduced.

We want to obtain the thermodynamic potential from this PF. We define the diagonal 
matrices and their respective characters for both fundamental and adjoint representations.
The finite dimensional diagonal matrix $R_{q ({\bar q})}$ in the basis of the color space 
represents the image~\cite{Weyl(book):TCG,Auberson:1986ft} of the group element in the 
irreducible representation of $SU(N_c)$ as
\begin{eqnarray}
{R}_q
&=& \mbox{diag}\left(
e^{i\theta_1}\,, e^{i\theta_2}\,, e^{i\theta_3} \right)\, ;  \hspace*{0.3in}
{R}_{\bar q}
= R^\dagger_q ,   
\label{eq:c4:char_fun}
\end{eqnarray}
with their respective characters
\begin{eqnarray}
 \chi_f=\mbox{tr}_c R_q=\sum_{i=1}^{N_c}e^{i\theta_i}; \hspace*{0.3in}
\chi_f^\dagger=\mbox{tr}_c R_q^\dagger=\sum_{i=1}^{N_c}e^{-i\theta_i}. \
\end{eqnarray}
Similarly, the character in adjoint representation is obtained as
\begin{eqnarray}
 \chi_{\mbox{adj}} &=& \chi_f\chi_f^\dagger-1=\mbox{tr}_cR_g=
\mbox{tr}_c \Big[ \mbox{diag}\big (1,1,e^{i(\theta_1-\theta_2)},
e^{-i(\theta_1-\theta_2)}, \nonumber \\
&& e^{i(2\theta_1+\theta_2)},e^{-i(2\theta_1+\theta_2)},e^{i(\theta_1+2\theta_2)},
 e^{-i(\theta_1+2\theta_2)}\Big)\Big]
\end{eqnarray}

We also define normalized characters by the respective dimension of the fundamental and 
adjoint representations as
\begin{eqnarray}
\Phi= \frac{1}{N_c} {\rm{tr}_c} R_q \,\, ,
\quad \quad 
\bar{\Phi} =\frac{1}{N_c}{\rm{tr}_c} R_q^\dagger\,\, ,
\quad \quad (N_c^2-1) \Phi_A &=& N_c^2\Phi{\bar \Phi}-1 \,.
\label{eq:c4:eq2}
\end{eqnarray}
We note here that the magnitude of the normalized character, $|\Phi|$, in fundamental representation is related to (i) the $Z(3)$ symmetry as shown in the section~\ref{sec:c4:nc_cs} and (ii) thermal expectation value of the PL as argued in the section~\ref{sec:c4:pl_nc}, in details.

With all these the Vandermonde term in (\ref{eq:pre:haar_measure}) can now be written in terms of $\Phi$ and ${\bar \Phi}$ as (vide appendix~\ref{app:vdm})
\begin{eqnarray}
 \prod_{i>j}^3 \left | e^{i\theta_i} - e^{i\theta_j} \right |^2 
&=& 27 [1-6\Phi {\bar \Phi} +4 (\Phi^3+{\bar \Phi}^3) 
-3(\Phi{\bar \Phi})^2 ]
= 27 \ H(\Phi,{\bar \Phi}) \, , \label{eq:c4:vdm}
\end{eqnarray}
for $SU(3)$. Further, this is in general not possible for $N_c>3$ as there are more than two 
independent parameters.
Now the Jacobian for variable transformation from $\{\theta_1,\theta_2\}$ to $\{\Phi , {\bar \Phi}\}$
can be obtained as
\begin{eqnarray}
 J(\Phi,{\bar \Phi})  = {(1/9)} \sqrt{27 H(\Phi,{\bar \Phi})}.
\end{eqnarray}
In the infinite volume $V$, one also needs to replace the discrete single particle sum by an integral as
$ \sum_\alpha \rightarrow (V/(2\pi)^3 \int d^3p \ .$

After performing the color trace of the matter and gauge parts, and expressing in terms of 
the characters of the fundamental and its conjugate, equation (\ref{eq:c4:par_fn_final}) becomes
\begin{eqnarray}
{\cal Z}_S&=& \int_{SU(3)} \ d\Phi \ d{\bar {\Phi}} \ e^{\Theta_q+\Theta_{\bar q}+\Theta_g+\Theta_H}; \nonumber \\
\Theta
%&=& \Theta_p+\Theta_H\nonumber \\
&=&\Theta_{q} + \Theta_{\bar q}+\Theta_g +\Theta_H \nonumber \\ 
&=& 2VN_f\int \frac{d^3p}{(2\pi)^3} 
\ln \Big [ 1 + e^{-3\beta\epsilon_q^{ +}} +  
N_c\left( \Phi + \bar{\Phi}e^{-\beta\epsilon_q^{+}}\right)e^{-\beta\epsilon_q^{ +}}
\Big] \nonumber \\
&& + 2VN_f\int \frac{d^3p}{(2\pi)^3}  
  \ln \left [ 1 +   e^{-3\beta\epsilon_q^{-}}
+ N_c\left( {\bar \Phi} + {\Phi}e^{-\beta\epsilon_q^{-}} \right) e^{-\beta\epsilon_q^{-}} \right] \nonumber \\
&& -2V\int \frac{d^3p}{(2\pi)^3}   
\ln\Big( 1 + \sum_{m=1}^8 a_m\, e^{-m\beta\epsilon_g}\Big ) +\frac{n}{2}\ln H  
.  \label{eq:c4:theta1}
\end{eqnarray}
with $\epsilon^\pm_q={\epsilon_q\mp\mu}$ and the coefficients $a_m$  are given by
\begin{eqnarray}\label{eq:c4:eq11}
a_1 &=& a_7
= 1 - N_c^2\bar{\Phi}\Phi\,,
\quad a_8
= 1\, , 
\nonumber\\
a_2
&=&
a_6
= 1 - 3 N_c^2 \bar{\Phi}\Phi
{}+ N_c^3 \left( \bar{\Phi}^3 + \Phi^3\right)\,,
\nonumber\\
a_4
&=&
2\left[
-1 + N_c^2 \bar{\Phi}\Phi - N_c^3\left( \bar{\Phi}^3 + \Phi^3\right)
{}+ N_c^4 \left( \bar{\Phi}\Phi \right)^2 \right ] \, , 
\nonumber\\
a_3
&=&
a_5
= -2 + 3 N_c^2 \bar{\Phi}\Phi
{}- N_c^4\left( \bar{\Phi}\Phi \right)^2\, .
\end{eqnarray}
We note here that the square of the Vandermonde determinant in (\ref{eq:pre:haar_measure}) enters at each point in space in the action. 
So, in the PF the logarithm of the product  of the Vandermonde term  should be proportional to $n$ as 
$\Theta_H = n\ln \sqrt H$, assuming $n$ is the number of points in the space. Also a constant normalization factor 
is dropped as it is subleading.

Now performing the integrations using the method of stationary points, one can write
\begin{eqnarray}
 {\cal Z}_S^0 \!\! &=& \!\! \left. e^{\Theta_q+\Theta_{\bar q}+\Theta_g +\Theta_{H}}
 \right |_{\Phi \rightarrow \Phi_0 \atop {\bar \Phi}\rightarrow {\bar \Phi_0} }
, \label{part_ic}
\end{eqnarray}
where the stationary values of $\Phi_0$ and 
${\bar \Phi_0}$ can be obtained from the extremum conditions as
\begin{eqnarray}
\frac{\partial \Theta}{\partial \Phi } &=&0 \, ; \, \, \,\,\,\, 
 \frac{\partial \Theta}{\partial {\bar \Phi} }  = 0\, . \label{extremum}
\end{eqnarray}
The color singlet thermodynamic potential density at the stationary points in infinite 
volume limit becomes
\begin{eqnarray}
{\Omega}_S^0  &=& -\frac{T}{V}\ln {\cal Z}_S^0 = 
-\frac{T}{V}\left[ \Theta_q+\Theta_{\bar q}+\Theta_g+\Theta_{\mbox{H}} \right ]_{\Phi \rightarrow 
\Phi_0 \atop {\bar \Phi} \rightarrow {\bar \Phi_0}}
 \nonumber \\
&=& \Big [ \Omega_q+\Omega_{\bar q}+\Omega_g
-\kappa T \ln H \Big ]_{\Phi \rightarrow 
\Phi_0 \atop {\bar \Phi} \rightarrow {\bar \Phi_0}} 
 =  {\Omega}^{\mbox{PL}} (\Phi_0,{\bar \Phi_0}). 
\label{eq:c4:potl} 
\end{eqnarray}
This exhibits that the $SU(3)$ color singlet ensemble of a quark-gluon gas is equivalent to that of PL with quarks. A close inspection of quark ($\Omega_q$), antiquark ($\Omega_{\bar q}$) and the Vandermonde terms in equation (\ref{eq:c4:potl}) reveals that they match with the respective terms in the thermodynamic potential in PNJL model as given in equation (\ref{eq:pnjl_pot}) in chapter~\ref{chapter:prelim}. The PL gauge potential is obtained here as $\Omega_g-\kappa T\ln H$, by considering the spatial gluons in the ensemble. We also note that the gluons 
as quasiparticles in PL model was also studied in some other context~\cite{Meisinger:2003id,Meisinger:2001cq,Meisinger:2001fi,Sasaki:2012bi}. But in PNJL model~\cite{Fukushima:2003fw,Fukushima:2003fm,Ratti:2005jh,Ghosh:2006qh,Mukherjee:2006hq,Roessner:2006xn,Ghosh:2007wy,Sasaki:2006ww,Bhattacharyya:2010wp,Bhattacharyya:2010jd,Bhattacharyya:2010ef,Asakawa:2012yv} the form of the gauge potential has usually been used by fitting pure gauge lattice data as discussed in equation (\ref{eq:pre:pot_gauge}). We emphasize that here the thermodynamic potential has been obtained phenomenologically by imposing $SU(N_c)$ color singlet restriction on a quark-gluon gas.

We note here that the calculation begun with a discussion of color 
neutrality effects in the free quark-gluon gas~\cite{Auberson:1986ft,Muller:1983ed,Redlich:1979bf,Turko:1981nr,Elze:1986db,Gorenstein:1982ua,Skagerstam:1983gv,Kusaka:1991ds,Zakout:2007nb,Ansari:1992qd,Mustafa:1995fv,Mustafa:1997ga,Mustafa:1993np,Mustafa:1992yv,Mustafa:1998zp,Abir:2009sh}. These works were concerned with global color neutrality
and the effect of which becomes irrelevant for a free gas in the limit where volume goes to infinity. Here, the extension to local 
neutrality is considered using Haar measure at every spatial point to address the effect of confinement. Nevertheless, the well-known 
calculations of the effective potential for PL to one loop in perturbation theory~\cite{Gross:1980br,Gocksch:1993iy,Weiss:1980rj,Weiss:1981ev} show that the local Vandermonde contribution  
due to the Haar measure is canceled out when spatially longitudinal gluon fields ($A_0(t,{\mathbf x})$) are integrated over. 
This becomes a problem to use local Haar measure as the basis for a fundamental theory of confinement and presently it is not known yet how
to use it. On the other hand, allowing PL field where $A_0(t,{\mathbf x})$ is constant (see next section) as ${\mathbf x}\rightarrow \infty$, 
the Vandermonde term contributes to the PF in (\ref{eq:c4:theta1}) and thus in the potential in (\ref{eq:c4:potl}). We further note that the contribution  from the Vandermonde term survives infinite volume limit since the constant, $\kappa=n/2V$, is the  ratio of two large numbers leading 
to a finite value which has to be determined from the lattice equation of state of pure $SU(3)$ gauge theory. This value is found as 
$\kappa\sim 0.0075$ GeV$^3$ in the literature~\cite{Asakawa:2012yv,Fukushima:2003fw,Fukushima:2003fm,Ratti:2005jh,Ghosh:2006qh,Mukherjee:2006hq,Roessner:2006xn,Sasaki:2006ww,Ghosh:2007wy,Bhattacharyya:2010wp,Bhattacharyya:2010jd,Bhattacharyya:2010ef}. Now, the potential in (\ref{eq:c4:potl}) has been considered~\cite{Fukushima:2003fw,Fukushima:2003fm,Ratti:2005jh,Ghosh:2006qh,Mukherjee:2006hq,Roessner:2006xn,Sasaki:2006ww,Ghosh:2007wy,Bhattacharyya:2010wp,Bhattacharyya:2010jd,Bhattacharyya:2010ef} as a
starting point for phenomenological models of quark-gluon thermodynamics and is also coupled to  chiral model~\cite{Nambu:1961tp,Nambu:1961fr,Kunihiro:1991qu,Hatsuda:1994pi} 
to study extensively the deconfinement and chiral dynamics together~\cite{Fukushima:2003fw,Fukushima:2003fm,Ratti:2005jh,Ghosh:2006qh,Mukherjee:2006hq,Roessner:2006xn,Sasaki:2006ww,Ghosh:2007wy,Bhattacharyya:2010wp,Bhattacharyya:2010jd,Bhattacharyya:2010ef}. 

Now, in the next section we discuss some of the phenomenological consequences of $SU(3)$ colorsingletness vis-a-vis PL that may lead to the formation of center domains~\cite{Pisarski:2000eq,Dumitru:2001xa,Dumitru:2002cf,Dumitru:2003hp,Deka:2010bc,Borsanyi:2010cw,Danzer:2010ge} in quark-gluon 
plasma produced in relativistic heavy-ion collisions.
 
\section{Polyakov Loop and normalized character in~${SU(3)}$}
\label{sec:c4:pl_nc}
The Polyakov loop ($L(\vec x)$) from the timelike Wilson line is 
given in subsection~\ref{sssec:pl} of chapter~\ref{chapter:prelim} through the equation 
(\ref{eq:pol_loop}). We recall here some of the important aspects of the Polyakov: 
the normalized trace over the Polyakov loop is known as Polyakov loop field $\Phi$ 
which transforms under the global $Z(N_c)$ symmetry as a field with charge one as $\Phi 
\rightarrow e^{{i2\pi j}/{N_c}}\Phi$ with $j=0,1,\cdots, (N_c-1)$. It is also
related with the free energy of a quark-antiquark pair and acts as an 
order parameter for confinement-deconfinement phase transition associated with
the spontaneous breaking of $Z(N_c)$ symmetry, the center symmetry of $SU(N_c)$.

% $\Phi$ is complex and transforms under the global $Z(N_c)$ symmetry as a field with charge one as $\Phi \rightarrow e^{{2\pi i}/{N_c}}\Phi$. Given the role of an order parameter for 
% pure gauge~\cite{Pisarski:2000eq,Dumitru:2001xa,Dumitru:2002cf,Dumitru:2003hp}, if $\Phi =0$ the $Z(N_c$) is unbroken and there is no ionization of $Z(N_c)$ charge, which is the confined phase below a certain temperature. At high temperature the symmetry is spontaneously broken,  $\Phi\ne 0$ corresponds to a deconfined phase of gluonic plasma and there are $N_c$ different equilibrium states distinguished by the phase $2\pi j/N_c$ with $j=0,\cdots (N_c-1)$. 

In the next section, it will be demonstrated in details that the magnitude of 
the normalized character in the fundamental representation of $SU(3)_c$ exhibits color localized and ionized domains of center symmetry, $Z(3)$ with three rotational angles (viz., $0$, $2\pi/3$, $4\pi/3$). This establishes that the normalized character in the fundamental representation of 
$SU(3)_c$ and the PL field are equivalent. We further note that
the temporal gauge field $A_0$ can be completely characterized~\cite{Sasaki:2012bi} by two diagonal 
generators\footnote{Diagonal in eigenvalues of $SU(N_c)$ group in terms of class parameter 
$\theta_i$. Since $\theta_i$ obeys $\sum_i^{N_c}\theta_i=0({\mbox{mod}}2\pi)$ ensuring 
that only $(N_c-1)$ parameters subgroups associated with two diagonal generators. For 
$SU(3)$ only two parameters $\theta_1$ and $\theta_2$ are sufficient to describe the
the finite dimensional diagonal matrix $R_{q ({\bar q})}$ 
in (\ref{eq:c4:char_fun}), which equivalent to PL matrix.} as 
$A_0=A_0^3\lambda_3+A_0^8\lambda_8$. Now, assuming a Gaussian 
approximation in equation (\ref{eq:c4:eq2}), one can 
write~\cite{Pisarski:2000eq,Dumitru:2001xa,Dumitru:2002cf,Dumitru:2003hp, Asakawa:2012yv}
\begin{eqnarray}
 |\Phi|&=& \left|\frac{1}{3} {\rm{tr}_c} R_q \right |= \sqrt{\Phi {\bar {\Phi}}}
%\nonumber \\
\approx
\exp\left [-\frac{g^2}{2T^2}{\mbox{tr}}_c  \langle A_0^2\rangle \right ]. \label{eq:c4:thfluc}
\end{eqnarray}
The dynamics of the magnitude of the normalized character, $|\Phi|$,  in fundamental representation is governed by the thermal average
of the square of the static temporal gauge field $A_0$. In the color confined phase $|\Phi|=0$ and the background temporal gauge field 
fluctuates with high amplitudes whereas in the color deconfined phase $|\Phi|=1$ and the fluctuations of the gauge field 
almost disappears. This background gauge field in the form of PL also interacts nonpertubatively with the quarks and gluons in 
the thermal medium as given in (\ref{eq:c4:potl}). 

\section{Normalized character, center symmetry and  consequences}
\label{sec:c4:nc_cs}
\begin{figure}
\begin{center}
\includegraphics[width=0.6\linewidth,height=0.5\linewidth, angle=0]{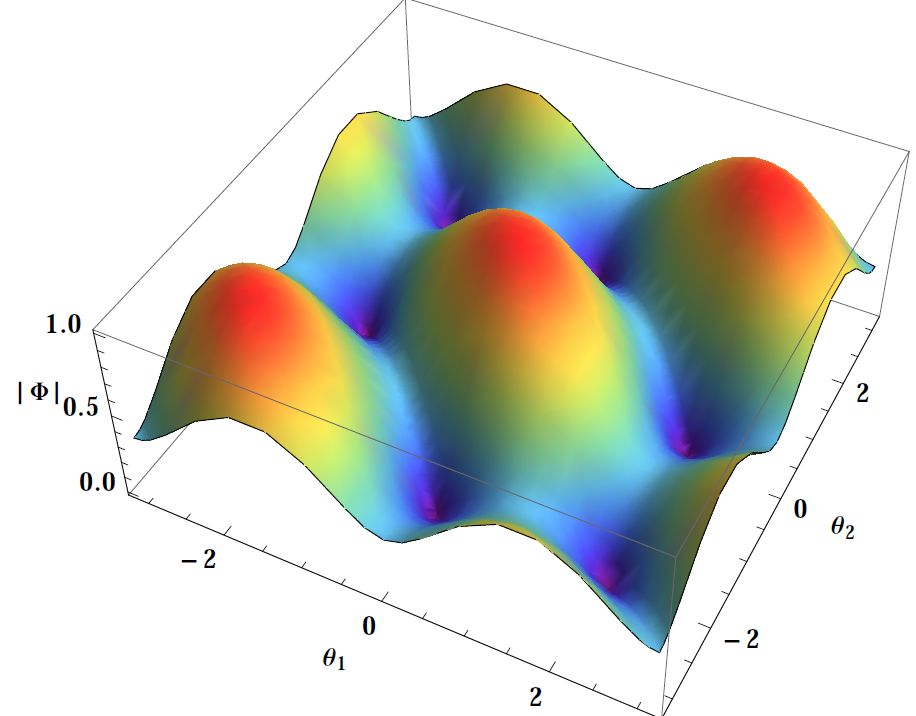}
\caption{A $3$D-plot of $|\Phi(\theta_1,\theta_2)|$, i.e.,
the normalized character in the fundamental representation of $SU(3)$ as given 
in \eqref{eq:c4:eq2} within $-\pi \leq \theta_1,\,\theta_2\leq \pi$.}
\label{fig:c4:phi_3d_f}
\end{center}
\end{figure}

In figure~\ref{fig:c4:phi_3d_f} a three dimensional view of 
the magnitude of the normalized character in fundamental 
representation of $SU(3)$,  $|\Phi(\theta_1,\theta_2)|$,
is shown within the domain $-\pi \leq \theta_1,\theta_2 \leq \pi$. 
It has three maxima for  $(\theta_1,\theta_2)= (0,0), \, (2\pi/3,2\pi/3), \, (-2\pi/3, -2\pi/3)$. 
There are also three minima (and three mirror images exist if
one interchanges $\theta_1\leftrightarrow\theta_2$) for $(\theta_1,\theta_2)= (0,2\pi/3), 
\, (0,-2\pi/3) \,{\mbox{and}} \, (2\pi/3, -2\pi/3)$. 
$|\Phi| $ has a
three fold degeneracy connected by a rotation of $2\pi/3$ in both 
$\theta_1$ and $\theta_2$. A Monte Carlo simulation of complex $\Phi$
is also displayed in figure~\ref{fig:c4:phi_comp_f} in a Argand plane with
 $-\pi \leq (\theta_1,\,\theta_2) \leq \pi$. 
This shows a three pointed star in a circle of unit radius, in which each point 
can be rotated by a phase $2\pi/3$ except the origin. This clearly indicates that $\Phi$  
in fundamental representation of $SU(3)$ has a center symmetry, $Z(3)$
with three rotational angles (viz., $0,\, 2\pi/3,\, 4\pi/3$ or $-2\pi/3$ ). This can also
be understood from the invariant Haar measure expressed in terms of $\Phi$ in (\ref{eq:c4:vdm}). Now, the 
three minima in figure~\ref{fig:c4:phi_3d_f}
uniquely correspond to the center of the circle at $\Phi=0$ in figure~\ref{fig:c4:phi_comp_f}, which
is $Z(3)$ symmetric phase or confined phase at low $T$. On the other hand, the three maxima in 
figure~\ref{fig:c4:phi_3d_f} correspond to the three pointed tips in figure~\ref{fig:c4:phi_comp_f} representing
the spontaneously broken phase or deconfined phase of $Z(3)$ at very high $T$. $\Phi$ can act as 
an order parameter for deconfinement phase transition.
In figure~\ref{fig:c4:phi_3d_a} a three dimensional plot of $\Phi_A$ is also displayed that exhibits
same features as figure~\ref{fig:c4:phi_3d_f} except minima appear in negative values, which could   
be understood from (\ref{eq:c4:eq2}). This clearly suggests that the magnitude of the normalized 
character in the fundamental representation of $SU(3)$ exhibits the center symmetry, $Z(3)$. 

\begin{figure}
\begin{center}
\includegraphics[width=0.55\linewidth,height=0.45\linewidth, angle=0]{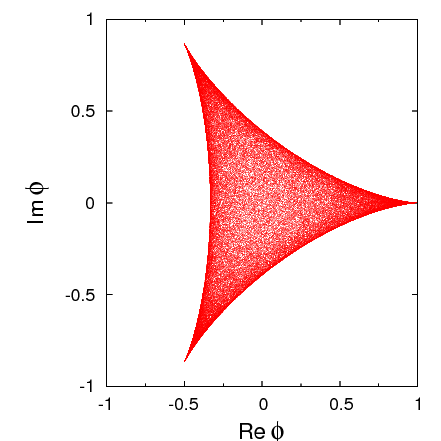}
\caption{A Monte Carlo simulation  of complex $\Phi(\theta_1,\theta_2)$
in Argand plane for which $\theta_1$ and $\theta_2$ are chosen randomly
in the domain $-\pi \le \theta \le \pi$.} 
\label{fig:c4:phi_comp_f}
\end{center}
\end{figure}

\begin{figure}
\begin{center}
\includegraphics[width=0.5\linewidth,height=0.4\linewidth, angle=0]{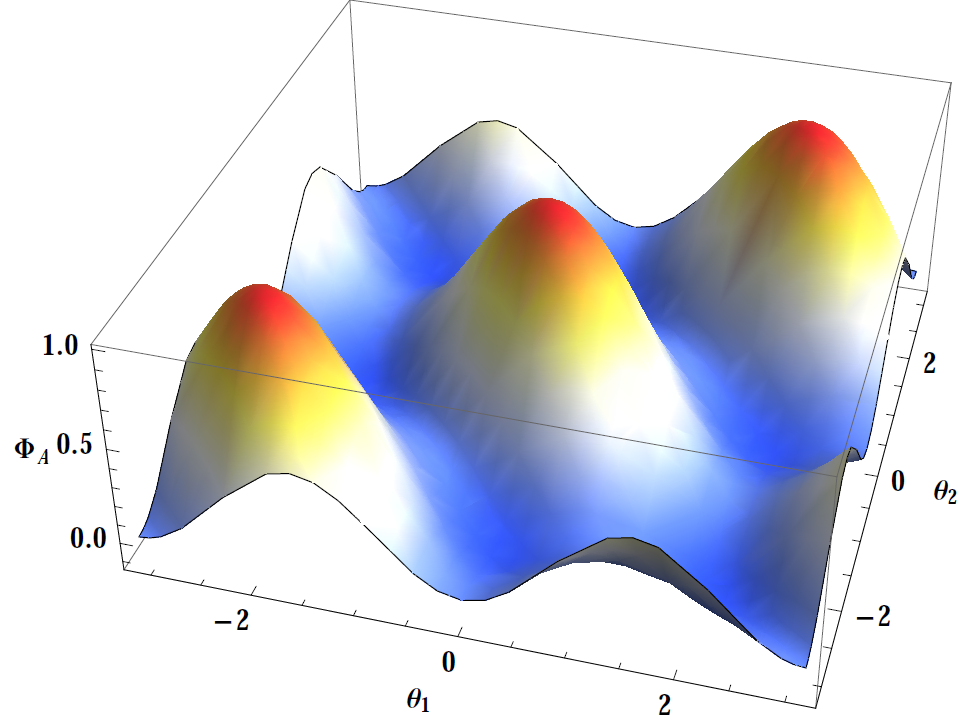}
\caption{Same as figure~\ref{fig:c4:phi_3d_f} but for   $\Phi_A(\theta_1,\theta_2)$.}
\label{fig:c4:phi_3d_a}
\end{center}
\end{figure}

\begin{figure}
\begin{center}
\includegraphics[width=0.5\linewidth,height=0.4\linewidth, angle=0]{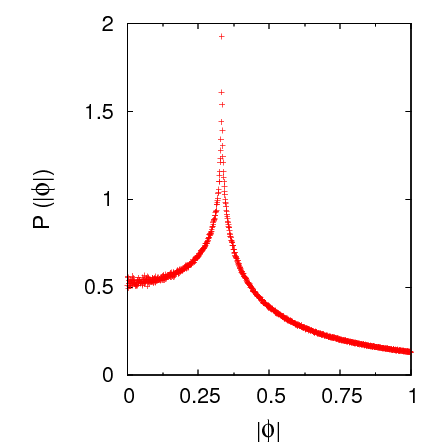}
\caption{A Monte Carlo simulation, $P(\Phi)$, corresponds to
the occurrence probability of $|\Phi (\theta_1,\theta_2)|$ in $SU(3)$ parameter
space $(\theta_1,\, \theta_2)$.
The values of $\theta_1$ and $\theta_2$ are chosen randomly within the domain 
$-\pi\leq \theta_1,\, \theta_2\leq \pi$ and then the obtained value of $|\Phi|$ 
is mapped in a given $\Phi$ bin, which is then normalized by the area of the bin.
}
\label{fig:c4:phi_prob}
\end{center}
\end{figure}

We also noticed some more interesting features of $\Phi$ in ($\theta_1, \, \theta_2$)-plane.
In figure~\ref{fig:c4:phi_prob} a Monte Carlo simulation of the occurrence probability, $P(|\Phi|)$, 
of $|\Phi(\theta_1,\theta_2)|$ is displayed in $SU(3)$ parameter space. This plot
indicates that the maximum probability  for $|\Phi|$ to occur  when $|\Phi|=1/3$
indicating a phase transition from a color confining phase to a color deconfining phase, which has also been 
observed in LQCD calculation~\cite{Cheng:2007jq}. This could be 
better viewed from figure~\ref{fig:c4:phi_eqva} which is a contour plot corresponding to figure~\ref{fig:c4:phi_3d_f}. 
As $T$ increases, $\Phi$ increases~\cite{Fukushima:2003fw,Fukushima:2003fm,Ratti:2005jh,Ghosh:2006qh,Mukherjee:2006hq,Roessner:2006xn,Sasaki:2006ww,Ghosh:2007wy,Bhattacharyya:2010wp,Bhattacharyya:2010jd,Bhattacharyya:2010ef} from zero in the confined phase and reaches unity for ideal gas. 
In the domain $ 0 < |\Phi| <1/3$, the color neutral states start decomposing
but prefer to reside in $Z(3)$ minima and its mirror images minima. Color charges (partons) with thermal momentum
in this domain cannot overcome the barriers provided by the large amplitude of the thermal fluctuations of the background 
gauge field in (\ref{eq:c4:thfluc}). This domain of $|\Phi|$ is shown by red dots ($|\Phi|\sim 0$) to purple triangles 
($|\Phi|\sim 0.3$) in figure~\ref{fig:c4:phi_eqva}. As long as such states are inside the domain of $Z(3)$ minima, a strong  color 
correlation exists among the color charges like a liquid~\cite{Asakawa:2012yv}, because the mean free path of the color charges is
of the order of size of the domain in $Z(3)$ minima. In this $|\Phi|$ domain, the normalized 
character in adjoint representation varies as $-1/8\leq \Phi_A < 0$ which is represented
in figure~\ref{fig:c4:phi_a_eqva}.

\begin{figure}
\begin{center}
\includegraphics[width=0.5\linewidth,height=0.45\linewidth, angle=0]{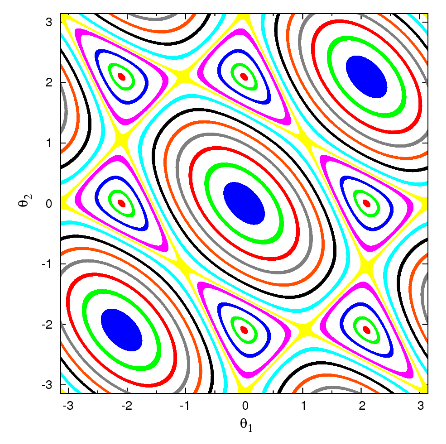}
\caption{A 2D projection of $|\Phi|$ in $\theta_1$ and $\theta_2$ plane in which 
each color corresponds to a equivalued $|\Phi|$. The red dots to purple triangles correspond to
$|\Phi|\sim0,\, 0.1,\, 0.2, \, 0.3$ whereas sea-blue lines to blue dots correspond to $|\Phi|\sim 0.4,\, 0.5,\,
0.6,\, 0.7,\, 0.8,\, 0.9, 1$. The equivalued mess connected by yellow triangles corresponds to $|\Phi|\sim 1/3$.}
\label{fig:c4:phi_eqva}
\end{center}
\end{figure}

\begin{figure}
\begin{center}
\includegraphics[width=0.55\linewidth,height=0.45\linewidth, angle=0]{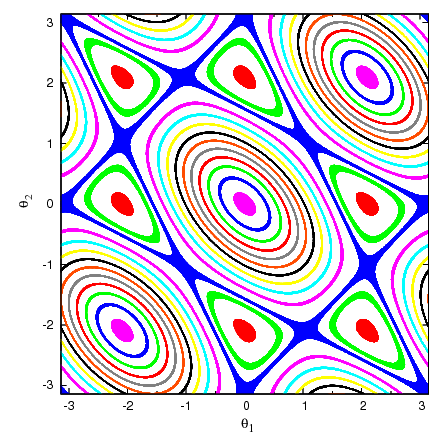}
\caption{Same as figure~\ref{fig:c4:phi_eqva} but for $\Phi_A$. 
The equivalued mess connected by blue triangles is for $\Phi_A=0$. From the blue mess to purple blobs, 
$\Phi_A$ increases by a step of $0.1$. The red dots are for $\Phi_A \sim -1/8$ whereas 
the green triangles $\Phi_A \sim -0.05$.}
\label{fig:c4:phi_a_eqva}
\end{center}
\end{figure}

Now for $|\Phi|=1/3$, the $Z(3)$ minima disappear and get connected to each other in $Z(3)$ space, which 
is represented by the yellow mess in figure~\ref{fig:c4:phi_eqva}. This causes $P(|\Phi|)$ to be 
maximum in figure~\ref{fig:c4:phi_prob} exhibiting a long range color correlation and the thermal fluctuations 
of the background gauge field attain a critical value as the separating barriers of minima become flat and wider.  
Here $\Phi_A=0$ as is also represented by the blue mess in figure~\ref{fig:c4:phi_a_eqva}.

When $1/3 < |\Phi| \leq 1$, the correlated $Z(3)$ domains start to get uncorrelated  and 
the ionization of $Z(3)$ color charges begin which is evident from equivalued $|\Phi|$ lines in figure~\ref{fig:c4:phi_eqva} starting 
from sea-blue ($|\Phi|\sim0.4$) to green lines ($|\Phi|\sim 0.9$) at a step of $0.1$. 
When $|\Phi|\sim 1$  a complete ionization of $Z(3)$ charges take place and they reside at those maxima in figure~\ref{fig:c4:phi_3d_f} which 
are also represented by blue blobs in figure~\ref{fig:c4:phi_eqva}. This ionization can also be seen in figure~\ref{fig:c4:phi_a_eqva} 
through purple equivalued $\Phi_A$ lines to purple blobs in the range $0.1\leq \Phi_A \leq 1$.  So, in the color 
deconfined phase ($1/3 < |\Phi| \leq 1$), there are formation of domains which are also separated by the nonperturbative 
interaction of the background gauge fields. These domains of ionized color charges can act as scattering centers 
in the deconfined phase lead to jet quenching.  A hard jet after losing energy through gluon emission by 
the scattering with those ionized domain of color charges~\cite{Asakawa:2012yv} in deconfined phase ($1/3 < |\Phi| \leq 1$) 
can enter the confining phase ($0< |\Phi| \leq 1/3$) and hadronize by recombination~\cite{Abir:2009sh,Fries:2003vb,Greco:2003xt}.

\begin{figure}
\begin{center}
\includegraphics[width=0.45\linewidth,height=0.3\linewidth, angle=0]{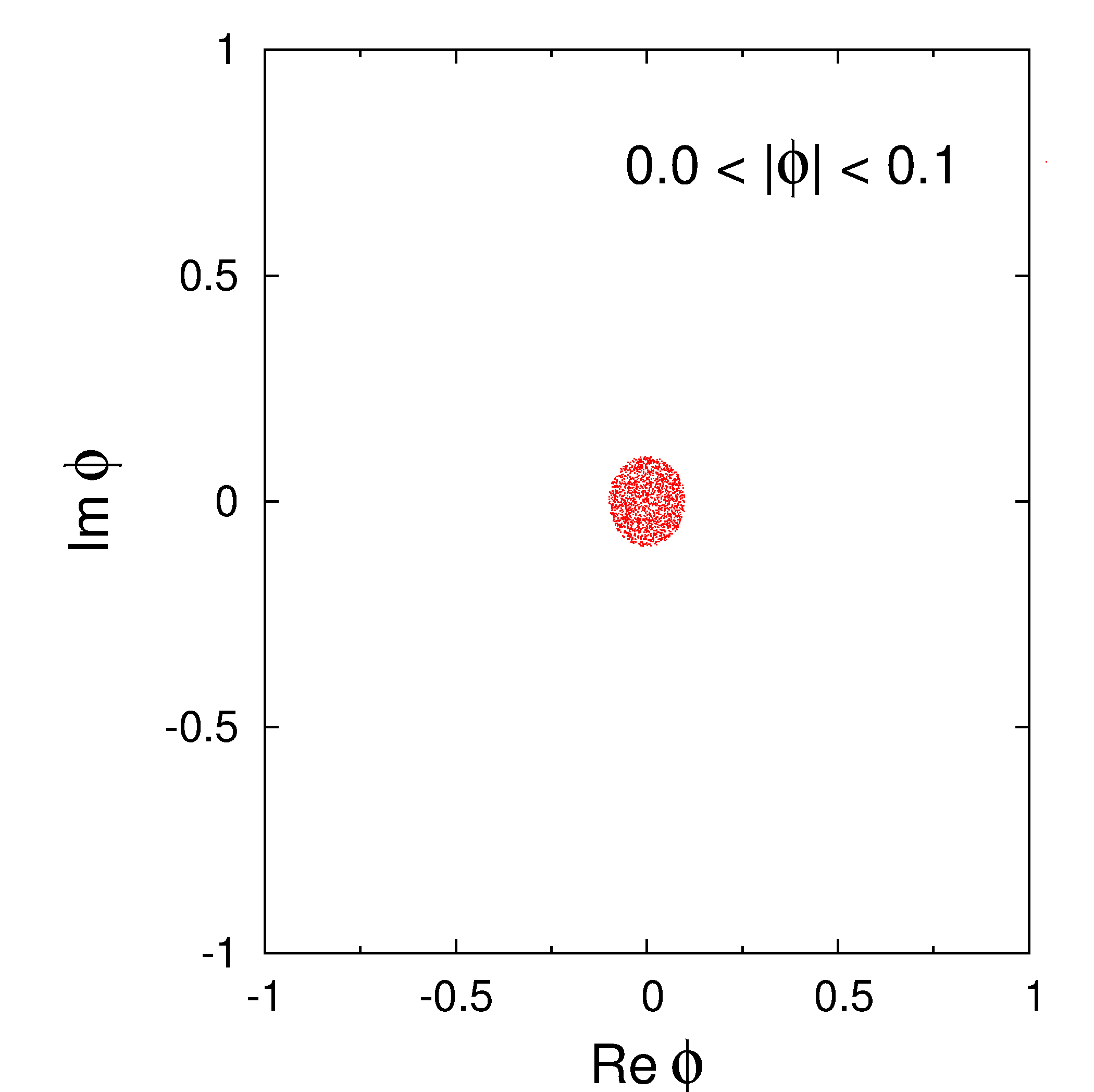}
\includegraphics[width=0.45\linewidth,height=0.3\linewidth, angle=0]{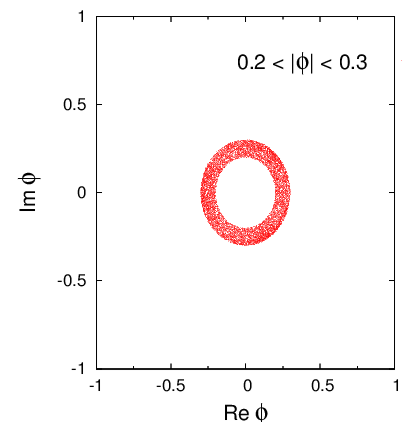}
\includegraphics[width=0.45\linewidth,height=0.3\linewidth, angle=0]{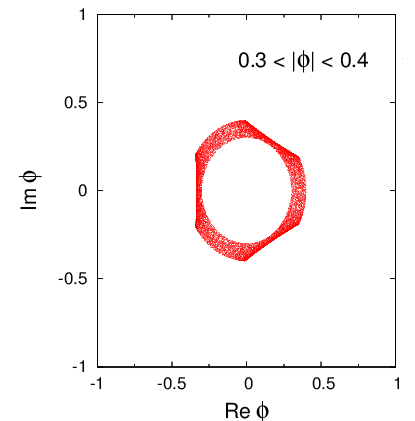}
\includegraphics[width=0.45\linewidth,height=0.3\linewidth, angle=0]{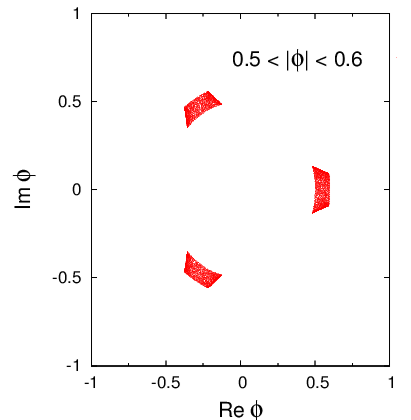}
\caption{The projection of $\Phi$ in Argand plane  for a fixed value, 
which are noted in the respective plots. This shows a strong color correlation in the 
range $0\leq \Phi \leq 1/3$ and ionization of color states in the range $1/3< \Phi \leq 1$.}
\label{fig:c4:phi_f_comp}
\end{center}
\end{figure}

All those features of center symmetry discussed above are also reflected in complex $\Phi$ plane in figure~\ref{fig:c4:phi_f_comp} indicating  
a snap shot of color localized and ionized domains of $Z(3)$. This clearly shows that there are center domains formation in the two
distinct regions of $|\Phi|$, which are  $0< |\Phi| \leq 1/3$ (confining domain) due to the center symmetry $Z(3)$, and  
$1/3 < |\Phi| \leq 1$ (deconfining or ionization domain) due to spontaneously breaking\footnote{Though also the $Z(3)$ symmetry is
explicitly broken with dynamical quarks unlike pure gauge sector, yet it can be regarded as an approximate symmetry and the PL 
expectation value is still useful as an order parameter as we will see later.}
 of the center symmetry $Z(3)$. 
The broken center domains formed in deconfining phases ($1/3 < |\Phi| \leq 1$ )
will be separated by domain walls as they are distinguished by the phases $2\pi j/3$ with $j=0,1,2$.
Nevertheless, the formation of the center domains, the path of ionization and the distribution of the color charges from confining 
phase to deconfining phase or vice-versa will also depend on the nature of the color singlet potential for pure gauge (where $Z(3)$ is 
spontaneously broken) and also that with matter field (where $Z(3)$ is explicitly broken), which we discuss below using (\ref{eq:c4:potl}).

\subsection{Gauge Sector}
\label{ssec:c4:gauge_sec}
The color singlet gauge potential is obtained in (\ref{eq:c4:potl}) as
\begin{eqnarray}
 \Omega^g_{S}&=&\Omega_g -\kappa T\ln H \nonumber \\
& =& 2  T \int\frac{d^3p}{(2\pi)^3} \ln
\left( 1 + \sum_{m=1}^8 a_m\, e^{-m\beta\epsilon_g}\right ) -\kappa T\ln H , \label{gpotl}  
\end{eqnarray}
where the first term describes the interaction of the spatial gluons with the PL 
(background temporal gauge field $A_0$ in (\ref{eq:c4:thfluc})) at finite $T$. 
The second term known as Vandermonde term comes from invariant Haar measure.
The $Z(3)$ domains are plentiful (viz., equation (\ref{eq:c4:eq11})) as the gluon dynamics are 
solely governed by the thermal fluctuation of the background gauge field $A_0$  in (\ref{eq:c4:thfluc}).

In figure~\ref{fig:c4:gauge_pot_fig} the color singlet gauge potential in a complex $\Phi$ plane is displayed for three temperatures. 
The left panel corresponds to 3-dimensional plots whereas right panel represents corresponding contour plots.
As can be seen the gauge potential, below $T< 270$ MeV,  has only one global minimum whereas for $T\ge 270$ MeV, it shows 
three minima representing a spontaneously broken $Z(3)$ phase for pure gauge.  The corresponding contour plots also display
the same features.  So, $T < 270$ MeV is color confining phase, where $Z(3)$ is unbroken as there is no ionization of
color charges. On the other hand $T\ge 270$ MeV there are ionization of color charges as the center symmetry is spontaneously broken
and those charges reside at the three minima in the potential in figure~\ref{fig:c4:gauge_pot_fig}  or three maxima in 
color space in Figs.~\ref{fig:c4:phi_3d_f} and \ref{fig:c4:phi_eqva} separated by distinct phases $2\pi j/3$ with $j=0,1,2$ and also  
by domain walls~\cite{Asakawa:2012yv}.  $T\sim (265-270)$ MeV, 
possibly indicates a phase transition for pure gauge and is in agreement with lattice result~\cite{Boyd:1996bx}.

\begin{figure}
\begin{center}
\includegraphics[width=0.45\linewidth,height=0.4\linewidth, angle=0]
{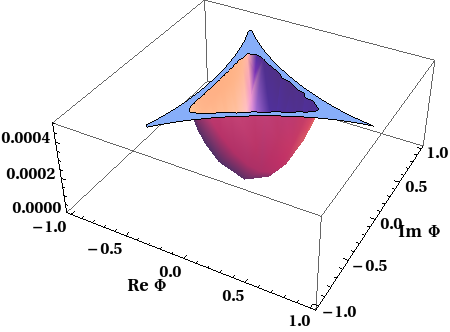}
\includegraphics[width=0.4\linewidth,height=0.35\linewidth, angle=0]
{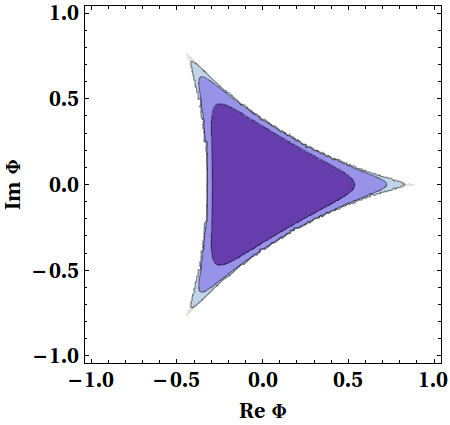}
\includegraphics[width=0.45\linewidth,height=0.4\linewidth, angle=0]
{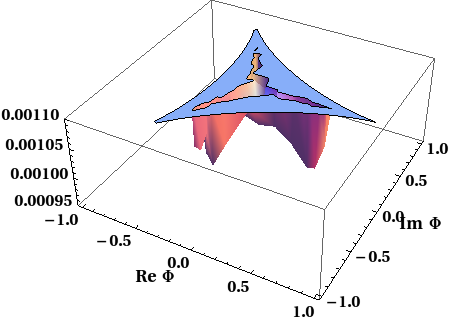}
\includegraphics[width=0.4\linewidth,height=0.35\linewidth, angle=0]
{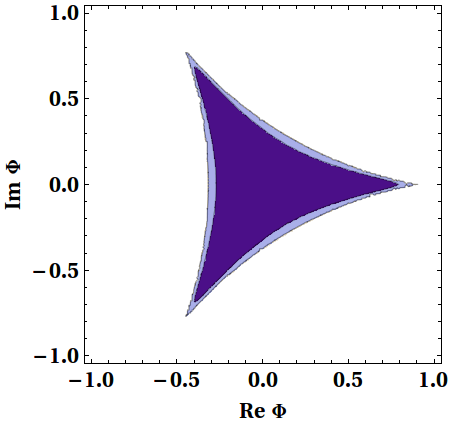}
\includegraphics[width=0.45\linewidth,height=0.4\linewidth, angle=0]
{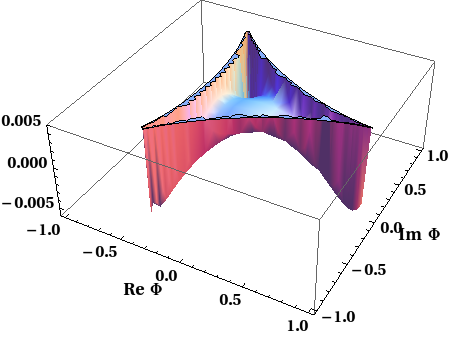}
\includegraphics[width=0.4\linewidth,height=0.35\linewidth, angle=0]
{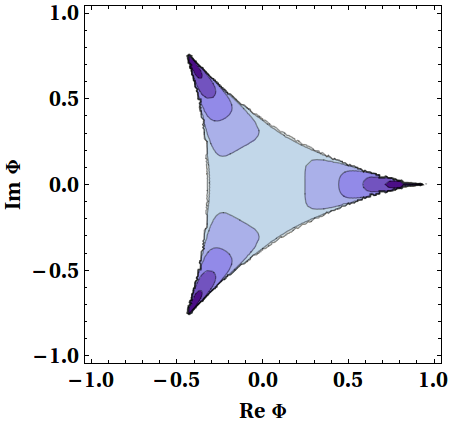}
\caption{{\it Left Panel:} A 3D plot of $\Omega_g\,  - \, \kappa T\ln \Theta_H$ in a 
complex $\Phi$ plane for $T=50,\, 270, \, {\mbox{and}} \,
350$ MeV and $\kappa=0.0075$ GeV$^3$.  {\it Right Panel:} Corresponding contour plots.}
\label{fig:c4:gauge_pot_fig}
\end{center}
\end{figure}

In asymptotically high temperatures ($T\gg T_c$),  $\Phi,{\bar\Phi} \to 1$, $\langle {A}_0^2\rangle \to 0$, one recovers 
free gluon gas from $\Omega_g^{\rm{PL}}$ as 
\begin{eqnarray}
\Omega_S^{g; \; \Phi,{\bar\Phi}\rightarrow 1}
\!\!\! &=& \!\! 2 (N_c^2-1) T \! \! \int\frac{d^3p}{(2\pi)^3}
\ln\left( 1 - e^{-\beta \epsilon_g} \right),
\label{eq12}
\end{eqnarray}
where $N_c=3$. The Vandermonde term due to the invariant Haar measure disappears and the spatial gluons are completely ionized.  

At low temperature ($T\ll T_c$),  the amplitude of $\langle A_0^2\rangle$ is high and $\Phi\rightarrow 0$, 
$\Omega_{S}^g$ becomes
\begin{eqnarray}
\Omega_S^{g;\;\Phi,{\bar\Phi}\rightarrow 0} 
= 2T \int\frac{d^3p}{(2\pi)^3} \left [
\ln\left( 1 - e^{-N_c\beta \epsilon_g} \right)^{2}\, \right.  
&& \nonumber \\ 
\left. \! +\! \ln\left( 1 - e^{2\pi i/N_c}e^{-\beta\epsilon_g} \right)
\! +\! \ln\left( 1 - e^{-2\pi i/N_c}e^{-\beta\epsilon_g} \right)\right ],
\label{eq:c4:eq13}
\end{eqnarray}
where the color charges are frozen through color-singlet states in the confining domain.
Now (\ref{eq:c4:eq13}) can be viewed in the following way: 

(i) The first term 
indicates that the  PL confines $N_c$ number of spatial gluons in a same energy 
state representing a glueball. There are two such copies which is consistent 
with $SU(3)$ gauge theory as $8\otimes 8\otimes 8$ generates only two singlet 
glueball states. Obviously, the $Z(3)$ charge is frozen in the color singlet 
glueballs through the localization of $Z(3)$ charge in the global $Z(3)$ 
minimum in figure~\ref{fig:c4:gauge_pot_fig}. The first term is negative which generates positive
pressure in QCD confined object.

(ii) The remaining two spatial gluons in the second and third terms are conjugate to each other but distinguished by $Z(3)$ phase. The potential, combining this two spatial gluons with $Z(3)$ phases,
can then be written as
\begin{eqnarray}
&& 2T \int\frac{d^3p}{(2\pi)^3}
\ln\left[ \left( 1 - e^{2\pi i/N_c}e^{-\beta\epsilon_g} \right)
\left( 1 - e^{-2\pi i/N_c}e^{-\beta\epsilon_g} \right)\right ] \nonumber \\
\nonumber \\ 
&& = 2T \int\frac{d^3p}{(2\pi)^3}
\ln\left( 1 + e^{-\beta \epsilon_g}+e^{-2\beta\epsilon_g} \right)\,,
\label{eq13c}
\end{eqnarray}
which is a positive definite quantity. Such states, may be condensates, are produced spontaneously in a nonabelian 
gauge theory as the pressure generated by this two spatial gluons in terms of the center group $Z(3)$ is negative 
compared to the confined object (first term in (\ref{eq:c4:eq13})). This two  gluons should not contribute directly 
to the thermodynamics in a confined phase. Rather they provide a nonperturbative ground state pressure which is negative and unbound 
from below that can be viewed as a general confining background of strong interaction. In LQCD calculations of equation of states 
this confining background is removed so that the pressure starts from zero or positive value in the confined phase (first term in (\ref{eq:c4:eq13})), 
i.e., in the  low temperature phase ($T\ll T_c$).

%%%%%%%%%%%%%%%%%%%%%%%%%%%%%%%%%%%%%%%%%%%%%%%%%%%%%
%\section{Matter Sector}
%\label{sec:Matter Sector}
%%%%%%%%%%%%%%%%%%%%%%%%%%%%%%%%%%%%%%%%%%%%%%%%%%%%

\begin{figure}
\begin{center}
\includegraphics[width=0.45\linewidth,height=0.3\linewidth, angle=0]
{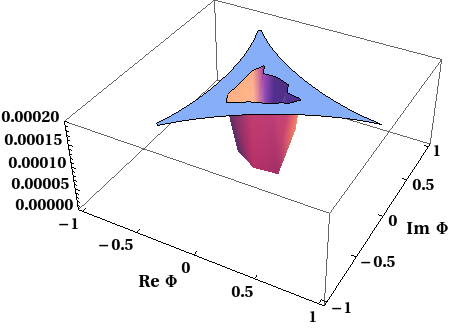}
\includegraphics[width=0.4\linewidth,height=0.3\linewidth, angle=0]
{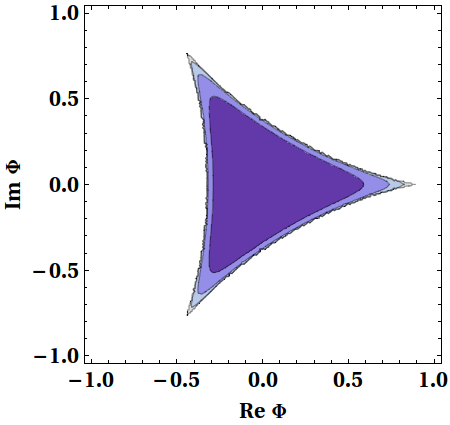}
\includegraphics[width=0.45\linewidth,height=0.3\linewidth, angle=0]
{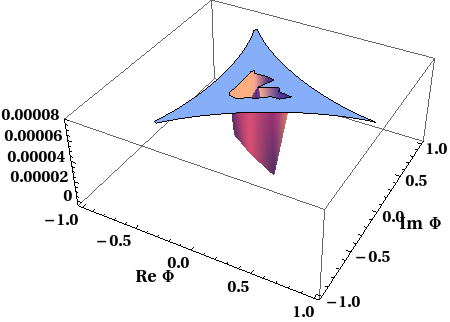}
\includegraphics[width=0.4\linewidth,height=0.3\linewidth, angle=0]
{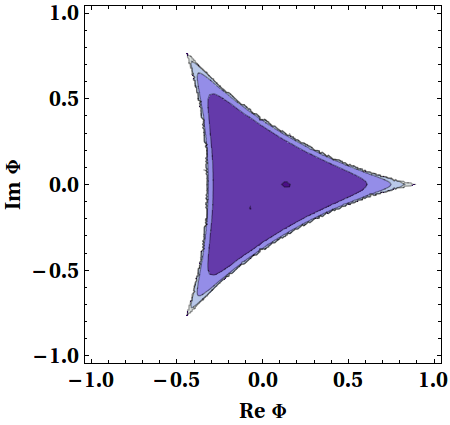}
\includegraphics[width=0.45\linewidth,height=0.3\linewidth, angle=0]
{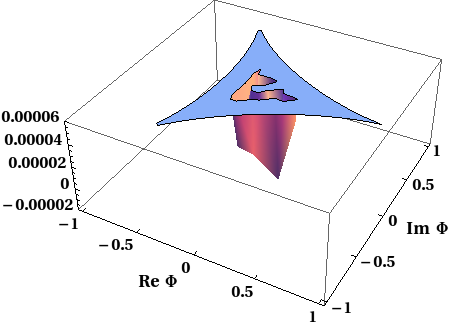}
\includegraphics[width=0.4\linewidth,height=0.3\linewidth, angle=0]
{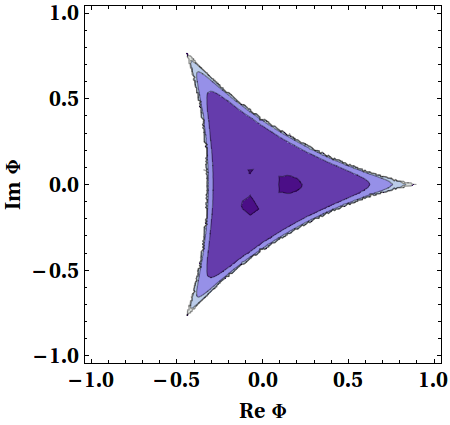}
\includegraphics[width=0.45\linewidth,height=0.3\linewidth, angle=0]
{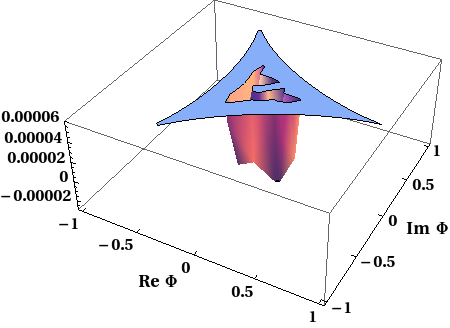}
\includegraphics[width=0.4\linewidth,height=0.3\linewidth, angle=0]
{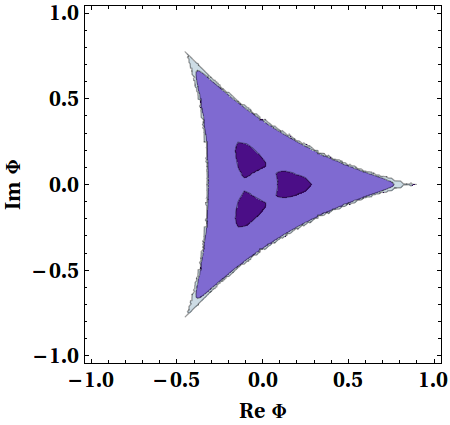}
\includegraphics[width=0.45\linewidth,height=0.3\linewidth, angle=0]
{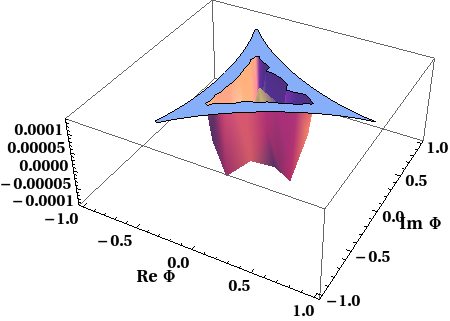}
\includegraphics[width=0.4\linewidth,height=0.3\linewidth, angle=0]
{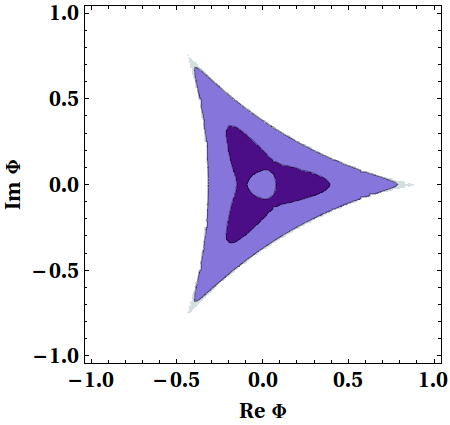}
\end{center}
\end{figure}

\begin{figure}
\begin{center}
\includegraphics[width=0.4\linewidth,height=0.3\linewidth, angle=0]
{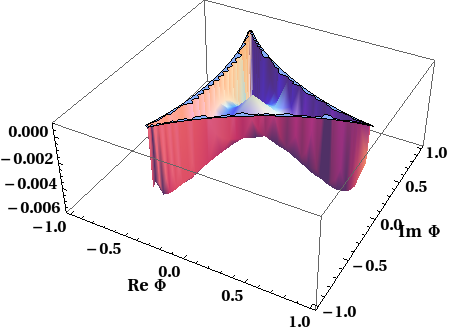}
\includegraphics[width=0.4\linewidth,height=0.3\linewidth, angle=0]
{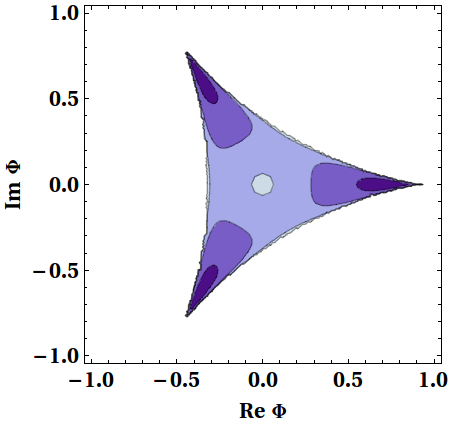}
\caption{{\it Left Panel:} A 3D plot of full potential $\Omega=\Omega_q+\Omega_{\bar q}+\Omega_g\,  - \, 
\kappa T \ln H$ in a complex $\Phi$ plane for $T=140,\, 149,\, 155, \, 160, \, 170 \, \, \, {\mbox{and}} \, \, \, 
250$ MeV with $\mu=0$ and $\kappa=0.0075$ GeV$^3$. 
{\it Right Panel:} Corresponding contour plots.}
\label{fig:c4:full_pot_fig}
\end{center}
\end{figure}

\subsection{Matter sector}
\label{ssec:c4:mat_sec}
The full color singlet potential in presence of matter field is given by (\ref{eq:c4:potl}) as
\begin{eqnarray}
\Omega^{S}=\Omega_q+\Omega_{\bar q}+\Omega_g-\kappa T\ln H . \label{eq:c4:mpotl}
\end{eqnarray}
The $Z(3)$ symmetry with quarks and antiquarks in the full potential is explicitly broken  
under the rotation of $Z(3)$ since they also carry the $Z(3)$ charge. This can be viewed from figure~\ref{fig:c4:full_pot_fig} that 
displays a plot of $\Omega^{S}$ in a complex $\Phi$ plane for a set of temperatures. The left panel in figure~\ref{fig:c4:full_pot_fig}
corresponds to 3-dimensional plots whereas right panel corresponds to contour plots in a complex $\Phi$ plane. 
In the range $ 0 <T < 140$ MeV, the potential has only one minimum (not shown here in figure~\ref{fig:c4:full_pot_fig}) that apparently 
represents $Z(3)$ global minimum. This is
because the  $Z(3)$ color charges are still frozen in hadrons in the confining domain. 
In the temperature domain $140 < T (\rm{MeV}) < 150$, there is still one minimum but has moved away from the $Z(3)$ center 
and does not remain symmetric under $Z(3)$ rotation $2\pi/3$. Here, the mean free path of the color charges is determined by
the effective size of the $Z(3)$ domains~\cite{Asakawa:2012yv}.  For $150 \le T(\rm{MeV}) < 170$,  
the potential shows one global minimum with larger depth and two local minima with smaller but uneven depth.  This implies that 
all three minima are not symmetric under the  $Z(3)$ rotation $2\pi/3$. This is unlike the pure gauge case in figure~\ref{fig:c4:gauge_pot_fig}, 
an indication of explicit $Z(3)$  symmetry breaking in presence of dynamical quarks. In this region $150 \le T(\rm{MeV}) < 170$ the mean free path 
of the color charges begin to increase as they tend to move from local minima to energetically favorable global minimum.  
 
Interestingly, the explicit symmetry breaking due to the presence of the matter fields leads to a {\it metastable state} in the temperature range 
($145\le T (\rm{MeV}) \le 170$) and beyond which the system crosses smoothly to
the deconfined phase. So, the concept of $T_c$  is not very well defined and LQCD calculations estimate it through 
various observable~\footnote{The analysis of HISQ/{\it tree} and {\it asqtad} 
action by HotQCD collaboration give a consistent results~\cite{Bazavov:2011nk,Petreczky:2012ct} in the continuum 
limit $T_c=(159\pm9)$ MeV from the peak of susceptibilities. The Wuppertal-Budapest collaboration~\cite{Borsanyi:2010bp} using {\it stout} action 
found $T_c=147(2)(3)$ MeV, $157(3)(3)$ MeV and $155(3)(3)$ MeV from  the peak position of susceptibility, and 
inflection points in chiral and renormalized chiral condensates.} and find in the range $(147-160)$ MeV. 
Our observation of $T_c\sim (165- 170)$ MeV where the instability of the metastable minima stabilizes due to 
the expansion of the domains as can be seen clearly from the contour plots in the right panel of figure~\ref{fig:c4:full_pot_fig}. 
Like gauge sector the depth of all three minima becomes almost symmetric for $T\ge 170 $MeV 
as the domains expand with temperature. The color charges, irrespective of their nature, begin to reside at those  minima in 
figure~\ref{fig:c4:full_pot_fig} for $T\ge 170$ MeV.
These domains are separated by non-perturbative domain walls even well above $T_c$ because the fluctuations of 
the background gauge field are still nonzero.  Moreover, the mean free path of the color charges becomes of the order of 
the effective size of these domains~\cite{Asakawa:2012yv}.  Since the domains expand, around $T\ge T_c$ the domain size is usually smaller 
that corresponds to shorter wavelength appropriate for hydrodynamics to be applicable
whereas the perturbative QCD may be applicable at high $T$ as the domain size increases. There will also be plenty of  domains
at high $T$ due to the fluctuations of the background gauge field.
In the color deconfined phase these domains as well domain walls act as scattering centers that cause high energy jet 
to lose energy through gluon radiations and get quenched~\cite{Asakawa:2012yv}. 

In the asymptotically high temperature ($T\gg T_c$),  equation (\ref{eq:c4:mpotl}) becomes
\begin{eqnarray}
\Omega_S^{\Phi,{\bar\Phi}\rightarrow 1}  
\!\!\! &=& \!\! -2N_fT \int\frac{d^3p}{(2\pi)^3}
\ln\left( 1 + e^{-\beta(\epsilon_q-\mu)} \right)^{N_c} \nonumber \\
&& -2N_fT \int\frac{d^3p}{(2\pi)^3}
\ln\left( 1 + e^{-\beta(\epsilon_q+\mu)} \right)^{N_c} \nonumber \\
&& + \Omega_g^{{\rm{PL}};\Phi,{\bar\Phi}\rightarrow 1}\, , 
\label{eq7}
\end{eqnarray}
where $N_c=3$. It represents the thermodynamic potential for
a free colored quark, antiquark and gluon gas at high temperature, i.e, the color charges are
completely ionized and reside at those minima in potential of figure~\ref{fig:c4:full_pot_fig} or equivalently at those 
maxima in color space of Figs.~\ref{fig:c4:phi_3d_f} and \ref{fig:c4:phi_eqva}. 

At low temperature ($T\ll T_c$), the potential with matter part can be written as
\begin{eqnarray}
\Omega^{\rm{PL};\Phi,{\bar\Phi}\rightarrow 0} 
&=&\!\! -2N_fT \!\! \int \!\! \frac{d^3p}{(2\pi)^3}
\ln\left( 1 + e^{-\beta N_c(\epsilon_q-\mu)} \right) \nonumber \\
&& -2N_fT \!\! \int \!\! \frac{d^3p}{(2\pi)^3}
\ln\left( 1 + e^{-\beta N_c(\epsilon_q+\mu)} \right) \nonumber \\
&&+ \Omega_g^{{\rm{PL}};\Phi,{\bar\Phi}\rightarrow 0} 
\,.
\label{eq:c4:eq8}
\end{eqnarray}
This represents the thermodynamic potential for a composite color singlet
object containing three quarks(antiquarks) in a same color state with the same energy. So 
is for gluons as discussed earlier. One can also combine appropriate terms in (\ref{eq:c4:eq8}) to get mesons, baryons, 
hybrid mesons, glueballs etc.
In other words, the $SU(3)$ color singlet restriction vis-a-vis PL dynamically confines three 
colored charges in a same energy state, which finally forms a color neutral
object like  baryon and glueball. This is because color charges get frozen in color singlet states like hadrons 
in the global minimum when $T\ll T_c$ . 
This essentially boils down to the fact that the $SU(3)$ color singlet restriction vis-a-vis PL dynamically 
provides the basis for the recombination of partons for hadronization
from quark-gluon plasma when it cools down  below $T_c$. In Refs.~\cite{Abir:2009sh} the colorsingletness has explicitly 
shown to provide the natural explanation of the scaling law (of the valence partons) of the elliptic flow of the identified 
hadrons in heavy-ion collisions, a direct evidence of deconfined phase~\cite{Adare:2006ti,Fries:2003vb,Greco:2003xt}.

\section{Conclusion}
\label{sec:c4:con}
We show that the color singlet ensemble of a quark-gluon gas  becomes equivalent to that of PL model within a stationary point approximation. The calculation is based on quantum statistical mechanics 
with a global $SU(3)$ symmetry but considering the Haar measure at each spatial points to take into account the confinement effect. The normalized character in fundamental representation of $SU(3)$ exhibits center symmetry, $Z(3)$, of $SU(3)$ akin to PL. In the process, we have also obtained pure gauge potential explicitly. 

The color singlet gauge potential shows center symmetry which is spontaneously broken in high temperature phase ($T\ge 270$ MeV). When matter field is added the center symmetry is found to be broken explicitly, which leads to a metastable state in the temperature domain $145 \le T(\rm{MeV}) \le 170$. The instability of the metastable state stabilizes for $T\ge 170$MeV and there are domains formed in the deconfined phase. We also discussed the phenomenological consequences of these center domains, both in pure gauge as well as with dynamical quarks, on color confining-deconfining phase transition or vice-versa in QCD, through the color singlet vis-a-vis PL potential. The center symmetry dictates that the confined phase appears as a color singlet object from the dynamical recombination of three partons as given in (\ref{eq:c4:eq8}), plus a confining background. This would solely describe the thermodynamic properties of color singlet structures like baryon, antibaryon, meson and glueball. Most of the effects of 
heavy-ion 
collisions: non-perturbative nature of the deconfined phase, fluid nature, jet quenching, recombination of hadronization etc can be understood in terms of the center domains. More calculations in this direction are required to make quantitative predictions on the consequences of center domains in heavy-ion phenomenology.

\chapter{Summary and prospect}
\label{chapter:sum_pros}

The phase diagram of strong interaction, as understood so far through QCD, is a complex one with many exotic phases. QGP, a strongly interacting hot and/or dense medium, is one of them which can be created with the currently available collider facilities known as HIC. In this dissertation we have studied some of the characteristics of this QGP phase. Such investigations have huge physical implications, since QGP is believed to have existed during the evolution of the early universe and is also presumed to be there in the core of neutron stars. Apart from that, knowledge of such strongly interacting phenomena also leads to a better understanding of other strongly correlated systems in nature.
 
The experimental evidence of asymptotic freedom and the success of standard model in the QCD sector, make us all believe that QCD is the theory of strong interaction. But still, there are some problems associated with the QCD. Because of its nonperturbative nature, the equations of motion arising from the Lagrangian of QCD cannot be solved exactly throughout the energy scale. The coupling constant, which runs with the energy inversely, becomes too large at low energy to be solved perturbatively. The failure of perturbative method poses serious problem in understanding the strong interaction phenomena at low energy scale \textendash~specifically the formation of hadrons from the quark degrees of freedom.

There is a first principle numerical method, known as lattice QCD (LQCD), which can solve the QCD Lagrangian exactly and can also describe the hadronic spectra starting with quarks and gluons. But this method is very costly and time consuming and also has some inherent problem at finite density, which is so far not resolved in a desired manner. So this prohibits the application of LQCD throughout the regime a hot and dense system. 

This is exactly where the applications of effective QCD models become useful. Since these are effective models they try to mimic QCD as closely as possible, particularly the hadronic properties are well reproduced by them. There are many such models. Here we work with the NJL model. The spontaneous chiral symmetry breaking is well taken care of by it. Apart from the chiral symmetry there is another important phase transition relevant for QGP \textendash~known as confinement-deconfinement transition. This confinement effect is not considered in the NJL model. To take it into account, NJL model is further augmented through the inclusion of a background gauge field in its Polyakov loop extended version, known as PNJL model. A major portion of the works done in this thesis uses these two models. PNJL model can further be extended by considering an entangled vertex and the incorporating model is known as entangled PNJL (EPNJL) model. In this model the coupling constants run with the temperature 
and chemical potential due to the entangled vertex. We use another effective model where we exploit the quantum statistical nature of QGP and the symmetries associated with it. Particularly the $SU(3)_C$ symmetry is emphasized through a group theoretical projection operator. This projection allows only the color singlet physical states to exist and we call such effective model as color singlet (CS) model.

Using the effective models we study the correlation function (CF) and the spectral properties associated with it. We know that the hadronic properties are embedded in the CFs. We have particularly calculated the vector meson current-current CF. The spatial part of the vector current CF is related with the dilepton multiplicity. The lepton pairs, having a longer mean free path, leaves the fireball with minimum interaction and thus carrying the undistorted information. This is why they have been considered as a reliable signal of QGP. Then we also exploit the temporal component of the vector CF and the associated conserved quark number density fluctuations. Thermodynamically this fluctuation is expressed in terms of susceptibility called as quark number susceptibility (QNS). QNS is also judged as an important probe for quark-hadron phase transition, which rises sharply in the transition region.

In chapter~\ref{chapter:JHEP} we explore the vector meson CF with and without the isoscalar-vector (I-V) interaction. We use a resummation scheme known as ring approximations for incorporating the I-V interaction, which becomes important specially for a system at finite quark density. First we study the behaviour of the mean fields both in absence and in presence of I-V interaction. Then we investigate the vector spectral function (SF) and extract the dilepton rate. Without the I-V interaction we observe that the SF and the associated dilepton rate in NJL model becomes quantitatively equal to that in free field theory. But in the PNJL model both quantities are enhanced in the deconfined state as compared to the free field theory or NJL model. This happens because of the presence of Polyakov loop field that suppress the color degrees of freedom in the deconfined phase just above $T_c$. We compare our findings with the available lattice data. Then the calculation is repeated again with the I-V interaction 
being included. It is observed that such inclusion further suppresses the color degrees of freedom up to a moderate value of the temperature above $T_c$ implying a stronger correlation among the color charges in the deconfined phase. This suppression is reflected in the corresponding dilepton plot with the increase in dilepton rate. As a prospect of this investigation we can further explore the photon rate in the ambit of these effective models. This will be interesting since photons are also considered as a trustworthy signal for QGP.

In chapter~\ref{chapter:PRD} we revisited the calculation done in chapter~\ref{chapter:JHEP} in an entangled environment where the chiral and the confinement dynamics are strongly correlated through some effective vertices. The choices for the effective vertices are done following some ansatzes which are guided by the symmetry principle, namely the chiral and $Z(3)$ symmetries. Through such vertices the coupling constants become dependent on PL fields and in turn on temperature and chemical potential. These dependence leads to the running of the coupling constants. We have investigated the implications of such running coupling constants on the SF and the corresponding dilepton rate and compare them with those in PNJL model. Because of the strong entanglement the color degrees of freedom in the deconfined phase gets enhanced as compared to the PNJL model. This in turn suppresses the strength of the vector SF and the peak is shifted to a higher energy compared to that of PNJL model but the strength is higher 
than 
the free one at low energy. This effect is expectedly reflected in the corresponding dilepton rate plots where the rate is suppressed in EPNJL model compared to the PNJL model, though it is higher as compared to the Born rate in the deconfined phase. This indicates that the entanglement effect leads to a relatively less production of lepton pairs at low energy.

Then we turn our attention to the Euclidean correlator in vector channel and the conserved density fluctuation associated with temporal correlator in chapter~\ref{chapter:Susc}. These quantities, particularly the QNS, are important to understand the properties of the deconfined nuclear matter. These quantities are explored both in presence and absence of the I-V interaction. In absence of the I-V interaction, the CF in NJL model become quantitatively equivalent to those of free field theory. In case of PNJL it is different from both free and NJL ones because of the presence of the PL fields that suppress the color degrees of freedom in the deconfined phase just above Tc, which suggests that some nontrivial correlation exist among the color charges in the deconfined phase. On the other hand, in absence of I-V interaction the QNS rises sharply in the crossover region and then saturates to the Stefan-Boltzmann limit at higher value of temperature. This happens for all the three effective models we have used \
textendash\, NJL, PNJL and EPNJL. The sharp rise remains there once we include the I-V interaction but it never reaches or shows the tendency to approach the Stefan-Boltzmann limit with the increase of temperature. It was expected that with the help of entangled vertex this problem will be eradicated. But we showed that with the available choices of ansatzes for the effective vertices this problem persists. This leaves us with the scope of guessing proper form of ansatzes that will eventually remove the problem. As we have discussed about the fluctuations it is also noteworthy that so far chiral susceptibility has not been explored in the PNJL model environment. There should be some attempt for this.

In the concluding chapter~\ref{chapter:JPG} we have made use of the quantum statistical nature of the QGP restricted by the $SU(3)_C$ symmetry. This symmetry restriction allows the existence of only color singlet physical states. We show here that the color singlet ensemble of a quark-gluon gas becomes equivalent to that of PL model within a stationary point approximation. It is also found that the normalized character in fundamental representation of $SU(3)_C$ exhibits center symmetry, $Z(3)$, akin to PL. We further explore the deconfinement phase transition in the ambit of CS model. What we found is that the pure gauge potential shows center symmetry which is spontaneously broken in high temperature phase ($T\ge 270$ MeV). As we add the quarks into the system the center symmetry is found to be broken explicitly, which leads to a metastable state in the temperature domain $145 \le T(\rm{MeV}) \le 170$. The instability of the metastable state stabilizes for $T\ge 170$MeV and there are domains formed in the 
deconfined phase. We 
discuss in details the implications of these domains both for pure gauge theory and the one with dynamical quarks. Many of the characteristics of the fireball created in HICs such as non-perturbative nature of the deconfined phase, fluid nature, jet quenching, recombination of hadronization etc can be understood in terms of these center domains.

\appendix
\chapter[Correlation Function]{Correlation function and spectral representation}
\label{app:cor_sp_rep}
%%%%%%%%%%%%%%%%%%%%%%%%%%%%%%%%%%%%%%%%%%%%%%%%%%%%%%%%%%%%%%%%%%%%%%%%%%%%%%%%%%%%%%%%%%%%%%%%

From the Kramers-Kronig relation it is known that the real and imaginary parts of a complex function are related with each other. Thus for a CF ${\cal G_{\mu\nu}}$ we can write
\begin{eqnarray}
{\mbox{Re}}\,{\cal G}_{\mu\nu}(\omega_n, \vec q) =
\frac{1}{\pi} P \int_{-\infty}^{\infty} d\omega \ 
\frac{{\mbox{Im}}\,{\cal G}_{\mu\nu}(\omega, \vec q)}{\omega-i\omega_n},
\end{eqnarray}
where $P$ stands for Cauchy principal value. Now with the definition of SF as $\sigma_{\mu\nu}(\omega, \vec q)=\frac{1}{\pi} {\mbox{Im}} \ {\cal G}_{\mu\nu}(\omega,\vec q)$ and writing ${\cal G}_{\mu\nu}$ in place of ${\mbox{Re}}\,{\cal G}_{\mu\nu}$ we have,
\begin{eqnarray}
{\cal G}_{\mu\nu}(\omega_n, \vec q) &=&
\int_{-\infty}^{\infty} d\omega \ \frac{\sigma_{\mu\nu}(\omega, \vec q)}{\omega-i\omega_n}
= \int_{-\infty}^{\infty} d\omega \ \frac{\sigma_{\mu\nu}(\omega, \vec q)}{\omega-i\omega_n}\  \frac{e^{\beta\omega}-1}{e^{\beta\omega}-1} \nonumber \\
&=& \int_{-\infty}^{\infty} d\omega \ {\sigma_{\mu\nu}(\omega, \vec q)}\ \big(1+n_B(\omega)\big)\  \frac{e^{-\beta\omega}-1}{i\omega_n-\omega},
\end{eqnarray}
where $n_B$ is the Bose-Einstein distribution function. For bosons the Matsubara frequencies are $\omega_n=2\pi n T$ and we can further write
\begin{eqnarray}
{\cal G}_{\mu\nu}(\omega_n, \vec q) &=& \int_{-\infty}^{\infty} d\omega \ {\sigma_{\mu\nu}(\omega, \vec q)}\ \big(1+n_B(\omega)\big)\  \frac{e^{(i\omega_n-\omega)\beta}-1}{i\omega_n-\omega}\nonumber \\
&=& \int_{0}^{\beta} d\tau'e^{i\omega_n\tau'} \ \int_{-\infty}^{\infty} d\omega \ {\sigma_{\mu\nu}(\omega, \vec q)}\ \big(1+n_B(\omega)\big)\ e^{\omega\tau'}.
\label{eq:app1:three}
\end{eqnarray}
Using ${\cal G_{\mu\nu}}(\tau,\vec q)=T\sum_{n=-\infty}^{\infty}{\cal G_{\mu\nu}}(\omega_n,\vec q)\ e^{-i\omega_n\tau}$ and $T\sum_{n=-\infty}^{\infty}e^{i\omega_n(\tau'-\tau)}=\delta(\tau'-\tau)$ we have from (\ref{eq:app1:three}), 
\begin{eqnarray}
{\cal G}_{\mu\nu}(\tau,\vec q) &=& \int_{-\infty}^{\infty} d\omega \ {\sigma_{\mu\nu}(\omega, \vec q)}\ \big(1+n_B(\omega)\big)\ \int_{0}^{\beta} d\tau'\delta(\tau'-\tau)e^{-\omega\tau'}\nonumber\\
&=&\int_{-\infty}^{\infty} d\omega \ {\sigma_{\mu\nu}(\omega, \vec q)}\ \big(1+n_B(\omega)\big)\ e^{-\omega\tau}.
\end{eqnarray}
Further breaking the limits in two regions and with some mathematical manipulations we have, 
\begin{eqnarray}
{\cal G}_{\mu\nu}(\tau,\vec q) &=&\int_{0}^{\infty} d\omega \ {\sigma_{\mu\nu}(\omega, \vec q)}\ \big(1+n_B(\omega)\big)\ e^{-\omega\tau}\nonumber\\
&-&\int_{\infty}^{0} d\omega \ {\sigma_{\mu\nu}(-\omega, \vec q)}\ \big(1+n_B(-\omega)\big)\ e^{\omega\tau}.
\end{eqnarray}
With the identity $1+n_B(-\omega)=-n_B(\omega)$ and also using the fact that the SF $\sigma(\omega,\vec q)$ is an odd function of the four momenta~\cite{Ding:2010ga} we have
\begin{eqnarray}
{\cal G}_{\mu\nu}(\tau,\vec q) &=&\int_{0}^{\infty} d\omega \ {\sigma_{\mu\nu}(\omega, \vec q)}\ \big(e^{-\omega\tau}+n_B(\omega)e^{-\omega\tau}+n_B(\omega)e^{\omega\tau}\big)\nonumber\\
&=&\int_{0}^{\infty} d\omega \ {\sigma_{\mu\nu}(\omega, \vec q)}\ \frac{\cosh[\omega(\tau-\beta/2)]}{\sinh[\omega\beta/2]}
\end{eqnarray}

\chapter[Density Fluctuation]{Response to conserved density fluctuation}
\label{app:flu_qns}
%%%%%%%%%%%%%%%%%%%%%%%%%%%%%%%%%%%%%%%%%%%%%%%%%%%%%%%%%%%%%%%%%%%%%%%%%%%%%%%%%%%%%%%%%%%%%%%%
We know the CF in terms of the SF. Writing down only the temporal part we have,
\begin{eqnarray}
{\cal G}_{00}(\tau,\vec q,T) &=&\int_{0}^{\infty} d\omega \ {\sigma_{00}(\omega, \vec q,T)}\ \frac{\cosh[\omega(\tau-\beta/2)]}{\sinh[\omega\beta/2]}.
\end{eqnarray}
In the limit of vanishing three momenta it becomes,
\begin{eqnarray}
{\cal G}_{00}(\tau T)\ \equiv\ {\cal G}_{00}(\tau,\vec 0, T) &=&-\int_{0}^{\infty} d\omega\ \chi_q\ \omega\ \delta (\omega)\ \frac{\cosh[\omega(\tau-\beta/2)]}{\sinh[\omega\beta/2]},
\label{eq:app2:two}
\end{eqnarray}
where $\sigma_{00}(\omega)=-\chi_q\ \omega\ \delta (\omega)$, the SF is represented by a delta function~\cite{Ding:2010ga}. Now we do a mathematical trick and write equation (\ref{eq:app2:two}) as
\begin{eqnarray}
{\cal G}_{00}(\tau T) &=&-\frac{1}{2}\int_{0}^{\infty} d\omega\ \chi_q\ \omega\ \delta (\omega)\ \frac{\cosh[\omega(\tau-\beta/2)]}{\sinh[\omega\beta/2]}\nonumber\\
&-&\frac{1}{2}\int_{0}^{\infty} d\omega\ \chi_q\ \omega\ \delta (\omega)\ \frac{\cosh[\omega(\tau-\beta/2)]}{\sinh[\omega\beta/2]}.
\end{eqnarray}
We change the variable, $\omega\rightarrow-\omega$ in the second term to obtain
\begin{eqnarray}
{\cal G}_{00}(\tau T) &=&-\frac{1}{2}\int_{0}^{\infty} d\omega\ \chi_q\ \omega\ \delta (\omega)\ \frac{\cosh[\omega(\tau-\beta/2)]}{\sinh[\omega\beta/2]}\nonumber\\
&-&\frac{1}{2}\int_{-\infty}^{0} d\omega\ \chi_q\ \omega\ \delta (\omega)\ \frac{\cosh[\omega(\tau-\beta/2)]}{\sinh[\omega\beta/2]}.
\end{eqnarray}
With few more mathematical manipulations we have,
\begin{eqnarray}
{\cal G}_{00}(\tau T)&=&-\frac{1}{2}\int_{-\infty}^{\infty} d\omega\ \chi_q\ \omega\ \delta (\omega)\ \frac{\cosh[\omega(\tau-\beta/2)]}{\sinh[\omega\beta/2]}\nonumber\\
&=&-\ T\ \chi_q(T).
\end{eqnarray}

\chapter[Temporal Component of Correlator]{Temporal component of correlator}
\label{app:correlator}
%%%%%%%%%%%%%%%%%%%%%%%%%%%%%%%%%%%%%%%%%%%%%%%%%%%%%%%%%%%%%%%%%%%%%%%%%%%%%%%%%%%%%%%%%%%%%%%%
The time-time (temporal) component of the vector correlator is 
\begin{align}
 \Pi_{00}(Q)=&\int{\frac{d^4P}{(2\pi)^4}{\mathrm {Tr}}\big[\gamma_0 S_f(K)\gamma_0 S_f(P)\big]},\\
 {\textrm {with}}~ S_f(P)=&~\frac{1}{\gamma_\mu P^\mu-M_f+\gamma_0\mu'}~;~\mu'_{NJL}=\tilde\mu,~{\textrm {and}}~\mu'_{PNJL}=\tilde\mu-iA_4,\nonumber
%\label{eq.pi00}
\end{align}
where the symbol (${\mathrm {Tr}}$) implies traces over Dirac, color and flavor spaces. In the medium we introduce the temperature and need to perform the Matsubara sum over the discrete set of energies,
\begin{align}
 \Pi_{00}(\omega,\vec q)=N_f\sum_n\frac{1}{\beta}\int\frac{d^3p}{(2\pi)^3}{\mathrm {Tr_{D,c}}}\big[\gamma_0S(k^0,\vec k)\gamma_0S(p^0,\vec p)\big]~{\textrm {with}}
 \label{eq:app:correlator:two}
 \end{align}
\begin{align}
 S(k^0,\vec k)=\frac{1}{\gamma_0(i\omega_n+\omega+\mu')-\vec\gamma\cdot\vec k-M}~{\textrm {and}}~S(p^0,\vec p)=\frac{1}{\gamma_0(i\omega_n+\mu')-\vec\gamma\cdot\vec p-M}.\nonumber
\end{align}
Now we want to perform the traces over Dirac space. Before that we simplify the expression of propagator and write
\begin{align}
 S(k^0,\vec k)=\frac{\gamma_0(i\omega_n+\omega+\mu')-\vec\gamma\cdot\vec k+M}{(i\omega_n+\omega+\mu')^2-(k^2+M^2)}~{\textrm {and}}~S(p^0,\vec p)=\frac{\gamma_0(i\omega_n+\mu')-\vec\gamma\cdot\vec p+M}{(i\omega_n+\mu')^2-(p^2+M^2)}.
\end{align}
Now we utilize properties of gamma matrices to perform the Dirac trace, which leads to the expression
\begin{align}
{\mathrm {Tr_{D,c}}}\big[\gamma_0S(k^0,\vec k)\gamma_0S(p^0,\vec p)\big]={\mathrm {Tr_{c}}}\frac{4\big[(i\omega_n+\omega+\mu')(i\omega_n+\mu')+\vec k\cdot\vec p+M^2\big]}{[(i\omega_n+\omega+\mu')^2-E_k^2][(i\omega_n+\mu')^2-E_p^2]},
\label{eq:app:correlator_four}
\end{align}
where $E_k^2=k^2+M^2$ and $E_p^2=p^2+M^2$. We now do the Matsubara sums for which we use the following identities 
\begin{align}
\sum_n\frac{1}{n+ix}\frac{1}{n+iy}=&~\frac{\pi}{x-y}\big(\textrm{coth}(\pi x)-\textrm{coth}(\pi y )\big),\nonumber\\ 
\sum_n\frac{1}{n-x}\frac{1}{n-y}=&~\frac{\pi}{y-x}\big(\textrm{cot}(\pi x)-\textrm{cot}(\pi y)\big)\nonumber\\
\textrm{and}~\textrm{tanh}({x})=&~1-2n_F(x);
\label{eq:app:correlator:five}
\end{align}
where $n_F(x)=\frac{1}{e^{\beta x}+1}$ is the Fermi-Dirac distribution function. We use the method of partial fraction in equation (\ref{eq:app:correlator_four}) and put it in equation (\ref{eq:app:correlator:two}) to obtain
\begin{align}
 \Pi_{00}(\omega,\vec q)=&~N_f\sum_n\frac{1}{\beta}\int\frac{d^3p}{(2\pi)^3}\frac{1}{E_k\,E_p}{\mathrm {Tr_{c}}}\Big[\frac{E_kE_p+M^2+\vec k\cdot\vec p}{D_k^-\,D_p^-}+\frac{E_kE_p-M^2-\vec k\cdot\vec p}{D_k^-\,D_p^+}\nonumber\\
 &+\frac{E_kE_p-M^2-\vec k\cdot\vec p}{D_k^+\,D_p^-}+\frac{E_kE_p+M^2+\vec k\cdot\vec p}{D_k^+\,D_p^+}\Big],
\label{eq:app:correlator:six}
\end{align}
with $D_k^\pm=(i\omega_n+\omega+\mu')\pm E_k$ and $D_p^\pm=(i\omega_n+\mu')\pm E_p$. As an example let's evaluate the first sum:
\begin{align}
\sum_n\frac{1}{(i\omega_n+\omega+\mu')-E_k}~\frac{1}{(i\omega_n+\mu')-E_p},
\label{eq:app:correlator:seven}
\end{align}
we put the Matsubara frequencies for fermions to get
\begin{align}
&\sum_n\frac{1}{\big(i\frac{(2n+1)\pi}{\beta}+\omega+\mu'\big)-E_k}~\frac{1}{\big(i\frac{(2n+1)\pi}{\beta}+\mu'\big)-E_p}\nonumber\\
&\!\!\!\!\!\!\!\!=-\sum_n\frac{1}{n+\frac{i\beta}{2\pi}\big(E_k-\omega-\mu'-i\frac{\pi}{\beta}\big)}~\frac{1}{n+\frac{i\beta}{2\pi}\big(E_p-\mu'-i\frac{\pi}{\beta}\big)}~\Big(\frac{\beta}{2\pi}\Big)^2\nonumber\\
&\!\!\!\!\!\!\!\!=\frac{\beta}{2}\frac{1}{(\omega-E_k+E_p)}\Big(\big(\textrm{coth}(\frac{\beta}{2}E_k-\omega-\mu')-i\frac{\pi}{2}\big)-\big(\textrm{coth}(\frac{\beta}{2}E_p-\mu')-i\frac{\pi}{2}\big)\Big)\nonumber,
\end{align}
then using the identities in equation (\ref{eq:app:correlator:five}) the frequency sum in (\ref{eq:app:correlator:seven}) reduces to
\begin{align}
&\sum_n\frac{1}{(i\omega_n+\omega+\mu')-E_k}~\frac{1}{(i\omega_n+\mu')-E_p}\nonumber\\
&\!\!\!\!\!\!\!\!=\beta~\frac{1}{\omega-E_k+E_p}\big(\ n_F(E_p-\mu')-n_F(E_k-\omega-\mu')\big)
\end{align}
Performing the frequency sums for the other three terms equation (\ref{eq:app:correlator:six}) becomes
\begin{align}
 \Pi_{00}(\omega,\vec q) =&\,\, N_f\int\frac{d^3p}{(2\pi)^3}\frac{1}{E_{p}E_{k}}\Bigg\{
\frac{E_{p}E_{k}+M_f^2+\vec{p}\cdot\vec{k}}{\omega+E_{p}-E_{k}} \nonumber\\
& \times \left[f(E_{p}-\tilde \mu)+f(E_{p}+\tilde \mu)
-f(E_{k}-\tilde\mu)-
f(E_{k}+\tilde \mu)\right] \nonumber \\
& + \left(E_{p}E_{k}-M_f^2-\vec{p}\cdot\vec{k}\right)
\left[\frac{1}{\omega-E_{p}-E_{k}}-
\frac{1}{\omega+E_{p}+E_{k}}\right] \nonumber\\
& \times \left[1-f(E_{p}+\tilde\mu)
-f(E_{k}-\tilde\mu)\right]\Bigg\}, 
\label{eq:app:correlator:nine}
\end{align}
which is the equation (\ref{eq:c1total_pi00}) in chapter~\ref{chapter:JHEP} apart from a $N_c$ factor, which comes out in a straightforward manner for NJL model and for PNJL model the overall thermal distribution gets modified as discussed in the subsection~\ref{ssec:resum_corr} of chapter~\ref{chapter:JHEP}. Now we want to calculate the real and imaginary parts of the temporal correlator. For that purpose we use the following identity
\begin{align}
 \lim_{\eta\rightarrow0}\frac{1}{x\pm i\eta}=\textrm{PV}\Big(\frac{1}{x}\Big)\mp i\pi\delta(x),
 \label{eq:app:correlator:ten}
\end{align}
where PV stands for principal value. We first analytically continue $\omega\rightarrow\omega\pm i\eta$ and then using the equation (\ref{eq:app:correlator:ten}) we obtain
\begin{align}
 \lim_{\eta\rightarrow0}\frac{1}{\omega+i\eta-E_k+E_p}=\textrm{PV}\Big(\frac{1}{\omega-E_k+E_p}\Big)-i\pi\delta(\omega-E_k+E_p).
 \label{eq:app:correlator:eleven}
\end{align}
Similarly the other terms can be separated into real and imaginary parts. Then from (\ref{eq:app:correlator:nine}) it is trivial to obtain equations (\ref{eq:c1:repi00}) and (\ref{eq:c1:im_pi}) given in chapter~\ref{chapter:JHEP}.

\chapter[QNS in RPA]{Quark number susceptibility in ring approximation}
\label{app:susc}
%%%%%%%%%%%%%%%%%%%%%%%%%%%%%%%%%%%%%%%%%%%%%%%%%%%%%%%%%%%%%%%%%%%%%%%%%%%%%%%%%%%%%%%%%%%%%%%%
We want to get the real part of the temporal resummed correlator. From equation (\ref{eq:c1:c00}) in chapter~\ref{chapter:JHEP} we have,
\begin{eqnarray}
 C_{00}(\omega,\vec q) = \frac{\Pi_{00}}{1+G_V(\frac{\omega^2}{{q}^2}-1)\Pi_{00}}.
 \label{eq:app:susc_c00}
\end{eqnarray}
Since we are interested in real part of $C_{00}$ we write the the above equation in a convenient way as
\begin{equation}
 C_{00}(\omega,\vec q)=\frac{{\rm Re \Pi_{00}}+{i\rm Im \Pi_{00}}}
 {1+ G_V \big(\frac{\rm Re\, {\cal S}+i\rm Im\, {\cal S}}{q^2}-1\big)+
 \big({\rm Re \Pi_{00}}+{i\rm Im \Pi_{00}}\big) ^2} 
 %\label{eq:c1:C00}
\end{equation}
with ${\cal S}=\omega^2$. Manipulating the denominator we have,
\begin{equation}
 C_{00}(\omega,\vec q)=\frac{{\rm Re \Pi_{00}}+{i\rm Im \Pi_{00}}}
 {1+ G_V\, {\rm DI}+ iG_V\, {\rm DII}} 
 %\label{eq:c1:C00}
\end{equation}
with ${\rm DI}=\frac{1}{q^2}\big(\rm Re\,{\cal S}\,\rm Re\Pi_{00}-\rm Im\,{\cal S}\,\rm Im\Pi_{00}\big)-\rm Re\Pi_{00}$ and ${\rm DII}=\frac{1}{q^2}\big(\rm Re\,{\cal S}\,\rm Im\Pi_{00}+\rm Im\,{\cal S}\,\rm Re\Pi_{00}\big)-\rm Im\Pi_{00}$. To separate the real and the imaginary part we write

\begin{equation}
 C_{00}(\omega,\vec q)= \frac{\big({\rm Re \Pi_{00}}+{i\rm Im \Pi_{00}} \big)\big(1+ G_V\, {\rm DI}-iG_V\,{\rm DII}\big)}{\big(1+G_V\,{\rm DI}\big)^2+\big(G_V\,{\rm DII}\big)^2}. 
 %\label{eq:c1:C00}
\end{equation}
From it the real part is separated out as
\begin{equation}
 {\rm Re}\,C_{00}(\omega,\vec q)= \frac{{\rm Re \Pi_{00}}\big(1+G_V\, {\rm DI}\big)+{\rm Im \Pi_{00}}\,G_V\,{\rm DII}}{\big(1+G_V\,{\rm DI}\big)^2+\big(G_V\,{\rm DII}\big)^2}. 
 %\label{eq:c1:C00}
\end{equation}
Now $\omega$ being a complex quantity we write it as $\omega\rightarrow\omega+i\eta$. Then ${\cal S} = \omega^2-\eta^2+2i\omega\eta$. Thus ${\rm Re}\,{\cal S}=\omega^2-\eta^2$ and ${\rm Im}\,{\cal S}=2\omega\eta$. To get the real part of the resummed temporal correlator that we will use, we take the limit, $\eta\rightarrow0$. This leads to the expression
\begin{align}
{\rm{Re}}C_{00}(\omega,\vec q) 
 = \frac{{\rm{Re}}\Pi_{00}(\omega, \vec q)+G_V\left (\frac{\omega^2}{ q^2}-1\right){\cal I}(\omega,\vec q)}
 {1+2G_V{\left (\frac{\omega^2}{ q^2}-1\right ){\rm{Re}}\Pi_{00}(\omega, \vec q)}
 +\left(G_V(\frac{\omega^2}{ q^2}-1)\right)^2{\cal I}(\omega,\vec q)}, 
 \label{eq_re_c00}
\end{align}
where ${\cal I}(\omega,\vec q)=\big({\rm{Re}}\Pi_{00}(\omega, \vec q))^2+({\rm{Im}}\Pi_{00}(\omega, \vec q)\big)^2$.

\chapter[Vandermonde term]{Vandermonde term}
\label{app:vdm}
%%%%%%%%%%%%%%%%%%%%%%%%%%%%%%%%%%%%%%%%%%%%%%%%%%%%%%%%%%%%%%%%%%%%%%%%%%%%%%%%%%%%%%%%%%%%%%%%
The Vandermonde term can be written as the product of differences of the eigenvalues as
\begin{eqnarray}
 H(\theta)&=& \prod_{i >j}\vert e^{i\theta_i}-e^{i\theta_j} \vert^2.
\end{eqnarray}
For any complex number $Z_i$ we can write
\begin{eqnarray}
 \prod_{i> j} (z_i-z_j) = \mathrm{det~}M = \begin{vmatrix}
1 & z_{1} & z_{1}^{2} & \dots & z_{1}^{N-1} \\ 
1 & z_{2} & z_{2}^{2} & \dots & z_{2}^{N-1} \\
\vdots & \vdots & \vdots & \ddots &\vdots \\
1 & z_{N} & z_{N}^{2} & \dots & z_{N}^{N-1} \\ 
\end{vmatrix}
\end{eqnarray}
To express it in a general way, $M_{ij}=z_i^{j-1}$ and $(M^\dagger)_{ij}=\bar{z}_j^{i-1}$. Here we have $z_k=e^{i\theta_k}$, then
\begin{eqnarray}
 H(\theta) = \prod_{i> j} \left(e^{i\theta_i}-e^{i\theta_j}\right)^\dagger \left(e^{i\theta_i}-e^{i\theta_j}\right) = (z_i-z_j)^\dagger (z_i-z_j) = \mathrm{det~}(M^\dagger M).
 \label{eq:app:vdm_three}
\end{eqnarray}
Now let us suppose that $\mathrm{det~}(M^\dagger M)=\mathrm{det~}X$, where 
\begin{eqnarray}
 X_{ij}=(M^\dagger M)_{ij}=\sum_k M^\dagger_{ik}M_{kj} = \sum_k \bar{z}_k^{i-1} z_k^{j-1} = \sum_k {z}_k^{j-i}.
\end{eqnarray}
Thus from (\ref{eq:app:vdm_three}), we have for $i,j\leq N=3$,
\begin{eqnarray}
 H(\theta)=\mathrm{det~}(M^\dagger M) = \begin{vmatrix}
\sum_k{z}_k^0 & \sum_k{z}_k & \sum_k{z}_k^{2} \\ 
\sum_k\bar{z}_k & \sum_k{z}_k^0 & \sum_k{z}_k \\
\sum_k\bar{z}_k^2 & \sum_k\bar{z}_k & \sum_k{z}_k^0 \\
\end{vmatrix},
\label{eq:app:vdm_five}
\end{eqnarray}
with $\sum_k {z}_k=z_1+z_2+z_3=e^{i\theta_1}+e^{i\theta_2}+e^{i\theta_3}$. Now rank of $SU(3)$ is two, so we have two independent parameters, let us say, $\theta_1$ and $\theta_2$ and $\theta_3$ can be expressed in terms of them two as $\theta_3=-(\theta_1+\theta_2)$. Expressing different elements of the matrix in equation (\ref{eq:app:vdm_five}) in terms of the normalized character of the fundamental representation of $SU(3)$, we get
\begin{eqnarray}
 H(\theta)&=& \begin{vmatrix}
3 & 3\Phi & (9\Phi^2-6\bar\Phi) \\ 
3\bar\Phi & 3 & 3\Phi \\
(9\bar\Phi^2-6\Phi) & 3\bar\Phi & 3 \\
\end{vmatrix}\nonumber\\
&=& 27\big(1-6\Phi\bar\Phi+4(\Phi^3+\bar\Phi^3)-3(\Phi\bar\Phi)^2\big),
\end{eqnarray}
where we have used $\sum_k{z}_k=3\Phi$,\, $\sum_k\bar{z}_k=3\bar\Phi$\, $\sum_k{z}_k^2=9\Phi^2-6\bar\Phi$ and $\sum_k\bar{z}_k^2=9\bar\Phi^2-6\Phi$.

\chapter[Partition function]{Partition function calculation}
\label{app:part}
%%%%%%%%%%%%%%%%%%%%%%%%%%%%%%%%%%%%%%%%%%%%%%%%%%%%%%%%%%%%%%%%%%%%%%%%%%%%%%%%%%%%%%%%%%%%%%%%
\section{Quarks}
\label{sec:app:part:quarks}
The partition function for quarks is
\begin{eqnarray}
{\cal Z}_q\!\!=\int_{SU(N_c)}{\hspace*{-0.15in}}\!\!\! {\rm d}\mu (g) \  e^{2 N_f\sum_\alpha {\rm{tr}_c}  \ln \left (1+R_q
e^{-\beta (\epsilon_{q}^\alpha -\mu_q)}
\right )},
\label{eq:app:part_one}
\end{eqnarray}
where the normalized characters for quarks and antiquarks are written as $\Phi= \frac{1}{N_c} {\rm{tr}_c} R_q$ and $\bar{\Phi} =\frac{1}{N_c}{\rm{tr}_c} R_{\bar q}$, with
\begin{eqnarray}
{R}_q
&=& \mbox{diag}\left(
e^{i\theta_1}\,, e^{i\theta_2}\,, e^{i\theta_3} \right)\, ;  \hspace*{0.3in}
{R}_{\bar q}
= R^\dagger_q.   
%\label{image_f}
\end{eqnarray}
Now, as mentioned in appendix~\ref{app:vdm} $\theta_3$ can be expressed in terms of $\theta_1$ and $\theta_2$ as $\theta_3=-(\theta_1+\theta_2)$. Thus we can write,
\begin{align}
{\rm{tr}_c}\ln\big(1+R_qe^{-\frac{(\epsilon_q-\mu)}{T}}\big) =&~{\rm{tr}_c}\ln\big(1+\mathrm{diag}\left(A,B,C\right)Q\big)\nn
 =&~\ln\big(1+(A+B+C)Q+(AB+AC+BC)Q^2+ABCQ^3\big),
 \label{eq:app:part_three}
\end{align}
with $A=e^{i\theta_1},\, B= e^{i\theta_2},\, C= e^{-i(\theta_1+\theta_2)}$ and $Q = e^{-\frac{(\epsilon_q-\mu)}{T}}$. Then expressing the normalized characters in terms of $\Phi$ and $\bar\Phi$ we have,

\begin{eqnarray}
A+B+C &=& N_c\Phi,\nn
AB+AC+BC &=& N_c\bar{\Phi},\nn
ABC &=& 1.
\end{eqnarray}
Putting these in equation (\ref{eq:app:part_three}) we have,
\begin{eqnarray}
{\rm{tr}_c}\ln\big(1+R_qe^{-\frac{\epsilon^+}{T}}\big) = \ln\left(1+N_c(\Phi+\bar{\Phi}~e^{-\frac{\epsilon^+}{T}})e^{-\frac{\epsilon^+}{T}}+e^{-\frac{3\epsilon^+}{T}}\right).
\label{eq:app:part_five}
 \end{eqnarray}
Thus in the infinite volume limit, using the equation (\ref{eq:app:part_five}) in equation (\ref{eq:app:part_one}), the partition function for the quarks can be written as
\begin{eqnarray}
{\cal Z}_q\!\!=2VN_f\int\frac{d^3p}{(2\pi)^3}\ln\left(1+N_c(\Phi+\bar{\Phi}~e^{-\frac{\epsilon^+}{T}})e^{-\frac{\epsilon^+}{T}}+e^{-\frac{3\epsilon^+}{T}}\right).
\end{eqnarray}
Similarly, the partition function for antiquarks can also be calculated.

\section{Gluons}
\label{sec:app:part:gluons}
Here we outline the partition function calculation for gluons. For gluons we have,
\begin{eqnarray}
{\cal Z}_g\!\!=\int_{SU(N_c)}{\hspace*{-0.15in}}\!\!\! {\rm d}\mu (g) \  e^{-2 \sum_\alpha {\rm{tr}_c}  \ln \left (1-R_g e^{-\beta (\epsilon_{g}^\alpha)}\right)},
%\label{eq:app:part_one}
\end{eqnarray}
where the character in the adjoint representation can be written as $\Phi_A = \frac{1}{N_c} {\rm{tr}_c} R_g$, with
\begin{eqnarray}
 R_g=\mbox{diag}\big (1,1,e^{i(\theta_1-\theta_2)},
e^{-i(\theta_1-\theta_2)}, e^{i(2\theta_1+\theta_2)},e^{-i(2\theta_1+\theta_2)},e^{i(\theta_1+2\theta_2)},
 e^{-i(\theta_1+2\theta_2)}\big).
\end{eqnarray}
Then we can write the trace as
\begin{align}
&{\rm{tr}_c}\ln(1-R_ge^{-\frac{\epsilon_g}{T}}) = {\rm{tr}_c}\ln\big(1-\mathrm{diag}(1,1,X,X^{-1},Y,Y^{-1},Z,Z^{-1})G\big)\nn
&\!\!=\ln\big((1-G)(1-G)(1-XG)(1-X^{-1}G)(1-YG)(1-Y^{-1}G)\nn
&(1-ZG)(1-Z^{-1}G)\big).
\label{eq:app:part_nine}
\end{align} 
After a lengthy but straightforward calculation of the trace the whole term can be written in terms of $\Phi$ and $\bar\Phi$, which when expressed the partition function for gluons in the infinite volume limit reads as,
\begin{eqnarray}
{\cal Z}_g\!\!=-2V\int \frac{d^3p}{(2\pi)^3}   
\ln\Big( 1 + \sum_{m=1}^8 a_m\, e^{-m\beta\epsilon_g}\Big ),
%\label{eq:app:part_one}
\end{eqnarray}
where the coefficients $a_m$ are given in equation (\ref{eq:c4:eq11}) of section~\ref{sec:c4:ther_pot} in chapter~\ref{chapter:JPG}

\onehalfspacing

}

\end{document}